%% file: main.tex
\documentclass[11pt]{style/thesis}
\makenomenclature

\makeatletter
\newcommand{\closenomencl}{%
  \closeout\@nomenclaturefile%
}
\makeatother

\newcommand{\writenomencl}[1]{%
  \closenomencl%
  \IfFileExists{#1.nlo}{%
    \write18{%
      makeindex -s nomencl.ist -o #1.nls -t #1.nlg #1.nlo%
    }%
  }{\typeout{Nothing there}}%
}

\AtEndDocument{\writenomencl{\jobname}}

\begin{document}

\thispagestyle{empty}
\begin{samepage}
\begin{center}











\vspace*{0.01cm}

\begin{adjustwidth}{-60pt}{-60pt}


\begin{center}
{\huge \bfseries How Can Datacenters Join the Smart Grid \\ to Address the Climate Crisis?  \par} 
\vspace{0.3cm}
{\Large \bfseries \textit{Using simulation to explore power and cost effects}} 

{\Large \bfseries \textit{of direct participation in the energy market} \par}
\end{center}
\end{adjustwidth}

\vspace*{1cm}

{\large
    \begin{tabular}{c}
    \small{By} \\ 
    \\ 
    \textbf{Hongyu He} \\
    \\ 
    \small{(VU: 2632195, UvA: 12958794)} \\
    \small \texttt{hongyu.he@vu.nl}
    \end{tabular}
}

\vspace*{1cm}

\centering
\begin{adjustbox}{width=1\textwidth}
{\normalsize 
    \begin{tabular}{rll}
    \textsl{1st supervisor:}   &  Prof.Dr.ir. Alexandru Iosup  & (VU Amsterdam) \\
    \textsl{daily supervisor:} & Fabian Mastenbroek  & (TU Delft) \\
    \textsl{3rd supervisor:} & Leon Overweel & (Dexter Energy Services B.V.) \\
    \textsl{2nd reader:} & Dr.ir. Animesh Trivedi & (VU Amsterdam) \\
    \end{tabular}
}
\end{adjustbox}

\vspace*{1.2cm}

\small{Submitted in fulfilment of the requirements for the degree \\ of Bachelor of Science (Honours) \\in Computer Science}

\vspace*{1cm}

\textsc{Vrije Universiteit Amsterdam} \\
Amsterdam, The Netherlands \\
July 2021

\end{center}
\end{samepage}

\newpage

\null\vfill
\thispagestyle{empty}
{\centering

\copyright\ \ 2021 \\
Hongyu He \\
All rights reserved \\
}

\renewcommand*\contentsname{Table of Contents}
\setcounter{tocdepth}{1} 

\frontmatter 

\input{chapters/0-front/0.1_dedication}
\input{chapters/0-front/0.2_acknowledgements}
\input{chapters/0-front/0.3_abstract}

\tableofcontents

\input{chapters/0-front/nomenclature}

\printnomenclature 
\listofalgorithms \addcontentsline{toc}{chapter}{List of Algorithms}
\listoffigures
\listoftables

\newpage

\mainmatter 

\input{chapters/1-intro/1_introduction}

\input{chapters/2-bg/2_background}

\input{chapters/3-degsin/3_design}

\input{chapters/4-impl/4_implementation}

\input{chapters/5-eval/5_evaluation}

\input{chapters/7-conclusion/7_conclusion}

\bibliographystyle{style/bib_nat} 
\renewcommand{\bibname}{References} 
\bibliography{references}

\end{document}

%% file: chapters/0-front/0.1_dedication.tex
\begin{dedication} 
\
\vfill
\centering

{\it
    To my mom, who always reminds me to shave my beard off,
    \\ and to my grandparents.
}
    
\vfill
\end{dedication}

%% file: chapters/0-front/0.2_acknowledgements.tex
\begin{acknowledgements} 


I would like to express my gratitude to my supervisors, Prof.Dr.ir. Alexandru Iosup, Fabian Mastenbroek, Leon Overweel, and Dr.ir. Animesh Trivedi, for their guidance through each stage of this research, and for inspiring my interest in the development of innovative technologies. Also, I appreciate all the training and support received from the AtLarge Research group. 

This thesis is the culmination of three-year education and research, marking the first milestone of my scientific contribution to the community and society.

\end{acknowledgements}

%% file: chapters/0-front/0.3_abstract.tex
\begin{abstracts} 





\enlargethispage{\baselineskip}

Amidst the climate crisis, the massive introduction of renewable energy sources has brought tremendous challenges to both the power grid and its surrounding markets. As datacenters have become ever-larger and more powerful, they play an increasingly significant role in the energy arena. With their unique characteristics, datacenters have been proved to be well-suited for regulating the power grid yet currently provide little, if any, such active response. This problem is due to issues such as unsuitability of the market design, high complexity of the currently proposed solutions, as well as the potential risks thereof. This work aims to provide individual datacenters with insights on the feasibility and profitability of directly participating in the energy market. By modelling the power system of datacenters, and by conducting simulations on real-world datacenter traces, we demonstrate the substantial financial incentive for individual datacenters to directly participate in both the day-ahead and the balancing markets. In turn, we suggest a new short-term, direct scheme of market participation for individual datacenters in place of the current long-term, inactive participation. Furthermore, we develop a novel proactive DVFS scheduling algorithm that can both reduce energy consumption and save energy costs during the market participation of datacenters. Also, in developing this scheduler, we propose an innovative combination of machine learning methods and the DVFS technology that can provide the power grid with indirect demand response (DR). Our experimental results strongly support that individual datacenters can and should directly participate in the energy market both to save their energy costs and to curb their energy consumption, whilst providing the power grid with indirect DR.

\keywords{Datacenter modelling, cloud simulation, smart grid, energy market, demand response, frequency scaling, DVFS scheduling, machine learning.}

\end{abstracts}

%% file: chapters/0-front/nomenclature.tex

\nomenclature{$f$}{Instant CPU frequency}
\nomenclature{$F$}{CPU capacity/frequency}
\nomenclature{$C$}{Capacitance}
\nomenclature{$V$}{Voltage}
\nomenclature{$t_\text{CPU}$}{CPU time}
\nomenclature{$t_\text{wall}$}{Wall time}
\nomenclature{$t_\text{idle}$}{Accumulative CPU idle time}
\nomenclature{$u$}{CPU usage}
\nomenclature{$u_\text{os}$}{CPU utilization}
\nomenclature{$f_\text{min}$}{Minimum scaling frequency}
\nomenclature{$f_\text{max}$}{Maximum scaling frequency}

\nomenclature{$E_\text{PSU}$}{Accumulative energy consumption of PSU}
\nomenclature{$N_\text{run}$}{Number of tasks that are being processed}
\nomenclature{$N_\text{queued}$}{Number of tasks in the run queue}
\nomenclature{$N_\text{blocked}$}{Number of tasks blocked by I/O}

\nomenclature{$P$  }{ Power draw }
\nomenclature{$P^\text{ total}$}{Power draw of datacenter}
\nomenclature{$P^\text{ IT}$}{ Power draw of IT infrastructure}
\nomenclature{$P^\text{ compute}$}{ Power draw of computing equipment}
\nomenclature{$U$}{Utilization of IT equipment}

\nomenclature{$G$}{Financial gain}
\nomenclature{$Q_s$}{Quantity of energy scheduled in the spot market}
\nomenclature{$Q_d$}{Quantity of energy delivered}
\nomenclature{$\zeta$}{Locational marignal pricing}

\nomenclature{$\pi$ }{ Nameplate loss coefficient }
\nomenclature{$\lambda$ }{ Tare loss coefficient }
\nomenclature{$\alpha$ }{ Proportional loss coefficient }
\nomenclature{$\beta$ }{ Square-law loss coefficient }
\nomenclature{$P^\text{\ in}$  }{ Inlet power }
\nomenclature{$P^\text{\ tare}$ }{ Tare power loss }
\nomenclature{$P^\text{\ loss}$ }{ Total power loss }
\nomenclature{$P^\text{\ rated}$ }{ Nameplate power }
\nomenclature{$P^\text{\ idle}$ }{ Idle power }
\nomenclature{$P^\text{\ max}$ }{ Maximum power }
\nomenclature{$P^\text{2nd}$ }{ Power draw of secondary power support}
\nomenclature{$N_\text{\ server}$ }{Number of active servers }
\nomenclature{$N_\text{\ PDU}$ }{ Number of attached PDUs }

\nomenclature{$p^S$ }{ Spot market price }
\nomenclature{$p^B_-$ }{ Shortage price in the balancing market }
\nomenclature{$p^B_+$ }{ Surplus price in the balancing market }
\nomenclature{$Q_-$ }{ Quantity of shortage energy of a BRP }
\nomenclature{$Q+$ }{ Quantity of surplus energy of a BRP }
\nomenclature{$Q+$ }{ Quantity of energy required by downwards regulations }
\nomenclature{$Q_\downarrow$ }{ Quantity of energy required by downwards regulations }
\nomenclature{$Q_\uparrow$ }{ Quantity of energy required by upwards regulations }
\nomenclature{$N_\text{BSP}$ }{ Number of participated BSPs in the balancing market }
\nomenclature{$N_\text{BRP}$ }{ Number of participated BRPs in the balancing market }

\nomenclature{$N_\text{ISP}$ }{ Number of ISPs}
\nomenclature{$p^S$ }{ Spot price in the day-ahead market }
\nomenclature{$p^F$ }{ Forecasted price}
\nomenclature{$p^f$ }{ Synthetically forecasted price}
\nomenclature{$E$ }{ Additive Gaussian random variable}

\nomenclature{$\tau$}{Rated output power of PSU}
\nomenclature{$\eta_l$}{Percentage of PSU load}
\nomenclature{$\eta_e$}{Percentage of PSU energy efficiency}
\nomenclature{$P^\text{ server}$}{Power draw of server}

\nomenclature{$\mathbb{Q}$}{Quantile function}
\nomenclature{$\mathbb{P}$}{Power function}
\nomenclature{$\mathds{1}$}{Indicator function}
\nomenclature{$a$}{Asymptotic Factor}
\nomenclature{$r$}{Approximation factor}
\nomenclature{$s$}{Scalar value}
\nomenclature{$c$}{Number of (logical) cores of a machine}
\nomenclature{$m$}{Size of memory unit}

%% file: chapters/1-intro/1_introduction.tex
\chapter{Introduction} \label{cha:intro}

Taking their toll at a rising speed over the past decades, environmental problems such as global warming have fast become a worldwide focal point \cite{Butler2018ClimateCH}, so much so that even the COVID-19 pandemic can barely arrest the development of this alarming trend \cite{zheng2020satellite}. To combat this exacerbating issue, remedies such as emission limits, carbon taxes, and perhaps most importantly, ambitious renewable energy targets adopted in 2015 \cite{paris_agreement} and enhanced in 2021 \cite{climate_summit}, have been introduced. As a result of such a push towards sustainability, the energy market has become increasingly volatile, bringing both challenges and opportunities to the ever-larger energy consumers, datacenters.

\begin{figure} [t!]
    \centering
    \includegraphics[width=0.6\textwidth]{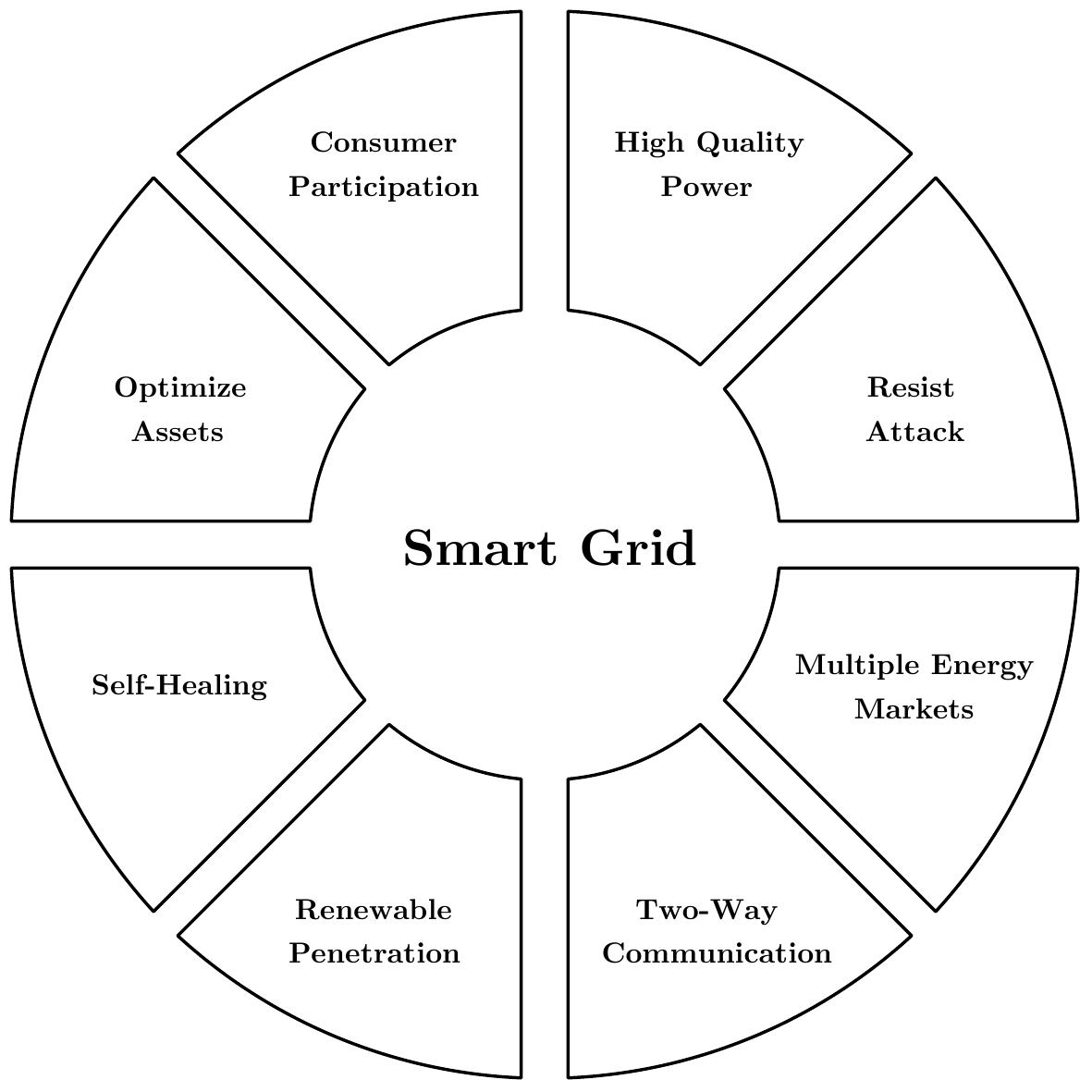}
    \caption[Functions of the smart grid]{Functions of the smart grid.}
    \label{fig:smart_grid}
\end{figure}

\paragraph*{Smart Grid.} Despite the fact that there is virtually no cost in the production of renewable energy as they are free of charge from nature, the substantial operational costs induced by its huge intermittency and stochasticity, however, greatly impedes the continuous penetration of renewable energy sources into the power grid \cite{Liang2017EmergingPQ}. Consequently, large numbers of expensive and carbon-intensive system operating reserves, which hinge on more reliable energy sources like petroleum or even diesel, are often required as hot/cold standby reserves to back up renewables in order to maintain the equilibrium of the power grid (\S\ref{sec:power_grid}). In addressing the above challenges, the power grid is becoming more and more intelligent (Figure \ref{fig:smart_grid}) --- the smart grid \cite{Fang2012SmartG}. Such recent advances in functions of the smart grid enable real-time, fluent interactions and coordination between energy producers and consumers, improving demand-side management (DSM) \cite{Wang2016ProactiveDR,Conejo2010RealTimeDR}.

\paragraph*{Demand Response.} The smartness of the grid, however, is hitting some limits due to the uncertainty caused by the massive introduction of renewable energy sources. One relatively recent, yet promising form of DSM is demand response (DR), which has been extensively explored in existing literature (\eg \cite{Hui20205GNI,Zhou2016DemandRC,Jindal2020GUARDIANBS,Khemakhem2019DoubleLH,zhang2009smart,stadler2008power,aalami2008demand}). In general, based upon the response time, DR programmes can be classified into two broad categories, direct and indirect control. The direct approach responds to requests and signals from the power grid quickly, which provides system operators with accurate and fast control over the power grid (different levels of DR are introduced in Section \ref{sec:balance_grid}). Thus, most of the ancillary services opt to direct DR control. Conversely, it is computationally and communicatively more intensive. In contrast, indirect DR is cheaper and more flexible but embodies a greater degree of uncertainty. As opposed to relying upon the direct requests from the system operators or upon constant monitoring of the power grid, it financially incentivizes prosumers to modify their energy consumption and/or production following signals from the energy markets and/or local utilities. Although unlike the frequency level of the power grid that directly reflects the status of the current demand, energy prices of the financial markets are good indicators and predictors of the demand-supply balance. Thus, indirect DR has the potential to effectively reduce the market volatility caused by wind and solar energy generation, mitigate network problems (\eg congestion, voltage), and respond to failures (\eg avoiding blackout), helping both the customers and the power grid dramatically reduce costs \cite{Sle2011DemandRF}. Although many grids currently do not have congestion problems, as more distributed solar generation and more erratic consumption resulted from mobility, \eg electrical vehicles (EVs), are introduced, a large fraction of the distribution grids is expected to suffer from congestion in the near future \cite{Hemmati2017StochasticPA}. To tackle this, intensive research has been conducted to explore the interrelationship between the mobility of EVs and DR, \eg using vehicle-to-grid (V2G) technology \cite{Wierman2014OpportunitiesAC,Kempton2005VehicletogridPF,Wang2017DistributedEM}. Throughout the years, people have also resorted to refrigerators \cite{marsh2009intermittent}, fans \cite{Vrettos2014RobustPO}, laundry machines and dishwashers \cite{Pipattanasomporn2014LoadPO,Paterakis2015OptimalHA}, or even boilers \cite{Demirbas2005PotentialAO}, searching for DR resources to ameliorate the intensive demand during peak times, During these periods, the energy delivery is much more costly since less efficient energy plants (usually powered by fossil fuels) are employed.

\begin{figure} [t!]
    \centering
    \includegraphics[width=\textwidth]{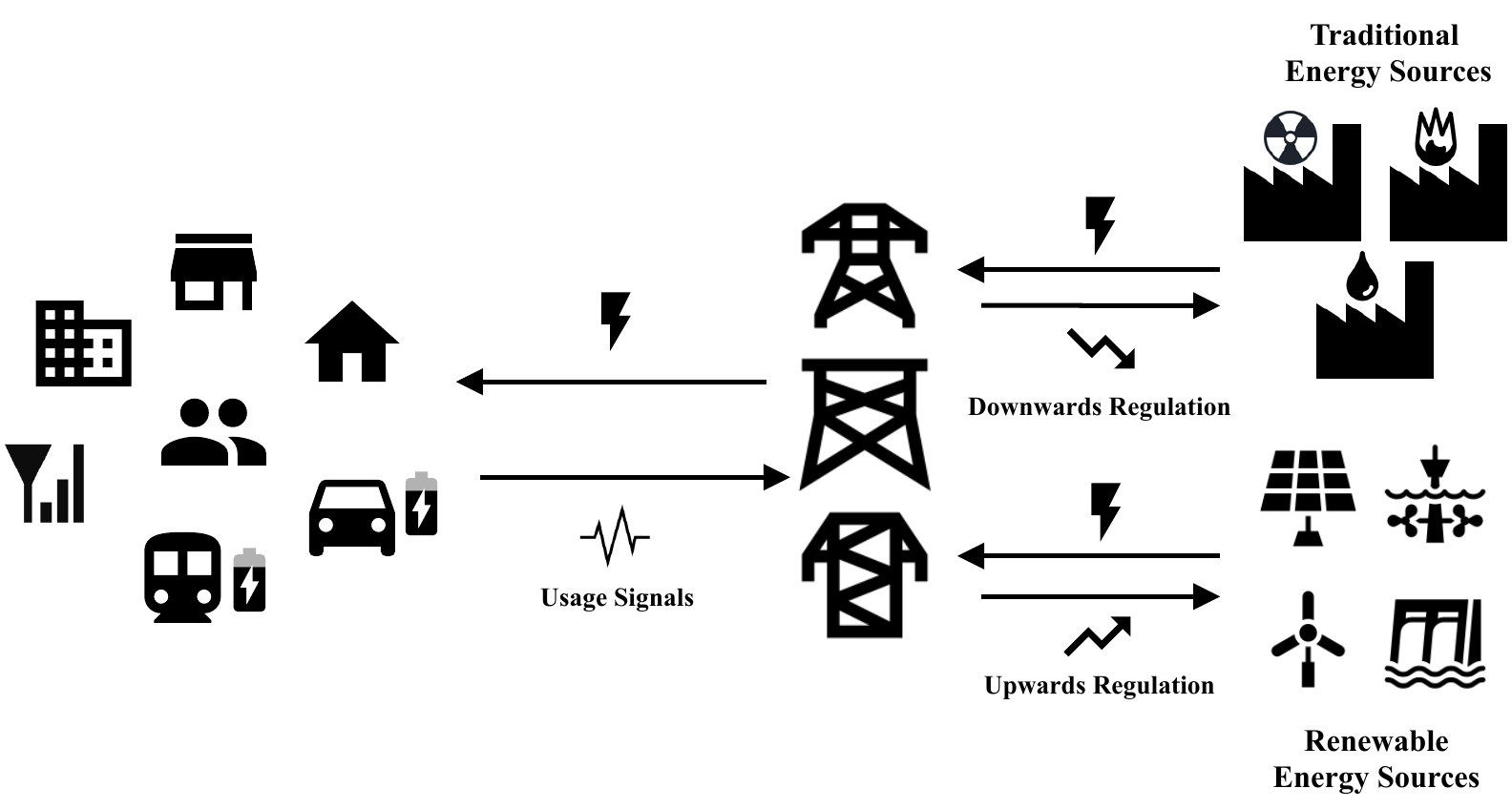}
    \caption[Demand response in the smart grids]{Demand response in the smart grids.}
    \label{fig:demand_response}
\end{figure}

\input{chapters/1-intro/1.1_problem_statement}
\input{chapters/1-intro/1.2_RQs}
\input{chapters/1-intro/1.3_methodology}
\input{chapters/1-intro/1.4_outline}

%% file: chapters/1-intro/1.1_problem_statement.tex
\section{Problem Statement} \label{sec:prob}

Over the past decades, datacenters have grown ever larger and more powerful \cite{masanet2020recalibrating}. According to recent reports, datacenters consume more than 1.5\% of global electricity use \cite{Li2019CSLdrivenAE,masanet2020recalibrating,Yadav2020AnAH}, and, with a growth rate of 10--12\% per year \cite{Vasques2019ARO,Brown2008ReportTC,Ghatikar2012DemandRO,koomey2011growth}, this figure is expected to grow to 3\% by 2025 \cite{vesa2020energy}. In the US, about 2.2--3.5\% electricity use can be attributed to datacenters \cite{koomey2011growth}. Similar phenomenons appear in the Netherlands, where the power consumption of datacenters amounts to 2.7 billion kWh by 2020, which is 2.3\% of the total Dutch energy use \cite{dda}. A decade ago (2011), Google datacenters already use almost 260 MW of power, exceeding the consumption level of Salt Lake City \cite{pedram2012energy}. Similarly, a single datacenter of  Microsoft in Washington DC consumes 48 MW of power, which is the equivalent of around 40k households in the US \cite{Wang2016ProactiveDR}. Due to their carbon footprints and energy consumption, datacenters have become a front-line target in the battle against climate change \cite{guyon2017much}. 

\subsection{Opportunities} \label{sec:opportunities}

\begin{figure} [t!]
    \centering
    \includegraphics[width=0.6\textwidth]{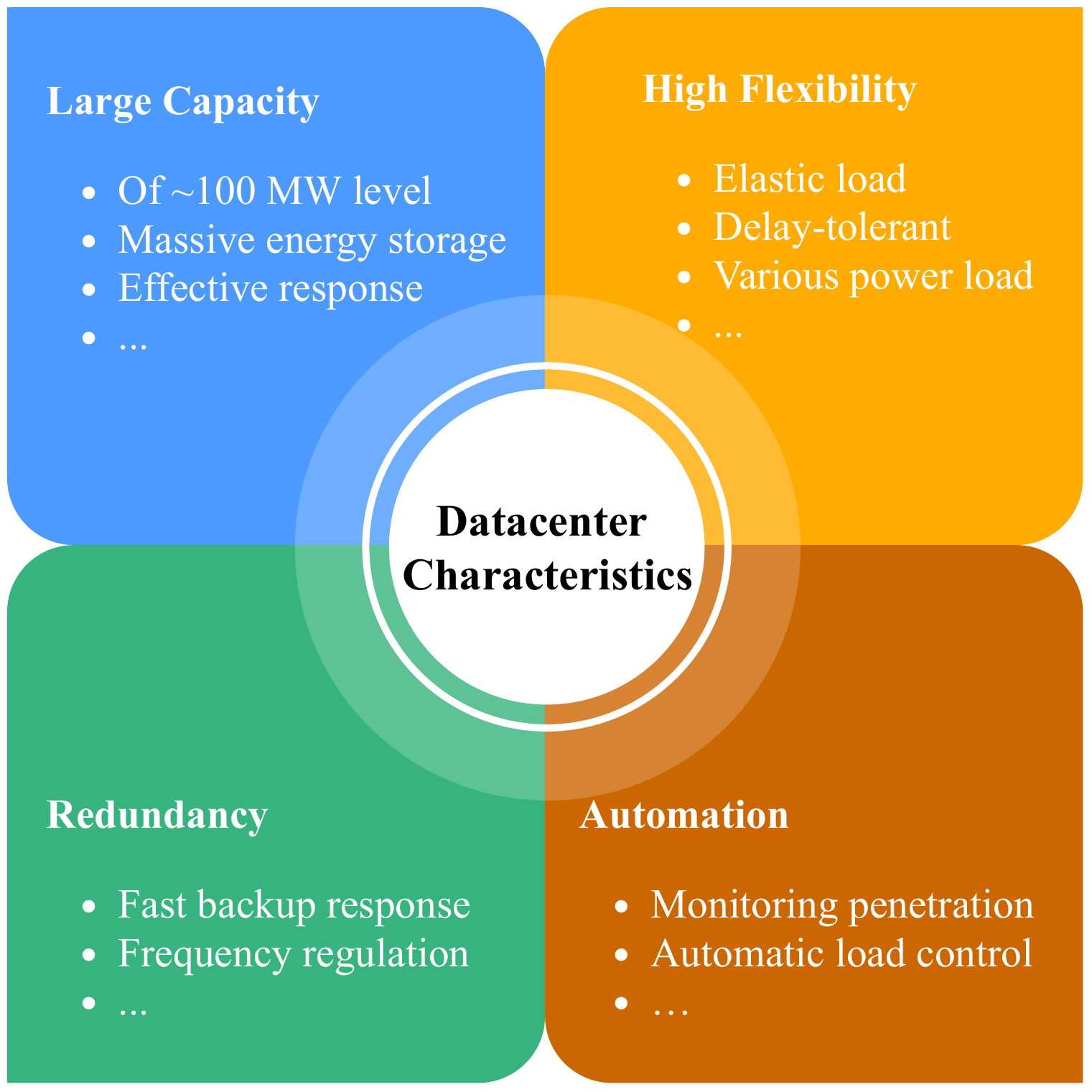}
    \caption[Characteristics of datacenters]{Characteristics of datacenters.}
    \label{fig:chars_dc}
\end{figure}

The increasing capacity of datacenters may well be a blessing in disguise for the power grid. A well-managed datacenter of 30 MW has approximately the same capability for regulating the power grid as huge energy storage of 7 MWh \cite{Wierman2014OpportunitiesAC}. Besides its capacity, unlike other subjects such as lighting and residential power, datacenters feature its elastic load. In other words, datacenters are capable of curbing their power demand without service degradation. Also, many Cloud service providers employ spot pricing mechanisms to manage demands (\eg \cite{aws_spot}). To put this into perspective, a study from Lawrence Berkeley National Laboratory (LBNL) shows that 15\% of the load can be shed within 15 minutes without adjusting temperatures or any other building managements \cite{Ghatikar2012DemandRO}. Additionally, the flexibility of datacenters can be leveraged at a finer-grained level through energy-saving techniques such as power capping \cite{chen2013dynamic} and dynamic voltage \& frequency scaling (DVFS) \cite{le2010dynamic}. Note that DR is not mean to save energy, but DVFS can. This enabling configuration will be described in detail later in Section \ref{sec:dvfs}. 
Furthermore, datacenters are built for extremely reliable and available services. For instance, two availability classes to which most modern datacenters belong are Tier \romnum{3} and Tier \romnum{4}. Datacenters that fall into the former category insure a 99.982\% availability, and for the latter, the figure is 99.995\% \cite{Mare2010DemandRA}. To accomplish promised uptime and performance guarantees, datacenters generally have a considerable amount of redundancy throughout their power systems, as well as large battery capacities in their primary power support. Lastly, datacenters are complex but highly automated systems. Ubiquitous monitors and controls empower datacenters’ participation in the DR programme. Therefore, datacenters are well-suited candidates for DR programmes \cite{Chen2019EnergyQAREQD,Wierman2014OpportunitiesAC}.

\subsection{Challenges} \label{sec:challenges}

Albeit with great potentials of taking part in DR programmes, datacenters nowadays provide the power grid with little, if any, response to the power grid \cite{Ghatikar2012DemandRO,glanz2012power,Liu2014PricingDC,Liu2013DataCD}. Firstly, the current market designs \cite{Johari2011ParameterizedSF,Xu2016DemandRW} and DR programmes  \cite{Baldick2018IncentivePO,Sle2011DemandRF} (i) are not particularly suitable for datacenters and (ii) can barely fully extract the flexibility of datacenters \cite{Liu2014PricingDC}. Secondly, datacenters may incur charges if no response is offered during DR programmes. No profit would be produced either if the peak in energy demand did not happen during the coincident periods. Thirdly, the proposed DR strategies for datacenters often require cooperation between the energy market, geographically distributed datacenters, and their utility companies \cite{Rao2010MinimizingEC,Liu2015GreeningGL,Ghamkhari2013EnergyAP,Wang2012DProDD,Zhang2015ATI,Wang2016ProactiveDR,Qureshi2009CuttingTE,Rao2010MinimizingEC,Liu2015GreeningGL,Lin2012OnlineAF,Ghamkhari2013EnergyAP,Wang2012DProDD,Zhang2015ATI}. In other words, they can hardly be carried out without structural changes to the energy market and/or substantial adjustments in datacenter operations. Hence, the complexity of the orchestration and the potential risks therein hinder the participation of datacenters in the power grid. Furthermore, experimenting, testing, and evaluating energy-aware techniques tend to be costly or sometimes even unrealistic in large-scale, modern datacenters. This could lend hesitancy for datacenter shareholders to embrace energy-saving techniques, \eg DVFS since the critical guarantee of performance and availability specified in their service-level agreement (SLA) always takes precedence over \textit{unknown} benefits brought by enabling energy-saving configurations. This challenge further imposes potential risks on datacenters when participating in the DR programmes. One way of bridging this gap is to measure energy consumption at the hardware level, for example, installing system/component power meters to monitor the power usage \cite{Chang2002EnergyDrivenSS,Flinn1999PowerScopeAT,Contreras2005PowerPF}. Although such a direct approach is becoming a common practice, it only applies to facilities that have already been built \cite{Economou2006FullSystemPA}. Consequently, it is not portable, scalable or informative for future planning and responsive scheduling. Another method is to employ software instruments, specifically, datacenter simulators, to model the energy consumption of thousands of servers as well as non-IT infrastructure, \eg cooling and ancillary equipment. In comparison to methods at the hardware level, datacenter simulators are more flexible, reproducible and cost-effective, playing an instrumental role in facilitating energy-aware decision-making \cite{iosup2017opendc}. As a result, datacenter simulation has been widely adopted for both academic research and industrial use \cite{Bambrik2020ASO,Ruepp2017CombiningHA}. 

%% file: chapters/1-intro/1.2_RQs.tex
\section{Research Questions} \label{sec:rqs}

\begin{figure} [t!]
    \centering
    \includegraphics[width=\textwidth]{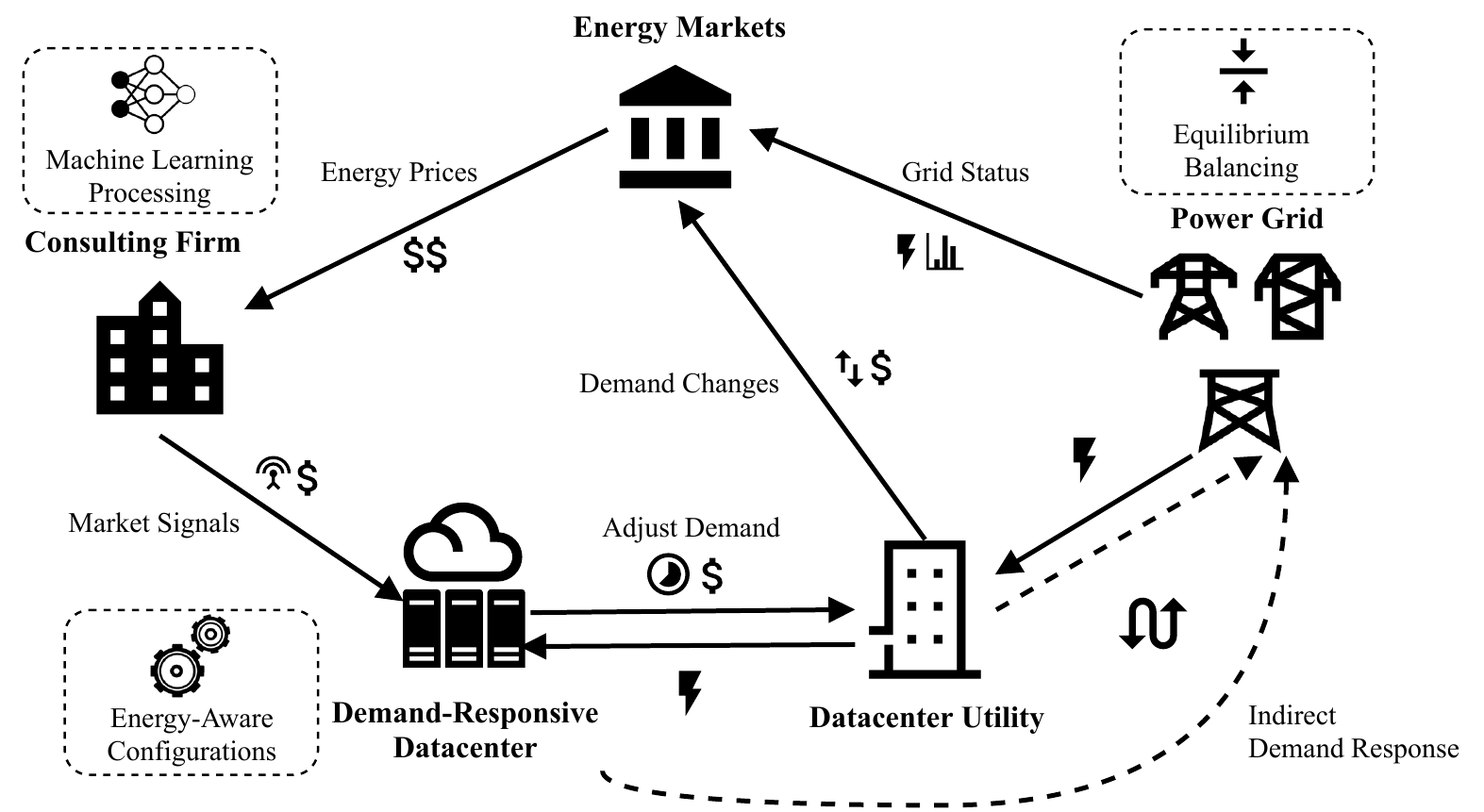}
    \caption{Aerial view of the project.}
    \label{fig:overview}
\end{figure}

As the first step towards addressing the aforementioned challenges, this work aims to provide individual datacenters with insights on the feasibility and profitability of directly participating in the energy market whilst offering indirect DR to the power grid (Figure \ref{fig:overview} demonstrates an aerial of this project). To achieve this objective, we put forwards the statement of this thesis:

\noindent\fbox{%
    \parbox{\textwidth}{%
        \noindent\textbf{\textit{Thesis Statement}} — Individual datacenters can and should directly participate in the energy market both to save their energy costs and to curb their energy consumption, whilst providing the power grid with indirect DR.
    }%
}

To offer evidence for the thesis statement, we raise the main research question (\ref{mrq}) followed by a sequence of research questions (\textbf{RQ}s).

\begin{enumerate} [label=\textbf{MRQ}]
 \item \label{mrq} How feasible and beneficial is it for individual datacenters to directly participate in the energy market whilst providing the power grid with indirect DR?
\end{enumerate}

\subsection{Research Question 1}

To answer the \ref{mrq}, we need to first estimate the energy consumption of datacenters. There are, however, a myriad of factors that influence the energy consumption of datacenters. To name a few such factors: different topologies of the datacenters, various types of hardware as well as many configurations thereof. In an effort to overcome this challenge, we resort to whole power system modelling in order to capture the heterogeneity of those factors. To this end, an energy resource chain, starting from the power source and ending at the IT infrastructure, is needed. This resource chain passes through several supporting subsystems and components of datacenters, for example, the automatic transfer switches (ATSs), the uninterrupted power supply (UPS) systems, and the floor/rack power distribution units (PDUs). Inside the servers, power supply units (PSUs) transform AC power to DC power (in a typical AC architecture). A detailed introduction to these various components can be found in Section \ref{sec:dc_grid}. Besides these hardware components and subsystems, fine-grained manoeuvres happen in the CPUs, adjusting the power consumption in real time. One of such techniques is the DVFS, an active power management technique whereby the frequency of a microprocessor can be automatically adjusted on spot based upon the computing loads. By exploiting DVFS, machines can save energy and, in turn, reduce operational costs (an elaborate introduction of DVFS can be found in section \ref{sec:dvfs}). Neither building the resource chain nor incorporating and synergizing various DVFS policies is a straightforward task, which gives rise to the first research question:

\begin{enumerate} [label=\textbf{RQ1}]
 \item \label{rq1} How to model the power system of datacenters?
\end{enumerate}

\subsection{Research Question 2}

IN \ref{rq1}, we estimate the energy consumption of datacenters through simulation. Now we are interested in the extent to which datacenters are able to benefit from directly participating in the energy market. We focus on benefits resulting from taking part in different markets, specifically, the day-ahead and the balancing markets. We do not, however, quantify the savings for the electricity grid resulted from a lower need to generate energy from fossil fuels and/or storing it for datacenter use, or the operational changes in the electricity grid (\eg supply curtailment \cite{Root2017UsingBE}). Should datacenters participate? If so, which market(s) should be given particular attention? To answer these questions, we post the second research question:

\begin{enumerate} [label=\textbf{RQ2}]
 \item \label{rq2} Is it beneficial for datacenters to participate in the energy market in the first place?
\end{enumerate}

\subsection{Research Question 3}

\ref{rq2} seeks answers to the question of why datacenters should participate in the energy market. The next step is to inquire about the most economical way of procuring energy in the energy market, \ie how to participate. Referring back to Section \ref{sec:opportunities}, reliability and availability are of paramount importance for datacenters. In fact, load forecast (\eg \cite{Bajracharya2018ForecastingDC,Uv2018EnergyMO,Fulpagare2017RackLF,Shaw2017UseOT,Karabinaoglu2017LoadFM}) is commonly resorted for energy-planning in datacenters. Therefore, load-forecast-based procurement strategies are of particular interest. In turn, we ask the third research question:

\begin{enumerate} [label=\textbf{RQ3}]
 \item \label{rq3} How to procure energy in the energy markets according to forecasted power load?
\end{enumerate}

\subsection{Research Question 4}

Leveraging different procurement strategies in \ref{rq3} may bring in substantial profits for datacenters, the energy consumption level, however, cannot be improved by only doing so. To provide indirect DR whilst reduce the carbon footprint, we turn to the novel orchestration between machine learning (ML) methods and the low-level energy-saving technique DVFS. We adjust the DVFS policies in accordance with the predicted market signals produced by ML models. Moreover, since almost every new advancement in computer science comes at a cost, mitigating the overhead introduced by using DVFS when responding to market signals is of the same importance. These challenges beg the fourth research question:

\begin{enumerate} [label=\textbf{RQ4}]
 \item \label{rq4} How to optimize energy consumption when participating in the energy market using DVFS, based upon ML methods?
\end{enumerate}

\subsection{Research Question 5}

There is much to explore when it comes to the interrelationship between datacenter simulation and the energy markets/power grid. However, as the old saying goes, “difference in the profession makes one feel worlds apart’’. To assist further explorations and bridge the gap in domain knowledge, we raise the last research question:

\begin{enumerate} [label=\textbf{RQ5}]
 \item \label{rq5} How to create an exploratory tool for problems in this domain, to be used by experts in both the IT and the energy industry?
\end{enumerate}

%% file: chapters/1-intro/1.3_methodology.tex
\section{Research Methodology} \label{sec:methodology}

To achieve \ref{rq1}, we employ quantitative research, specifically, system modelling and simulation \cite{Kheir1995SystemsMA}. We design \cite{Iosup2019TheAV,Peffers2008ADS} and prototype \cite{Hamming1997TheAO} the energy modelling and power management subsystem of OpenDC in which a set of power models and a number of generic DVFS algorithms are built and integrated. As a result, these components should be integrated and work in concert. On the basis of these models and management techniques, we model and simulate the power system of a typical datacenter. 

\noindent To answer \ref{rq2}, we carry out experimental research \cite{Jain1990TheAO}, conducting discrete-event simulations on real-world datacenter workload, quantifying and comparing the energy costs of participating in different markets. Note that the prediction of the workloads in production will be more precise as the time approaches the energy delivery period. However, we do not take into account such an increasing precision in workload predictability since it is the controlled variable in this work.

\noindent To address \ref{rq3}, we conduct a case study, collaborating with partners and experts in the energy market. We obtain the predicted energy prices of the intraday market as well as details concerning trading and operations in the energy market. We regard the predicted prices of the balancing market as indicators of energy demand, with which we adjust the fractions of energy bought in during the day-ahead market period. By doing so, we leverage energy cost whilst still sufficiently satisfy energy needs in the balancing market.

\noindent To attack \ref{rq4}, we further employ various frequency scaling algorithms in response to the market signals, developing a DVFS scheduling algorithm powered by ML methods. By designing workload-level benchmarks \cite{benchmarkcrimes,Ousterhout2018AlwaysMO}, we conduct various experiments using trace-based simulations, focusing on the impact of various datacenter phenomena, such as specialized/mixtures of scenarios and correlated forms of performance variability.

\noindent To tackle \ref{rq5}, we honour open-science guidelines \cite{bezjak2018open,Wilkinson2016TheFG} and build open-source scientific software, following rigorous software engineering methods and PR-review software development cycle \cite{Yu2014ReviewerRO}. Agreeing on various specifications with our partners and experts, we adhere to the standard format of market data from official websites and containerize the deployment of our research instruments. We make our OpenDC datacenter simulator together with its market extension an out-of-box tool that is ready to be used by experts in both the energy and the IT industry. 

%% file: chapters/1-intro/1.4_outline.tex
\section{Thesis Contribution} \label{sec:contribution}

By addressing all research questions with our best efforts, we endeavour to transfer our knowledge and experience/lessons learned as well as to deliver developed software and experimental results to the community as much as possible without reservation. In this section, we list our scientific contributions (\textbf{TC}) as well as technical contributions (\textbf{TC}). Also, we identify potential societal impact (\textbf{SI}) and economic impact (\textbf{EI}) of this research. Additionally, we honour the FAIR data principles\cite{Wilkinson2016TheFG}  (\textbf{FAIR}), exercising the best practices for sharing data. The contribution of this work is nine-fold:

\begin{enumerate}
    \item We are the first to demonstrate the substantial financial incentive for individual datacenters to directly participate in both the day-ahead and the balancing market \{\textbf{SC}, \textbf{EI}\}.
    \item We suggest a new short-term, direct scheme of energy market participation for individual datacenters, in place of the current long-term, inactive market participation \{\textbf{SC}\}.
    \item We develop a novel proactive DVFS scheduling algorithm that is able to both reduce the energy consumption \{\textbf{SI}\} and save the energy cost of datacenters \{\textbf{EI}\}. 
    \item We propose an innovative combination of machine learning methods and the energy-management technique DVFS \{\textbf{SC}\} that can provide the power grid with indirect DR in an effort to overcome the increasing challenges brought by renewable energy sources \{\textbf{SI}\}.
    \item We are the first to achieve whole power system modelling in datacenter simulation \{\textbf{TC}\}.
    \item We create a user-friendly and ready-to-use tool for experts in both the IT and the energy industry to further explore the research potential lies at the intersection between the two fields \{\textbf{TC}\}.
    \item We publish our code as open-source projects, facilitating future scientific explorations and collaborations \{\textbf{TC}, \textbf{FAIR}\}. \footnote{\url{https://github.com/atlarge-research/opendc}}$^,$ \footnote{\url{https://github.com/hongyuhe/opendc-eemm}}
    \item Alongside the code, we publish our datasets and raw experimental results, which, in turn, are findable, accessible, and do not subject to any ethical, legal or contractual restrictions \{\textbf{FAIR}\}.
    \item Besides the datasets, we build documentation and make tutorials with examples to ensure reproducibility, interoperability, and reusability of both the software and the data \{\textbf{TC}, \textbf{FAIR}\}. \footnote{\url{https://opendc-eemm.rtfd.io}}$^,$ \footnote{\url{https://opendc.org}}
\end{enumerate}

\clearpage

\section{Plagiarism Declaration}

I confirm that this thesis work is my own work, is not copied from any other source (person, Internet, or machine), and has not been submitted elsewhere for assessment.\footnote{\url{https://www.vu.nl/en/about-vu-amsterdam/academic-integrity/index.aspx}}

\section{Thesis Structure} \label{sec:structure}

Firstly, key terms pertinent to this work are covered in Chapter \ref{cha:background}. Then, in Chapter \ref{cha:degsin} we present the design of the energy modelling and management system, including its subsystems and the market extension. In addition, it also introduces the development pipeline and requirement engineering process. Next, we detail the implementations of the system in Chapter \ref{cha:impl}. After that, in Chapter \ref{cha:eval} we employ the developed infrastructure and tools to conduct experiments. In the last chapter, we answer research questions, summarizing our key findings and results. Lastly, we identify limitations and envision future research (Chapter \ref{cha:conclusion}).

%% file: chapters/2-bg/2_background.tex
\chapter{Background} \label{cha:background}

An overview of related subjects is laid out in this chapter. Key terms covered include the metrics used for evaluating energy consumption and for the saturation of critical loads (\S\ref{sec:metrics}), energy proportionality and existing solutions (\S\ref{sec:energy_prop}), frequency scaling (\S\ref{sec:dvfs}), and the power grid (\S\ref{sec:power_grid}), especially, the architecture of the power system in datacenters that we simulate is also presented (\S\ref{sec:dc_grid}). Last but not least, we discuss the energy modelling for datacenters in Section (\S\ref{sec:modelling}). 

\input{chapters/2-bg/2.1-2_metrics_prop}

\newpage
\input{chapters/2-bg/2.3_dvfs}

\newpage
\input{chapters/2-bg/2.4_power_grid}

\newpage
\input{chapters/2-bg/2.5_modelling}

%% file: chapters/2-bg/2.1-2_metrics_prop.tex
\section{Metrics} \label{sec:metrics}

\paragraph*{PUE \& CPE. } Two fundamental energy metrics are commonly used in datacenters, a) Power Usage Effectiveness (PUE) first proposed by \citet{malone2006metrics} (Equation \ref{eq:pue}), which can be used for both benchmarking and energy estimation, and b) Compute Power Efficiency (CPE) for measuring the computational efficiency of datacenters (Equation \ref{eq:cpe}).

\begin{align}
    \text{PUE} &= \dfrac{P^\text{ total}}{P^\text{ IT}}  \label{eq:pue} \\
    \text{CPE} &= \dfrac{U}{\text {PUE}}
               = \frac{U \cdot P^\text{ IT}}{P^\text{ total}},  \label{eq:cpe}
\end{align}

\noindent where $P^\text{ total}$ is the total facility power, $P^\text{ IT}$ is the power draw of the IT infrastructure, and $U$ denotes the utilization of the IT equipment.


Reflecting on the continuous growth of energy consumption from 1992 to 2014, and the raw performance per watt that is radically doubling every year, \citeauthor{malone2006metrics} point out that this trend is unlikely to stop in the next years. Also, they suggest a significant change in the past decades that the major cost of the datacenters is moving from operation-driven \cite{Xu2013MinimizingTO} to an infrastructure- and energy-oriented model. 

The average PUE value of datacenters worldwide is still close to 2.0 in 2021 \cite{uptime_pue}. This is a clear indication of widespread energy-inefficiency in the sense that to produce 1 W of computational power, around 2 W of power is consumed by supporting infrastructures such as cooling and other ancillary facilities. In 2008, only 0.4 improvements of PUE can result in about 350,000\$ reduction in energy cost, which is equivalent to about 430,000\$ in 2019. To better capture the power usage of IT infrastructure, the authors propose the adoption of CPE that directly reflects the actual fraction of energy used for computing. Note that a slight increase in PUE can boost the CPE significantly. For example, a typical PUE of 2.0 for a well-managed datacenter has a CPE of 10\%, whilst a PUE of 1.6 corresponds to a CPE of 50\%. Thus, CPE is a relatively more sensitive and intuitive indicator of energy losses for both the IT and supporting infrastructure.

\paragraph*{TUE \& ITUE. } \citet{patterson2013tue} developed IT-power usage effectiveness (ITUE) and total-power usage effectiveness (TUE) to improve PUE and CPE respectively, by taking into account the power distribution and cooling losses inside IT equipment \cite{patterson2013tue}. These metrics aim to tackle the challenge of estimating and tracking the total efficiency of the entire energy stack for datacenters. The computation of the metrics are captured by Equation \ref{eq:itue} and \ref{eq:tue}:

\begin{align}
    \text{ITUE} &=\dfrac{P^\text{ total}}{P^\text{ compute}} \label{eq:itue} \\
    \text{TUE} &= \text{ITUE} \cdot \text{PUE}, \label{eq:tue} 
\end{align}

\noindent where $P^\text{ compute}$ is the power draw of only the equipment related to computing service.


\paragraph*{Resource Utilization Metrics.} \label{sec:cpu_metrics} The debate of performance metrics is known as one of the “Rat Holes” in systems research \cite{Jain1991TheAO}. Gauging the saturation of the CPUs is never a trivial task, and over time, a number of metrics have been widely used in various contexts. In this section, we focus on three commonly employed metrics, namely, CPU load, CPU usage, and CPU utilization. First and foremost, it is worth noting that the definition of CPU load/usage/utilization varies with different use cases, platforms, and organizations \cite{vmware_vm,vmware_metrics,ibm,wiki}. Sometimes they are even used interchangeably by many people, albeit quite different. This is partially due to the fact that there is no single, universally standard way of defining these metrics. Nevertheless, the CPU load ($l$) is generally used in the context of the Linux scheduler, in which the run queues of the processors are accessible. The calculation of $l$ along the lines of Equation \ref{eq:cpu_load}; computing the average CPU load ($l_a$) is also a common practice (Equation \ref{eq:cpu_load_avg}). 

\begin{align}
    l =&\ N_\text{run} + N_\text{queued} + N_\text{blocked}  \label{eq:cpu_load} \\
    l_a =&\ \dfrac{l} {c}\ , \label{eq:cpu_load_avg}
\end{align}

\noindent where $N_\text{run}$ is the number of tasks that are being processed, $N_\text{queued}$ is the number of tasks in the run queue, $N_\text{blocked}$ is the number of tasks blocked by I/O, and $c$ is the (logical) core count of the machine. 


When it comes to CPU utilization/usage, the definitions often become a bit more nebulous. In this work, we distinguish these two in accordance with different viewpoints in the context of server architecture. With regard to the host, neither the run queue nor scheduler is visible at the firmware level. Therefore, from the standpoint of bare-metal machined, it is impractical to evaluate any metrics other than the ratio of the current CPU speed to the CPU capacity. In our case, we simulate Type-\romnum{1} hypervisors directly on top of the physical interface of the host machines. Hence, we employ the concept of the CPU usage ($u$) illustrated by Equation \ref{eq:cpu_usage} throughout our energy modelling and management system.

\begin{equation}
    u = \dfrac{f} {F}\ , \label{eq:cpu_usage}
\end{equation}

\noindent where $f$ is the instant CPU frequency, and $F$ is the CPU capacity.


In contrast, from the virtual machine (VM) or operating system (OS) point of view, the run queue, the scheduler and the wall time for every task are available. In turn, evaluating the CPU load is a feasible work. Thus, in the case of this work, we specify the CPU utilization \textit{at the VM/OS level} following Equation \ref{eq:cpu_util}. In addition, the average CPU load (also known as Per-entity Load Tracking \cite{linux_plt}) is used in the \texttt{schedutil} scaling governor, and the concept of CPU utilization $u_{os}$ (called “CPU load” in the context of the Linux kernel) is used in other governors.

\begin{align} \label{eq:cpu_util} 
    u_\text{os} 
        &= \dfrac{t_\text{CPU}}{t_\text{wall}} \\
        &= 100\% - \dfrac{t_\text{idle}}{t_\text{wall}} \ , \nonumber
\end{align}

\noindent where $t_\text{CPU}$ denotes the CPU time, $t_\text{wall}$ denotes the wall time, and $t_\text{idle}$ is the accumulative time in which the CPU is idle.


Besides the available abstraction offered by the current architecture of our infrastructure, another reason for using the CPU usage rather than the other two metrics is that the time measurement in simulation is generally at a coarser granularity than that of run-time systems in the real world. In other words, datacenter simulations are of a high abstraction of the real-world scenarios in that the traces, on which the experiments are conducted, usually have an interval of several seconds or even minutes between records. Conversely, timing is critical in gauging the CPU utilization, so much so that pitfalls often occur if the interrupts were to happen at undesirable points in time \cite{linux_load}. In addition, when assessing the CPU utilization, the Linux kernel does not take account of the total CPU capacity and the actual CPU speed since both of which are out of reach at the OS level; they are, nevertheless, available through the hardware interface in our simulation infrastructure. Hence, we decide to take advantage of the resources available in our instrument, basing the following development of our energy modelling and management system upon the CPU usage.

\section{Energy Proportionality} \label{sec:energy_prop}

The inclinations demonstrated in previous reports \cite{58670,koomey2007estimating} show that the rising energy consumption steadily dominates the total cost of ownership (TCO) including both computing and infrastructure costs. \citet{Barroso2007TheCF} claim that, in order to prevent energy footprints of datacenters from exploding, the improvement of energy efficiency should keep up with the growth of computing power. From the result of their study \cite{malone2006metrics}, datacenters’ raw performance has been growing five times faster than their performance-watt ratio. Moreover, \citeauthor{malone2006metrics} demonstrates a fatal mismatch of datacenters in which the most used working mode often runs in the least energy-aware way with fairly low CPU utilization. To combat this challenge, energy proportionality, \ie datacenters consume nearly no energy when standing by and gradually raise power consumption as workloads increase, is proposed as an ultimate design goal. Consequently, operating datacenters at the near-peak performance level with a high utilization is preferred as the higher utilization, the better energy efficiency.

That been said, energy proportionality is not an easily achievable objective. By the nature of distributed systems, not only computation but data is allocated amongst hundreds of, if not thousands of, geographically distributed nodes. The purpose of such an architecture is to create duplications, increasing service availability and, in turn, reducing risks caused by disastrous situations. One of many services that hinge on such an architecture is the Google File System (GFS) \cite{Ghemawat2003TheGF}. Conversely, such settings come at a cost --- distributed servers are expected to be always up and running, even if no heavy workloads are hosted \cite{malone2006metrics}. As a result, servers in a datacenter are often neither fully idle nor operates at their maximum utilization. Instead, they mostly operate in the 10 to 50\% utilization range \cite{ryckbosch2011trends}. Moreover, as \citeauthor{malone2006metrics} have suggested, because of the need for performing constant, small operations in the background, networked nodes can hardly enter deep sleep states. When severs are running at their lowest operational states, they consume more than half of their full power \cite{malone2006metrics} (approximately 70\% of their full-speed power \cite{Raghavendra2008NoS,Kusic2008PowerAP,Verma2008pMapperPA,Gandhi2009OptimalPA}). In addition, the impact of the energy loss incurred during the wake-up stage is not as significant compared to the energy consumption under normal utilization level. These characteristics can rarely be found in other systems, such as mobile and embedded systems. Besides the peak-power range, energy efficiency, therefore, should be optimized at all frequency steps.


With the growing energy consumption of modern datacenters and the increasing concern of global warming, more and more studies are being conducted in the search for sustainable solutions. With regard to the energy efficiency of modern datacenters and cloud systems, a full taxonomy has been built by \citeauthor{Beloglazov2010ATA}. Methods for energy management are generally categorised as two major types, a) static power management (SPM) and b) dynamic performance scaling (DPS) \cite{Beloglazov2010ATA,iosup2017opendc}. In each of them, both the hardware and the software solutions have their important role to play \cite{Lim2009MDCSimAM}. The major objectives thereof are two-fold: (1) improving hardware design such as energy-efficient computing as well as models of cooling systems \cite{Gupta2011GDCSimAT,Soeleman1999UltralowPD,Seta199550AS}, and (2) creating better resource management algorithms, including workload scheduling \cite{Bag2006EnergyET,Abbasi2010ThermalAS,Mukherjee2009SpatiotemporalTJ}, policies of power management \cite{Mastroleon2005AutomaticPM}, etc.

\citeauthor{Beloglazov2010ATA} investigated the solutions at a more fine-grained level. They separate the case studies into four levels namely, the hardware and firmware level, the OS level, the virtualization level and the datacenter level. To achieve a better TOC and to speed up Returns On Investments (ROI), and most importantly, to mitigate the carbon footprints of datacenters and cloud services, an integrated approach detailed at each of the four levels is proposed \cite{Beloglazov2010ATA}. Similarly,  \citeauthor{Chien2015ZeroCarbonCA} proposed a set of computing models, the Zero-Carbon Cloud (ZCCloud) that features a bottom-up approach from the selection of sustainable sites to high-level infrastructure design \cite{Chien2015ZeroCarbonCA}.

%% file: chapters/2-bg/2.3_dvfs.tex
\section{DVFS as Mechanism to Manage Energy} \label{sec:dvfs}

In computing machines, different hardware components usually operate at various states, and generally speaking, binary states (\eg active and inactive) are applicable to most of these components. The idle state and inactive state should be clearly distinguished, as they are rather different from each other. When a machine is idle, it operates at its lowest active state without undertaking any useful work, whereas, in the inactive states, it is in one of the standby or in sleep states \cite{Heinrich2017PredictingTE}. Note that, in some literature, the inactive states are regarded as idle states whilst, in this work, the idle state is referred to as the lowest available working state. Also, CPU frequency scaling introduced in this section is an active power management technology. 

\subsection{Frequency Scaling}

Operational states differ from component to component. For example, memory could be in states such as precharging, refreshing, writing, reading, etc., whilst for network switches, the ports therein can operate at various rates, which correspond to another set of states different from that of the memory. Each of these operational states corresponds to a different energy state,\ie their energy efficiency differs in accordance with the type of the operations and/or the speeds of that operations are carried out. This, in turn, makes the energy management of the CPUs relatively more intricate. 

\begin{equation}
    P \propto C \cdot V^2 \cdot F + P^\text{ idle}
    \label{eq:ptof}
\end{equation}

Modern CPUs are capable of running at various speeds/frequencies, each of which has a certain rate of power consumption; Equation \ref{eq:ptof} illustrates this well. The power consumption ($P$) of a CPU is proportional to the capacitance ($C$), the voltage ($V$) and the clock frequency ($F$) of the CPU, whereas its idle power $P^\text{ idle}$ is an addictive term. Clearly, the frequency is linearly correlated with the power consumption, whilst the voltage has a quadratic correlation with it. However, this might give the impression that halving the CPU frequency will make the running time of the hosted workloads twice as long, whilst they still consume the same amount of energy. If so, the best option would always be race-to-halt,\ie execute all tasks (threads in the view of the kernel) as fast as possible and then, put CPUs into sleep. 

This impression is not actually the case, due to the fact that the CPU frequency and the voltage applied to the CPU move together. In other words, by the law of physics, it is impossible to acquire a higher CPU frequency without increasing the voltage because boosting the voltage level requires frequency uplift. Therefore, adjusting CPU frequency has a quadratic impact on its energy consumption. In turn, race-to-halt is not ideal since the $P^\text{ idle}$ only has a linear effect that is not able to offset the impact of voltage variation. Hence, if possible, the CPUs should be reduced to a lower speed with a lower voltage in order to save power, which is where dynamic frequency and voltage scaling (DVFS in short) comes into play. Furthermore, the purpose of using DVFS is twofold \cite{dvfs_twofold}: besides saving power, DVFS helps processors reduce the peak thermal load. The effectiveness of the cooling system relies upon the peak power instead of the average power. Thus, capping the peak power decreases the cost and the size of the cooling equipment. 

\begin{figure}[!t]
  \begin{center}
    \includegraphics[width=\textwidth]{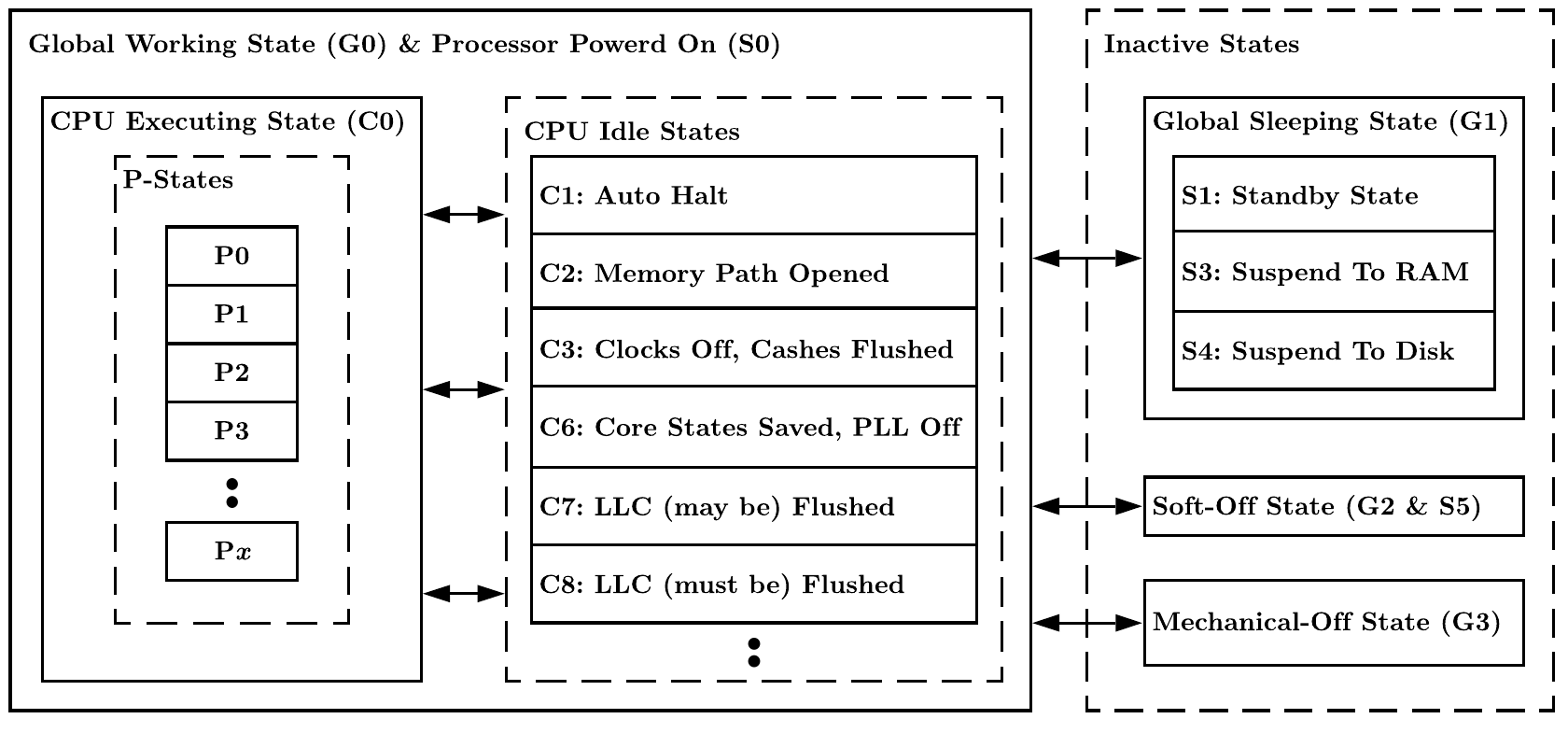}
  \end{center}
  \caption[Transitions between the power states defined in the ACPI standard]{Transitions between the power states defined in the ACPI standard.}
  \label{fig:acpi}
\end{figure}

\subsection{Power States} 

\begin{figure}[!ht]
  \begin{center}
    \includegraphics[width=0.7\textwidth]{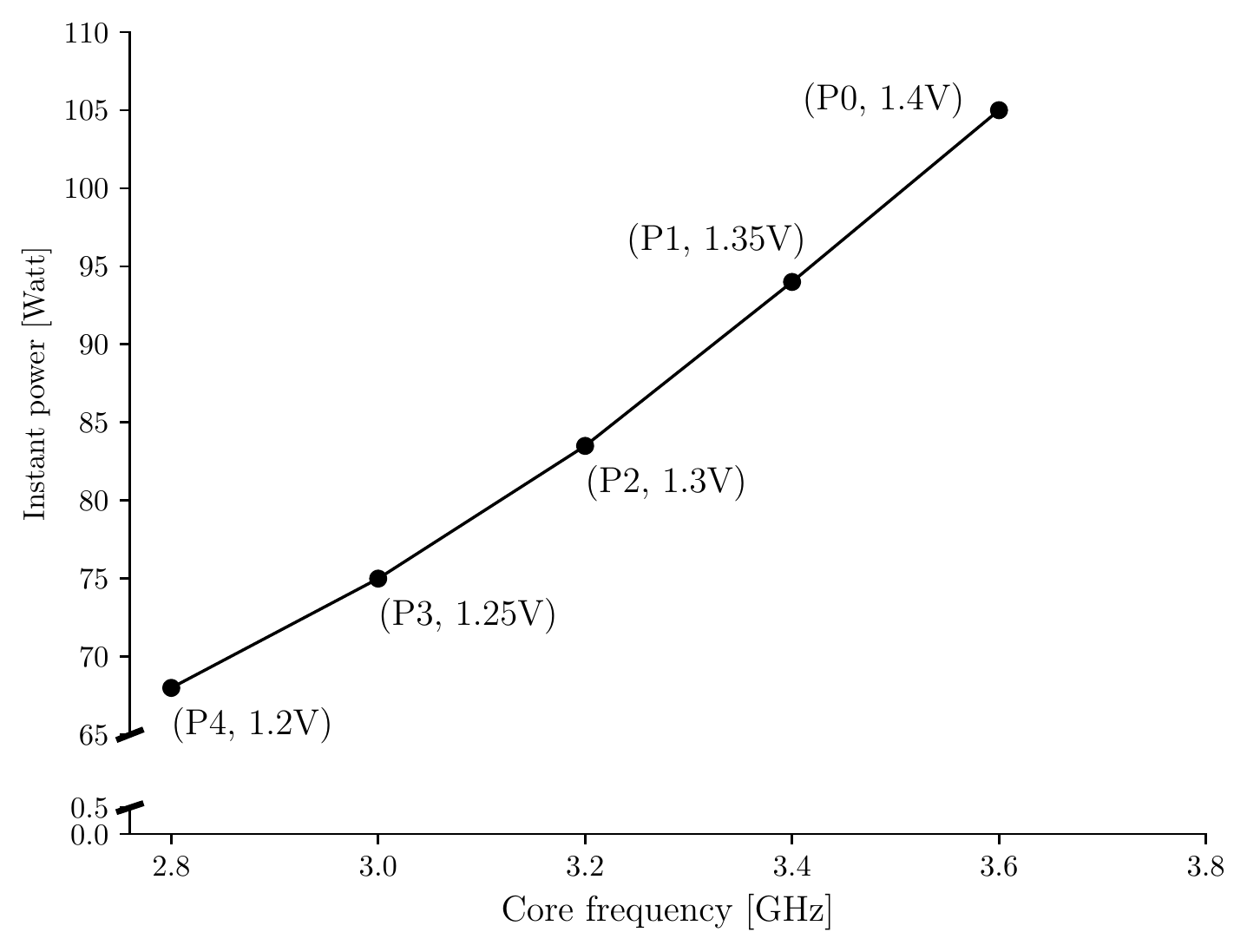}
  \end{center}
  \caption[P-states and conresponding power consumption levels] {P-states and corresponding power consumption levels (data source: \cite{intel_powerarch}).}
  \label{fig:pstates}
\end{figure}

To achieve DVFS, Operating Performance Points were introduced, representing several levels of voltage and clock frequency at which the CPUs are able to operate. P-states (performance states) is the terminology used for Operating Performance Points in the Advanced Configuration and Power Interface (ACPI) standards \cite{acpi}. As shown in Figure \ref{fig:pstates}, the higher the P-state, the lower the core frequencies and voltage, in turn, ultimately saving more energy. Besides P-states, ACPI also defines a set of other types of states, namely, G-states for global system states, D-states for device power management, C-states for the CPU power states, and S-states that entail a number of sleep states. In respect of the CPU, P-states and C-states (CPU-states) are of particular interest. Figure \ref{fig:acpi} further illustrates how P-states and C-states work with other power states. In the context of \texttt{G}0 -- Global Working State, it refers to the active mode in which the CPU is executing instructions, whilst from \texttt{C}1 to \texttt{C}8, the power usage of the CPU is gradually reduced in order to sequentially save more energy. In the context of \texttt{C}0 -- CPU Executing State, the processor can operate at different P-states to further curb the energy consumption. As explained above, P-states modulate both the CPU frequency (in MHz) and the voltage at the same time, and the P-state \texttt{P}\textit{x} depends on the number of frequency steps available in different platforms. By the virtue of the quadratic relationship demonstrated by Equation \ref{eq:ptof}, noticeable energy saving can be achieved by enabling P-state scaling. In addition, in some machines of older generations that predate C-States and P-States, throttling states (T-States) are used under thermal emergency where the processor is overheating. This is achieved by gating the CPU clock --- the higher the temperature, the higher the T-State and, in turn, more CPU cycles will be gated. As T-States are outdated technology, they are not considered in this study.

\subsection{DVFS in Linux} 

P-states are effective in saving power, they, however, are handled differently by different platforms. In this work, we focus on the DVFS implementation of Linux. The \texttt{CPUFreq} subsystem is responsible for CPU frequency scaling in the Linux kernel, offering a basic infrastructure and user-mode interface for all devices that support P-states. It not only provides other components with a framework in which they operate but also gives the opportunity to implement various frequency-scaling mechanisms in accordance with the (estimation of) CPU capacity demanded by workloads. To this end, a number of generic scaling \texttt{Governors} and \texttt{Drivers} are provided by the Linux kernel \cite{linux_kernel}. A \texttt{Governor} is a piece of software in which the algorithms/policies for adjusting the CPU frequencies are implemented. The scaling rules thereof are based on the estimation of the required CPU capacity. Each of the \texttt{Governor}s implements one set of frequency scaling rules, and these policies are located under the directory \texttt{/sys/devices/system/cpu/cpufreq/}. The scaling \texttt{Governor}s are independent of specific CPU architectures. A \texttt{Driver} is another piece of software that is responsible to interact with hardware directly. They offer available \texttt{Governor}s a set of machine-specific P-states and apply the frequencies proposed by \texttt{Governor}s to the machine via hardware-dependent interfaces. If the scaling algorithm implemented by a \texttt{Governor} is per-policy as opposed to system-wide/global, \texttt{Driver}s will find the corresponding tunable attributes (\texttt{sysfs}) in the subdirectory of the policies (\texttt{/sys/devices/system/cpu/cpufreq/policy\{\#\}}). In spite of the existence of driver-specific properties, \texttt{Driver}s and \texttt{Governor}s are designed to be orthogonal, \ie they are supposed to be used in any combinations. This design is achieved via a set of \texttt{struct cpufreq\_policy} objects, each of which is associated with one or several CPUs. The type of \texttt{Governor}s can be altered during runtime, and in turn, several \texttt{Governor}s attached to the CPUs can share the same policy object by setting the \texttt{scaling\_governor} attribute in \texttt{sysfs}. 

\begin{figure}[!t]
  \begin{center}
    \includegraphics[width=0.7\textwidth]{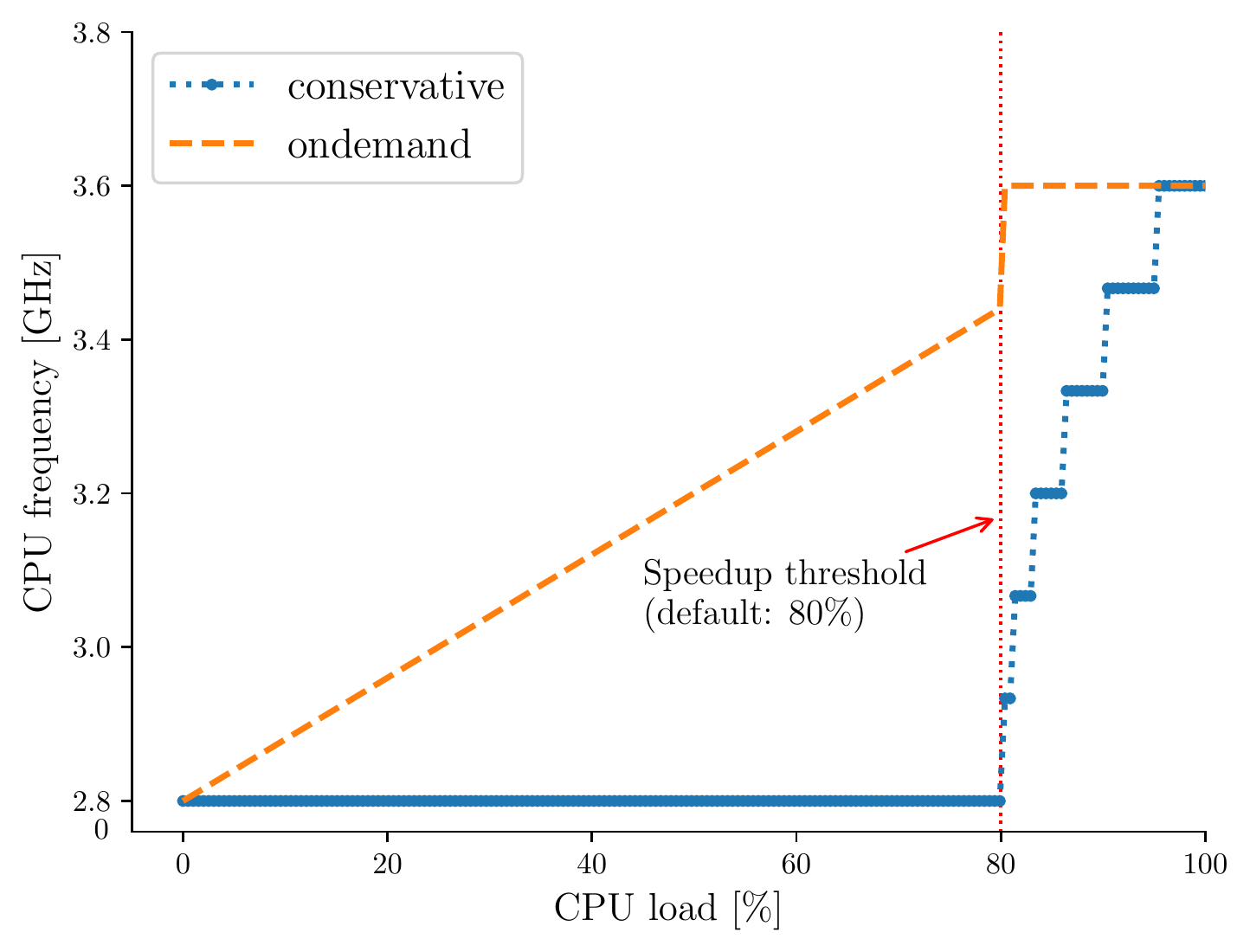}
  \end{center}
  \cprotect \caption[The \texttt{ondemand} \& \texttt{conservative} governors in the Linux kernel] {Different behaviours of the \texttt{ondemand} and the \texttt{conservative} governors in the Linux kernel.}
  \label{fig:governors}
\end{figure}

\subsection{Governors \& Drivers} 

Six generic \texttt{Governor}s are available in the Linux kernel: 1) the \texttt{perf\-ormance} governor, which immediately request the highest frequency within the limit specified by the \texttt{scaling\_max\_freq} attribute of each policy, 2) the \texttt{powersave} governor, which proposes the lowest frequency above the threshold specified by the \texttt{scaling\_min} \texttt{\_freq} of each policy, 3) the \texttt{userspace} governor, which does nothing aside from allowing the \texttt{scaling\_setspeed} attribute of each policy to be set in user mode, 4) the \texttt{schedutil} governor, which runs in the context of the Linux scheduler and uses the scheduler for estimating the next frequency to which the CPUs ought to be adjusted, 5) the \texttt{ondemand} governor, which uses the CPU load as the metric for selecting the CPU frequency, and 6) the \texttt{conservative} governor, whose policy resembles the two-stage frequency scaling process of the \texttt{ondemand} governor but requests changes in small steps. The \texttt{schedutil} governor is developed to tackle the challenge of estimating the requested CPU capacity. It employs data obtained from the scheduler instead of only the data from the CPUs since the scheduler is more informed by the run queue, information of I/O blocking, etc. Such a mechanism can be of great help in DVFS because, for instance, a task that is not running but waiting or blocked by I/O also contribute to the load of the system, whereas a long-running task that accumulatively consumes a large number of resources may not be as demanding at the moment. In addition, the difference between the \texttt{ondemand} governor and the \texttt{conservative} governor are subtle but important. Both of them are running in the process context asynchronously, which causes little overhead to the scheduler but generates more context switches. The interrupts triggered by them for updating the P-states can be irregular, and the idle time of the CPU is, thereby, reduced. As illustrated in Figure \ref{fig:governors}, before reaching the (configurable) speedup threshold, the \texttt{ondemand} governor will propose the next frequency proportional to the current CPU load \footnote{\url{https://github.com/torvalds/linux/blob/master/drivers/cpufreq/cpufreq\_ondemand.c}} (Equation \ref{eq:ondemand}). Conversely, as for the \texttt{conservative} governor, no changes in frequency will be requested at the first stage \footnote{\url{https://github.com/torvalds/linux/blob/master/drivers/cpufreq/cpufreq\_conservative.c}}.

\begin{align}
    f = &\ f_\text{min} + l \cdot \dfrac{ f_\text{max} - f_\text{min}}{100}\ ,
    \label{eq:ondemand}
\end{align}

\noindent where $f$ is the next frequency to propose, $f_\text{max}$ and $f_\text{max}$ denote the maximum and minimum frequency specified in the scaling policy, respectively. 


Once the threshold is met, the \texttt{ondemand} governor will jump straight to the maximum frequency limit (\texttt{scaling\_max\_freq}), whilst the \texttt{conservative} governor will request frequency changes continuously (both increase and decrease) in small steps in order to avoid significant frequency fluctuations over short periods of time. This mechanism is particularly useful when drastic changes in CPU frequencies are not supported or suited for the machine. The default threshold in the Linux kernel is 80\%. In other words, if the idle time to wall time ratio is less than 20\%, the two governors will start boosting core frequencies. In addition, the minimum step size of the \texttt{conservative} governor is 5\% of the maximum frequency limit. 

Note that the P-state scaling provided by Intel (\texttt{intel\_pstate} \cite{intelp}) is rather different from that of the generic policies. It comes with its own algorithms and bypasses the built-in drivers of the Linux kernel. Specifically, the Intel SpeedStep$^\text{®}$ and the Speed Shift$^\text{®}$ technologies \cite{intel_manual} are available by the time of this study. The former switches P-states based on certain algorithms that are not open-sourced, and the latter is an improved version of the former. Instead of changing P-states in a discrete manner, Speed Shift$^\text{®}$ enables a full multiplier range or narrow window. It is able to fully ramp up a core speed in response to a lower P-state faster (30-35 ms) compared to that of the SpeedStep$^\text{®}$ (100-150 ms). However, it is limited to the Skylake architecture and needs support from operating systems. Thus, \texttt{intel\_pstate} is not considered in this work.

\subsection{Multicore} 

\label{sec:multicore} As elaborated above, although a \texttt{struct cpufreq\_policy} object can be assigned to multiple CPUs, a single CPU is able to occupy a policy object itself. However, in old-generation machines, all (logical) cores of a package are managed under the same power domain \cite{Siddha2007ProcessSC}, which means they are all in the same P-states at any time based upon their maximum load \cite{intel_manual}. This is akin to the mechanism of the multicore C-states. Furthermore, under the OS C-states, there are 1) CC-states, which offer a set of idle states for each physical core, and 2) PC-states for the idles states at the package level covering various shared resources. Both the CC-states and the PC-states are set by taking the minimum level of their respective state (which has the highest frequency value) over all its components. Notably, in Intel $\times$86 processors that support hyper-threading \cite{marr2002hyper}, per-thread C-states can be obtained, managed in the same manner as that of the PC-states and CC-states. For P-states, however, only until recent generations (after Haswell \cite{Hackenberg2015AnEE}), processors have just started supporting P-states for each (logical) core, known as multicore-aware P-state coordination \cite{intel_manual}. Nevertheless, few datacenter simulators support this feature (\S\ref{sec:modelling}), as it is too detailed for high-level datacenter simulation and may introduce substantial overhead as well \cite{Casanova2014VersatileSA}. Also, these datacenter simulators or their respective energy extensions \cite{Heinrich2017PredictingTE,Silva2019AccuratelySE,Casta2013Emc2AF} reuse the core-level infrastructure previously built to achieve this feature so that backward compatibility is better maintained. As we do not have such an issue in our OpenDC simulator, we take package-level decisions when switching between P-states to circumvent the unnecessary overhead (\S\ref{sec:sys_impl}).

%% file: chapters/2-bg/2.4_power_grid.tex
\begin{figure} [t!]
    \centering
    \includegraphics[width=0.5\textwidth]{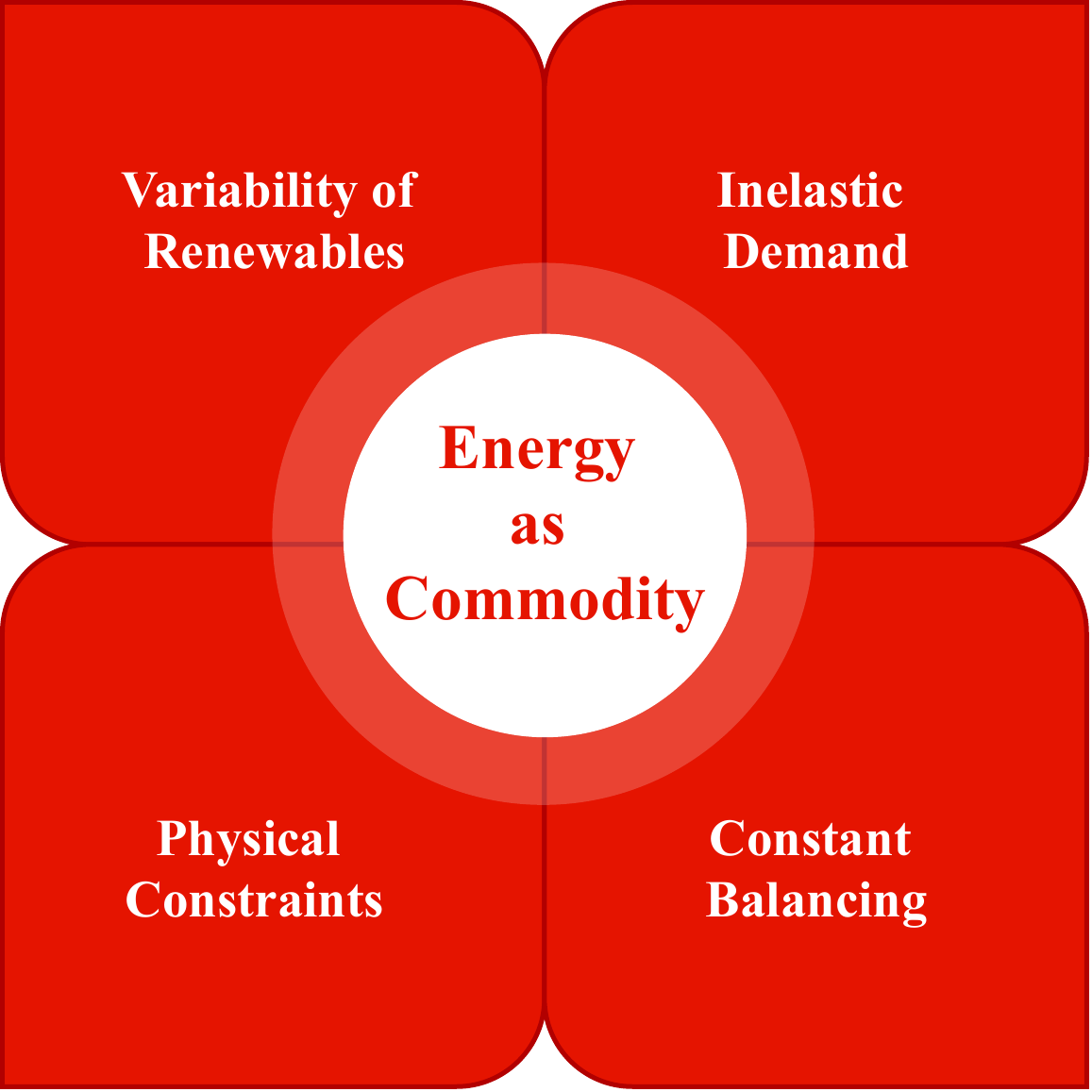}
    \caption[Challenges faced by energy transmission and trading]{Challenges faced by energy transmission and trading.}
    \label{fig:energy_commodity}
\end{figure}

\section{Power Grid} \label{sec:power_grid}

Energy, or specifically, electricity, is a special type of commodity. This section covers the basics of the power grid (in the EU) with regard to both its financial perspectives and real operations. As of the time of this work, once it has been generated, it is neither practical nor economical to store them on a large scale for a long time. To function well and to avoid issues such as blackout and power outage, a balance between power generation and consumption must be kept at all times. Its transportation and distribution are performed on a power network, governed by specific physical rules of mother nature. Also, it features inelastic demand as the power load can affect the energy prices but not the other way around. The majority of the end-users is of rubric of the society (\eg residential, production, hospital, etc.), whilst the roots of the generated energy are not differentiable, \ie the power produced by burning coal and the power produced by solar panels have no difference in the eyes of the consumer. Nevertheless, the origin of energy makes a huge difference in the producer side as well as to the energy market. Determined to tackle the climate crisis, the EU has turned its market into a massive entry of renewable energy sources. However, this comes at costs that greatly diminish the competence of renewable energy compared to fossil fuel.

\begin{figure} [t!]
    \centering
    \includegraphics[width=0.4\textwidth]{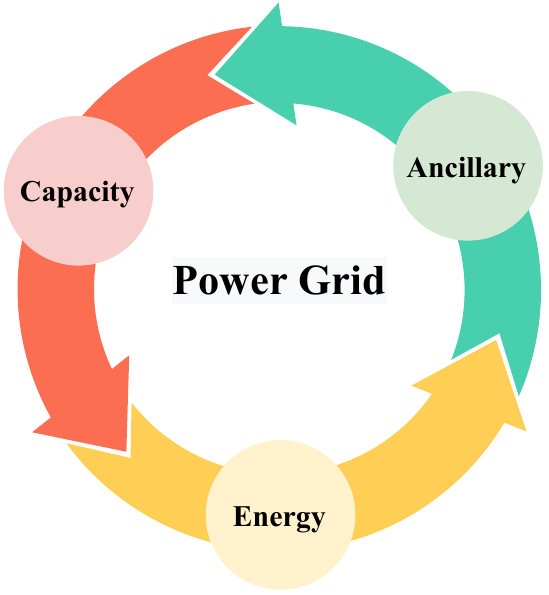}
    \caption[Markets around the power grids]{Markets around the power grids.}
    \label{fig:grid_markets}
\end{figure}

\subsection{Markets Around the Power Grid} \label{sec:grid_markets}

\begin{figure} [t!]
    \centering
    \includegraphics[width=0.6\textwidth]{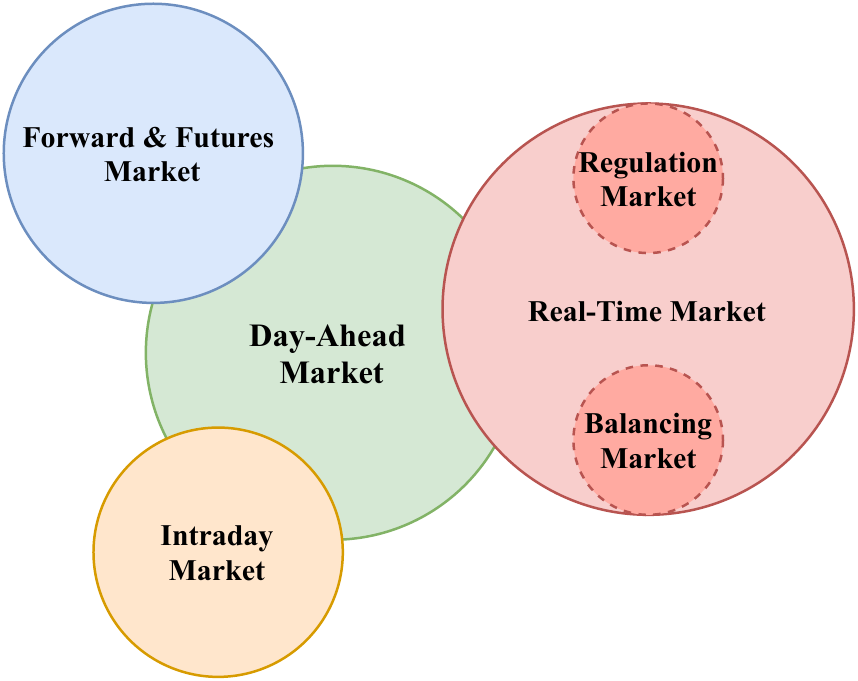}
    \caption[Types of energy markets in the EU]{Types of energy markets in the EU.}
    \label{fig:energy_markets}
\end{figure}

As shown in Figure \ref{fig:grid_markets}, when it comes to the power grid, there are three types of markets in the EU. In the capacity market, system operators ensure sufficient capacity of power generation is retained for sustaining competitive prices and reliable operation in the coming years. They also provide many services such as  Primary/Secondary/Tertiary reserves and voltage control, of which the ancillary market is comprised. In this work, we focus on the energy market in which optimal scheduling and power exchanges take place. In respect of the energy market, four major parts arranged in a sequence interact with one and another (Figure \ref{fig:energy_markets}), keeping a delicate balance for both the financial markets and the actual operations in the power grid. Firstly, contracts for physical delivery of energy, price hedging and risk management are made in the Forward \& Futures markets. Secondly, treating the current trending as the spinal core for predicting the matching of everyday demand and supply, participants buy or/and sell energy for the next 24 hours in a closed auction held in the Day-Ahead market, which is the most important market in the EU. In other words, according to the previous market-clearing outcomes (price and volumes for each market time unit), market operators will dispatch themselves concerning the pattern of power generation and consumption of the next day (often in a 15-minute increment).

After this blind auction, a spot price is settled at the intersection where the demand meets the supply. This can be regarded as the first balance reached in terms of the financial market of the power grid, it, however, does not imply any obligation towards prosumers, \ie no one is forced to produce or consume the promised quantities. Then, as a continuation, the Intraday market provides prosumers with a bilateral trading platform by which they can adjust their self-dispatched units in the Day-Ahead market based on the newest status (\eg updated weather/market predictions) before the actual operation/delivery commence. Finally, the Real-Time market serves as the final guard during physical operations, which is where the system operators (in the EU) take over the market, offering regulation to counteract the remaining imbalance and charging the prosumers based upon actual figures in the power meters against the contracted volume.

\subsection{Balancing the Power Grid} \label{sec:balance_grid}

\begin{figure} [t!]
    \centering
    \includegraphics[width=0.9\textwidth]{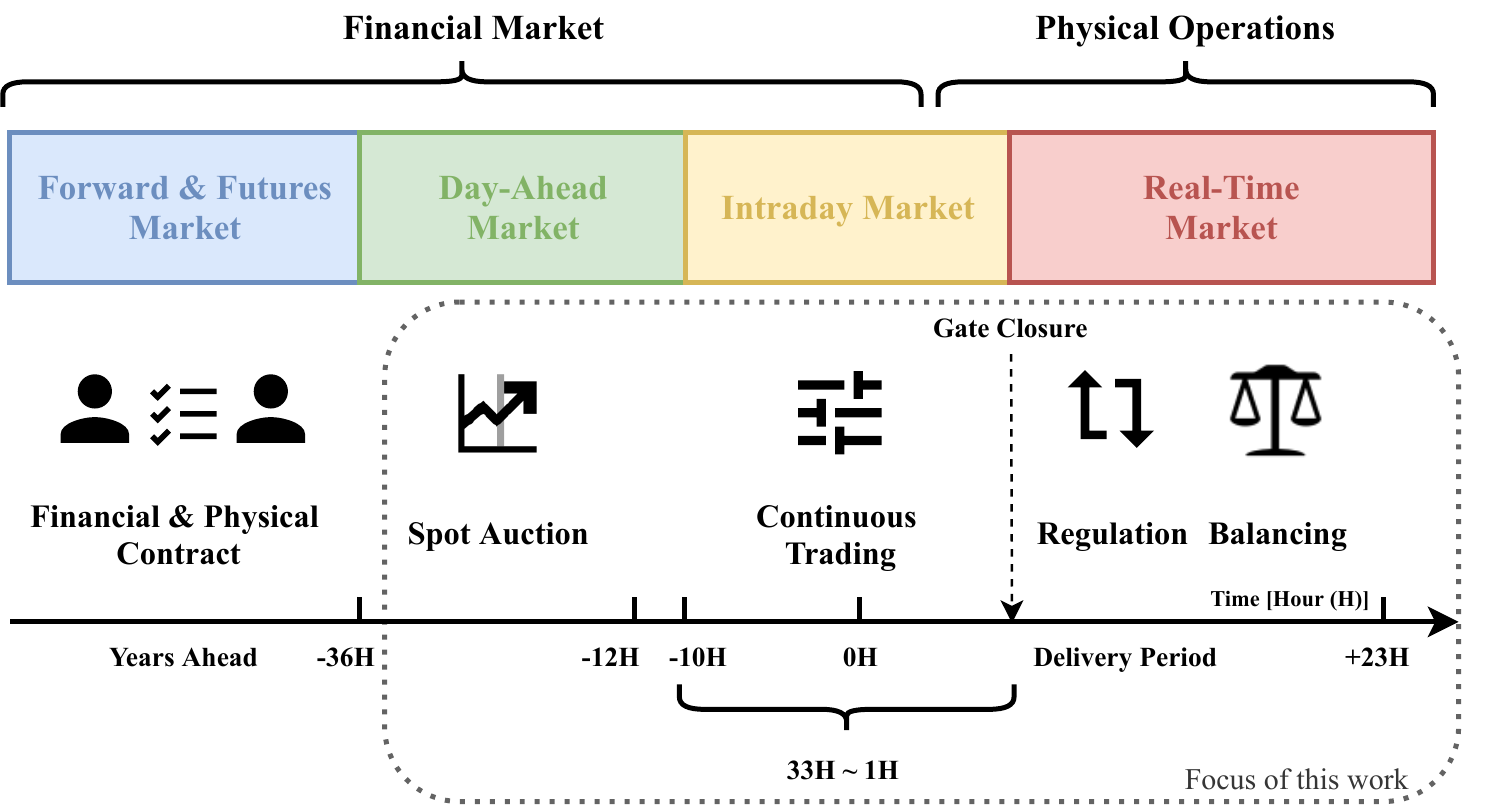}
    \caption[Sequence of energy markets]{Sequence of energy markets.}
    \label{fig:market_sequence}
\end{figure}

There are two levels of (im)balance in the power grid, the positive/negative/no imbalance at the overall system level and that of the prosumer level. Even if not all prosumers have reached a balance, it is possible that the overall system of the power grid can still strike a delicate balance. Although balancing the grid is the central duty of the system operators, it is neither solely pertinent to the system operators nor only take place in the Real-Time market. Instead, it is relevant to every participant and is performed at almost every stage of the market sequence demonstrated in Figure \ref{fig:market_sequence}. Firstly, as described in the previous section, in the EU, the day-ahead auction is the most important market whose clearing yields a spot price at which demand and supply meet for the first time. This, however, does not imply any actual obligation towards prosumers. In other words, they are free to (intentionally or accidentally) break the contract during real-time operations. 

As shown in Figure \ref{fig:market_sequence}, there is still a maximum 33-hour interval in time between the clearing of the spot auction and the start of the delivery. Market operators can take advantage of this period to adjust their initial commitments in the Intraday market until one hour before the delivery phase, hedging the risks of under-/over-production. This can be treated as the second balancing procedure of the power grid. Last but certainly not least, after the physical operations commence, the system operators take control to ensure the grid is balanced during energy transmission. The Real-Time market consists of two sub-markets, namely, the obligation market and the balancing market. Prior to the start of transmission operations, prosumers who participated in the regulation market can offer to buy and/or sell regulation power to the system operators. In other words, these are the participants who are capable of helping balance the grid, declaring to the system operators the willingness of altering their set-points on the consumption side and/or the production quantity. Also, the system operators can actually purchase the regulation resources from neighbouring countries. This is usually in the form of commercial entities buying, selling and transporting electricity across a border, and participating in markets in both countries as any other energy prosumer would. BritNed, the cable between the United Kingdom and the Netherlands, is an example of this. In contrast, the prosumers who do not respect their original agreements and, in turn, induce imbalance in the power grid will be charged by the system operators on the basis of the difference between their original schedules and the actual readings from the power meter. On the contrary, if a participant is helping the system operator balance the power grid, \eg a consumer is in positive imbalance and the system is in positive imbalance as well, then the prosumer would be rewarded by the system operator as they are helping balance the grid (in spite of not following their schedule). Via these settlements and other energy reserves, the balance of the power grid is maintained in real-time operations by the system operators.

Note that the elaboration above is tailored to the EU energy market in which the transmission system operator (TSO) has the ultimate responsibility to keep its transmission system in balance. For other places such as the USA that relies on the independent system operator (ISO), the energy market, especially, the balancing mechanisms are different. Nevertheless, the market sequence is shown in Figure \ref{fig:market_sequence} still applies and the objectives of the balancing phase stay the same. An excellent load/price forecast and flexible operational responses to the market will bring substantial benefits to the prosumer. For example, if it is predicted that the demand/price will be lower during the Real-Time market, then one can sell (short) in the Day-Ahead market and buy in more energy in the Real-Time market. This can facilitate energy-aware scheduling in industry production. As a result, the prosumer will receive the profits along the lines of Equation \ref{eq:balance}:

\begin{equation}
    G = (Q_{a} - Q_{s}) \cdot \zeta\ , \label{eq:balance}
\end{equation}

\noindent where $G$ denotes the financial gain, $Q_a$ represents the actual quantity of energy transmission, $Q_s$ is the scheduled quantity in the spot market, and $\zeta$ denotes the locational marginal pricing.

Bidding in the Day-Ahead market is going to push the prices up; good forecast and flexible responses to the market foster smooth pricing in the Real-Time market, which will bring the prices down to what they ought to be. Consequently, it can bring the prices of today and tomorrow closer, and ultimately help the power grid strike a balance that would greatly reduce the electrical system charges and, in turn, lower the grid balance cost.

\input{resources/tables/tab_reserves}

Furthermore, the imbalance energy trading happens in the regulation \& balancing markets are of particular interest of this work. During the trading, Balancing Service Providers (BSPs) offer bids to counter the imbalance in power supply and consumption during the delivery hour. The imbalance is introduced by Balance Responsible Parties (BRPs), which are the prosumers responsible for the deviation between their actual delivery and their self-dispatched volume concluded in the day-ahead/intraday clearings. Specifically, in the Netherlands, these self-dispatched commissions are in the form of ``E-programs’’ in a 15-minute resolution. Having been reported the readings of the meters, TSO will ascertain the deviation for every BRP. Then, BRPs are obliged to pay their energy deficit or surplus by participating in the imbalance settlement. In terms of timing, BRPs can change their ``E-programs’’ until one hour before the start of the delivery through, \eg the Intraday market, whilst BSPs are able to change their bids 30 minutes before the start of a delivery hour. The 15-minute interval of an ``E-programs’’ is known as an Imbalance Settlement Period (ISP), and every day has 96 ISPs. As summarized in Table \ref{tab:reserves}, various mechanisms are employed by TSO in the imbalance system, including Frequency Containment Reserve (FCR), automatic Frequency Restoration Reserve (aFRR), manual Frequency Restoration Reserve (mFRR), and Replacement Reserve (RR). They differ in their rate of ramping up. For example, the activation time of an aFRR is required to be no longer than 15 minutes with the minimum bid size of 1 MW. In this work, we do not distinguish the types of means by which the BSPs provide balancing energy to the grid since they are paid from the same ground irrespective of the type of reserves activated.

\subsection{Challenges of Renewables}

\begin{figure} [t!]
    \centering
    \includegraphics[width=0.8\textwidth]{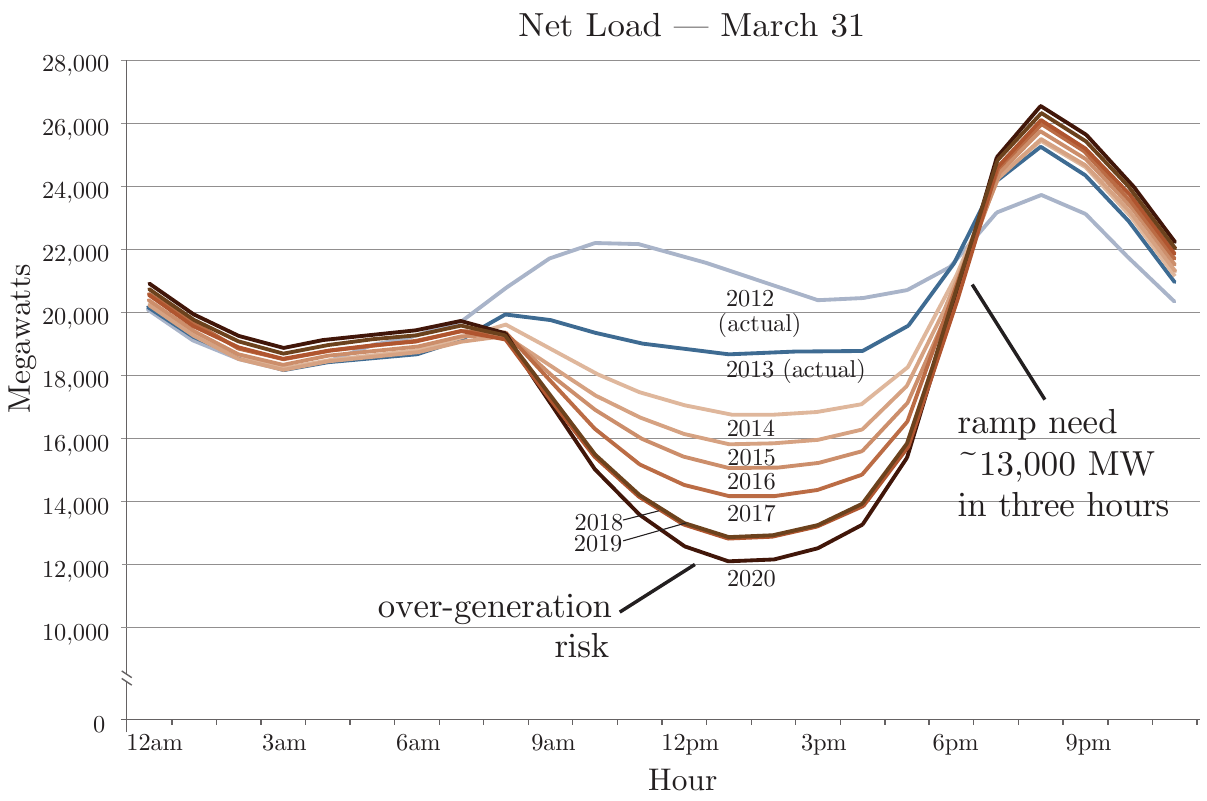}
    \caption[Duck curve showing the net load of the power grid] {Duck curve showing the net load of the power grid for 11 January from 2012 to 2020 in California (figure source: \cite{cal_iso}).}
    \label{fig:duck_curve}
\end{figure}

As elaborated in the previous section, balancing the power grid is a win-win for both the grid system and the prosumers. However, it faces increasing challenges brought by the introduction of renewable energy sources, for example, solar, wind and tidal power (its current contribution is diminutive to the grid). This is due to the fact that these origins of energy are erratic and, thus, hard to predict. This brings risks to both the prosumers and our environment. For prosumers, the energy market incurs more fluctuation and, in turn, is more precarious that even the Intraday market will not give enough operational headroom. Concerning the environment, renewable energy generation needs constant backup from traditional energy sources, which exacerbate the reliance on fossil fuels. One of many quintessential examples for variable generation resources is the duck-curve phenomenon (Figure \ref{fig:duck_curve}) first introduced by California independent system operator (ISO). As the degree to which the power grid is dependent on renewable sources grows from 2012 to 2020, the daily variation of the load within increasingly boosts over the years. 

\begin{figure} [t!]
    \centering
    \includegraphics[width=0.85\textwidth]{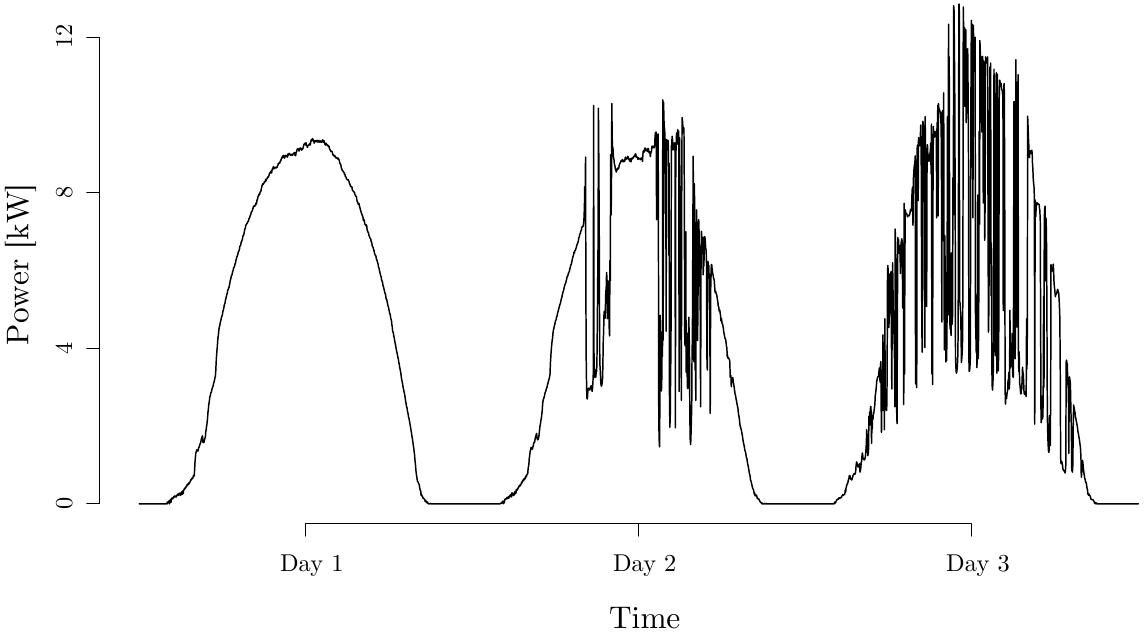}
    \caption[Solar power generation over three consecutive days] {Solar power generation at the University of Queensland over three days in September 2018 (data source: \cite{cal_iso}).}
    \label{fig:solar_power}
\end{figure}

To further illustrate the challenges brought by renewables, Figure \ref{fig:solar_power} shows the power generated by the solar panels at the University of Queensland, St Lucia campus, Prentice building in three consecutive days. When there is no sunshine at all, a drastic drop will occur in solar power generation; if the weather is overcast, the reflection of light on the clouds may actually bring more solar power. The third day in this graph is obviously rather cloudy. The challenges brought by renewable energy sources in balancing the grid put further emphasis on the paramount importance of flexible and quick demand-side responses to the requests from the system operators and price/load of the energy market. This will not only enable prosumers to stay competent and profitable but also provide the power grid with sufficient energy reserves in the long run.

\subsection{Demand Side Management} \label{sec:dsm}

Intensive work has been conducted to explore the datacenter’s participation in DSM programmes. A synergistic control strategy and a flexibility factor have been developed to increase the capacity of frequency regulation in a recent study \cite{Fu2020AssessmentsOD}. \citeauthor{Fu2020MultimarketOO} presented a real-time, multi-market optimization framework for the datacenters \cite{Fu2020MultimarketOO}, facilitating datacenters to participate in both the energy market and the regulation operations. Taking into account the energy cost, demand costs and regulation revenues, the framework optimizes the bid in each hour, helping datacenters meet energy and demand goals whilst retain minimum cost through obtaining maximum regulation revenues. However, this work focuses on enlarging the FR capacity of datacenters, which is of rarity in the industry \cite{Bates2014ElectricalGA}.

Via dynamic pricing, \citeauthor{Li2013TowardsDP} proposed a collaborative framework for optimizing the overall costs of several datacenters in their position paper \cite{Li2013TowardsDP}. The framework employs collaborative efforts across multiple geographically distributed datacenters that communicate via dedicated network fabrics to negotiate mutually optimal energy prices. In this collaborative system, an optimization platform \cite{Liu2012CologneAD} is used to enable constraint optimization problems (COPs). However, to benefit from these schemes, multiple datacenters and their utility companies have to be involved in order to optimize the price updates in the energy market.

In respect of market design, two categories of programmes for DR are commonly available. The first category entails biding/supplying a certain amount of demand flexibility into the market. In other words, consumers bid their flexibility via supply functions that are parameterized (\eg \cite{Johari2011ParameterizedSF,Xu2016DemandRW}). As for the second category, consumers buy in or respond to published prices that were selected according to predictions on (potentially) available flexibility; examples of the second sort include \cite{Conejo2010RealTimeDR,
Li2011OptimalDR,Rad2010OptimalRL,Liu2014PricingDC}. Specifically, programmes such as the Coincident Peak Pricing (CPP) \cite{Baldick2018IncentivePO} and price-based incentive programmes \cite{Sle2011DemandRF} are available for datacenters to participate in. These programmes charge much higher prices for electricity during peak hours (usually over 200$\times$ higher than the base price \cite{Liu2014PricingDC}). Such peak-hour costs can account for 23+\% of the consumers’ energy bill \cite{peak_bill}, which is a strong motive for consumers to reduce their energy usage during peak hours. However, when it comes to datacenters, \citeauthor{Liu2014PricingDC} argued that the market designs and models in the second category outperform those in the first. Also, the current available DR programmes, such as CPP, are not particularly suitable for datacenters since they can barely fully extract the flexibility of datacenters due to many reasons. Chief amongst them is that although datacenters may incur charges if no response is offered, no profit would be produced either if the peak in energy load does not happen during the coincident periods. As a result, datacenters nowadays provide little, if any, response to market signals \cite{Ghatikar2012DemandRO,glanz2012power,Liu2014PricingDC,Liu2013DataCD}. In addition, a large amount of effort has been devoted to the optimization of workload management of datacenters (\eg \cite{Chen2010IntegratedMO,Gandhi2011MinimizingDC,Heo2011OptiTunerOP,Lin2011DynamicRF,Meisner2011PowerMO,Xu2012CostED,Yao2012DataCP,Zhang2012DynamicEC}), especially, via workload distribution \cite{Rao2010MinimizingEC,Liu2015GreeningGL,Ghamkhari2013EnergyAP,Wang2012DProDD,Zhang2015ATI,Wang2016ProactiveDR}. When such optimization is considered, time for shedding the load is able to be reduced and, in turn, the flexibility of datacenters can be exploited further. Moreover, by leveraging the regional difference of the energy cost, datacenters can leverage their distributed nature to further optimized \cite{Qureshi2009CuttingTE,Rao2010MinimizingEC,Liu2015GreeningGL,Lin2012OnlineAF,Ghamkhari2013EnergyAP,Wang2012DProDD,Zhang2015ATI}. In fact, these management strategies require cooperation between the power grid and geographically distributed datacenters together with their utility companies. Thus, they are far more complex to orchestrate than what is introduced in this work, which, in turn, could potentially hinder their adoption. In this work, we develop a straightforward, short-term scheme, whereby individual datacenters can participate in the energy market, both saving their energy cost and curbing their energy consumption, whilst providing the power grid with indirect DR.

\input{chapters/2-bg/2.4.x_dc_and_grid}

%% file: resources/tables/tab_reserves.tex
\begin{table}[h!]
    \centering
    \begin{adjustbox}{width=1.1\textwidth,center=\textwidth}
    \begin{tabular}{c|c|c}
    \toprule
        \textbf{Network Code} & \textbf{Definition} & \textbf{Activation Time} \\
    \midrule
        FCR  &  Primary control (automatic activation) & $>$ 15 minutes \\
        aFRR & Secondary control (automatic activation) & 30 seconds -- 15 minutes \\
        mFRR & Tertiary control (semi-automatic or manual activation) & $\geq$ 15 minutes \\
        RR   & Optional control (semi-automatic or manual activation) & $\geq$ 15 minutes \\
    \bottomrule
    \end{tabular}
    \end{adjustbox}
    \caption[Balancing reserves in the EU]{Balancing reserves in the EU.}
    \label{tab:reserves}
\end{table}

%% file: chapters/2-bg/2.4.x_dc_and_grid.tex
\subsection{Datacenters and the Power Grid} \label{sec:dc_grid}

\begin{figure}[!ht]
    \centering
    \includegraphics[width=0.7\textwidth]{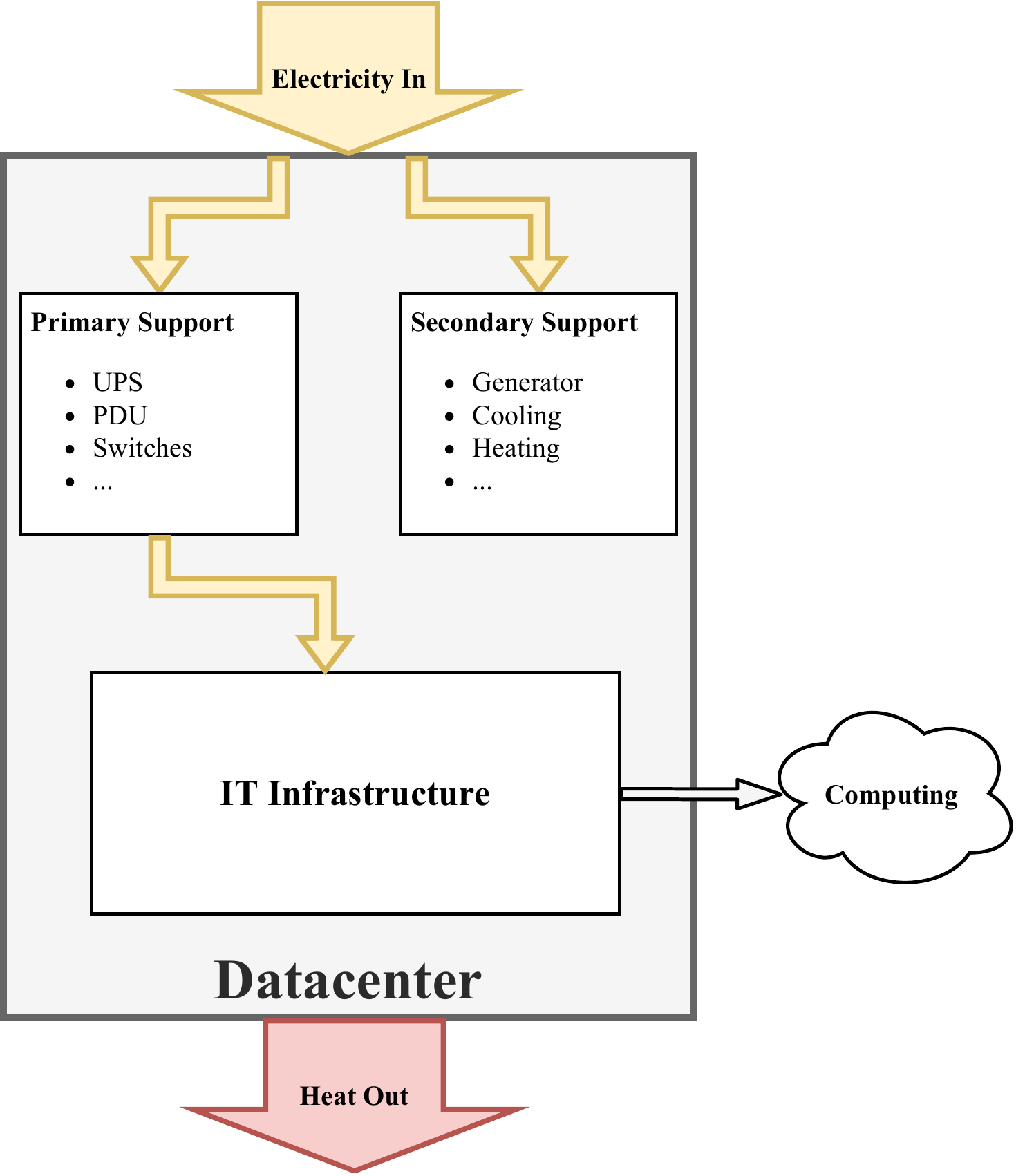}
    \caption[Energy flow in datacenters]{Energy flow in datacenters.}
    \label{fig:energy_flow}
\end{figure}

Whilst system-level challenges have been brought into the grid by renewable energy sources, increasing numbers of opportunities are being created in the meantime. Especially, as the proportion of worldwide electricity attributed to datacenters is sky-rocketing, these challenges could well be a blessing in disguise for datacenters. Over recent years, more and more studies regarding the participation of datacenters in the power grid have been carried out. Several identified characteristics therein of datacenters underpin the increasingly important role played by datacenters. To name a few, firstly, the capacity of datacenters are huge in the sense that the nameplate load of a single datacenter is able to reach as large as 50+ MW \cite{Brown2008ReportTC}. Moreover, for large cloud providers, the critical power of a single datacenter can even exceed 100 MW \cite{Barroso2018TheDA}. To put this into perspective, \citeauthor{Wierman2014OpportunitiesAC} suggest that a well-managed datacenter of 30 MW has approximately the same capability for regulating the power grid as massive energy storage of 7 MWh. Thus, overlooking datacenters in balancing the grid means missing out on a huge amount of capacity. Secondly, datacenters are built for extremely reliable and available services. For instance, two availability classes to which most modern datacenters belong are Tier \romnum{3} and Tier \romnum{4}. Datacenters that fall into the first category insure a 99.982\% availability, and for the latter, the figure is 99.995\% \cite{Mare2010DemandRA}. To accomplish promised uptime and performance guarantees, a considerable amount of redundancy is introduced in the power system of datacenters. 

As demonstrated in Figure \ref{fig:energy_flow}, a large portion of the energy goes into the power support system that has a high degree of redundancy, and almost all energy ultimately turns into heat in the end. To ensure a secure power supply, datacenters usually have two energy sources. As accessing to two utility power sources is not often feasible, most datacenters use a backup generator as the second power source. These generators are powered by gas, diesel or flywheel, providing power when the utility power fails. In the event of power failure, the automatic transfer switch (ATS) will start the generator to power the uninterruptible power supply (UPS) load. The UPS system supports servers, data communication systems and other equipment during a sudden power failure or voltage drop. It provides clean power to sensitive data equipment by eliminating power surges, noise, spikes, etc. The UPS system also constantly conditioning and monitoring utility power to protect the load. Batteries therein are constantly been charged for emergency support during a utility outage. Note that one of the most important factors to ensure proper UPS performance is the battery quality in the sense that one bad battery is able to bring the entire system down during a power interruption. Serving as the last layer of adjusting the electricity, Power distribution units (PDUs) provide the ability to control and monitor how the power is distributed to the IT infrastructure. The facilities in the mechanical yard are critical for environmental controls (\eg heating, cooling and humidity). They maintain a proper environment for electronic equipment by tolerating fluctuations in moisture and temperature. One of the most efficient forms of cooling for datacenters is a close-coupled in-row, chilled water cooling system, also known as a computer room air conditioner (CRAC). Furthermore, the appropriate placement of air returns and the use of perforated floor tiles/sensors can help eliminate hot spots and gain efficiencies in the building. All these equipment are built for facilitating the computing services provided by the server farm, in which racks designed specifically for datacenters offer a modern enclosure with strength and stability for any server environment. Components such as the backup generators, UPS, transformers, chillers/Computer Room Air Conditioner units (CRACs), Computer Room Air Handler units (CRAHs), etc., can all be regarded as redundant equipment.

Further, perhaps what counts more is the significant flexibility of datacenters, making them extremely elastic power loads for the grid. For example, the wide range of temperatures under which the datacenters can operate results in various power loads \cite{ashrae2015thermal}. Also, many workloads of modern datacenters are delay-tolerant. In other words, the schedule of these workloads can be shifted in response to the energy market and requests from the system operators to optimize profit as elaborated in previous sections. Power management techniques such as frequency scaling \cite{Chen2015OnTI}, power capping \cite{Zhang2012DynamicEC,Chen2013DynamicSP,Lin2011DynamicRF} and different levels of energy-saving configurations \cite{Heo2011OptiTunerOP} further boost the flexibility of datacenters. Lastly, datacenters are complex but highly automated systems. Ubiquitous monitors and controls empower datacenters’ participation in the power grid.

In general, datacenters participate in the power grid in two ways, Demand Response (DR) and Frequency Regulation (FR) \cite{Bates2014ElectricalGA}. Note that FR can be categorized as a special means of DR \cite{Wang2019FrequencyRS} as the overlap between the two is substantial. LBNL proposes potential DR resources that reside at different components of datacenters \cite{Mare2010DemandRA}, including supporting equipment such as backup generators and UPS, programmable power managements such as DVFS and power capping, server consolidation by virtualization, and load (re)scheduling \& migration. Note that some of these methods are not particularly environment-friendly, for example, the using the backup generators \cite{ashrae2015thermal}. But in terms of effectiveness, even methods like power capping are relatively coarse-grained, it can potentially enable datacenters to tuck about 25\% more servers into the same amount of space \cite{samson2009power}. In respect of FR, its resources have a huge overlap with that of the DR. The main difference between the two is that FR requires a way faster timescale, usually at the second level. Moreover, both the supply side and the demand side need to constantly work together to automatically lubricate the small frequency fluctuations in the power grid. By the virtue of fast charging and discharging, UPS becomes a good candidate in FR; the ``UPS-as-a-Reserve” \cite{ups_as_reserve} is one pilot project of such. Similarly, because of their quick response, power management such as DVFS \cite{Chen2013RealtimePC,Chen2014TheDC} and dummy workload \cite{Wang2019FrequencyRS} are also well-suited resources for FR. 

FR enables datacenters to take part in the real-time market as well as the ancillary market. Nevertheless, DR is far more common than FR because, for datacenters, FR is too fast and risky to visualize and control \cite{Bates2014ElectricalGA}. In this work, we focus on datacenters’ participation in the day-ahead and the intraday markets. Thus, DR is the focal point of this study.

%% file: chapters/2-bg/2.5_modelling.tex
\section{Energy Modelling for Datacenters} \label{sec:modelling}

Methods for modelling energy consumption in datacenters can be broadly classified into two categories: measuring energy usage at the hardware level \cite{Chang2002EnergyDrivenSS,Flinn1999PowerScopeAT,Contreras2005PowerPF} and modelling energy consumption using simulation \cite{Bellosa2000TheBO,Brooks2000WattchAF,Cignetti2000EnergyET,Gurumurthi2002UsingCM,Rawson2004MEMPOWERAS,Vijaykrishnan2000EnergydrivenIH,Wang2002OrionAP,Zedlewski2003ModelingHP}. Hardware measurement has a huge advantage over the whole-system simulators in terms of speed, so much so that the latter can hardly be used for long-term applications and very large dataset without applying reduction configurations \cite{Economou2006FullSystemPA}. However, online energy monitoring and metrics collection systems are of rarity incurring substantial costs in practice. To our knowledge, the work from Fan et al. \cite{Fan2007PowerPF} is the first to use theoretical energy models for very large-scale datacenter power provisioning on live, production workloads. 

\begin{figure}[!t]
  \begin{center}
    \includegraphics[width=0.8\textwidth]{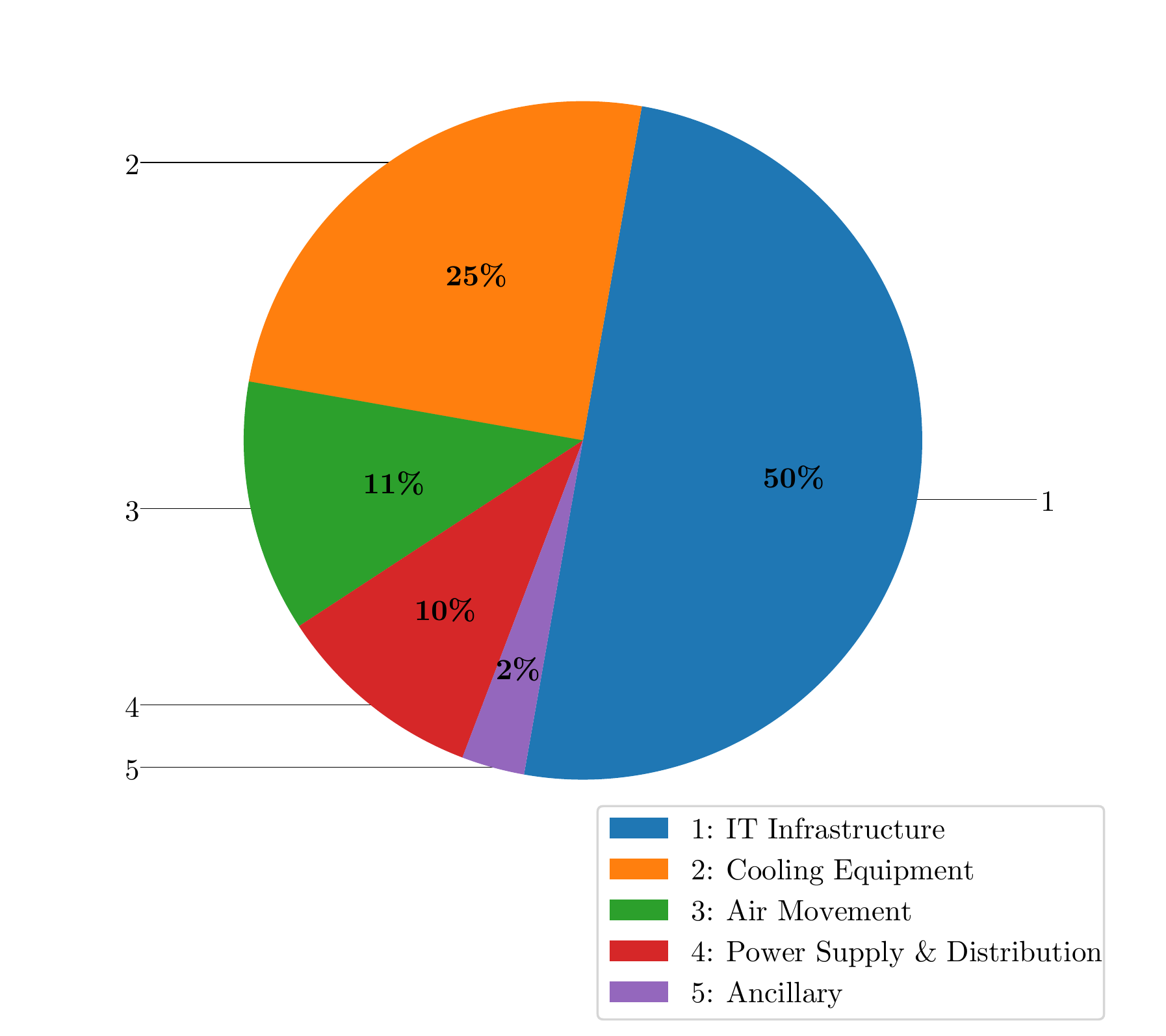}
  \end{center}
  \caption[Approximate distribution of energy usage in a datacenter] {Approximate distribution of energy usage in a datacenter with PUE value of 2.0 (data source: \cite{Barroso2008TheDA}).}
  \label{fig:distro_total}
\end{figure}

\begin{figure}[!t]
    \begin{adjustbox}{width=1.2\textwidth,center=\textwidth}

    \includegraphics[width=\textwidth]{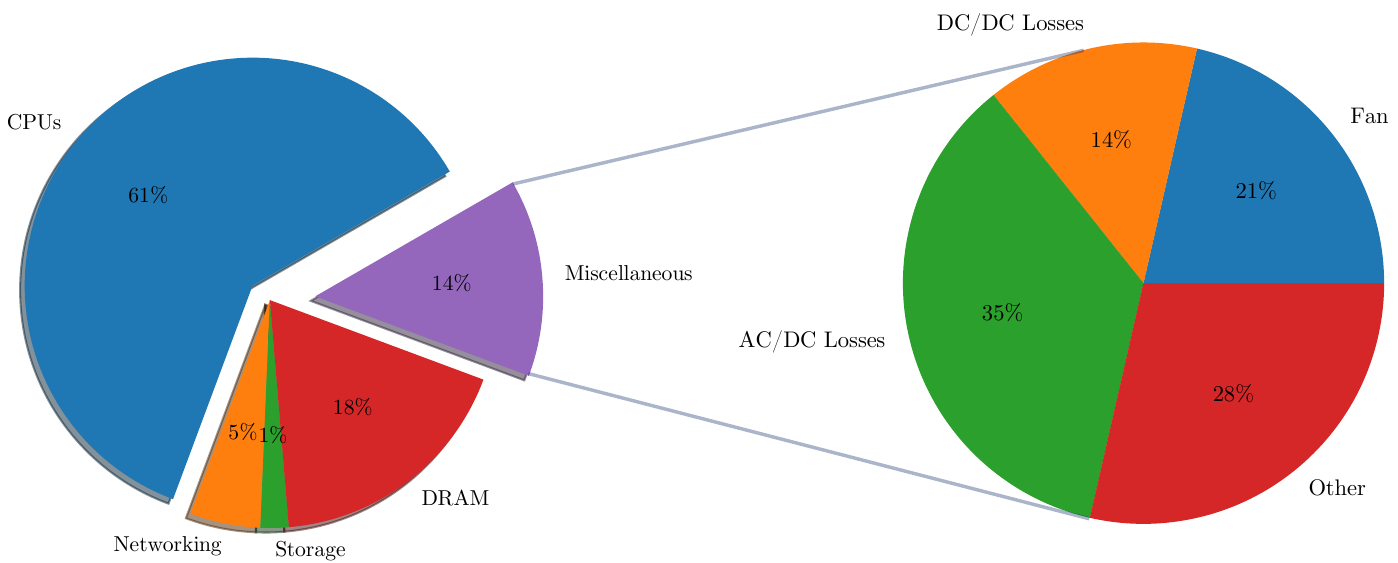}
    \end{adjustbox}
  \caption[Approximate distribution of energy losses in IT equipments] {Approximate distribution of energy losses in IT equipments (data source: \cite{Barroso2008TheDA,intel_powerarch}).}
  \label{fig:distro_IT}
\end{figure}

Similar to the approach from \citet{Economou2006FullSystemPA}, instead of considering the full system with fine-grained models, \citeauthor{Fan2007PowerPF} take a helicopter view using metrics such as CPU utilization and I/O activity at a coarser granularity to estimate energy consumption. The authors focus on critical power without taking into account energy losses and cooling power consumption at the datacenter level. \citet{Economou2006FullSystemPA} suggest that the power consumption of non-IT infrastructure is by no means negligible because they amount to 30-50\% of the total energy consumption. Figure \ref{fig:distro_total} gives an approximate distribution of energy usage in a datacenter with a PUE value of 2.0 \cite{Barroso2008TheDA}, showing that about 50\% of the total energy consumption is used by IT infrastructure. Figure \ref{fig:distro_IT} offers an overview of the rough proportions of the energy losses in terms of various IT components operating at peak power \cite{intel_powerarch,Barroso2018TheDA}.

Although non-IT components account for around half of the total power usage, Fan et al. \cite{Fan2007PowerPF} argue that in modern datacenters, critical power can accurately capture the energy consumption for other non-IT facilities in the sense that the dynamic power of the non-IT components can be modelled as a static tax proportional to the critical power in modern datacenters. This estimation can be further facilitated by proper calibrations. The authors proposed a power model (Equation \ref{eq:base_1} and \ref{eq:base_2}) where $u$ is the CPU utilization that is obtained via the operating systems, which averages across all CPUs, and $r$, which was set to 1.4 in the original experience, is the calibration parameter chosen to optimize the mean square error (MSE). This model, albeit simple, has been widely adopted in simulating energy consumption in datacenters.

In this model, the CPU utilization is employed as the single indicator for estimating the power consumption of the critical load. Fan et al. demonstrated that CPU utilization is able to serve as an extremely accurate signal providing veracious results for thousands of machines. Consequently, measurements for additional loads, e.g., hardware performance meters, are complementary yet unnecessary. Furthermore, a sub-/super-linear correlation between power and the CPU frequency is assumed in this power model. Studies \cite{Beloglazov2012EnergyawareRA,Raghavendra2008NoS,Kusic2008PowerAP,Verma2008pMapperPA,Gandhi2009OptimalPA} have illustrated that validity of this assumption lies in the fact that DVFS is only applied to the CPU, not to other components, and the number of states defined in DVFS for the CPU is finite. 

\begin{subnumcases}{ }
    \mathbb{P}(u) &= $P^{\text{ idle}}+\left(P^{\text{ max}}-P^{\text{ idle}}\right) {u}$ \label{eq:base_1}\\
    \mathbb{P}_\text{MSE}(u) &= $P^{\text{ idle}}+\left(P^{\text{ max}}-P^{\text{ idle}}\right)\left(2 {u}-{u}^{r}\right)$  \label{eq:base_2}
\end{subnumcases}

In addition to this, whilst advocating the usage of actual peak power as opposed to the nameplate power (because the latter is so conservative that 80-130\% more machines can be deployed in the case that the nameplate power is targeted), Fan et al. \cite{Fan2007PowerPF} emphasize the importance of energy management overall power range, \eg DVFS. To support such optimization, the authors set a predefined threshold for the CPU utilization and half the power provisioning of the CPU (whilst leaving others unchanged) whenever the utilization drops below the threshold. Despite such DVFS strategy used in their simulation is rather simple, in a cautious measurement, this strategy results in about 30\% reduction in peak power and is capable of saving around 23\% system energy. In addition, servers are idle for a large fraction of the running time, which consumes about 50-60\% of the actual peak power, whilst they are barely completely inactive, i.e., sleep or standby. Therefore, in terms of large-scale datacenters consisting of thousands of machines, different sleep states (C-states) have a limited impact on the energy consumption at the datacenter level.

\enlargethispage{\baselineskip}

\input{resources/tables/tab_simulators}

Furthermore, over the last decade, many datacenter simulators \cite{Tian2015ATF,Calheiros2011CloudSimAT,Gurout2013EnergyawareSW,tighe2012dcsim,Mrkus2017CostAwareIE,Kecskemeti2017ModellingLP,Kecskemeti2015DISSECTCFAS,Gupta2011GDCSimAT,Boru2013EnergyefficientDR, Toosi2011ResourcePP,Kliazovich2010GreenCloudAP,Casta2013Emc2AF,Nez2012iCanCloudAF,Heinrich2017PredictingTE,Silva2019AccuratelySE,Casanova2014VersatileSA,Kurowski2013DCwormsA,Malik2017CloudNetSimAG,Louis2015CloudSimDiskES} have been developed. Some of them embed energy models for different components, including both IT and non-IT infrastructure. Such advancements foster the development of energy-saving algorithms and energy-aware decision-making acknowledging the benefits brought upon by simulation, such as simplicity, reproducibility, and cost-friendliness. Table \ref{tab:simulators} gives an overview of nice state-of-the-art datacenter simulators in which the power consumption has been captured to various degrees. As shown in the table, we are the first to achieve whole power system modelling in datacenter simulation.

%% file: resources/tables/tab_simulators.tex
\begin{table}[H]
\centering

\begin{adjustbox}{width=1.2\textwidth,center=\textwidth}
\begin{tabular}{l|c|c|c|c|c|c}
\toprule
\multirow{2}{*}{\textbf{Simulator}} & \multicolumn{2}{c|}{\textbf{IT Infrasturcture}} & \multicolumn{2}{c|}{\textbf{Primary Support}} & \multirow{2}{*}{\textbf{Secondary Support}}  & \multirow{2}{*}{\textbf{Energy Market Integration}} \\
\cline{2-5}
& \textbf{Critical Load} & \textbf{DVFS} & \textbf{UPS} & \textbf{PDU} & & \\
\midrule


\textit{{DCSim}} \cite{Gupta2011GDCSimAT}  & \yes & \no & \no & \no & \no & \no \\

\textit{{CloudSim}} \cite{Calheiros2011CloudSimAT} & \yes & \yes & \no & \no & \no & \no \\

\textit{{GDCSim}} \cite{Gupta2011GDCSimAT} & \yes & \no & \no & \no & \yes$^+$ & \no \\

\textit{{CloudSched}} \cite{Tian2015ATF} & \yes & \yes & \no & \no & \no & \no \\

\textit{{DISSECT-CF}} \cite{Mrkus2017CostAwareIE,Kecskemeti2017ModellingLP,Kecskemeti2015DISSECTCFAS} & \yes & \yes & \no & \no & \yes$^+$ & \no \\

\textit{{GreenCloud}} \cite{Boru2013EnergyefficientDR, Toosi2011ResourcePP,Kliazovich2010GreenCloudAP} & \yes$^+$ & \yes$^+$ & \no & \no & \no & \no \\

\textit{{iCanCloud/E-mc$^2$}} \cite{Casta2013Emc2AF,Nez2012iCanCloudAF} & \yes$^+$ & \yes$^+$ & \no & \no & \no & \no \\

\textit{SimGrid} \cite{Casanova2014VersatileSA,Heinrich2017PredictingTE,Silva2019AccuratelySE} & \yes$^+$ & \yes$^+$ & \no & \no & \no & \no \\

\hline

\textit{\textbf{OpenDC}} \cite{iosup2017opendc,mastenbroek2021opendc} & \yes & \yes$^+$ & \yes$^+$ & \yes$^+$ & \yes & \yes$^+$ \\

\bottomrule
\end{tabular}
\end{adjustbox}

\caption[Overview of the nine surveyed datacenter simulators] {Overview of the nine surveyed datacenter simulators, where the \textcolor{darkgreen}{\cmark} symbol means that the corresponding energy model is available, the \textcolor{red}{\xmark} symbol means that it is unavailable, and $^+$ represents advanced support.}
\label{tab:simulators}

\end{table}

%% file: chapters/3-degsin/3_design.tex
\chapter{Energy Modelling \& Management} \label{cha:degsin}

In this chapter, we reason and describe our design of the energy modelling and management system. We begin by introducing the development pipeline closely followed in this work (\S\ref{sec:pipeline}). Then, in Section \ref{sec:req_eng}, we detail the requirement engineering process carried out in this work. Finally, we present the architecture of the entire system as well as its subsystems in Section \ref{sec:architecture}.

\section{Development Pipeline} \label{sec:pipeline}

\begin{figure}[!ht]
    \centering
    \includegraphics[width=\textwidth]{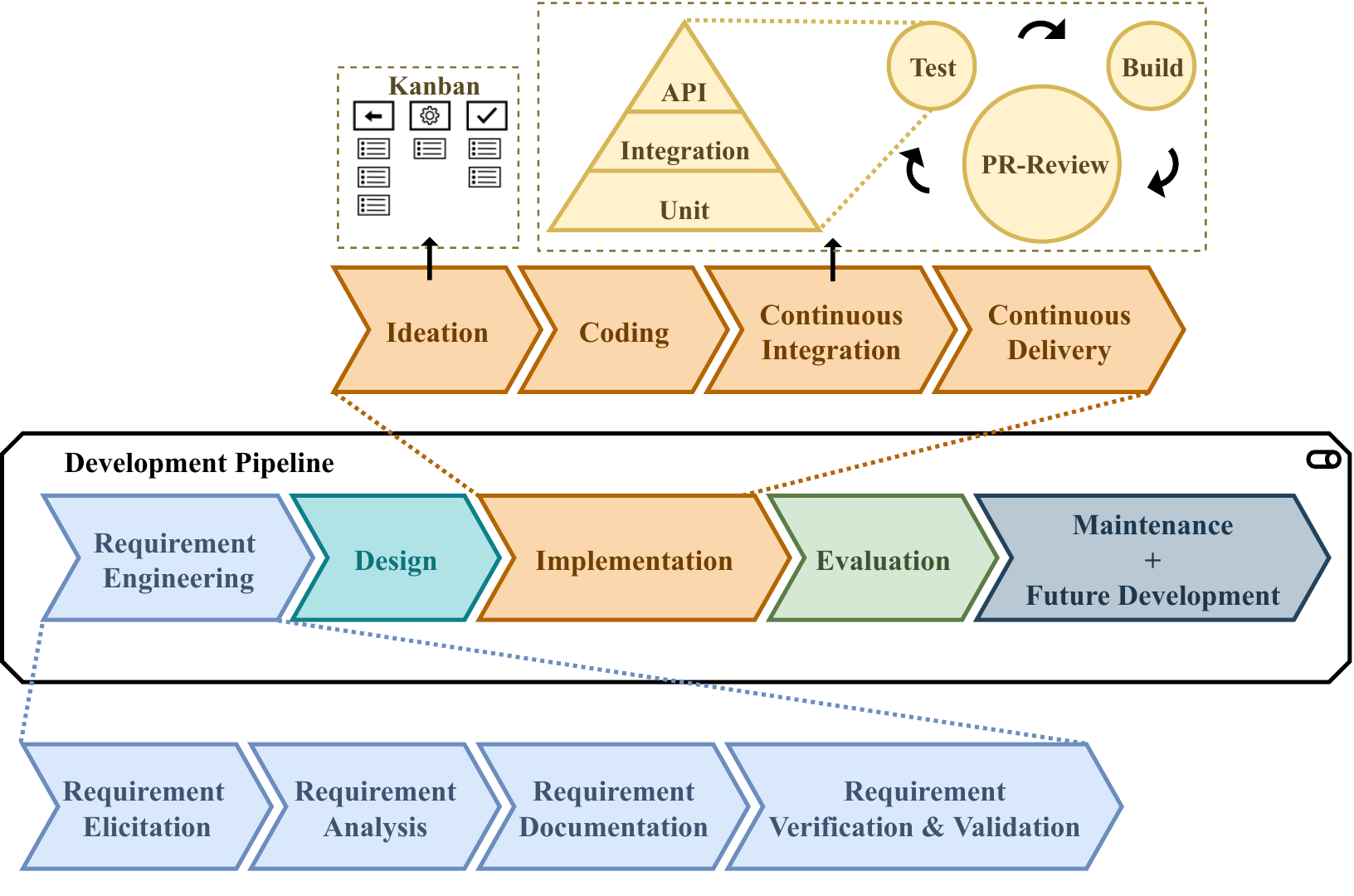}
    \caption[Development pipeline] {Development pipeline.}
    \label{fig:pipeline}
\end{figure}

To achieve a coherent design and reproducible results, we rigorously follow a development pipeline shown in Figure \ref{fig:pipeline} from the outset, adhering to the code of conduct in the AtLarge research group. 

The first stage in this pipeline is requirement engineering, which will be described in detail in the next section (\S\ref{sec:req_eng}). Based on the requirements elicited and documented in the first stage, we then draft our design, outlining more elaborate specifications. During the implementation stage, we put up and organize ideas into different categories using Kanban. These ideas are constantly being reviewed and updated together with the initial design. Furthermore, instead of developing the system in a separate branch, we practice Continuous Integration and Delivery (CI/CD), which is commonly employed in modern software development. During CI/CD, we develop the system in small, measurable steps, every integration of which is reviewed by at least one senior engineer from our research group. Testing, building, and deployment are all instrumented and automated so that the system is a useable instrument at all times. Next, we conduct regular evaluations of the system, which provides frequent feedback for previous stages. At last, any changes to the codebase are carefully documented, facilitating maintenance and future development.   

\input{chapters/3-degsin/3.2_req_eng}

\newpage

\input{chapters/3-degsin/3.3_architecture}

%% file: chapters/3-degsin/3.2_req_eng.tex
\section{Requirement Engineering} \label{sec:req_eng}

In this section, we detail the first stage of the development pipeline -- Requirement Engineering, which is further modulated into several steps. We first begin with eliciting requirements from a variety of facets in the views of stakeholders using the Six Thinking Hats technique \cite{de2017six} (\ref{sec:req_elicit}). Next, we conduct analysis on the elicited requirements through use case modelling in Section \ref{sec:req_analysis}. In Section \ref{sec:req_doc}, we then formally document the analysed requirements, categorized into function requirements (FRs) and non-function requirements (NFRs). Last, but certainly not least, we verify and validate the documented requirements in Section \ref{sec:req_veva} with experts in both datacenters and the energy market.

\subsection{Requirement Elicitation} \label{sec:req_elicit}
In this section, we first identify and classify potential stakeholders, and then, employ the Six Thinking Hats approach \cite{de2017six} to explore system requirements. 

\subsubsection{Stakeholders.} The importance of stakeholders is paramount, we, therefore, start with identifying some potential stakeholders in Table \ref{tab:id_stakeholders}.

\input{resources/tables/tab_id_stakeholders}

Having identified potential stakeholders, we classify them into two categories in Table \ref{tab:cat_stakeholders}: active stakeholders of whom the successfulness of this research is the predominant interest, and passive stakeholders who more care about whether the outcome of this work abides by their agreements and rules.

\input{resources/tables/tab_cat_stakeholders}

\subsubsection{Change of Perspective.} 

Now, we employ the Six Thinking Hats method \cite{de2017six} seeking system requirements. With the purpose of creating new ideas, the colour sequence of hats applied in the following elicitation is: blue, white, red, green, yellow, black, and blue.

\paragraph*{Blue Hat.} The role of the blue hat is to manage and control the elicitation. Thus, we commence the process by explaining the objectives. We aim to find possible requirements from various perspectives so that the design of this work caters for the needs of identified stakeholders.

\paragraph*{White Hat.} Firstly, we summarize facts (\textbf{F}s) regarding the power grid and datacenter energy management from previous sections (\S\ref{sec:prob}, \S\ref{sec:power_grid}).

\begin{enumerate} [label=\textbf{F\arabic*}]
    \item \label{f1} By virtue of smart grid functions, activities from energy prosumers are able to have bidirectional influence in regulating and balancing the power grid.
    \item \label{f2} The capability of the power grid is hitting some limits due to the massive introduction of renewable energy sources, which features intermittency and stochasticity caused by a range of sporadic environmental factors.
    \item \label{f3} Datacenters are well suited for regulating and balancing the power grid because of their unique characteristics such as large capacity, high flexibility and redundancy, etc.
    \item \label{f4} Datacenters nowadays rarely actively participate in the energy market, providing little regulation capacity such as DR to the power grid.
\end{enumerate}

\paragraph*{Red Hat.} According to the perspective of the red hat, we next express emotions and feelings of the initiatives. Referring back to Chapter \ref{cha:intro}, the environmental crisis is taking its toll at a worrying pace and consequently, the calling for a lower carbon footprint is at an all-time high. Datacenters play an essential role in our day-to-day life, whilst with their ever-increasing energy consumption, they are at the front line of curbing environmental issues. Also, the operational cost resulted from energy consumption incurs heavy bills on datacenters. People from both the society and the computing industry are longing for further exploration regarding datacenters’ participation in the energy market.

\paragraph*{Green Hat.} Chief amongst the spirits of the green hat is creativity, promoting innovative solutions and new ideas. With this initiative in mind, we propose the following potential solutions (\textbf{S}s) to the aforementioned challenges.

\begin{enumerate} [label=\textbf{S\arabic*}]
    \item \label{s1} We can model the whole datacenter power system, providing datacenters managers and operators the trending of energy consumption by means of datacenter simulation.
    \item \label{s2} We can estimate the energy costs in different markets given the (predicted) workload, supporting the active participation of datacenters in the energy market.
    \item \label{s3} We can take advantage of available resources such as price forecasts produced by machine learning methods, facilitating the decision-making in the energy market.
    \item \label{s4} We can simulate fine-grained energy management configurations such as the DVFS technique, assisting datacenters in optimizing operational scheduling and further enabling proactive demand response.
\end{enumerate}

\paragraph*{Yellow Hat.} Optimism and positivity of the proposed solutions are brought to the table by the yellow hat, which puts emphasis on the possible advantages and opportunities. Firstly, regarding \ref{s1}, we believe knowing is power -- providing the status of the datacenters’ energy consumption visually and quantitatively will empower datacenter managers and operators. For \ref{s2}, diversifying the types of markets in which datacenters can participate can give market operators more options when certain constraints occur in one market, such as the available bids in the intraday market cannot meet the current energy needs. Also, \ref{s2} can leave datacenters with more leeway under situations where, for example, the workload is not as delay-tolerant. \ref{s3} can prompt enabling cooperation between datacenters and consulting firms, creating a level playing field in the energy market for datacenters to participate. Lastly, \ref{s4} can further add values to the synergetic collaborations in \ref{s3} by employing proactive, fine-grained optimizations through low-level energy configurations.

\paragraph*{Black Hat.} To provide early criticism and judgment before the requirement verification \& validation stage (\S\ref{sec:req_veva}), the black hat plays devil’s advocate, bringing up potential difficulties and risks that the proposed solutions could face. With regard to \ref{s1}, when it comes to datacenter energy modelling, there is a myriad of factors that have substantial impacts on the energy consumption of datacenters, for example, the heterogeneity of supporting equipment and machines, the various topologies, etc. Different from detailed emulations, simulations incur less overhead but can hardly capture many of those factors. Concerning \ref{s2}, load prediction is by no means trivial and can have a great effect on the participation of datacenters in different markets. Thirdly, the performance of the machine learning methods from the consulting firm can be both beneficial, if the inference is accurate, and detrimental if it is not. Thus, it is hard for datacenter managers and operators to embrace \ref{s3} without bounded estimations of the performance impact. In addition, fine-grained energy management techniques such as DVFS may be too low-level that could barely make a sizable influence. Further, such optimization could introduce additional overhead to the system. Lastly, experts in the energy industry and the IT industry may well have drastically different familiarity with datacenter simulation tools. Consequently, the usability of the developed tool could significantly vary amongst different user groups. 

\paragraph*{Blue Hat.} To conclude, we first reiterate four major observations (\ref{f1} -- \ref{f4}) in the stage of the white hat. Then, we add on to the facts the emotional elements entailed by the perspective of the red hat. Next, via the green hat, we propose four possible solutions (\ref{s1} -- \ref{s4}) to overcome the challenges. From the perspective of the yellow hat and that of the black hat, we reflect on the potential pros and cons of the proposed solutions respectively. As a result, we recognize the potential benefits the proposed solutions can bring to the stakeholders whilst, in the meantime, be aware of the possible side effects and risks that come along with the positive facet. 

\begin{figure}[H]
    \centering
    \caption[Use case diagram of the system]{Use case diagram of the system.}
    \includegraphics[width=\textwidth]{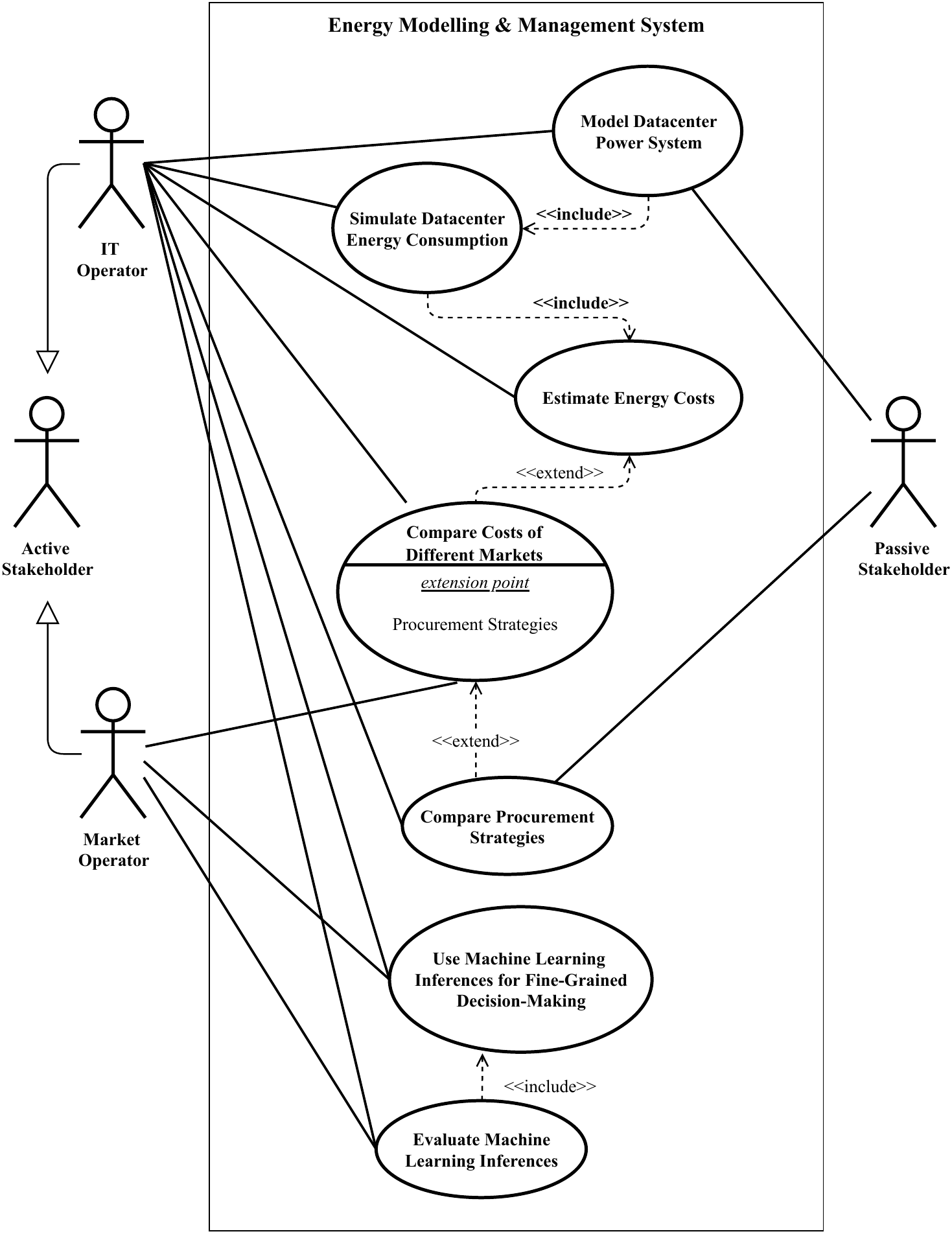}
    \label{fig:use_case}
\end{figure}

\subsection{Requirement Analysis} \label{sec:req_analysis}

In this section, we take results from the previous stage, requirement elicitation, and put them into the context of the energy modelling and management system that we are trying to build. Figure \ref{fig:use_case} is a generalized use case diagram, showing the interactions and communications between the system and various actors. The IT operator represents the active stakeholders in the IT industry summarized in Table \ref{tab:id_stakeholders} and \ref{tab:cat_stakeholders}. As a primary actor, the IT operator should be able to initiate all four major use cases in the system, namely, modelling datacenter power system, simulating datacenter energy consumption, estimating energy costs, comparing costs of different energy markets, making decisions based on ML inferences, and evaluating the ML model performance. Another primary actor is the energy market player, including the active stakeholders in the energy industry (Table \ref{tab:id_stakeholders} and \ref{tab:cat_stakeholders}), and can initiate the last three use cases. The associations between the first three use cases are “include” since the system should only be able to simulate the power consumption given a specific power system model and to provide cost estimations based on simulation results. Similarly, the association between the last two is also “include” as the performance of the ML inferences should always be measured. In the case those procurement strategies are provided, the system should allow users to compare them based on a load prediction in order to mitigate the concern raised by the {Black Hat} regarding \ref{s2}. Unlike active stakeholders, passive stakeholders, such as legislators and cloud architects, serve as the secondary actor who is primarily concerned with, for example, the validity of the power system model and the legitimacy of the energy procurement strategies.

\subsection{Requirement Documentation} \label{sec:req_doc}

In this section, we extract the marrow of the previous analysis and formally document the requirements into functional requirements (FRs) and non-functional requirements (NFRs).

Based on the use case analysis in the previous section, we summarize \textbf{FR}s as follows:

\noindent(we use “the system” to indicate the energy modelling and management system)

\begin{enumerate} [label=\textbf{FR\arabic*}]
    \item \label{fr1} The system should enable users to model the power system of datacenters.
    \item \label{fr2} The system will simulate the datacenter energy consumption given a power system model when hosting user workloads.
    \item \label{fr3} The system will estimate the energy cost of datacenter operations based on the results of the workload simulation.
    \item \label{fr4} The system shall demonstrate to users the potential costs in participated markets. 
    \item \label{fr5} The system shall provide users with the ability to compare various procurement strategies.
    \item \label{fr6} The system will empower fine-grained decision-makings for users according to ML inferences.
\end{enumerate}

\noindent Now, referring back to the concerns raised by the Black Hat, we further specify a list of NFRs below:

\begin{enumerate} [label=\textbf{NFR\arabic*}]
    \item \label{nfr1} When modelling the power system of datacenters, the system should incorporate heterogeneous topologies, hardware components, etc.
    \item \label{nfr2} When comparing various load-forecast-based procurement strategies, the system should be able to assess their impact to an extensive extent.
    \item \label{nfr3} When employing inferences from machine learning methods, the system should conduct bounded evaluations on the effectiveness of using the predicted prices.
    \item \label{nfr4} When developing the user interface of the tool, the system should respect inclusive design, \eg experts from the energy and the IT industry should be able to run the system with less than four steps/click and with minimum prerequisite knowledge.
\end{enumerate}

\begin{samepage}
\subsection{Requirement Verification \& Validation} \label{sec:req_veva}

Last, but certainly not least, we request experts in both the IT and the energy industry to examine our requirements and corresponding analyses, seeking any flaws and conflicts. Table \ref{tab:experts} lists the experts involved in the verification and validation process. 

\begin{table}[H]
\centering
\begin{adjustbox}{width=0.9\textwidth}
    \begin{tabular}{c|l|l}
    \toprule
        \textbf{Industry} & \textbf{Role} & \textbf{Size of Infrastructure Dealt w/}\\
        \midrule
        Energy            & Machine Learning Engineer & Large\\
        IT                & Engineer and Scientist    & Medium\\
        IT                & Researcher                & Medium \& Small \\
    \bottomrule
    \end{tabular}
\end{adjustbox}
\caption[Experts involved in requirement verification and validation]{Experts involved in requirement verification and validation.}
\label{tab:experts}
\end{table}

\end{samepage}

%% file: resources/tables/tab_id_stakeholders.tex
\begin{table}[H]
\centering
\begin{adjustbox}{width=\textwidth}

    \begin{tabular}{l|l}
        \toprule
        \textbf{Industry} & \textbf{Stakeholders}\\
        \midrule
        IT                & \makecell[l]{datacenter managers, datacenter operators, datacenter technicians, \\ cloud architects, cloud tenants} \\
        \hline
        Energy            & \makecell[l]{consulting firms, energy market operators, power grid system operators,\\ renewable energy suppliers} \\
        \hline
        Others            & legislators, end-users of cloud services    \\
        \bottomrule
    \end{tabular}
\end{adjustbox}
\caption[Potential stakeholders]{Potential stakeholders.}
\label{tab:id_stakeholders}
\end{table}

%% file: resources/tables/tab_cat_stakeholders.tex
\begin{table}[H]
\centering
\begin{adjustbox}{width=\textwidth}

    \begin{tabular}{l|l}
    \toprule
        \textbf{Category}   & \textbf{Stakeholders}\\
        \midrule
        Active stakeholder  & \makecell[l]{datacenter managers, datacenter operators, consulting firms, \\ energy market operators, renewable energy suppliers} \\
        \hline
        Passive stakeholder & \makecell[l]{datacenter technicians, cloud architects, power grid system operators, \\ legislators, end-users of cloud services}\\
    \bottomrule
    \end{tabular}
\end{adjustbox}
\caption[Stakeholder classification]{Stakeholder classification.}
\label{tab:cat_stakeholders}
\end{table}

%% file: chapters/3-degsin/3.3_architecture.tex
\begin{figure} [H]
\centering
    \caption[Overview of the architecture of the entire system] {Overview of the architecture of the entire system.}
    \begin{adjustbox}{width=1.2\textwidth,center=\textwidth}
    \includegraphics[width=\textwidth]{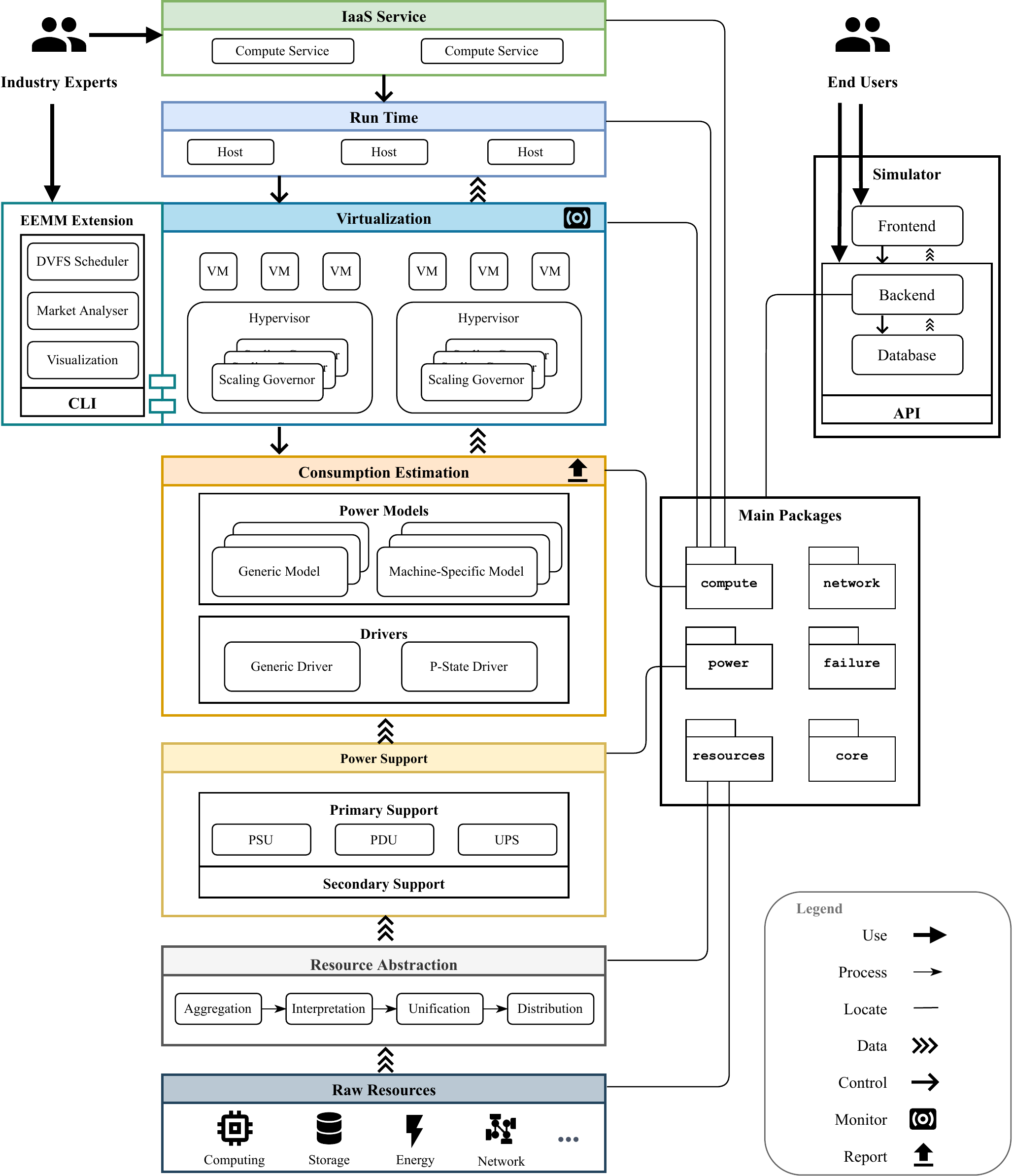}
    \end{adjustbox}
    \label{fig:overall}
\end{figure}

\section{System Architecture} \label{sec:architecture}

In this chapter, we elaborate on the design of the system architecture, teasing it down layer by layer in the following sections. Firstly, we give an overview of the architecture of the entire system in Section \ref{sec:sys_overview}. Then, in Section \ref{sec:power_support}, we present the blueprint of the power support subsystem. Finally, the market extension EEMM is described in the last section of this chapter (\S\ref{fig:eemm}). 

\subsection{System Overview} \label{sec:sys_overview}

Referring back to Chapter {\ref{cha:intro}}, datacenter simulators have been widely adopted in both academia and the industry. OpenDC is one of such simulation tools, which is easy-to-use with a wide range of state-of-the-art features, for example, capacity planning \cite{Andreadis2021CapelinDC}, modelling serverless computing and hosting ML workloads \cite{mastenbroek2021opendc}. This work develops advanced power models and a unified energy resource chain integrated into the infrastructure of the simulator. 

Figure \ref{fig:overall} dissects the system in a layered manner. Most of the high-level functionalities are readily available to end-users through the frontend UI and the code API. IT professionals and experts in the energy industry can access more functionalities with fine-grained control over the simulator by directly invoking the infrastructure as a service (IaaS) interface. Such services are provided by one of six packages, the \texttt{compute} package, which resides in the backend of the simulator. Via the IaaS interface, users are able to specify detailed simulation setup. The simulator supports heterogeneous hardware types, topologies, and scenario portfolios. Moreover, this work further enables users to configure the energy modelling and management system in a flexible way.

Experts can enable platform-dependent energy models and energy-saving configurations, such as a set of frequency scaling governors in the Linux kernel and various power estimation models (either generic or machine-specific), as well as different scaling drivers with customizable P-states. On top of that, users can configure the topology of the power support subsystem, which locates in the \texttt{power} package and is underpinned by a unified resource chain penetrated throughout the system stack. This power support subsystem will be detailed in Section \ref{sec:power_support}.

The bottom layer of the system represents the physical resources assigned by users. Instead of using their raw form directly, we abstract all resources into one single representation with a common unit. This process is realized by first aggregating the resources from all sources (\eg computing, storage, energy and so on), next interpreting different resources to unify them into a common unit, and finally, distributing the resources to support subsystems. Such an abstraction happens in the \texttt{resources} package, forming the backbone of the power modelling and management system. In this way, various physical resources with rather different units can be utilized throughout the system with one common view.

Furthermore, working towards accomplishing native support for functionalities related to the energy market, this work provides an extension of the energy modelling and management system (EEMM), which will be introduced in Section \ref{sec:eemm}.

\begin{figure} [H]
\centering
    \caption[Architecture of the modelled power system of datacenters] {Architecture of a quintessential AC power system of datacenters modelled in this work.}
    \begin{adjustbox}{width=1.3\textwidth,center=\textwidth}
    \includegraphics[width=1.3\textwidth]{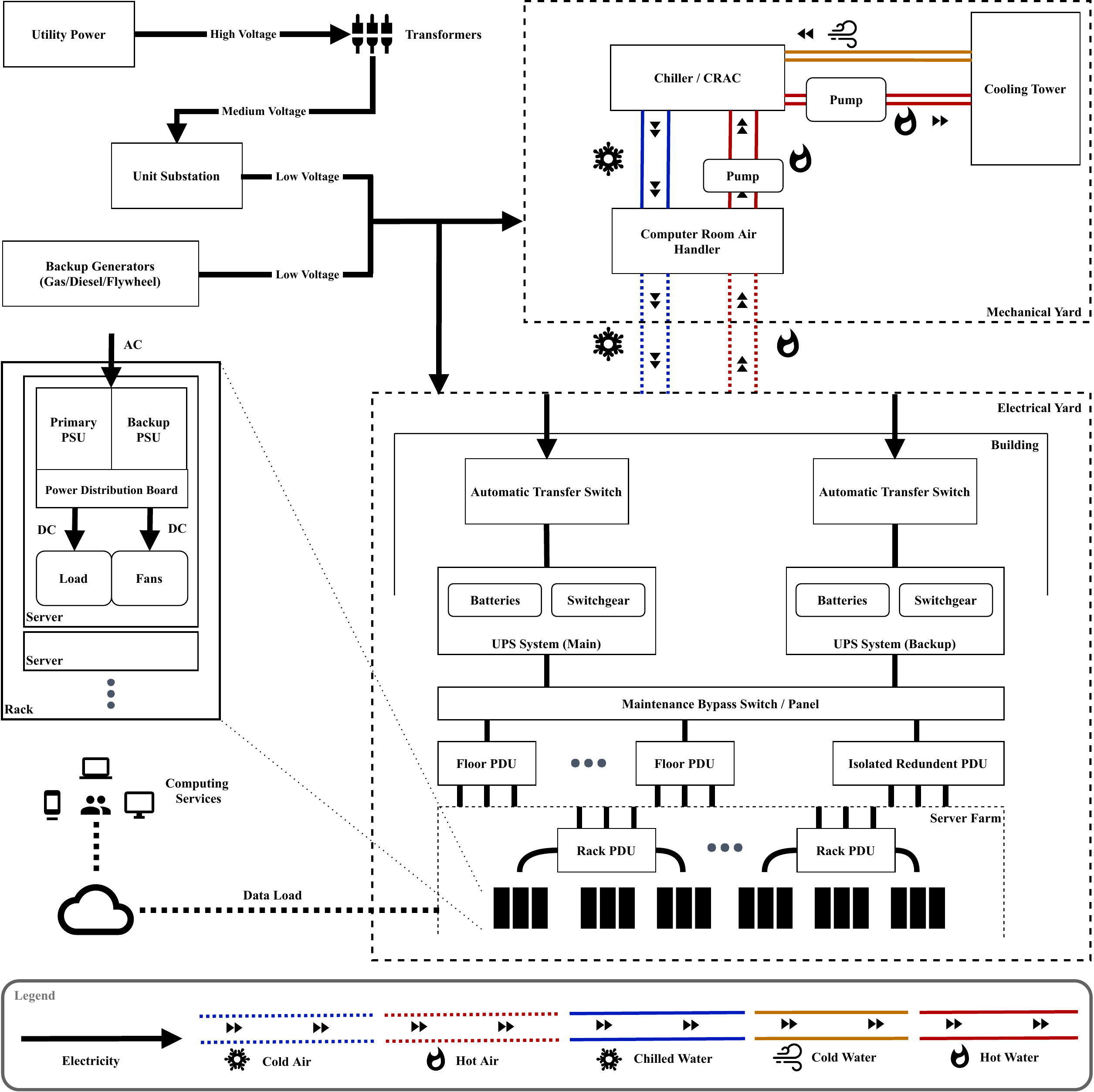}
    \end{adjustbox}
    \label{fig:dc_power}
\end{figure}

\subsection{Power Support Subsystem} \label{sec:power_support}

To achieve the \textbf{FR}s documented in Section \ref{sec:req_doc}, the very first premise is reliable power modelling and flexible energy management, underpinning \ref{fr1} and \ref{nfr1}. This section describes the core design of the power support subsystem.

First and foremost, according to the study presented in Section \ref{sec:dc_grid}, we construct the skeleton of a quintessential datacenter power system, upon which the power modelling subsystem is built. Figure \ref{fig:dc_power} demonstrates the AC power system modelled in this work. First, the electricity comes into the datacenter from the utility power station. Then, the energy is distributed to two parts, the primary support in the electrical yard and the secondary support in the mechanical yard. Inside the datacenter building, the energy passes through the transfer switches and reaches the UPS systems. The UPS systems will next distribute the energy to several floor PDUs. These floor PDUs deal with a higher voltage than the rack PDUs which are directly connected with the servers. Inside of the server, the PSU transform AC power to DC and power both the computing load and the internal cooling system. Note that almost all input electricity will ultimately be exiled in the form of heat.
We recognize that such an architecture may well vary from datacenters to datacenters. With this caution in mind, we design the power support subsystem that is highly customizable.

\begin{figure} [!t]
\centering
 \begin{adjustbox}{width=1\textwidth,center=\textwidth}
    \includegraphics[width=1\textwidth]{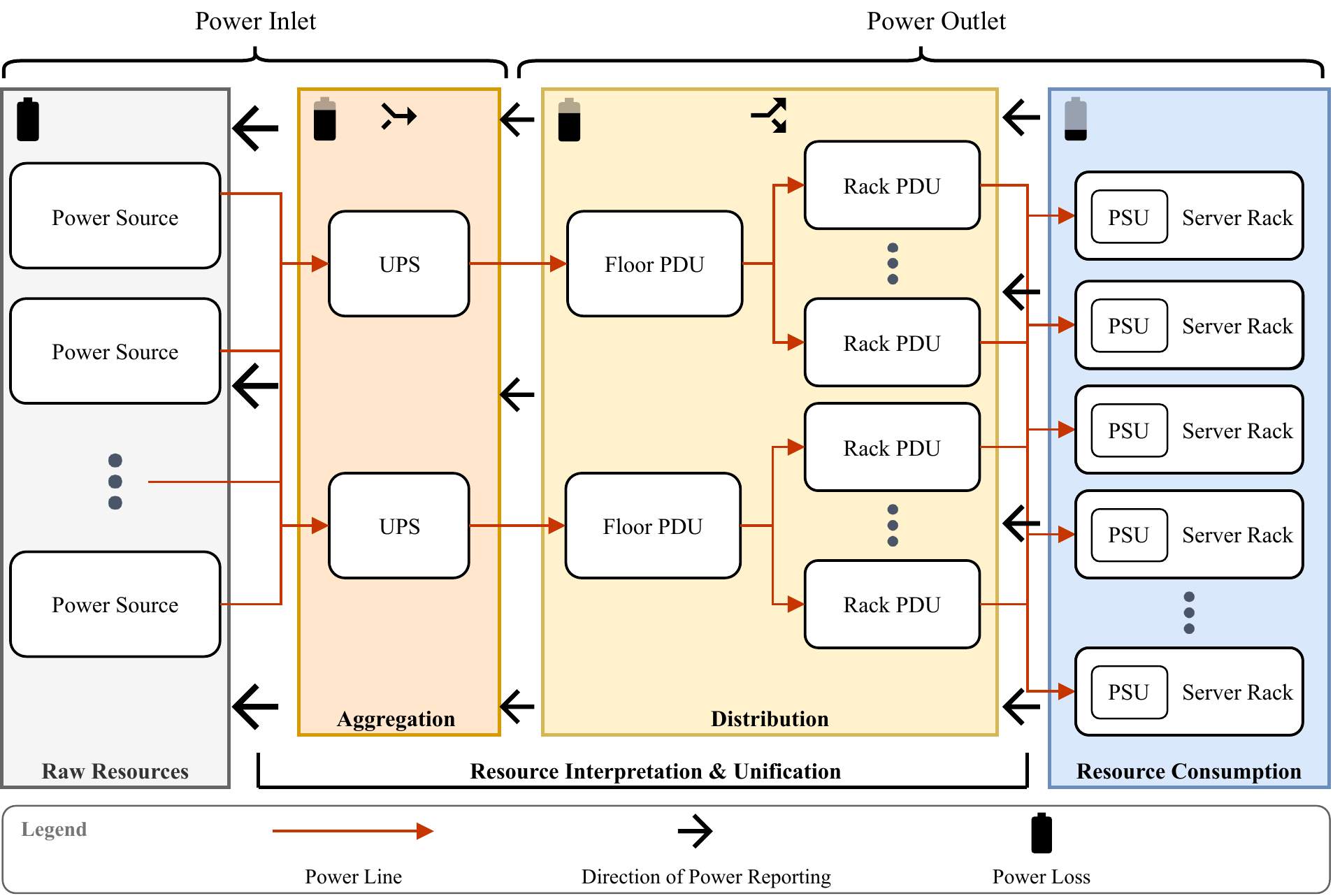}
    \end{adjustbox}
    \caption[Architecture of the power support subsystem] {Architecture of the power support subsystem.}
    \label{fig:power_support}
\end{figure}

Figure \ref{fig:power_support} shows the design of the power support subsystem. Components of the system form four layers, namely, the raw resource layer, the aggregation layer, the distribution layer, and the resource consumption layer. To support various power system topologies, the number of components at each layer can be customized by users. The middle two layers are responsible for the core abstraction of resource interpretation and unification elaborated in Section \ref{sec:sys_overview}. Moreover, from a higher point of view, components in the first two layers serve as power inlets that generate and aggregate energy, and the last two serve as power outlets that distribute and consume energy. 

From left to right, the power chain therein starts at various power sources in the first layer and ends at the IT infrastructure, the server farm. On the contrary, the reporting of energy use follows exactly the opposite direction. This architecture facilitates the energy monitoring and data collection mechanisms one abstraction above, directly addressing \ref{fr2}. The implementation of the power support subsystem described in Chapter \ref{cha:impl} closely follows this design.

\subsection{Market Extension} \label{sec:eemm}

In this section, we turn our attention to addressing the market-related requirements, \ie \ref{fr3} to \ref{fr6}, and \ref{nfr2} to \ref{nfr4}.

\begin{figure} [H]
\centering
    \caption[Architecture of EEMM] {Architecture of the market extension of the energy modelling and management system.}
    \begin{adjustbox}{width=0.7\textwidth,center=\textwidth}
    \includegraphics[width=\textwidth]{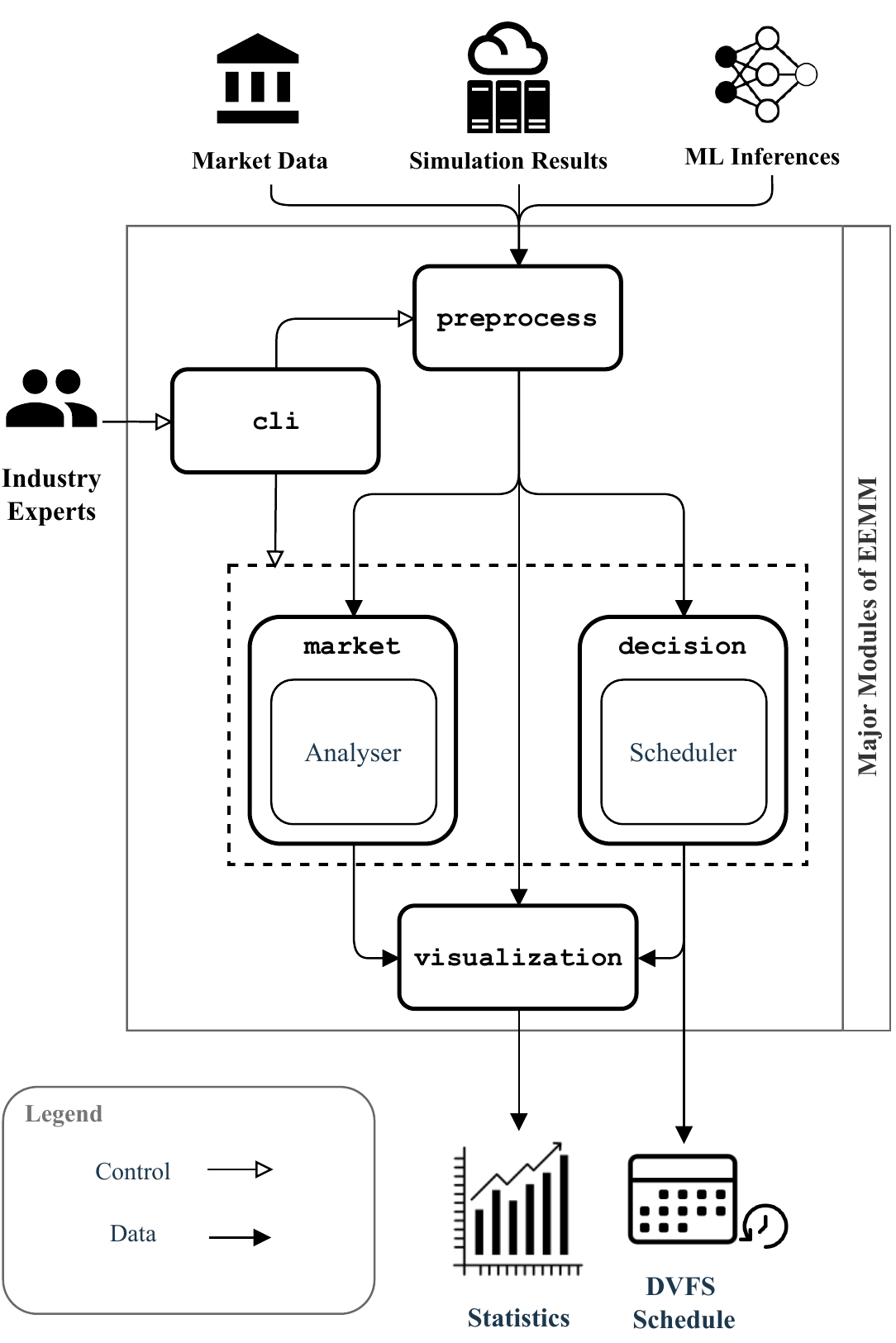}
    \end{adjustbox}
    \label{fig:eemm}
\end{figure}

Upon the basis of the simulation infrastructure presented in the previous two sections, we build an extension of the energy modelling and management system (EEMM). Figure \ref{fig:eemm} demonstrates the architecture of the extension, in which five core modules work in concert.

Industry experts can directly interact with the extension via the command-line interface provided by the \texttt{cli} modules in a Unix system. By providing the extension with the simulation results and market data, including the energy prices in various markets, the \texttt{preprocess} module will convert data and feed them into the \texttt{market} module. Note that this step does not require any extra manual data processing from users other than the standard market data from corresponding official websites \footnote{\url{https://transparency.entsoe.eu/dashboard/show}}\footnote{\url{https://www.tennet.org/english/operational_management/system_data_relating_processing/settlement_prices/index.aspx}}, which aim at fulfilling \ref{nfr4}.

The \texttt{market} module estimates the energy consumption and provides insights of the energy costs in different markets, addressing \ref{fr3} and \ref{fr4}. Moreover, it contains an analyser that inspects the market data and simulates procurement strategies. Next, these analyses will be passed to the \texttt{visualization} modules that visualizes the statistics for users to compare different strategies (\ref{fr5}).

Furthermore, to address \ref{fr6}, the \texttt{decision} module is developed to process the ML inferences as market signals. It embodies a proactive DVFS scheduler, supporting users to make fine-grained decisions in response to the market signals.

%% file: chapters/4-impl/4_implementation.tex
\chapter{Implementation} \label{cha:impl}

In this chapter, we take a deep dive into the implementation of the system. Firstly, the development of the energy modelling and management system is introduced in Section \ref{sec:sys_impl}. Then, we move on to the implementation of the market extension EEMM in Section \ref{sec:eemm_impl}. 

\input{chapters/4-impl/4.1_sys_impl}

\input{chapters/4-impl/4.2_eemm_impl}

\newpage

%% file: chapters/4-impl/4.1_sys_impl.tex
\begin{subnumcases} {}
    \texttt{ConstantPowerModel} & $\mathbb{P}(s) = s$ \label{eq:odc_const}\\
    \texttt{LinearPowerModel} & $\mathbb{P}(u)=P_{\text {idle}}+\left(P_{\max }-P_{\text {idle}}\right) u$ \label{eq:odc_linear}\\
    \texttt{SquarePowerModel} & $\mathbb{P}(u)=P_{\text {idle }}+\left(P_{\max }-P_{\text {idle }}\right) u^{2}$ \label{eq:odc_squre}\\
    \texttt{CubicPowerModel} & $\mathbb{P}(u)=P_{\text {idle }}+\left(P_{\max }-P_{\text {idle }}\right) u^{3}$ \label{eq:odc_cubic}\\
    \texttt{SqrtPowerModel} & $\mathbb{P}(u)=P_{\text {idle }}+\left(P_{\text {max }}-P_{\text {idle }}\right) \sqrt{u}$ \label{eq:odc_sqrt}\\
    \texttt{MsePowerModel} & $\mathbb{P}(u) = P_{ {idle }}+\left(P_{ {max }}-P_{ {idle }}\right)\left(2 \mathrm{u}-\mathrm{u}^{r}\right)$ \label{eq:odc_mse} \\
    \texttt{InterpolationPowerModel} & $\mathbb{P}(u)=P\left(u_{1}\right)+\left(P\left(u_{2}\right)-P\left(u_{1}\right)\right) \frac{u-u_{1}}{u_{2}-u_{1}}$
    \label{eq:odc_intp} \\
    \texttt{AsymptoticPowerModel} & $\mathbb{P}(u)=P_{\text {idle}}+\frac{\left(P_{\text {max}}-P_{\text {idle}}\right)}{2}\left(1+u-e^{-\frac{u}{a}}\right)$ \label{eq:odc_asym} \\
    (\texttt{AsymptoticPowerModelDvfs}) & $\mathbb{P}(u)=P_{\text {idle}}+\frac{\left(P_{\text {max}}-P_{\text {idle}}\right)}{2}\left(1+u^{3}-e^{-\frac{u^3}{a}}\right)$ \label{eq:odc_asymdvfs}
\end{subnumcases}

\section{Energy Modelling \& Management System} \label{sec:sys_impl}

\begin{figure}[H]
    \centering
    \caption[UML diagram of the energy modelling and management system]{Simplified UML diagram of the energy modelling and management system.}
     \begin{adjustbox}{width=1.25\textwidth,center=\textwidth}
    \includegraphics[width=\textwidth]{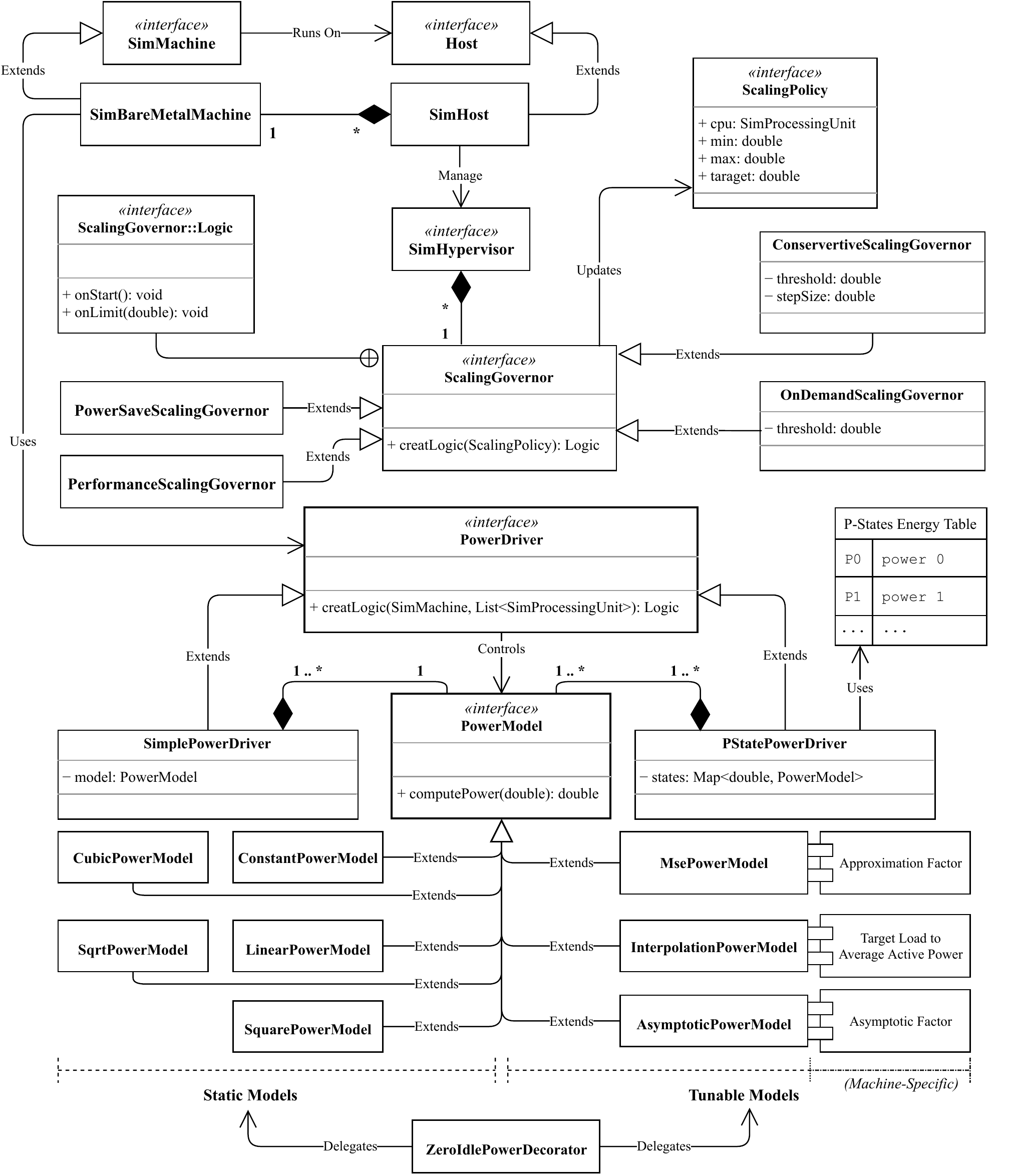}
    \end{adjustbox}
    \label{fig:uml_power_model}
\end{figure}

Closely following the design described in Section \ref{sec:architecture}, we implement and integrate the energy modelling and management system into the OpenDC simulator. Figure \ref{fig:uml_power_model} is a simplified UML diagram of the system. 

Power models occupy the lower part, of the diagram. These models include both the generic models and the machine-specific ones that can be further tuned towards a particular computing platform. These models align with their mathematical formulation shown in Equation \ref{eq:odc_const} to \ref{eq:odc_asymdvfs}. Note that these models are implemented during the Honours research of the author (please see the report \cite{hongyu_hp} for more details) but are reorganized and improved during this work. These power models are controlled by two power drivers, and the \texttt{PStatePowerDriver} utilizes the P-states provided by users to adjust power estimation in discrete steps.  

Algorithm \ref{algo:pstate_driver} illustrates the detailed mechanism of this driver. As described in Section \ref{sec:multicore}, we take package-level decisions in line \ref{line:decision_start} to \ref{line:decision_end}. CPU saturation is measured by the metric specified by Equation \ref{eq:cpu_usage} (\S\ref{sec:cpu_metrics}) in line \ref{line:util}.

\input{resources/algos/algo_drive}

Moreover, four generic scaling governors are developed, corresponding to the four governors found in the Linux kernel (\S\ref{sec:dvfs}), namely, the \texttt{powersave}, 
the \texttt{performance}, the \texttt{conservative}, and the \texttt{ondemand} governors. Note that the \texttt{schedutil} governor is not included since it is scheduler-dependent, and the governor \texttt{userspace} is also excluded as it is nothing but a static governor that can be easily realized by either the \texttt{PowerSaveScalingGovernor} or the \texttt{PerformanceScalingGovernor}.

Algorithm \ref{algo:ondemand} shows the scaling mechanism realized in the \texttt{OnDemandScalingGovernor}, and that of the \texttt{ConservativeScalingGovernor} is presented in Algorithm \ref{algo:conservative}. 

\input{resources/algos/algo_ondemand}

\input{resources/algos/algo_conservative}

\newpage

The \texttt{ScalingPolicy} is implemented to model the \texttt{struct cpufreq\_policy} object in the Linux kernel, as introduced in Section \ref{sec:dvfs}. It contains essential information concerning the associated CPU, as well as the target frequency modulated by the scaling governors. This object is initialized by the \texttt{SimHypervisor}s which operates scaling governors thereof. To achieve a flexible initialization for various types of CPUs as an effort to address \ref{nfr1}, the inner interface \texttt{ScalingGovernor::Logic} is introduced, serving as a factory to assist this initialization process. \texttt{SimHypervisor}s are managed at runtime by \texttt{SimHost}s that run on a \texttt{SimBareMetalMachine}. A bare-metal machine is able to invoke \texttt{PowerDiver}s at the physical layer for energy estimation.

\begin{figure}[H]
    \centering
    \caption[UML diagram of the power support subsystem]{Simplified UML diagram of the power support subsystem.}
     \begin{adjustbox}{width=1.3\textwidth,center=\textwidth}
    \includegraphics[width=\textwidth]{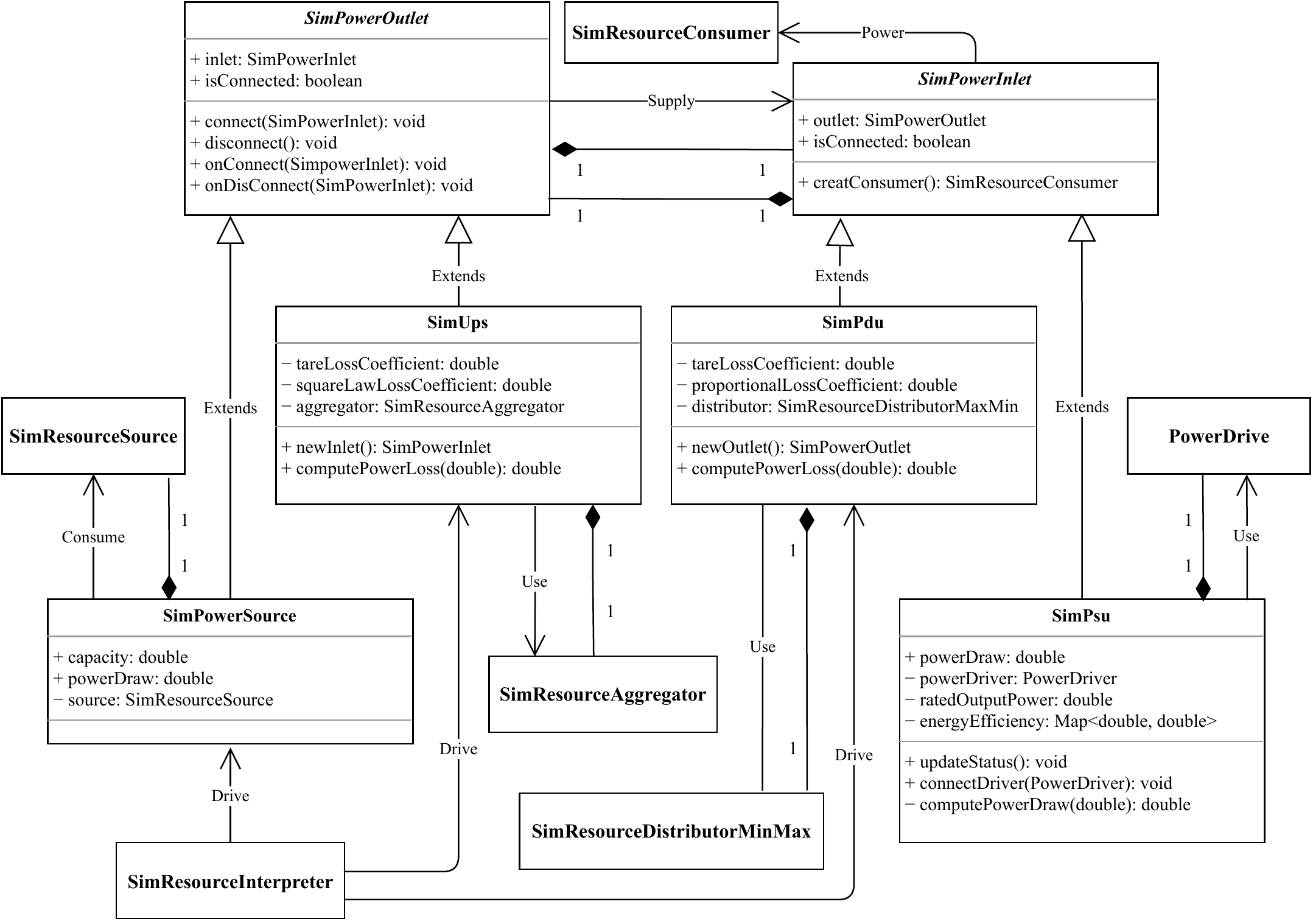}
    \end{adjustbox}
    \label{fig:uml_power_support}
\end{figure}

With regard to the implementation of the power support subsystem, its development adheres to the UML diagram shown in Figure \ref{fig:uml_power_support}. Referring back to the design described in Section \ref{sec:power_support}, the \texttt{SimPowerOutlet} and the \texttt{SimPowerInlet} classes represent the two top categories in Figure \ref{fig:power_support}, inherited by various component classes at different layers. A \texttt{SimPsu} that resides in a bare-metal-machine directly interacts with the aforementioned \texttt{PowerDriver}s. Furthermore, power distribution is achieved by a simple max-min fair-sharing policy, illustrated by Algorithm \ref{algo:maxmin}. Note that the system has been integrated with OpenDC, which is fully containerized and can be easily run via \texttt{docker} (\ref{nfr4}).

\input{resources/algos/algo_maxmin}


%% file: resources/algos/algo_drive.tex
\begin{algorithm}[H]
\KwInput{
    \newline 
    A list of scaling context $C$ associated with the CPUs of $machine$; \newline
    A map $M$ that contains a set of P-states as keys and power models at each respective frequency level as values;}
\KwOutput{\newline The next P-state;}

\KwData{A power consumption table $T$ for each corresponding P-state in $M$;}

\tcc{Initializing $M$ according to $T$.}
  
\ForEach {$state,powerLevel \in T.entries$} {
	Choose a power model $m$ \tcp*{Model types can vary from level to level.}
	Instantiate $m$ based on $powerLevel$\;
	$M.put(state,\ m)$\;
}
  
\tcc{Updating the current P-state.}
\textbf{initially} $target \gets 0$\;
\textbf{initially} $currentUsage \gets 0$\;
\textbf{initially} $isUpdated \gets false$\;
\If{\underline{$\exists c \in C: c\ $\upshape has been updated by governors}} 
{
	$isUpdated \gets true$\;
	\ForEach {$c \in C$} { \label{line:decision_start} 
		$target \gets \max(c.requested,\ target)$ \tcp*{Take package-level decisions.}
		$currentUsage \gets c.cpu.speed + currentUsage$\; \label{line:decision_end}
	}
}

\tcc{Locating the appropriate P-state.}
\textbf{initially} $pstate \gets 0$\;
\If{\underline {isUpdated}} 
{
    \tcp{The following can be simplified via a tree map instead of a normal hash map.}
	$upperBound \gets \max(M.getKeys())$\;
	$target \gets \min(upperBound, target)$\;
	$levels \gets$ sort$(T.getKeys())$ \tcp*{Sort in ascending order.}
	\ForEach {$l \in levels$} {
		\If{\underline {$level \geq target$}} {
			$pstate \gets level$\;
			\textbf{break}\;
		}
	}
} 
\Else
{
	$pstate \gets$ P-state from the last update\;
	\ForEach {$cpu \in machine.cpus$} {
		$currentUsage \gets cpu.speed + currentUsage$\;
	}
}



\tcc{Computing the instant power consumption.}
$model \gets M.get(pstate)$\;
$u \gets \cfrac{currentUsage} {pstate * C.size}\ $\; \label{line:util}
$model.computePower(u)$\;

\Return $pstate$\;

\caption[P-state scaling algorithm] {P-state scaling algorithm applied in the \texttt{PStatePowerDriver}.}
\label{algo:pstate_driver}
\end{algorithm}

%% file: resources/algos/algo_ondemand.tex
\begin{algorithm}[H]
\SetAlgoLined
\KwInput{
    \newline
    The load threshold $t$ with default value being $0.8$;
    \newline
    The scaling policy $P$;
}
\KwOutput{\newline \texttt{void};}

\If{\underline {the associated CPU has not been initialized}} 
{
    $P.target \gets P.min$\;
    \Return \texttt{void}\;
}

\If{\underline{$l > t$}}
{
    \tcc{Proportional scaling.\footnote{\url{https://github.com/torvalds/linux/blob/master/drivers/cpufreq/cpufreq_ondemand.c}}}
    $P.target \gets P.min + l * \cfrac{P.max - P.min}{100}$\;
}
\Else
{
    $P.target \gets P.min$\;
}

\Return \texttt{void}\;

\cprotect \caption[Scaling algorithm in the \texttt{OnDemandScalingGovernor}] {Scaling algorithm in the \texttt{OnDemandScalingGovernor}.}
\label{algo:ondemand}
\end{algorithm}

%% file: resources/algos/algo_conservative.tex
\begin{algorithm}[H]
\SetAlgoLined
\KwInput{
    \newline
    The load threshold $t$ with default value being $0.8$;
    \newline
    The step size $s$ with defaut value being $-1.0$;
    \newline
    The previous CPU load $o$;
    \newline
    The current CPU load $l$;
    \newline
    The scaling policy $P$;
}
\KwOutput{\newline \texttt{void};}

\If{\underline {the associated CPU has not been initialized}} 
{
    $P.target \gets P.min$\;
    \Return \texttt{void}\;
}

\If{\underline{$s \le 0$}}
{
    $s \gets P.max * 0.05$ \tcp*{Set the step size to the default value in the Linux kernel.\footnote{\url{https://github.com/torvalds/linux/blob/master/drivers/cpufreq/cpufreq_conservative.c}}}
}
\Else
{
    $s \gets \min(s,\ P.max)$\;
}

\textbf{initially}, $step \gets -1$\;
\If{\underline{$l > t$}}
{
    \tcc{Checking for load increase.}
    \uIf{\underline{l > o}}
    {
        $step \gets +s$\;
    }
    \uElseIf{\underline{l < o}}
    {
        $step \gets -s$\;
    }
    \Else
    {
        $step \gets 0.0$\;
    }
}

$P.target \gets \min(\max((P.target + step),\ P.min),\ P.max)$\;

$o \gets l$\;
\Return \texttt{void}\;

\cprotect \caption[Scaling algorithm in the \texttt{ConservativeScalingGovernor}] {Scaling algorithm in the \texttt{ConservativeScalingGovernor}.}
\label{algo:conservative}
\end{algorithm}

%% file: resources/algos/algo_maxmin.tex
\begin{algorithm}[H]
\SetAlgoLined
\KwInput{
    \newline
    A list of demands $D$;
    \newline
    The capacity $c$ of the power sources;
}
\KwOutput{\newline The remaining capacity of the power source;}

$D \gets \text{sort}(D)$\tcp*{Sort $D$ in ascending order.}
$n \gets \text{sizeof}(D)$\;

\textbf{initially}, $ration \gets \cfrac{c}{n}$\ \;
\textbf{initially}, allotments $A \gets$ an empty list of size $n$\;

\tcc{Initial assignment.}
\For{$i \gets 0$ \KwUntil $n$ \KwBy $1$}
{
    $A[\ i\ ] \gets r$\;
}

\tcc{Handling overloaded demands.}
\For{$i \gets 0$ \KwUntil $n$ \KwBy $1$}
{
    $ration \gets A[\ i\ ]$\;
    \textbf{initially}, $share \gets 0.0$\;
    
    \If{\underline{$i < (n - 1)$\ and\ $ration \ge D[\ i\ ]$}}
    {
        $share \gets \cfrac{ration - D[\ i\ ]}{n - (i+1)}$\tcp*{Fair-share over-supplied capacity.}
        \For{$j \gets i$ \KwUntil $n$ \KwBy $1$}
        {
            $A[\ j\ ] \gets A[\ j\ ] + share$\;
        }
        $A[\ i\ ] \gets D[\ i\ ]$\;
    }
    \Else
    {
        $A[\ i\ ] \gets \min(ration,\ D[\ i\ ])$\;
    }
    
}

\Return $c - \sum_i^{n-1} A[\ i\ ]$\;

\cprotect \caption[Max-min fair-sharing power distribution algorithm] {Max-min fair-sharing power distribution algorithm.}
\label{algo:maxmin}
\end{algorithm}

%% file: chapters/4-impl/4.2_eemm_impl.tex
\section{Market Extension} \label{sec:eemm_impl}

he core of the extension EEMM is the DVFS scheduler. Referring back to Section \ref{sec:dvfs}, the major drawback of enabling DVFS is the prolonged execution time. By the virtue of the dominating quadratic relationship shown in Equation \ref{eq:ptof}, the benefit of the power saving brought by DVFS will ultimately outweigh the linearly scaled overhead, \ie the longer duration. That been said, such an overhead incurred by using DVFS is by no means negligible and should be further mitigated. Therefore, we take this factor into account when developing the scheduler.

When hosting traces of the virtual machines (VMs), our simulation currently does \textit{not} prolong the execution time of the jobs in the case that the capacity of the host is reduced. Instead, the overhead of DVFS is reflected by the over-commission of the CPUs. To elaborate on this further, when the frequency of the CPUs are restrained to a much lower level, the total capacities of all VMs may well (temporarily) exceed the maximum capacity of their host. We do not scale the VMs downwards to accommodate the reduced hosting capacity but capture the over-commissioned CPU cycles instead. Similar circumstances could also occur when new VMs are spawned, which can lead to exceeding the total capacity of the host. In contrast, as part of the capacity of the host is released (\eg some VMs have finished the execution or the frequency of the CPU is increased), the scaling driver in our simulator does not stick to the proposed CPU frequencies by the scaling governor and level down the resource processing speed accordingly. 

Thus, the DVFS scheduler should juggle both the energy consumption affected by switching the scaling governor and the overhead of using DVFS, the CPU over-commission. To this end, we introduce the concept of \textit{damping factor}, which is inspired by the Google PageRank algorithm \cite{rogers2002google} with different implications and implementations. The damping factor, in this case, is not a probability value but a threshold that restrains the increase in the level of CPU over-commission. The lower the damping factor, the more frequent the scheduler ameliorates the constraint on the CPU frequency by switching to a more performant scaling governor. Algorithm \ref{algo:schedule} illustrates such a scheduling strategy implemented in the EEMM. We will further elaborate on the two decision points (line \ref{line:1st_decision} and \ref{line:2nd_decision}) in Chapter \ref{cha:eval}.

In an effort to address \ref{nfr4}, we develop the market extension as a Python library, which can be easily installed by using the following command in a Unix system.

\begin{lstlisting}
$ pip install opendc-eemm
\end{lstlisting}

\input{resources/algos/algo_scheduler}

%% file: resources/algos/algo_scheduler.tex
\begin{algorithm}[H]
\SetAlgoLined
\KwInput{
    \newline
    The spot price $p^S$ of the next ISP;
    \newline
    The forecasted imbalance shortage price $p^F$ of the next ISP;
    \newline
    A list of availabel scaling governors $G$;
    \newline
    The damping factor $d$;
    \newline
    The current damping factor counter $c$;
}
\KwOutput{\newline The next scaling governor to use;}

\KwData{A serise of datacenter traces $T$ up util now;}

\textbf{initially} $prev \gets$ get previous over-commission level from $T$\;

\textbf{initially} $curr \gets$ compute current over-commission level from $T$\; 

\textbf{initially} $governor \gets null$\;

\tcc{Gauging the over-commission status.}
\If{\underline {$curr > prev$}} 
{
    $c++$ \tcp*{Record the increase of the current over-commission level.}
}
\Else
{
    $c--$\;
}
$prev \gets curr$\;

\tcc{The first decision point.}
\If{\underline {$p^F \le 0$}} 
{   \label{line:1st_decision}
    $governor \gets G.performance$\;
}
\Else
{
    \tcc{The second decision point.}
    \If{\underline {$p^F > p^S$}} 
    {   \label{line:2nd_decision}
        $governor \gets G.powersave$\; 
    }
    \Else
    {
        \If{\underline {$c \ge d$}}
        {
            $governor \gets G.conservative$\;
			$c \gets 0$ \tcp*{Relax the over-commission meter.}
        }
        \Else
        {
            $governor \gets G.ondemand$\;
        }
    }
}

\Return $governor$

\caption[DVFS scheduling algorithm] {DVFS scheduling algorithm implemented in the market extension.}
\label{algo:schedule}
\end{algorithm}

%% file: chapters/5-eval/5_evaluation.tex
\chapter{Evaluation} \label{cha:eval}

In this chapter, we elaborate on the experiments conducted to answer \textbf{RQ}s described in Section \ref{sec:rqs}, as well as to meet the \textbf{FR}s and \textbf{NFR}s listed in Section \ref{sec:req_doc}. We start with detailing the setup in Section \ref{sec:setup}. Then, in Section \ref{sec:market}, we analyse the results regarding the participation of datacenters in the energy market. Lastly, in Section \S\ref{sec:scheduling}, we evaluate the performance of the proactive DVFS scheduler powered by ML methods.

\input{chapters/5-eval/5.1_setup}

\newpage

\input{chapters/5-eval/5.2_market}

\newpage

\input{chapters/5-eval/5.3_dvfs_schedule}
\newpage

%% file: chapters/5-eval/5.1_setup.tex
\section{Experiment Setup} \label{sec:setup}

In this section, we describe the simulation model employed in the experiments, specifically, the market model (\S\ref{sec:market_model}), the specifications of machines simulated (\S\ref{sec:machine_model}), and the energy models employed for estimating the energy consumption of the datacenter (\S\ref{sec:energy_model}). 

\subsection{Market Model} \label{sec:market_model}

Firstly, referring back to Section \ref{sec:power_grid}, there are two financial markets prior to the start of the real-time delivery, the day-ahead market and the intraday market. The energy prices settled in these two markets are often referred to as the spot price. The major difference between the two is that the day-ahead market has one single settlement for all participants, whereas the intraday market consists of continuous bilateral trading carried out through the trading day and, in turn, has no common settlement. As described in Section \ref{sec:balance_grid}, the intraday market is where prosumers conduct the final adjustments to their self-dispatched quantities. Besides the varying prices, the number and types of products are highly dependent on the surplus/deficit situation of every participant. Therefore, in our experiment, we do not make any assumption in the participation of the intraday market, \ie we do not make any adjustment in the quantity of energy ordered in the day-ahead market before the start of the actual delivery period.

Secondly, in respect of the day-ahead market and the balancing market, we focus on the energy market of the Netherlands. In other words, all prices used in the following experiments are the energy prices of the Netherlands only. Similarly, although the energy trading system introduced in Section \ref{sec:power_grid} applies across the EU, there are still nuances between different countries. For this matter, we also only focus on the trading system of the Netherlands, but the experiment results do not lose their generality as these nuances are rather minute.

Lastly, datacenters normally either have long-term contracts with their utility companies or buy energy in an on-demand scheme. Since it is not feasible to obtain/disclose the energy prices of the bilateral, long-term contracts, we only focus on the on-demand scheme. Note that, in practice, datacenters generally do not have significant discounts through long-term contracts due to the lack of intermittency, but we, nevertheless, want our experiment results to be inclusive and representative. Therefore, we consider three on-demand energy prices from low to high summarized in Table \ref{tab:od_prices}.

\input{resources/tables/tab_od_prices}

\subsection{Machine Model} \label{sec:machine_model}

DVFS technology is one of the major interests of this work. Although the frequencies and voltages of P-states in different CPUs are commonly available, their corresponding consumption levels, however, need to be specially measured by either software or hardware power meters. That said, since neither developing an instrument for measuring the P-state power nor testing the accuracy of the measurements is within the scope of this work, we resort to the existing literature for this matter. 

\input{resources/tables/tab_pstates}

For the consumption levels of P-states, the latest reputable reference that we found is \cite{Gurout2013EnergyawareSW}. However, the CPU therein is old and, therefore, might not be representative in terms of its power consumption. To understand the impact of using this old machine model, we also include a recent machine with a type of CPU released this year (2021) from the SPEC benchmark \cite{Bucek2018SPECCN}. These two machines models are summarized in Table \ref{tab:machines}, and the P-states consumption levels of the old machine model are in Table \ref{tab:pstates}.

\input{resources/tables/tab_machines}

In the experiments, we adopt a set of Business Critical Workload (BCW) traces \cite{traces} from the Dutch IT service provider Solvinity, containing records monitored over a course of one month. To host the traces properly without overloading/underutilizing the hosts, it is necessary to carefully calculate the resources needed by each of the two machine models in order to set up the compute service at the IaaS layer (Figure \ref{fig:overall}). The number of hosts ($N_\text{Hosts}$) for each machine model is computed following Equation \ref{eq:hosts}, where $\Psi_d$ denotes the maximum instant CPU demand of the traces, the $\Psi_f$ represents the maximum frequency of the machine, and $c$ is the core count of the CPU package. Similarly, the number of memory units $N_\text{Units}$ populated in each machine is calculated by Equation \ref{eq:units}, where $\Psi_m$ is the maximum instant memory request of the traces, and $m$ denotes the size of the memory unit.

\begin{align}
    N_\text{Hosts} &= \left\lceil\ \left\lceil \dfrac{\Psi_d}{\Psi_f} \right\rceil\ \middle/\ c\ \right\rceil \label{eq:hosts}\\ 
    N_\text{Units} &= \left\lceil\ \left\lceil \dfrac{\Psi_m}{N_\text{Hosts}} \right\rceil\ \middle/\ m\ \right\rceil \label{eq:units}
\end{align} 

Note that, although the results are rounded up at each step, CPU over-commission can still occur since the original traces contain requests that exceed the capacity of the VMs in the first place \cite{Shen2015StatisticalCO}. Also, the computing node used in the SPEC benchmark of the new machine model contains two packages; we set the number of cores per host to the total number of logical cores, which determine the actual capacity, as opposed to the number of dies/chips.  The detailed setup for each machine model is summarized in Table \ref{tab:hosts}. 

\input{resources/tables/tab_hosts}

\subsection{Energy Model} \label{sec:energy_model}

\paragraph*{Critical Load.} According to the results of the author's Honours research \cite{hongyu_hp}, when no particular computing platform is assumed, the linear power model (Equation \ref{eq:odc_linear}) and the square root power model (Equation \ref{eq:odc_sqrt}) are able to properly bound the total energy consumption of the critical load for a specific machine. Therefore, we use these two power models for the old machine model, which are referred to as “LINEAR” and “SQRT” in the following experiments. 

In the case that platform-specific data is available, machine-specific models are preferred over generic ones. Since we have the load-to-power data of the new machine model from the SPEC benchmark, we use the interpolation power model (Equation \ref{eq:odc_intp}) for the new machine model, which is referred to as “INTERPOLATION” in the experiments. The load-to-power data of the new machine model is presented in Table \ref{tab:spec}.

\input{resources/tables/tab_spec}

With regard to the PSU, we employ the power model from another state-of-the-art datacenter simulator, iCanCloud/E-mc$^2$ \cite{Casta2013Emc2AF,Nez2012iCanCloudAF}. To model the power losses of PSU, we first compute the percentage of load ($\eta_l$) following Equation \ref{eq:icancloud_load} based on the power draw of the server ($P_\text{server}$) and then, maps it to the corresponding energy efficiency ($\eta_e$) by Equation \ref{eq:icancloud_eff}, where both the rated output power ($\tau$) and the mapping for the ($\eta_e$) are specified in manufacturers’ datasheets. In this work, the energy efficiency of the PSU adhere to the 80 Plus Titanium standard, and the rated output power $\tau$ is set to a value 870 W, which is arbitrary but can be commonly found in PSU products of datacenter servers.

\begin{equation}
    \eta_l=\frac{100}{{ \tau }} \cdot P_\text{server}
    \label{eq:icancloud_load}
\end{equation}

\begin{equation}
    \eta_e = \left\{\begin{array}{ll}
        90 \% & \text { if } 0 \leqslant \eta_l \leqslant 10 \% \\
        94 \% & \text { if } 10 < \eta_l \leqslant 20 \% \\
        96 \% & \text { if } 20 \% < \eta_l \leqslant 50 \% \\
        91 \% & \text { if } 50 \% \leqslant \eta_l < 100 \%
    \end{array}\right.
    \label{eq:icancloud_eff}
\end{equation}

\noindent Having computed the energy efficiency, we then accumulate the energy consumption of the PSU ($E_\text{PSU}$) following Equation \ref{eq:icancloud_psu}.

\begin{equation}
    E_\text{PSU}=\int_{t_{0}}^{t_{\text {PSU }}} \frac{\left( P_\text{server} \cdot 100\right)}{\eta_e}-P_\text{server} d t
    \label{eq:icancloud_psu}
\end{equation}

\paragraph*{Primary Support.} To estimate the energy consumption of the primary support, specifically, the UPS systems and the PDUs, we employ the power model proposed by \citeauthor{Rasmussen2007ElectricalEM} \cite{Rasmussen2007ElectricalEM}, which has been widely adopted in estimating the power consumption of the UPS and PDU over the years. \citeauthor{Rasmussen2007ElectricalEM} suggests that it is insufficient to model the consumption of components of the power support system by using only the single parameter, the nameplate loss coefficient, specified in the datasheets from the manufactures. Instead, their energy consumption should be captured by two values, the tare loss coefficient ($\pi$) that is independent of the load, and a polynomial loss coefficient that is load-dependent. In the case of PDU, the load-dependent part of consumption is captured by the proportional loss coefficient ($\alpha$), and that of the UPD is captured by the square-law loss coefficient ($\beta$). The coefficients are summarized in Table \ref{tab:coefficient}, based on the work from \citeauthor{Rasmussen2007ElectricalEM}.

\input{resources/tables/tab_coefficients}

\begin{equation}
    \begin{cases}
    \alpha = \pi_\text{\ UPS} - \lambda_\text{\ UPS} \\
    \\
    P^\text{\ tare}_\text{\ UPS} = \lambda_\text{\ UPS} \cdot P^\text{\ rated}_\text{\ UPS} \\
    \\
    P^\text{\ loss}_\text{\ UPS} = P^\text{\ tare}_\text{\ UPS} + \alpha \cdot (\sum^{N_\text{\ PDU}}_{i} P^\text{\ in}_{\text{\ PDU}_i})
    \end{cases}
    \label{eq:ups}
\end{equation}

\begin{equation}
    \begin{cases}
    \beta = \pi_\text{\ PDU} - \lambda_\text{\ PDU} \\
    \\
    P^\text{\ tare}_\text{\ PDU} = \lambda_\text{\ PDU} \cdot P^\text{\ rated}_\text{\ PDU} \\
    \\
    P^\text{\ loss}_\text{\ PDU} = P^\text{\ tare}_\text{\ PDU} + \beta \cdot (\sum^{N_\text{\ server}}_{i} P^\text{\ in}_{\text{\ server}_i})^2
    \end{cases}
    \label{eq:pdu}
\end{equation}

We compute the energy consumption of the UPS and PDU using Equations \ref{eq:ups} and \ref{eq:pdu}, where $P^\text{\ in}$ denotes the inlet power, $P^\text{\ tare}$ denotes the tare power loss, $P^\text{\ loss}$ denotes the total power loss, $P^\text{\ rated}$ denotes the nameplate power, $N_\text{\ server}$ denotes the number of active servers, and $N_\text{\ PDU}$ denotes the number of attached PDUs.

\paragraph*{Secondary Support.} Without presuming the consumption rate of every single piece of equipment, such as the cooling tower, the CRAC, the backup generator, etc. (Figure \ref{fig:dc_power}), which can differ greatly from datacenter to datacenter, we make use of the PUE value to estimate the energy consumption of the entire datacenter as a whole. We compute the power consumption of the secondary support ($P^\text{2nd}$) using Equation \ref{eq:2nd}, based upon the power draw of the servers ($P_{\text{\ server}_i}$), the UPS systems ($P_{\text{\ server}_i}$), and the PDUs ($P_{\text{\ UPS}_i}$). 

\begin{equation}
    P^\text{ 2nd} = \text{PUE} \cdot \sum^{N_\text{\ server}}_{i} P_{\text{\ server}_i} - \left(\sum^{N_\text{\ PDU}}_{i} P_{\text{\ PDU}_i} + \sum^{N_\text{\ UPS}}_{i} P_{\text{\ UPS}_i} \right)
    \label{eq:2nd}
\end{equation}

%% file: resources/tables/tab_od_prices.tex
\begin{table}[H]
\centering
\begin{adjustbox}{width=0.9\textwidth}

    \begin{tabular}{c|c|l}
    \toprule
        \textbf{Price Level}   & \textbf{Price [€/MWh]} & \textbf{Source}\\
        \midrule
        Low  &  38.0 & NieuweStroom B.V. (2021, average) \cite{low_od_price}\\
        Medium & 56.5 & PricewaterhouseCoopers (2017, average) \cite{med_od_price}\\
        High & 80.4 & Essent N.V. (2021, fixed) \cite{high_od_price}\\
    \bottomrule
    \end{tabular}
\end{adjustbox}
\caption[On-demand energy prices considered in the experiments]{On-demand energy prices considered in the experiments.}
\label{tab:od_prices}
\end{table}

%% file: resources/tables/tab_pstates.tex
\begin{table}[H]
\centering
\begin{adjustbox}{width=0.7\textwidth}

    \begin{tabular}{r|rrrrr}
    \toprule
    \textbf{Frequency Steps [MHz]} & 1600  & 1867  & 2113  & 2400  & 2670  \\
    \hline
    \textbf{Idle Power [Watt]}            & 82.70 & 82.85 & 82.95 & 83.10 & 83.25 \\
    \hline
    \textbf{Max. Power [Watt]}            & 88.77 & 92.00 & 95.50 & 99.45 & 103.0 \\
    \bottomrule
    \end{tabular}
\end{adjustbox}
\caption[P-states consumption levels of the old machine model]{P-states consumption levels of the old machine model.}
\label{tab:pstates}
\end{table}

%% file: resources/tables/tab_machines.tex
\begin{table}[H]
\centering
\begin{adjustbox}{width=1.2\textwidth, center=\textwidth}
    \begin{tabular}{c|c|l|c|r|r|r}
    \toprule
    \textbf{Machine Model} & \textbf{Year of Release} & \textbf{CPU}& \textbf{Base Frequency} & \textbf{Cache} & \textbf{\#Cores} & \textbf{\#Threads} \\
    \midrule
    Old & 2007      & Intel$^\text{®}$ Core™2 Quad Q6700   & 2.66 GHz & 8 MB & 4 & 4   \\
    New & 2021      & Intel$^\text{®}$ Xeon$^\text{®}$ Platinum 8380 & 2.30 GHz & 60 MB          & 40& 80 \\
    \bottomrule
    \end{tabular}
\end{adjustbox}
\caption[Machine models]{Machine models used in the experiments.}
\label{tab:machines}
\end{table}

%% file: resources/tables/tab_hosts.tex
\begin{table}[H]
\centering
\begin{adjustbox}{width=\textwidth}

    \begin{tabular}{crrrr}
    \toprule
    \textbf{Machine Model} & \textbf{\#Cores/Host} & \textbf{\#Host} & \textbf{Size of Memory Unit {[}MB{]}} & \textbf{\#Memory Units/Host} \\
    \midrule
    Old & 4  & 284             & 4,000               & 4         \\
    New & 160& 9               & 3,200               & 48      \\
    \bottomrule
\end{tabular}
\end{adjustbox}
\caption[Host setup for each machine model]{Host setup for each machine model.}
\label{tab:hosts}
\end{table}

%% file: resources/tables/tab_spec.tex
\begin{table}[ht!]
\centering
\begin{adjustbox}{width=1.2\textwidth,center=\textwidth}
\scriptsize
\begin{tabular}{l|rrrrrrrrrrr}
\toprule
\backslashbox{\textbf{Model}}{\textbf{Load}}  & \textbf{0\%}  &\textbf{ 10\%}  & \textbf{20\%}  & \textbf{30\%}  & \textbf{40\%}  & \textbf{50\%}  & \textbf{60\%}  & \textbf{70\%}  & \textbf{80\%}  & \textbf{90\%}  & \textbf{100\%} \\
\hline
New  & 118.0 & 188.0 & 224.0 & 258.0 & 273.0 & 298.0 & 333.0 & 380.0 & 430.0 & 492.0 & 633.0 \\
\bottomrule
\end{tabular}
\end{adjustbox}
\caption[Load-to-power data of the new macine model]{Target loads to average active power values (in Watts) for the new machine model.}
\label{tab:spec}
\end{table}

%% file: resources/tables/tab_coefficients.tex
\begin{table}[H]
\centering
\begin{adjustbox}{width=0.5\textwidth}

    \begin{tabular}{ccccc}
    \toprule
    \textbf{Component} & $\lambda$   &  $\alpha$     &   $\beta$    &  $\pi$     \\
    \midrule
    \textbf{UPS} & 0.040 & 0.050 & ---   & 0.090 \\
    \textbf{PDU} & 0.015 & ---   & 0.015 & 0.030 \\
    \bottomrule
    \end{tabular}
\end{adjustbox}
\caption[Coefficient values of the UPS and PDU power model]{Coefficient values of the UPS and PDU power model.}
\label{tab:coefficient}
\end{table}

%% file: chapters/5-eval/5.2_market.tex
\section{Energy Market} \label{sec:market}

In this section, we present the results regarding the participation of datacenters in the energy market. First and foremost, for \ref{rq2} we ascertain whether it is financially beneficial for datacenters to participate in the energy market in the first place and if so, which market to participate in (\S\ref{sec:power_loads} and \S\ref{sec:energy_costs}). Next, in Section \ref{sec:new_machine} we investigate the implications of using the old machine model. Then, in Section \ref{sec:procurement} we research the impact of different energy procurement strategies (\ref{rq3}), \ie how to participate in the day-ahead and the balancing market? Finally, we investigate why ML methods could be of help in further leveraging profits during market participation.

\subsection{Power Loads} \label{sec:power_loads}

The concept of power loads plays a key role in the following experiments. Power load is the quantity of energy consumed by all equipment in a datacenter. It can be further categorized into two sub-classes, the base load and the peak load. Base load is the bare minimum active power required to keep the datacenter up and running, whereas peak load is demand-dependent. In other words, the base load is the constant power draw at all times, and peak load can be perceived as the margin between the power draw at peak demands and the constant base load. 

\begin{figure}[!t]
    \centering
    \begin{adjustbox}{width=1.35\textwidth,center=\textwidth}
        \begin{subfigure}[b]{.55\textwidth}
          \centering
          \includegraphics[width=\linewidth]{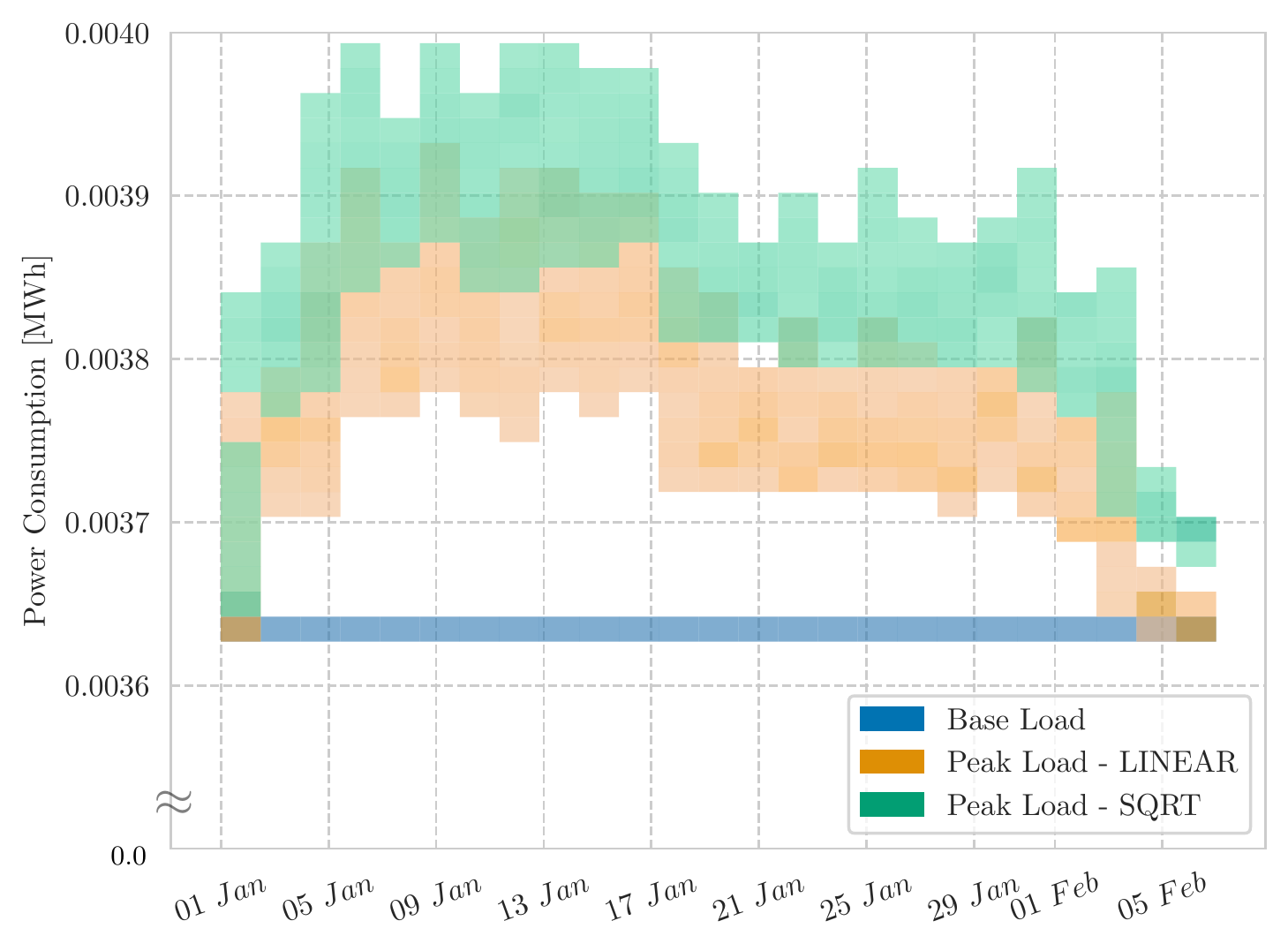}
          \caption{Instant power loads.}
          \label{fig:loads_model}
        \end{subfigure}%
        \begin{subfigure}[b]{.45\textwidth}
          \centering
          \includegraphics[width=\linewidth]{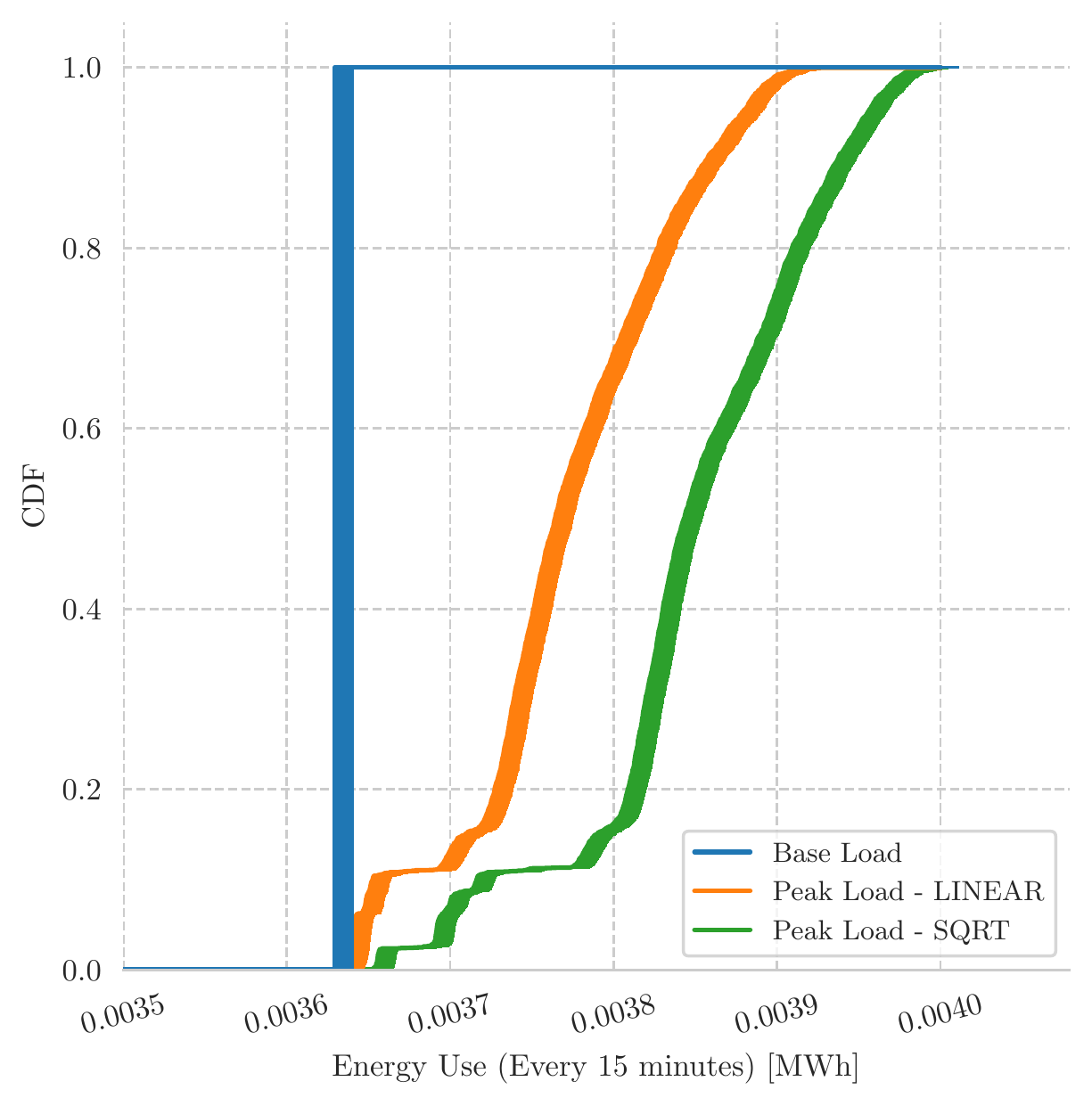}
          \caption{CDF of the power loads.}
          \label{fig:loads_cdf}
        \end{subfigure}
    \end{adjustbox}
    \caption[Different power loads]{Different power loads.}
    \label{fig:loads}
\end{figure}

Specifically, we define the active idle power of the datacenter as the base load, and the proportion of power draw catering for the demands as the peak load. Figure \ref{fig:loads} demonstrate these two loads estimated by the two power models, where Figure \ref{fig:loads_model} shows the instant power loads, and Figure \ref{fig:loads_cdf} shows their cumulative distribution functions (CDF). Note that we add on the base load to the peak load to demonstrate their relationship, the peak load would be below the base load otherwise. As described in Section \ref{sec:setup}, the two generic power models, the LINEAR and the SQRT are able to serve as the lower and upper bounds for a specific computing platform. Therefore, the actual power load should lie between the two curves.

\subsection{Energy Costs} \label{sec:energy_costs}

\begin{figure}[!ht]
    \centering
    \begin{adjustbox}{width=1.37\textwidth,center=\textwidth}
        \includegraphics[width=\textwidth]{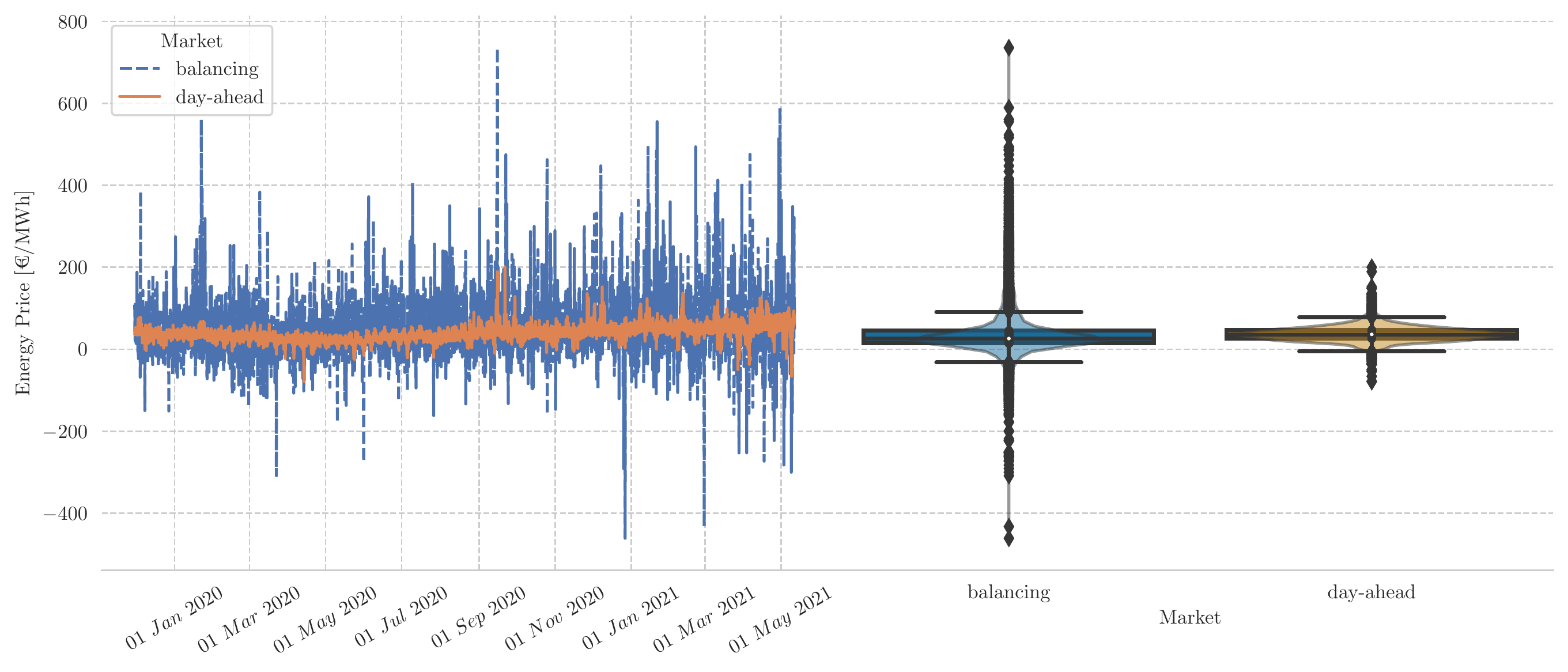}
    \end{adjustbox}
    \caption[Distributions of day-ahead prices and imbalance prices]{Distributions of day-ahead prices and imbalance prices.}
    \label{fig:prices_dist}
\end{figure}

Now, we investigate the energy costs of the power loads in the day-ahead and the balancing markets. The market data analysed runs from 30 November 2019 to 11 May 2021, which is about 1.5 years in duration.  Figure \ref{fig:prices_dist} illustrates the distribution of the energy prices of the two markets. Note that the positive price is the price that a datacenter, as a balance responsible party (BRP), would have to pay for its energy use. On the contrary, if the price is negative, datacenters will be paid by the energy market or the system operator to consume energy (mostly with the purpose of balancing the power grid). We can see from Figure \ref{fig:prices_dist} that the balancing market generally has a much higher positive price than that of the day-ahead market. However, the level of negative price in the balancing market is much lower than that of the day-ahead market to a similar extent. Albeit larger variation, the median of the energy price in the balancing market is slightly below the mean of the energy price in the day-ahead market. Hence, we conclude that the balancing market demonstrates a {coexistence} of \textit{high risks} (greater positive price and larger variation) and \textit{high profitability} (lower negative price and lower median value). Having said that, is it beneficial in the long run to take part in the day-ahead and/or the balancing market?

\begin{figure}[H]
    \centering
    \caption[Energy Costs of the two power loads in different markets]{Energy Costs of the two power loads in different markets.}
        \begin{subfigure}[b]{\textwidth}
          \centering
           \begin{adjustbox}{width=1.3\textwidth,center=\textwidth}
            \includegraphics[width=\linewidth]{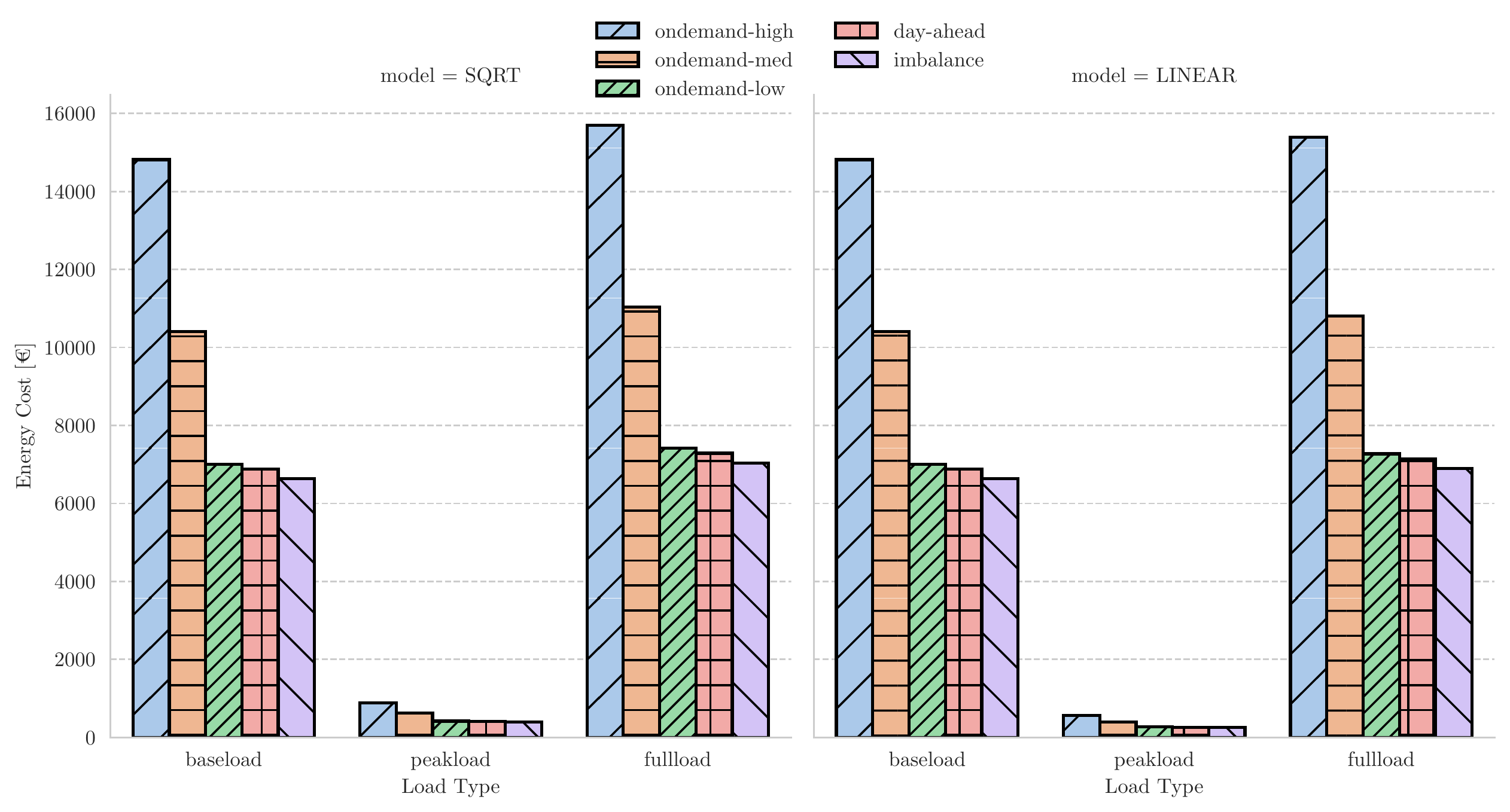}
            \end{adjustbox}
          \caption{Energy costs estimated by two power models.}
          \label{fig:cost_models}
        \end{subfigure}
        
        \begin{subfigure}[b]{\textwidth}
          \centering
            \begin{adjustbox}{width=0.9\textwidth,center=\textwidth}
             \includegraphics[width=\linewidth]{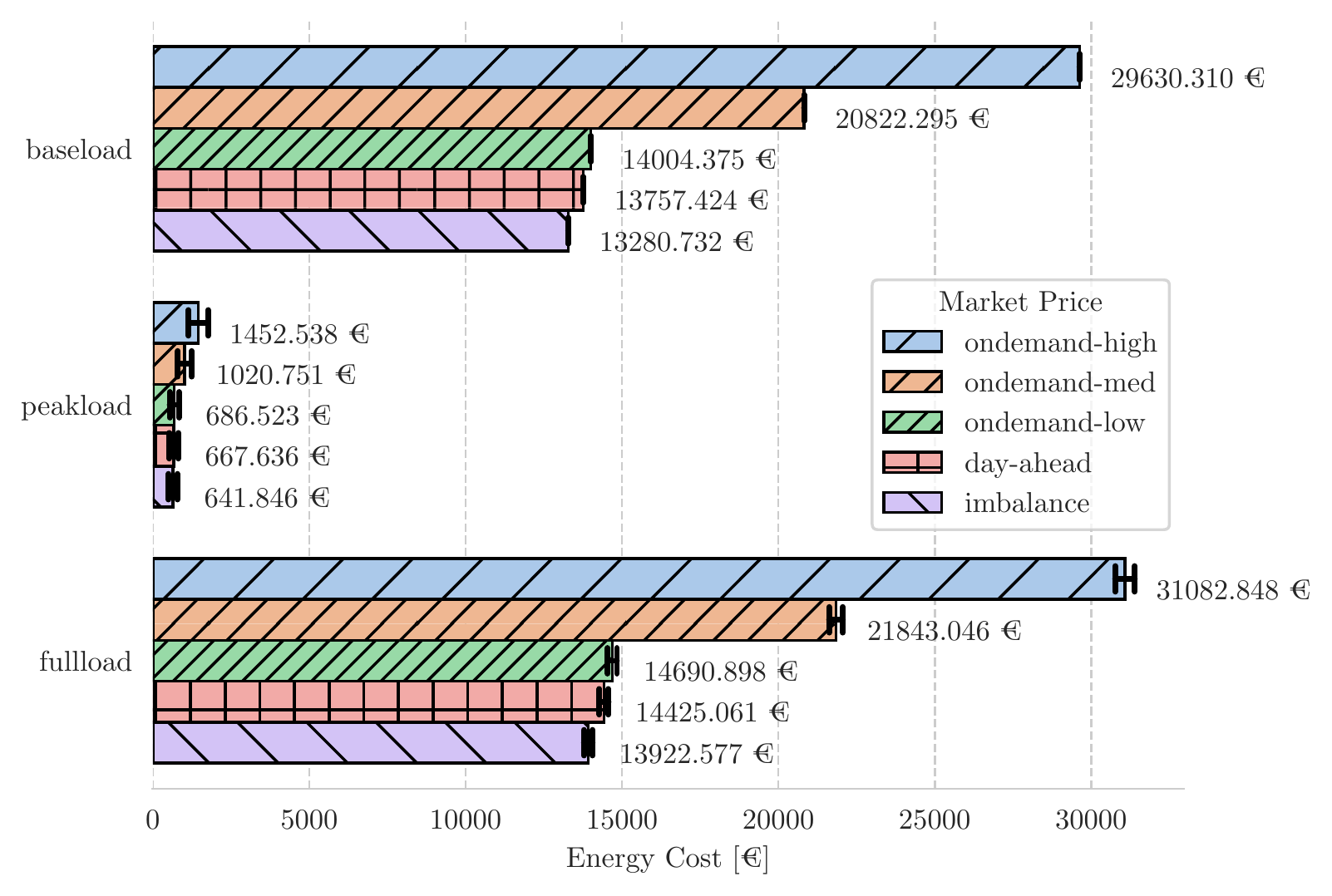}
            \end{adjustbox}          
            \caption{Combined cost estimation from two power models.}
          \label{fig:cost_combine}
        \end{subfigure}
    \label{fig:costs}
\end{figure}

To answer the aforementioned question, we use the EEMM extension to compute the energy costs for the two pow loads of the \textit{old} machine model under different market prices. Figure \ref{fig:cost_models} shows the costs of the power loads estimated by the two power models in different markets (\ref{fr4}), and Figure \ref{fig:cost_combine} shows the combined estimation (\ref{fr3}), in which the error bars capture the variation therein. As we can see that participating in either the day-ahead market or the balancing market results in lower energy cost even compared to the energy cost of the on-demand scheme of the lowest price. Also, the cost in the balancing market is even lower than that in the day-ahead market. Therefore, we conclude that it is financially beneficial to participate in both the day-ahead market and the balancing market (note that, referring back to Section \ref{sec:balance_grid}, it is even compulsory for a BRP to resolve its unbalance in the balancing market).

\subsection{Newer Machine Model} \label{sec:new_machine}

\begin{figure}[!t]
    \centering
    \begin{adjustbox}{width=0.8\textwidth,center=\textwidth}
        \includegraphics[width=\textwidth]{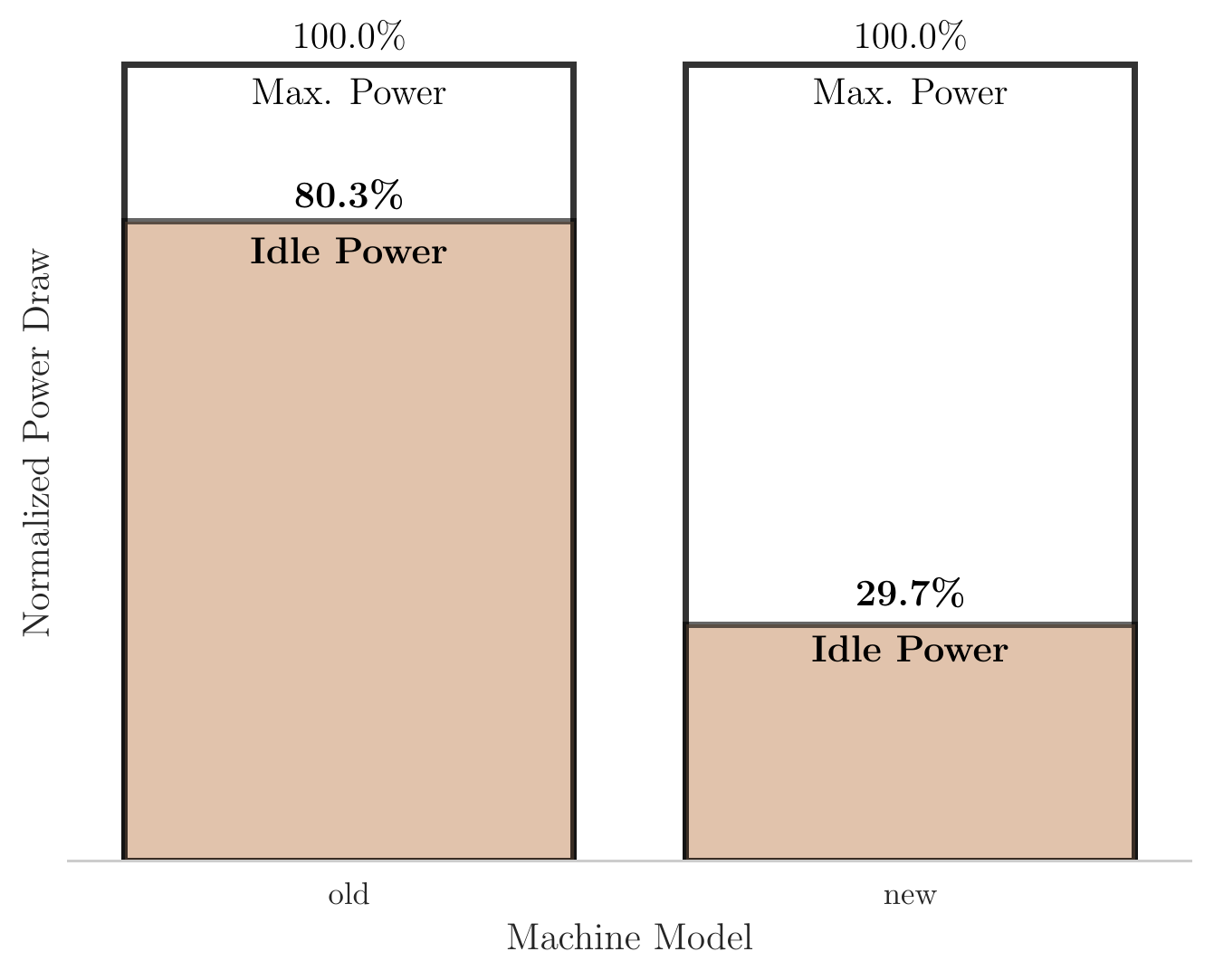}
    \end{adjustbox}
    \caption[Comparison of CPU energy efficiency of two machine models]{Comparison of CPU energy efficiency of two machine models.}
    \label{fig:new_cpu}
\end{figure}

Having ascertained the incentive of participating in both the day-ahead and the balancing market for \ref{rq2}, we now investigate the implementations of using the old machine model.

Figure \ref{fig:new_cpu} illustrates the difference in the energy efficiency of the two machine models in terms of the idle-maximum power ratio. The idle power of the old machine model takes up more than 80\% of its total power capacity. In contrast, the idle power of the new machine model only accounts for less than 30\% of the maximum power. Thus, the new machine is drastically more energy efficient compared to the old one.

\begin{figure}[!t]
    \centering
    \begin{adjustbox}{width=1.35\textwidth,center=\textwidth}
        \includegraphics[width=\textwidth]{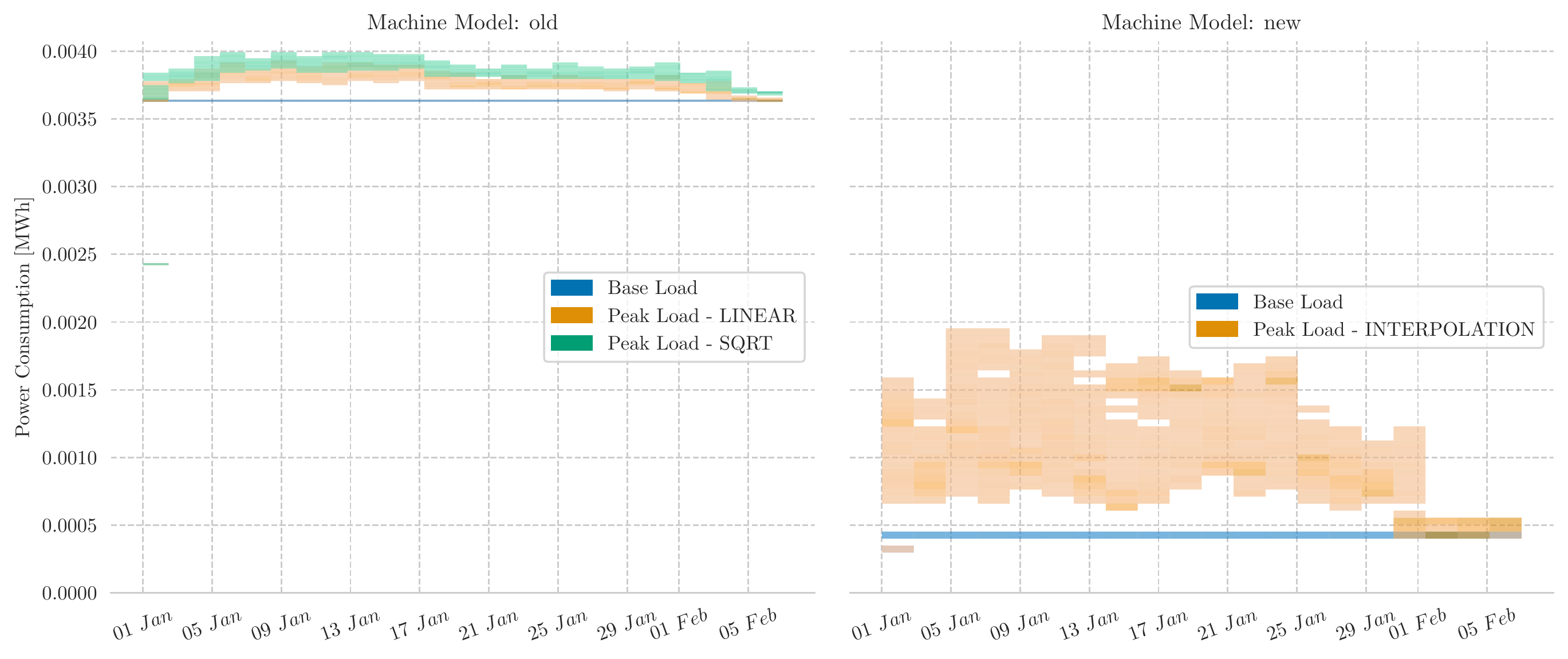}
    \end{adjustbox}
    \caption[Comparison of power loads of the two machine models]{Comparison of power loads of the two machine models.}
    \label{fig:new_loads}
\end{figure}

\begin{figure}[!ht]
    \centering
    \begin{adjustbox}{width=1.35\textwidth,center=\textwidth}
        \includegraphics[width=\textwidth]{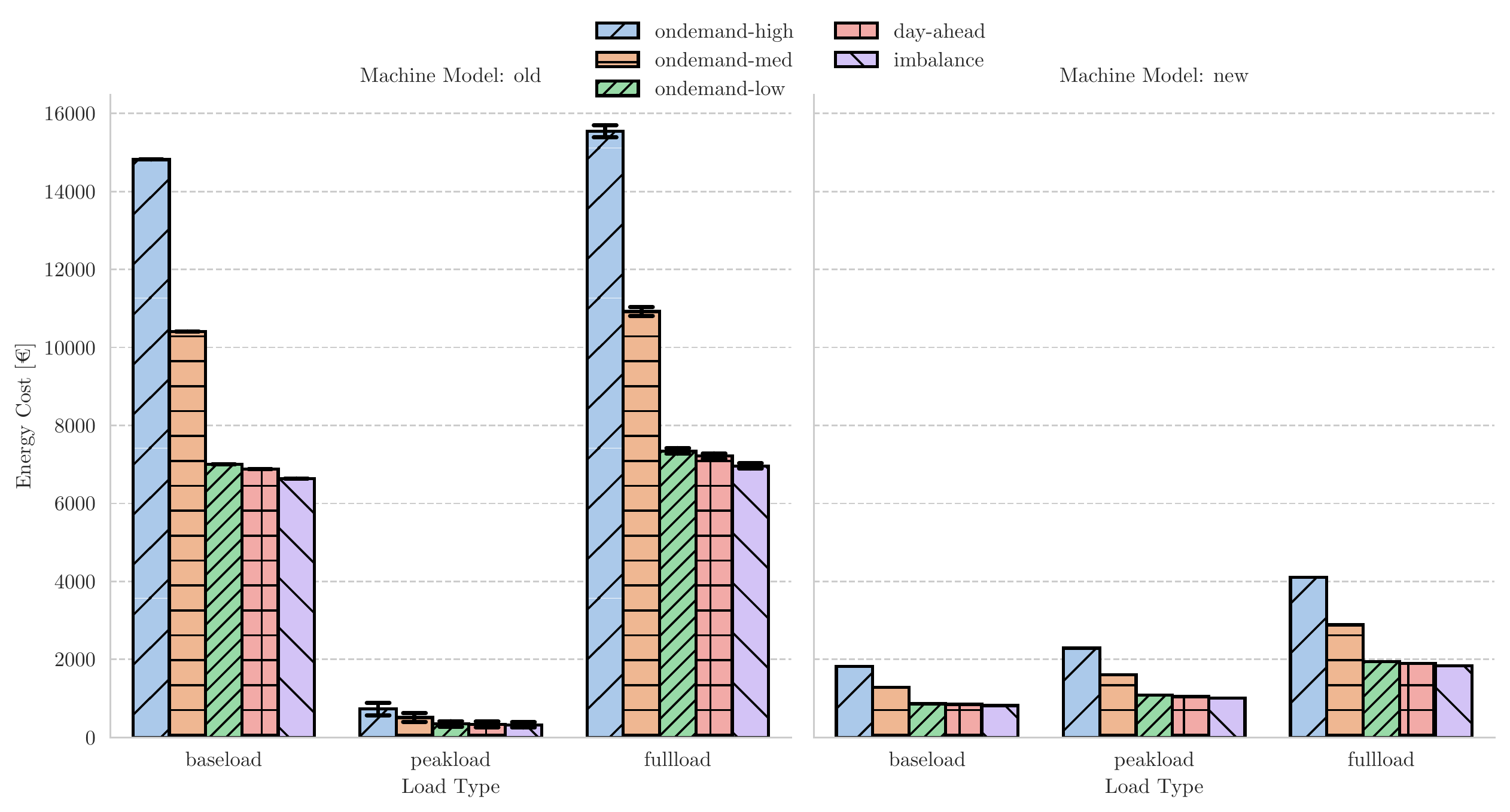}
    \end{adjustbox}
    \caption[Comparison of energy costs of the two machine models]{Comparison of energy costs of the two machine models.}
    \label{fig:new_costs}
\end{figure}

When it comes to the power loads, the new machine model yields a much lower level of energy consumption whilst, in the meantime, exhibits a much greater variation in the peak load (Figure \ref{fig:new_loads}). Consequently, the new machine model results in a $\sim$73.6\% total cost-saving, as shown in Figure \ref{fig:new_costs}. Most importantly, the portion of energy costs resulted from the peak load of the new machine model is about $2.5\times$ higher than that of the old machine model. Nevertheless, the conclusion from Section \ref{sec:energy_costs} still hold, \ie participating in both the day-ahead and the balancing markets is beneficial in terms of the energy cost. These observations are important for generalizing the conclusions in later sections. From the next section onwards, we come back to the old machine model, through which the core experiments, the DVFS optimization, is conducted.

\subsection{Energy Procurement Strategy} \label{sec:procurement}

\begin{figure}[!ht]
    \centering
    \begin{adjustbox}{width=1.2\textwidth,center=\textwidth}
        \includegraphics[width=\textwidth]{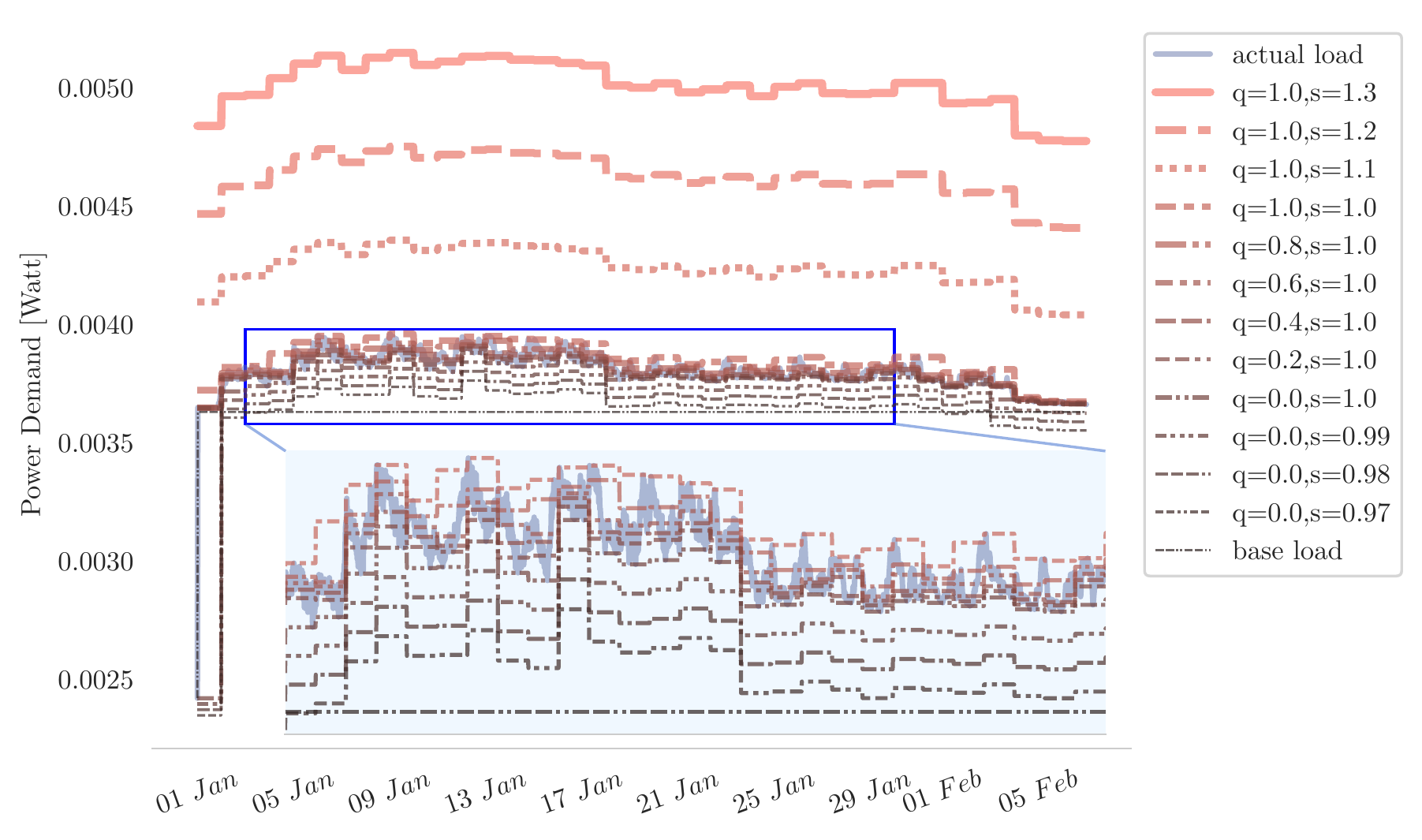}
    \end{adjustbox}
    \caption[Simulated load schedules in the day-ahead market]{Simulated load schedules in the day-ahead market.}
    \label{fig:load_schedules}
\end{figure}

\begin{figure}[!t]
    \centering
    \begin{adjustbox}{width=1.2\textwidth,center=\textwidth}
        \includegraphics[width=\textwidth]{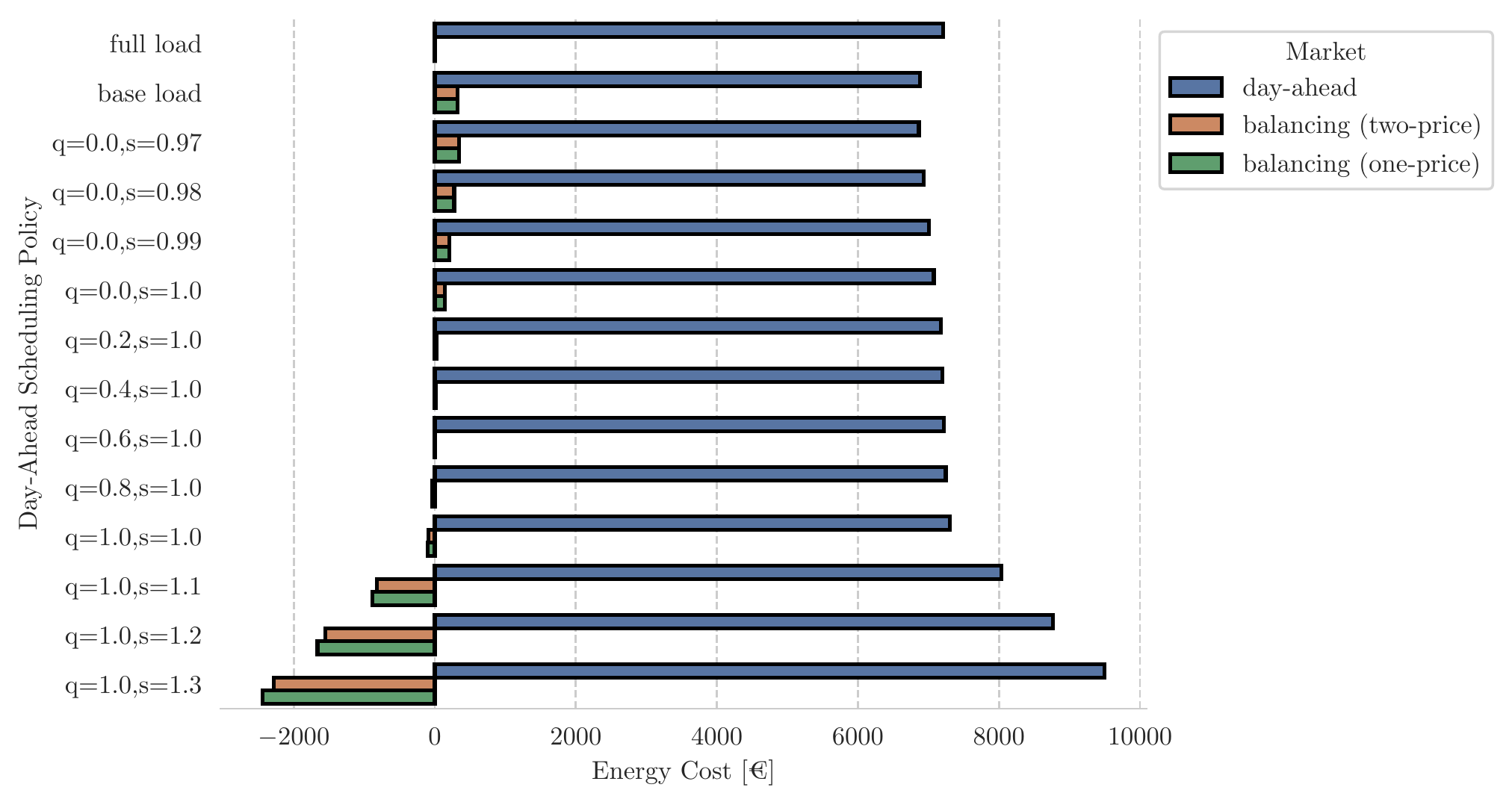}
    \end{adjustbox}
    \caption[Comparison of energy costs of simulated load schedules]{Comparison of energy costs of simulated load schedules.}
    \label{fig:no_stack_load_schedules}
\end{figure}

Results from previous sections (\S\ref{sec:energy_costs}and \S\ref{sec:new_machine}) illustrate the reason as to why datacenters should participate in the energy market. In the following sections, we investigate the impact of employing different load-forecast-based procurement strategies for \ref{rq3}, \ie how datacenters should participate? To answer this question, we start with stating the assumptions (\textbf{A}s) made for the experiments.


\begin{enumerate} [label=\textbf{A\arabic*}]
    \item \label{a1} Datacenter operators purchase energy in the day-ahead market based on the load forecast of the corresponding \textit{day}.
    \item \label{a2} The load forecast is \textit{perfect}. In other words, the load predictions always precisely match the actual loads of the corresponding day.
    \item \label{a3} Datacenter operators do not deliberately schedule even less energy than the bare-minimum quantity --- the base load.
    \item \label{a4} Datacenters' participation is abided by the two-price balancing system.
\end{enumerate}

To elaborate on the two assumptions further, one should recall from Section \ref{sec:grid_markets} that market participants self-dispatch the quantity of energy that they are expected to produce/consume in \textit{each hour} of the corresponding day during the day-ahead market (before the gate closure of the spot market). In turn, \ref{a1} orientates the experiment to the average scenario, as the load forecast can be more frequent (\eg on an hour-basis) or be in a less proactive manner (\eg weekly or even monthly). Concerning \ref{a2}, the performance of the load forecasting of a datacenter is the controlled variable here. Hence, we do not presume its level of accuracy/precision. As for \ref{a3}, datacenter operators are not expected to intentionally introduce a large energy deficit in the first place. Concerning \ref{a4}, the reasoning will be explained in detail in Section \ref{sec:pricing_systems}.

On the basis of \ref{a1} to \ref{a3}, we define the quantity of energy to schedule in the spot market ($Q^S$) using Equation \ref{eq:da_load}:

\begin{equation}
\mathlarger{
    Q^S = \mathbb{Q}_q(l_f) \cdot s,
}
\label{eq:da_load}
\end{equation}

\noindent where $l_f$ denotes the load forecast of the next day, $q$ is the quantile of the quantile function $\mathbb{Q}$, and $s$ is a scalar to apply. 

In Figure \ref{fig:load_schedules}, we use the extension EEMM to demonstrate the resulting load schedules in the spot market (\ref{fr5}), where $q \in [0, 1]$,  and $\ s \in [0.97, 1.30]$. Specifically, $q < 1.0$ models the situation in which datacenter operators schedule less power than the forecasted load of the day (under-scheduling), and $s > 1.0$ models the over-scheduling scenario in which the operators schedule more energy than the forecasted load. In addition, the “base load” represents the strategy that in the day-ahead market, the operator only procures the bare-minimum energy whose quantity is \textit{certain}. As shown in the figure, the portion where $s > 1.0$ is sufficiently high (\ref{nfr2}), whilst the part where the $0.0 \ge q < 1.0$ is still lower-bounded by the base load.

Having obtained the load schedules of the day-ahead market, we use the EEMM extension to compute the \textit{total} energy cost for every schedule (\ref{fr5}), where the energy surplus/deficit is resolved in the balancing market. Figure \ref{fig:no_stack_load_schedules} shows an unstacked comparison of the energy costs for the simulated schedules. It illustrates that the more energy ordered in the day-ahead/spot market, the more datacenters have to pay. In contrast, as the amount of scheduled energy increases, the energy cost in the balancing market decreases from positive to negative values. As described in Section \ref{sec:market}, datacenters will be paid back when the energy price is negative; therein lies the question: can datacenters deliberately (or even maliciously) schedule arbitrarily large amounts of energy in the day-ahead market so that they eventually would gain profit during the imbalance settlement? To answer this question, we need to take a deep dive into the balancing mechanism of the (EU) energy market. 

\subsubsection{Imbalance Pricing Systems} \label{sec:pricing_systems}

\input{resources/tables/tab_sym_im}

As indicated in Figure \ref{fig:no_stack_load_schedules}, there are two balancing systems: one-price and two-price systems \cite{1_2_price}. In the case of the one-price system, if the imbalance of a prosumer is (unintentionally) helping balance the grid, the prosumer will in effect earn extra monetary rewards during the imbalance settlement. On the contrary, under the two-price balancing system, such inadvertent assistance is not encouraged since the price level of the compensation is the same as that of the spot market. The mechanism of the one-price system follows Equation \ref{eq:one_price}, and the two-price systems adheres to Equation \ref{eq:two_price}; Table \ref{tab:sym_im} shows the meanings of the symbols therein.

\begin{align}
    \sum^{N_\text{BSP}}_i \left(  Q_{\downarrow_i} \cdot p^B_\downarrow \right) + 
        \sum^{N_\text{BRP}}_j \left(  Q_{-_j} \cdot p^B_- \right)
&=
    \sum^{N_\text{BSP}}_i \left(  Q_{\uparrow_i} \cdot p^B_\uparrow \right) + 
        \sum^{N_\text{BRP}}_j \left(  Q_{+_j} \cdot p^B_+ \right) 
\label{eq:one_price} \\
    \sum^{N_\text{BSP}}_i \left(  Q_{\downarrow_i} \cdot p^B_\downarrow \right) + 
        \sum^{N_\text{BRP}}_j \left(  Q_{-_j} \cdot p^B_- \right)
&=
    \sum^{N_\text{BSP}}_i \left(  Q_{\uparrow_i} \cdot p^B_\uparrow \right) + 
        \sum^{N_\text{BRP}}_j \left(  Q_{+_j} \cdot p^S \right)
\label{eq:two_price}
\end{align}

\begin{figure}[!t]
    \centering
    \begin{adjustbox}{width=1\textwidth,center=\textwidth}
        \includegraphics[width=\textwidth]{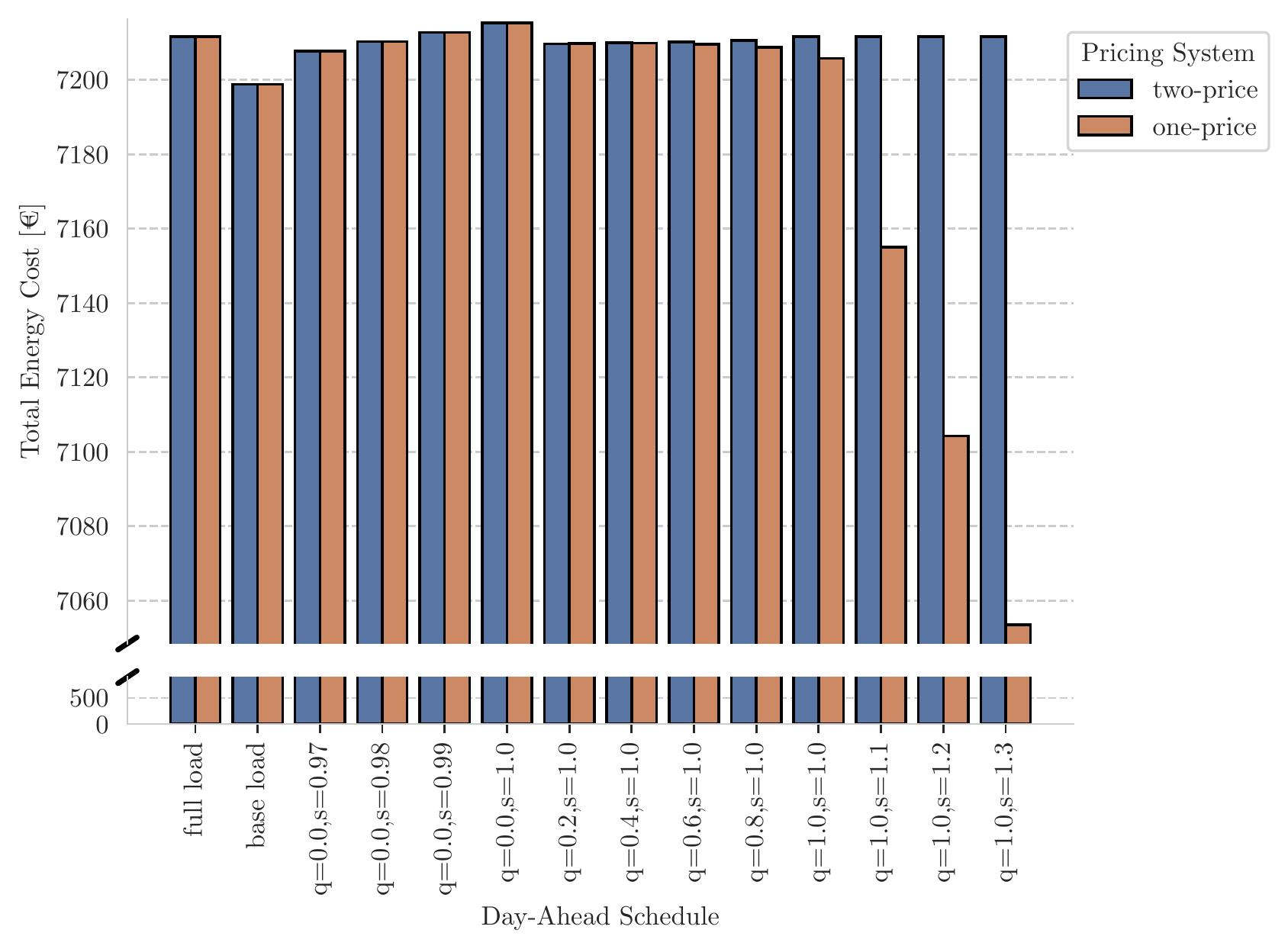}
    \end{adjustbox}
    \caption[Comparison of energy costs of the two imbalance pricing system]{Comparison of the two imbalance pricing system.}
    \label{fig:pricing_system}
\end{figure}

Figure \ref{fig:pricing_system} demonstrates the difference between the two pricing systems by comparing their resulting total energy costs. As expected, datacenters cannot gain extra profit by scheduling much larger quantities of energy than actually needed under the two-price system, but they can in the case of the one-price system, which will be prohibited by the system operator. Therefore, datacenters are expected to comply with the two-price balancing system, and \textit{do not} intensionally introduce imbalance to the power grid. In turn, the fourth assumption (\ref{a4}) is valid. 

\begin{figure}[!t]
    \centering
    \begin{adjustbox}{width=0.95\textwidth,center=\textwidth}
        \includegraphics[width=\textwidth]{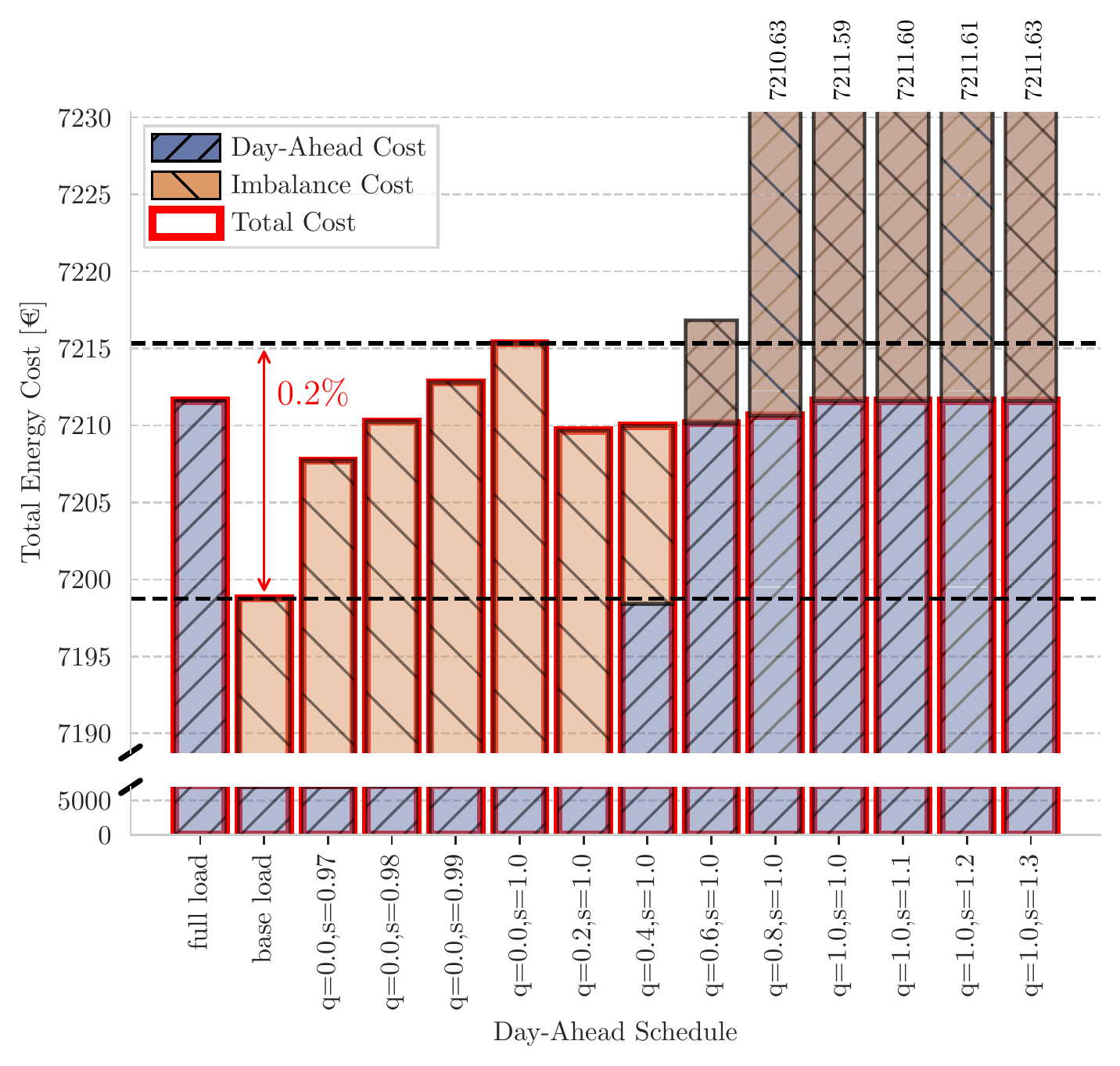}
    \end{adjustbox}
    \caption[Stacked comparison of energy costs of the imbalance pricing systems]{Stacked comparison of the two imbalance pricing systems.}
    \label{fig:stack_load_schedules}
\end{figure}

Having clarified all assumptions (\ref{a1} -- \ref{a4}), we now conduct the final comparison for the energy costs of the simulated schedules (Figure \ref{fig:load_schedules}). To this end, we stack the energy costs of the two markets in Figure \ref{fig:stack_load_schedules}, where the positive imbalance costs are added on top of the day-ahead costs, whilst the negative ones intrude into the day-ahead costs from the top, forming the overlaps. As highlighted in red, the variation between the total energy costs of the simulated schedules is about 0.2\% with the minimum being the base-load strategy. In other words, although the margin is not significant, only scheduling the bare-minimum energy in the day-ahead market is the preferred strategy. Also, as more and more energy is scheduled in the day-ahead market ($s > 1.0$), the total energy cost demonstrates a small yet gradual and steady increase. Note that this conclusion also applies to the new machine model because it is not subjected to load variation due to the assumption of having a perfect load forecast (\ref{a2}).

\subsection{Relationship between Prices of the Two Markets} \label{sec:price_corr}

\begin{figure}[!t]
    \centering
    \includegraphics[width=\textwidth]{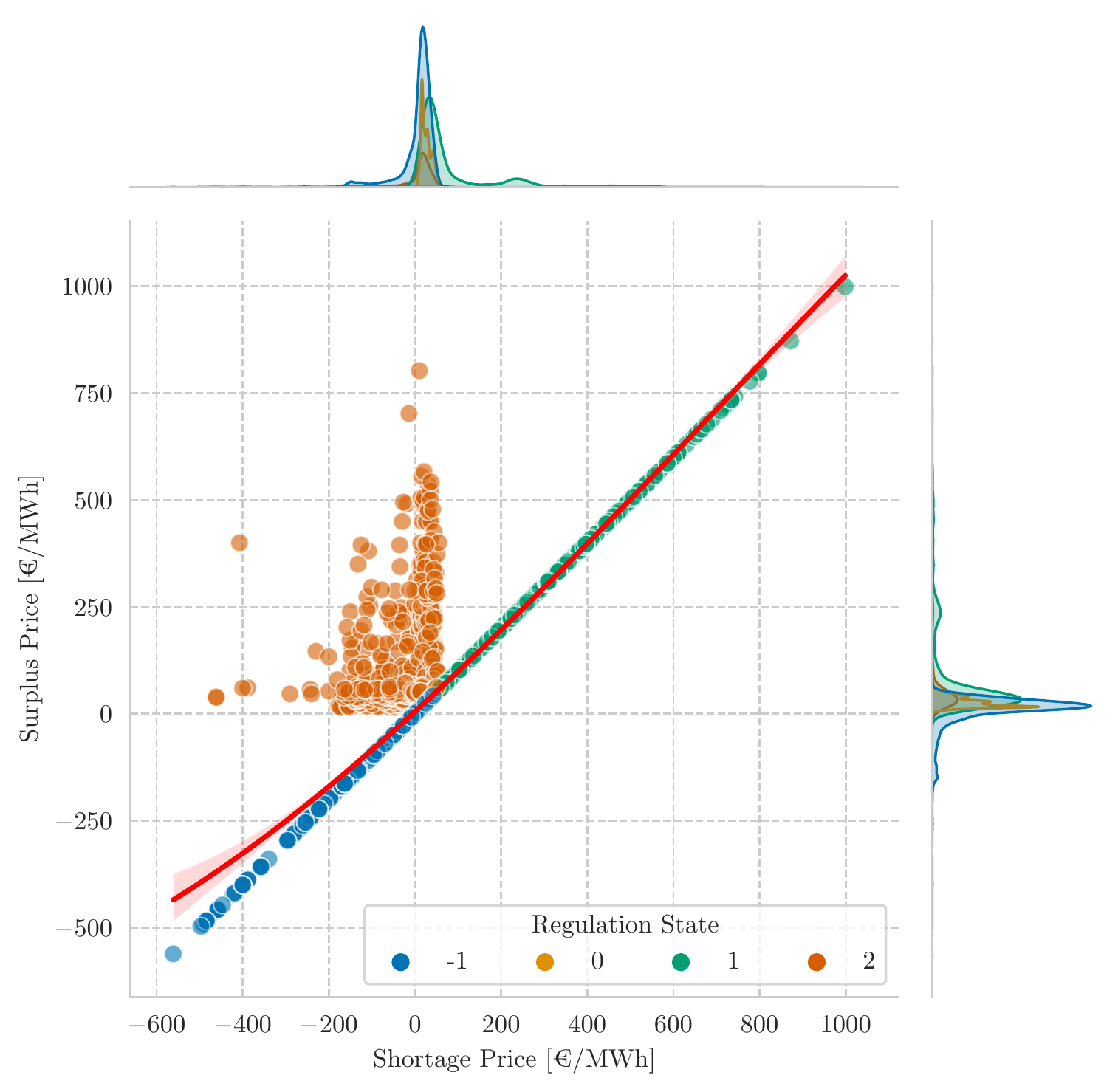}
    \caption[Correlation between imbalance prices]{Correlation between imbalance prices with second order interpolation.}
    \label{fig:corr_im}
\end{figure}

\begin{figure}[!t]
    \centering
    \includegraphics[width=0.85\textwidth]{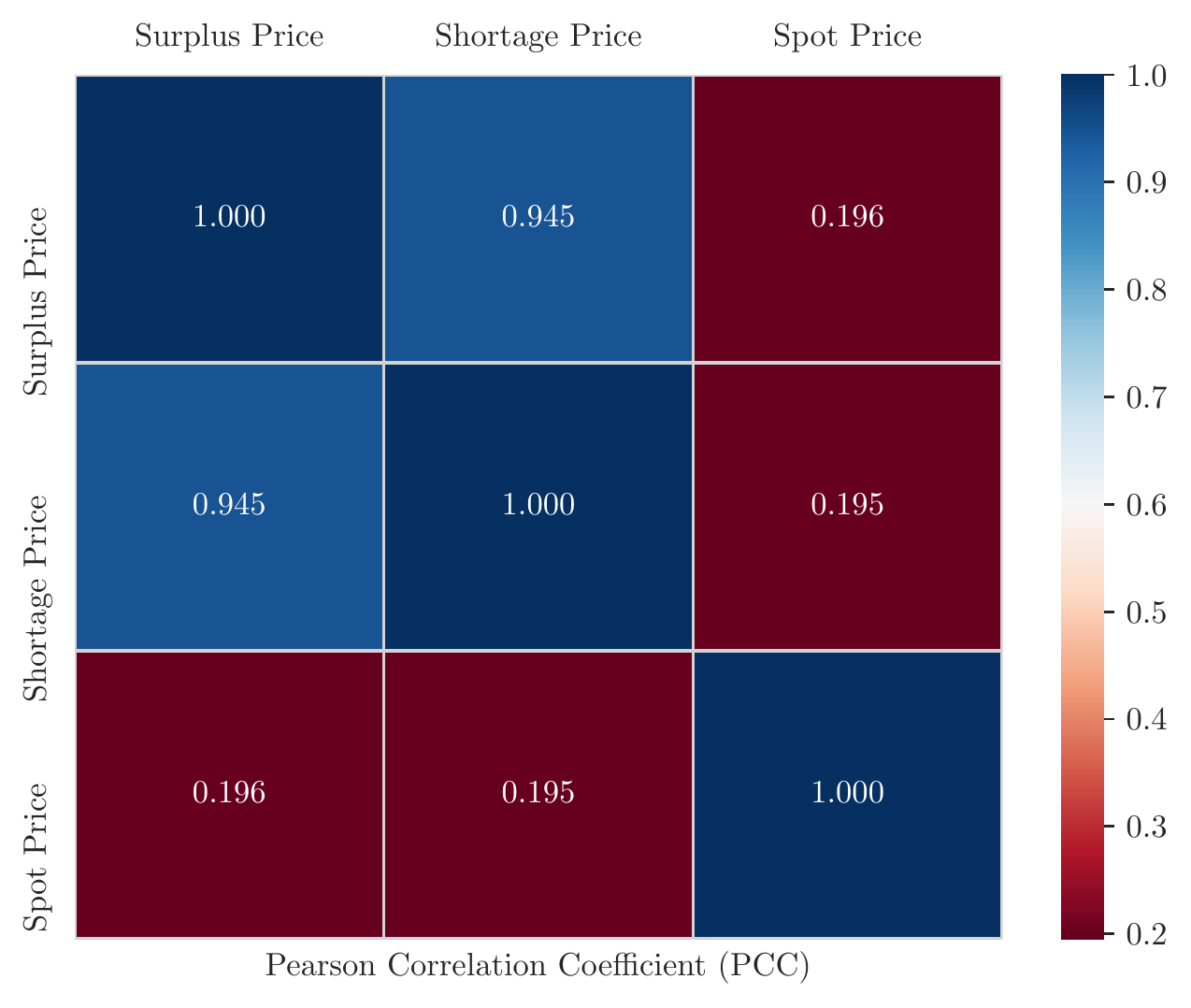}
    \caption[Correlation between day-ahead prices and imbalance prices]{Pearson Correlation Coefficients between day-ahead prices and imbalance prices.}
    \label{fig:corr_im_da}
\end{figure}

\begin{figure}[!t]
    \centering
    \begin{adjustbox}{width=0.8\textwidth,center=\textwidth}
        \includegraphics[width=\textwidth]{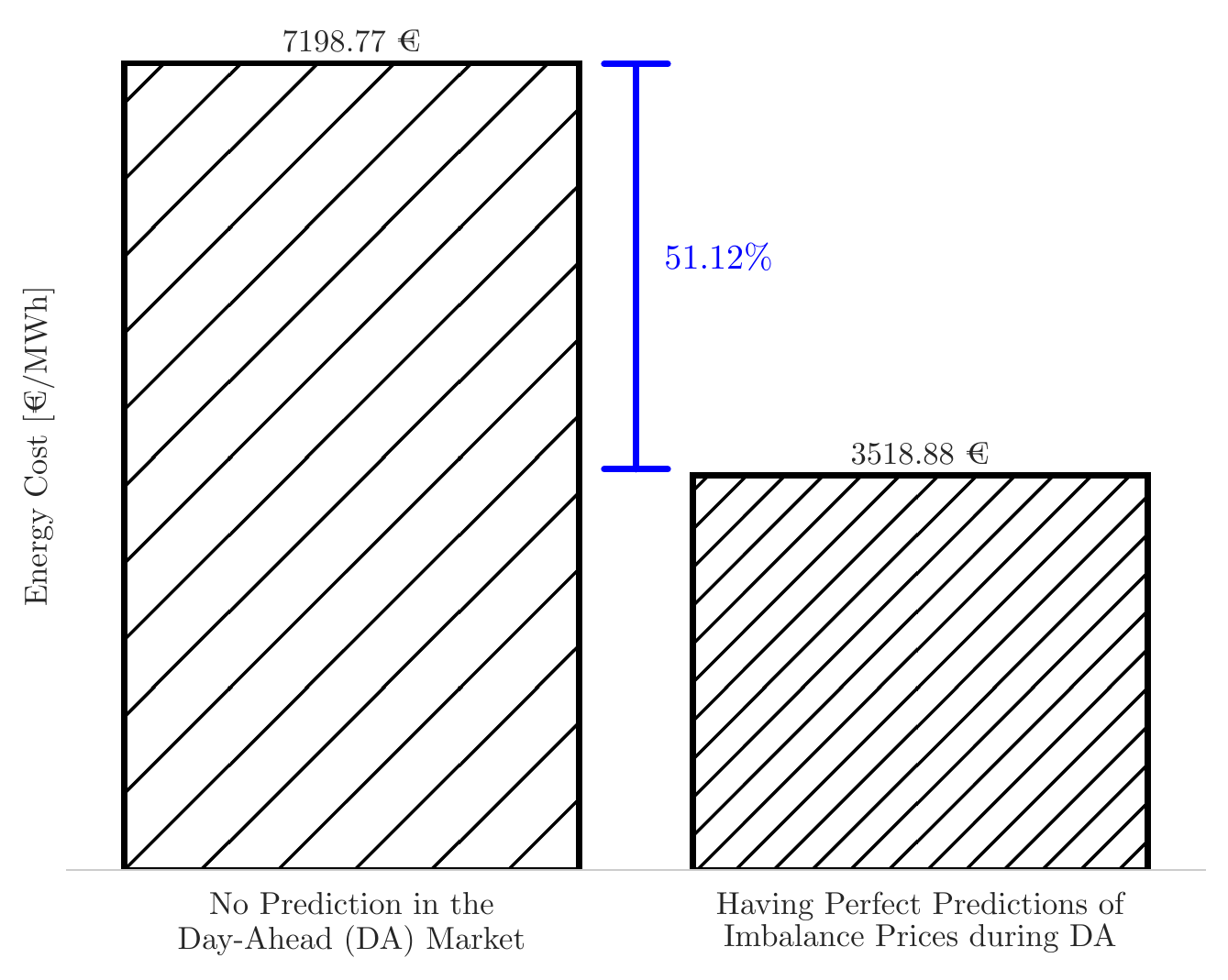}
    \end{adjustbox}
    \caption[Potential benefit of obtaining predictions for imbalance prices early]{Potential benefit of obtaining predictions for imbalance prices during the day-ahead market.}
    \label{fig:pred_im_in_da}
\end{figure}

In previous sections, we answered the question of why and how to participate in the energy market (\ref{rq2}, \ref{rq3}). In this section, we seek answers to the question of why ML methods can be of help for datacenters in terms of leveraging the profit when taking part in the energy market. To this end, we search for potential correlations between energy prices in order to ascertain whether it is reasonable to make decisions based on simple heuristics.

Firstly, we demonstrate the correlation between the imbalance prices (shortage and surplus prices) with corresponding regulation states in Figure \ref{fig:corr_im} (the latest, detailed descriptions of the regulation states in the Netherlands can be found in \cite{tennet}). The probability distribution functions (PDFs) shown at the top and the right part illustrate that a substantial number of values concentrate between $-$200 and 200. Furthermore, from the second-order interpolation between the two prices, the red curve, we can clearly see a strong linear correlation between the two imbalance prices. Also, when the regulation state is 2, surplus price is generally higher than shortage price.

Conversely, when it comes to the inter-market relationship, there is little if any correlation between the prices of the two markets. Figure \ref{fig:corr_im_da} shows the Pearson Correlation coefficients (PCC) of the prices. The PCC between the spot price and the imbalance prices is as little as $\sim 0.19$, which would provide datacenters with little to no help in making heuristics for leveraging profits. Thus, we conclude that it is not feasible to optimize profits when juggling the two energy markets by making simple heuristics.

Furthermore, currently, our ML inferences of the imbalance prices can only be obtained during the balancing market, which will be described in detail in Section \ref{sec:scheduling}. Nevertheless, to illustrate the promising potential of employing ML methods, we assume for a moment that (1) the predictions can be obtained during the day-ahead market, and (2) the predictions are perfect. Then, one straightforward procurement strategy could be: purchasing the (forecasted) full load in the day-ahead market if $p^S < p^B_-$, otherwise, using the base-load strategy, \ie scheduling the base load in the day-ahead market and settling the peak load in the balancing market. This straightforward policy together with the assumption of having the perfect prediction will provide an upper bound for the potential benefit of using the early obtained ML predictions for energy procurement.
As shown in Figure \ref{fig:pred_im_in_da}, if such an early forecast can be achieved, we can dramatically reduce the energy cost further by up to 51.12\% compared to the preferred base-load strategy by using the above mentioned straightforward procurement policy.

\subsection{Summary}

To summarize, in Section \ref{sec:energy_costs} we shed light on the coexistence of high risks and high profitability when taking part in the energy market. Then, for answering \ref{rq2} we ascertain the high financial incentive for datacenters to take part in both day-ahead and the balancing market. Next, we demonstrate the higher energy efficiency and the large variation of the power load featured by the new machine model (\S\ref{sec:new_machine}), which is important for later generalizing the experimental results. Next, in Section \ref{sec:procurement} we illustrate proved that scheduling only the bare-minimum energy, the base load, is the preferred procurement strategy (\ref{rq3}). In Section \ref{sec:price_corr}, we show that it is infeasible to leverage market participation simply by making heuristics and that substantial profit could be derived by employing early ML predictions.

%% file: resources/tables/tab_sym_im.tex
\begin{table}[!h]
    \centering
    \begin{adjustbox}{width=0.7\textwidth,center=\textwidth}
    \begin{tabular}{ll}
    \toprule
        $p^S$ & Spot market price \\
        $p^B_-$ & Shortage price in the balancing market \\
        $p^B_+$ & Surplus price in the balancing market \\
        $Q_-$ & Quantity of shortage energy of a BRP \\
        $Q+$ & Quantity of surplus energy of a BRP \\
        $Q+$ & Quantity of energy required by downwards regulations \\
        $Q_\downarrow$ & Quantity of energy required by downwards regulations \\
        $Q_\uparrow$ & Quantity of energy required by upwards regulations \\
        $N_\text{BSP}$ & Number of participated BSPs in the balancing market \\
        $N_\text{BRP}$ & Number of participated BRPs in the balancing market \\
    \bottomrule
    \end{tabular}
    \end{adjustbox}
    \caption[Symbols used in defining the two balancing systems]{Symbols used in defining the two balancing systems.}
    \label{tab:sym_im}
\end{table}

%% file: chapters/5-eval/5.3_dvfs_schedule.tex
\section{DVFS Scheduling} \label{sec:scheduling}

\begin{figure}[!t]
    \centering
    \begin{adjustbox}{width=1.3\textwidth,center=\textwidth}
        \includegraphics[width=\textwidth]{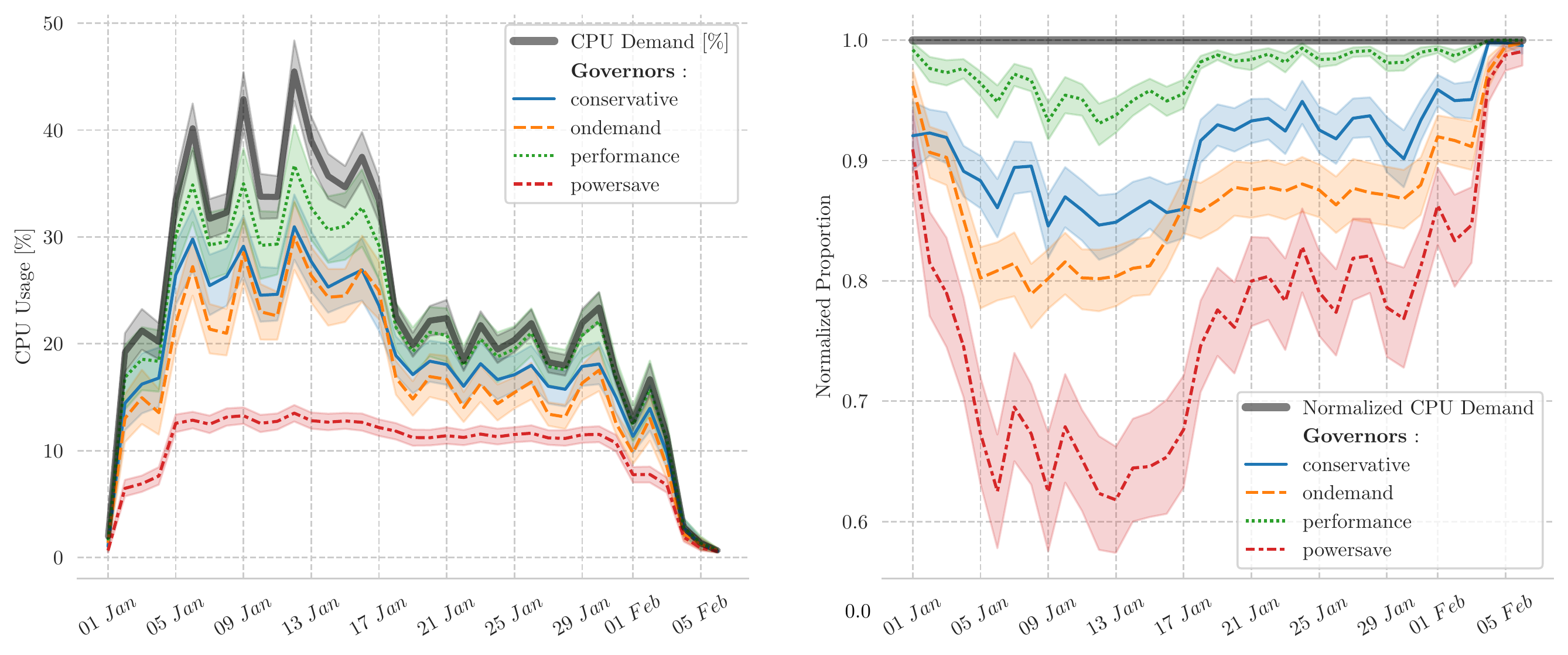}
    \end{adjustbox}
    \caption[Relationships between CPU demand and usage of DVFS]{Relationships between CPU demand and CPU usage of the four scaling governors.}
    \label{fig:dvfs_demand_usage}
\end{figure}

In this section, we present the results regarding the performance and effect of the developed proactive DVFS scheduler, in which ML inferences are employed, to answer \ref{rq4}. Firstly, we demonstrate the behaviours of the DVFS implemented in our system (\S\ref{sec:dvfs_behaviour}). Then, in Section \ref{sec:damping_factor} we investigate the effect of the hyperparameter of the scheduler, the damping factor. After that, we conduct bounded performance estimation, comparing the ML methods with synthetic estimators in Section \ref{sec:bounded}. 

\subsection{DVFS Behaviour} \label{sec:dvfs_behaviour}

\begin{figure}[!t]
    \centering
           \begin{adjustbox}{width=1.35\textwidth,center=\textwidth}
            \includegraphics[width=\linewidth]{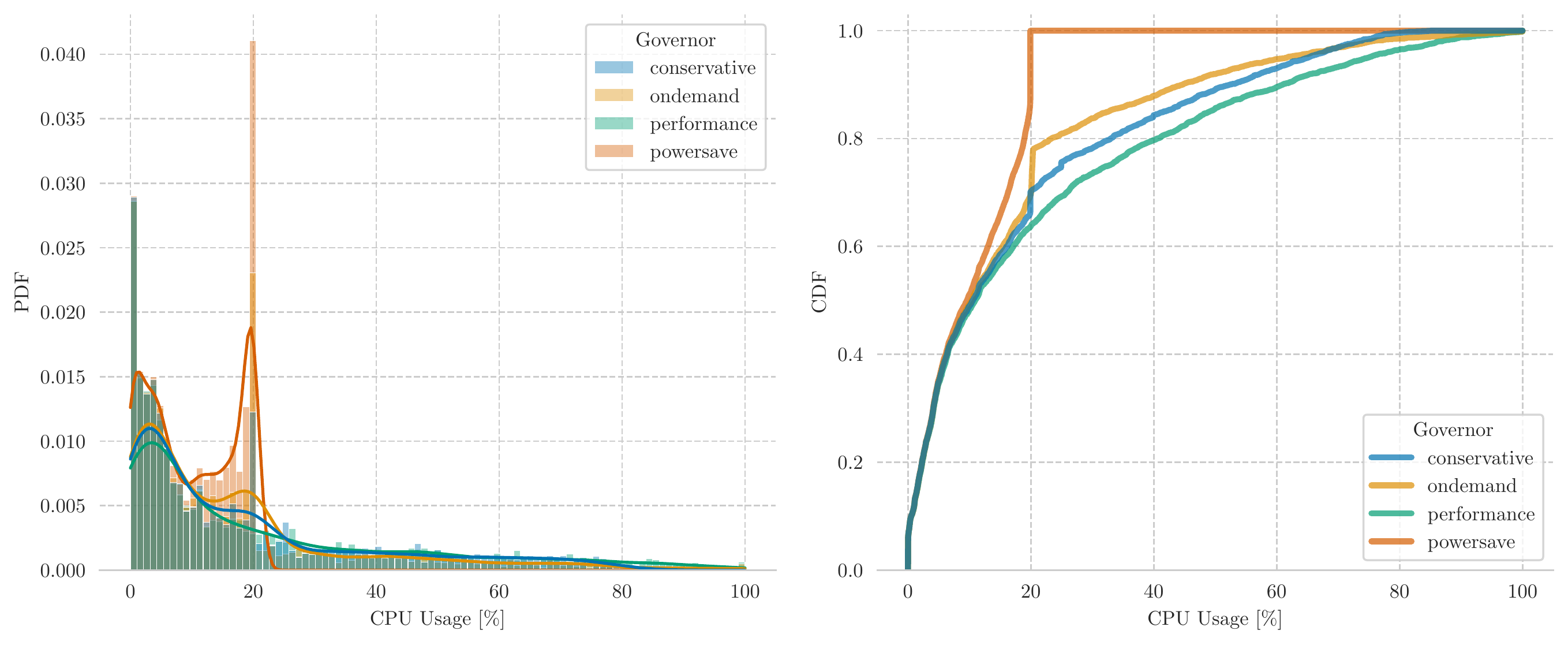}
            \end{adjustbox}
          \caption[PDF and CDF of the CPU usage of different governors]{PDF and CDF of the CPU usage of different scaling governors.}
          \label{fig:usage_pdf_cdf}
\end{figure}

\begin{figure}[!ht]
    \centering
           \begin{adjustbox}{width=1.3\textwidth,center=\textwidth}
             \includegraphics[width=\linewidth]{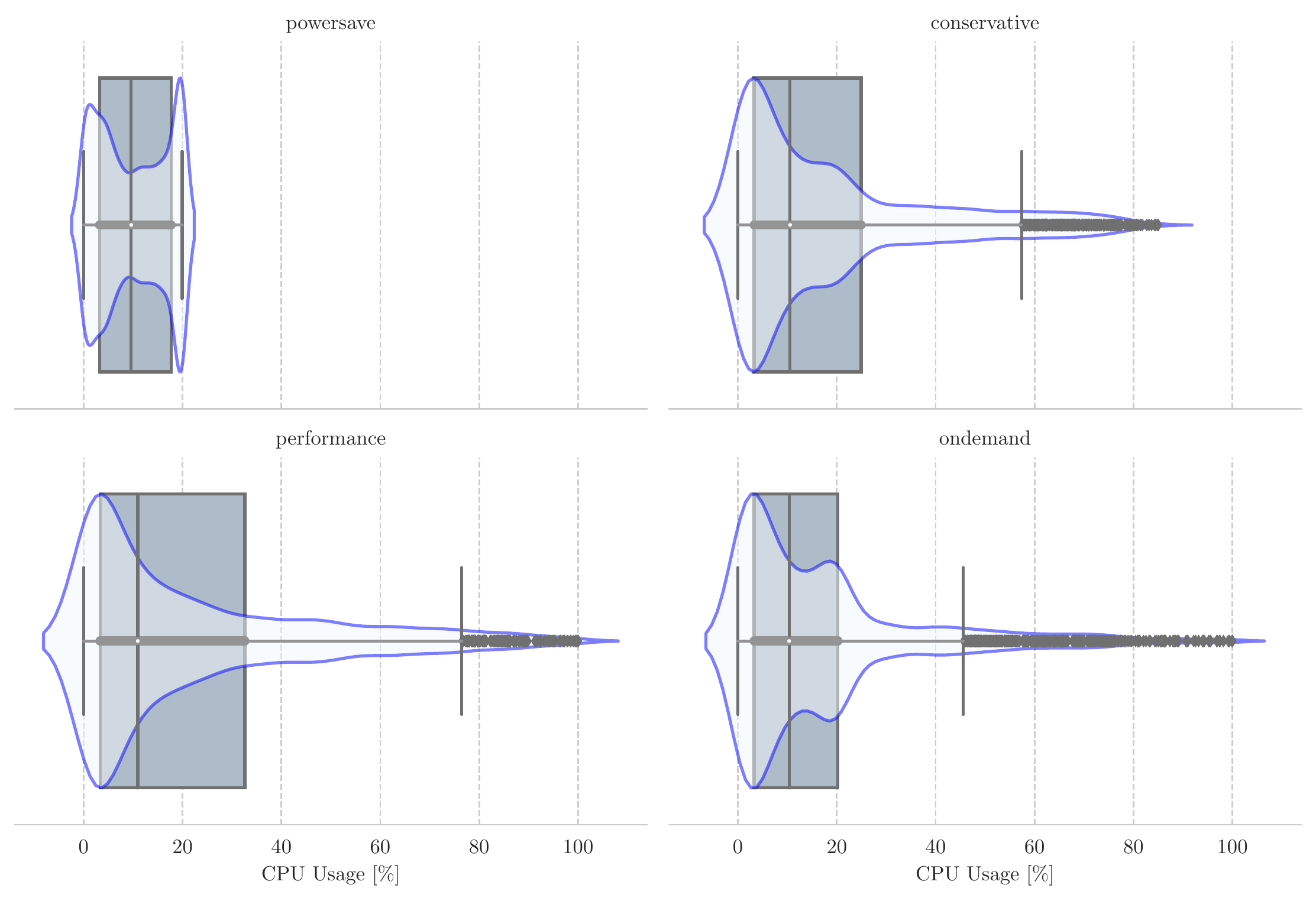}
            \end{adjustbox}          
            \caption[Detailed distributions of CPU usage of different governors]{Detailed distributions of CPU usage of different scaling governors.}
          \label{fig:usage_box}
\end{figure}

\begin{figure}[!ht]
    \centering
    \begin{adjustbox}{width=1.2\textwidth,center=\textwidth}
        \includegraphics[width=\textwidth]{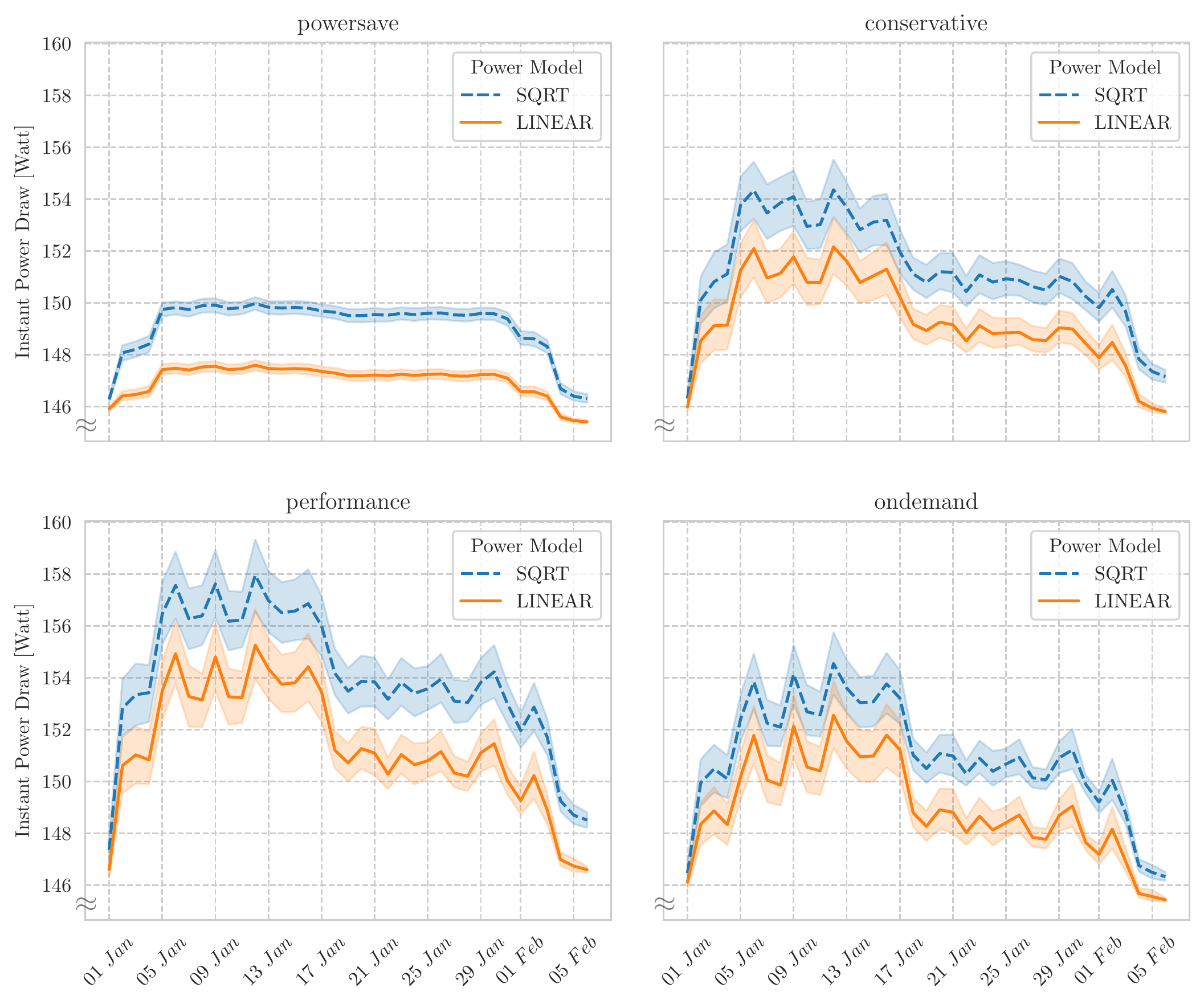}
    \end{adjustbox}
    \caption[Comparison of instant power draw of different governors]{Comparison of instant power draw of different scaling governors estimated by two power models.}
    \label{fig:dvfs_instant_power}
\end{figure}

First of all, we demonstrate the behaviours of the basic DVFS mechanism described in Section \ref{sec:sys_impl}. The left plot of Figure \ref{fig:dvfs_demand_usage} shows the relationships between the CPU demand and the actual CPU usage of the four scaling governors, namely, the \texttt{conservative}, the \texttt{ondemand}, the \texttt{powersave}, and the \texttt{performance} governors. As shown by the dotted green curve, the \texttt{performance} governor tries its best to meet the required computation power, whilst the CPU usage of the \texttt{powersave} is capped under 20\% because of the frequency limit imposed on the CPU. The figure on the right offers a scaled view of that of the right, where the CPU usage is normalized by the CPU demand. The distribution of their CPU usage is further demonstrated by Figure \ref{fig:usage_pdf_cdf}, where the corresponding PDFs and CDFs are presented. As shown in Figure \ref{fig:usage_box}, other than the \texttt{powersave} governor, the other three governors have about the same median value of CPU usage. Also, the \texttt{ondemand} governor exhibits more concentrated CPU usage compared to that of the \texttt{conservative}. Conversely, the CPU usage of the \texttt{conservative} has never reached 100\% since it proposes frequency changes in consecutive small steps as described in Section \ref{sec:dvfs} and \ref{sec:sys_impl}. With regard to the instant power draw shown in Figure \ref{fig:dvfs_instant_power}, the \codeword{conservative} governor exhibits large fluctuations, again because of the gradual adjustments in CPU frequency. In contrast, the power consumption of the \codeword{ondemand} governor forms narrow lanes due to its drastic changes in CPU frequency. As suggested in Figure \ref{fig:dvfs_instant_power}, the behaviours of their instant power draw exhibit similar patterns to that of the CPU usage shown in Figure  \ref{fig:dvfs_demand_usage}.

\newpage

\begin{figure}[!ht]
    \centering
    \begin{adjustbox}{width=1.2\textwidth,center=\textwidth}
        \includegraphics[width=\textwidth]{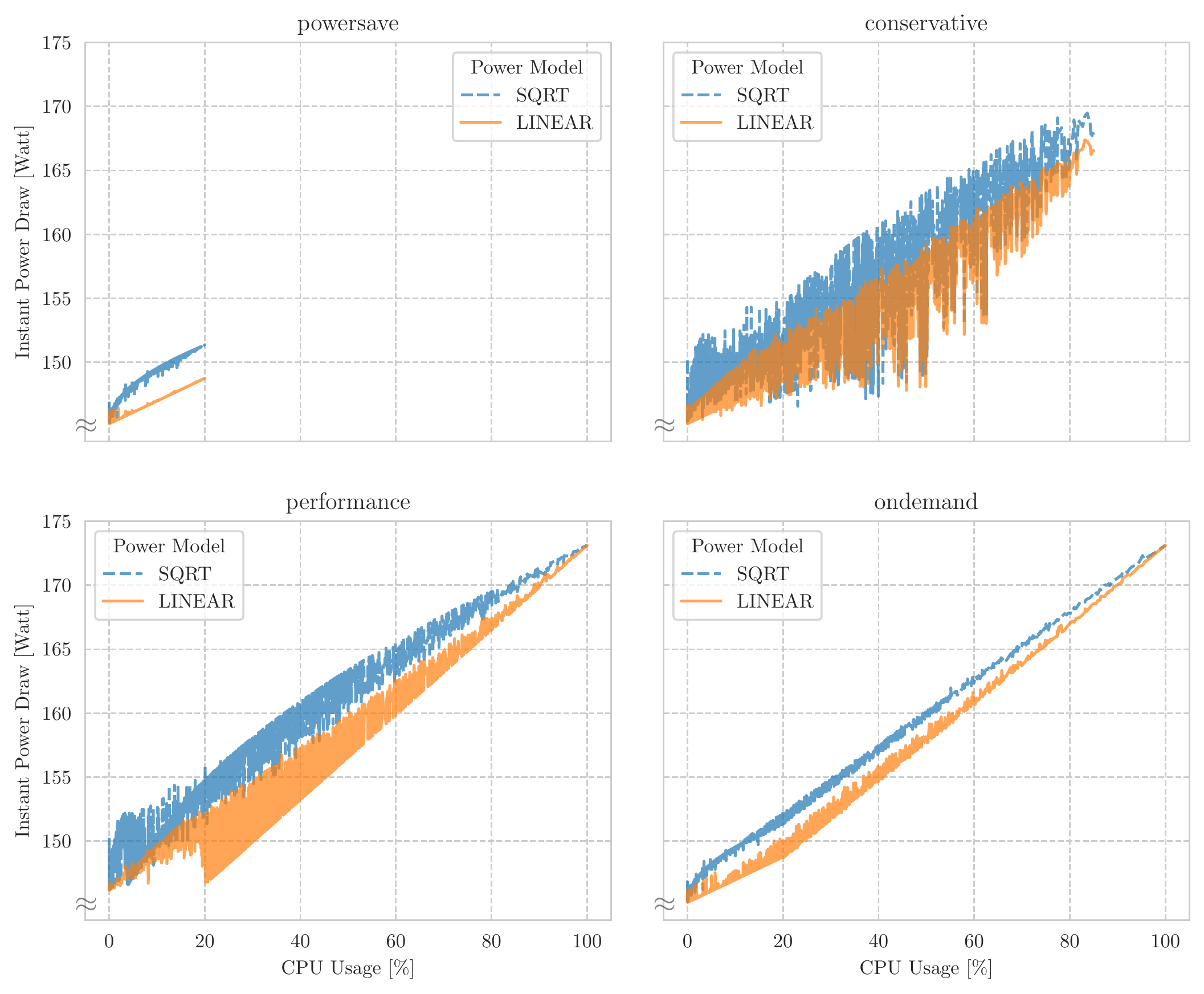}
    \end{adjustbox}
    \caption[Comparison of power estimation of the two models for governors]{Comparison of power estimation of the two models for the four scaling governors.}
    \label{fig:dvfs_usage_power}
\end{figure}

\newpage

\newpage

\begin{figure}[!t]
    \centering
    \begin{adjustbox}{width=0.8\textwidth,center=\textwidth}
        \includegraphics[width=\textwidth]{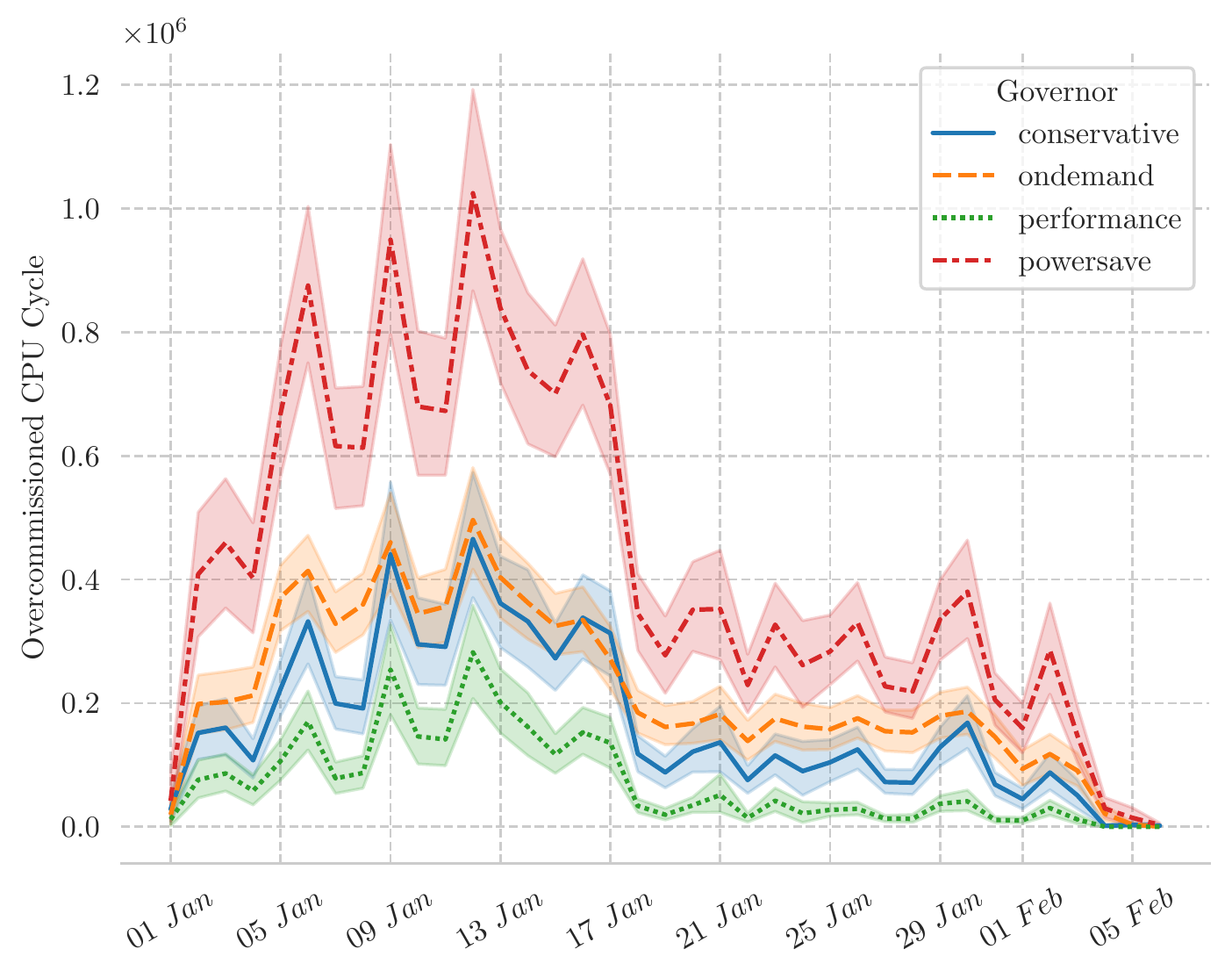}
    \end{adjustbox}
    \caption[Instant CPU over-commission level of different scaling governors]{Instant CPU over-commission level of the four scaling governors.}
    \label{fig:dvfs_oc}
\end{figure}

When it comes to CPU over-commission, however, the behaviour pattern is exactly the opposite of that of the CPU usage. As shown in Figure \ref{fig:dvfs_oc}, the \codeword{powersave} governor has a much higher CPU over-commission compared to the others. In contrast, the \codeword{performance} governor has the lowest level since it does not stress the CPU speed whatsoever. Moreover, to further understand the behaviour of CPU over-commission, we examine its relationship with other factors. 

\begin{figure}[!ht]
    \centering
    \caption[Over-commission at different CPU usage levels]{Over-commission at different CPU usage levels of the four scaling governors.}
    \begin{adjustbox}{width=1.35\textwidth,center=\textwidth}
        \includegraphics[width=\textwidth]{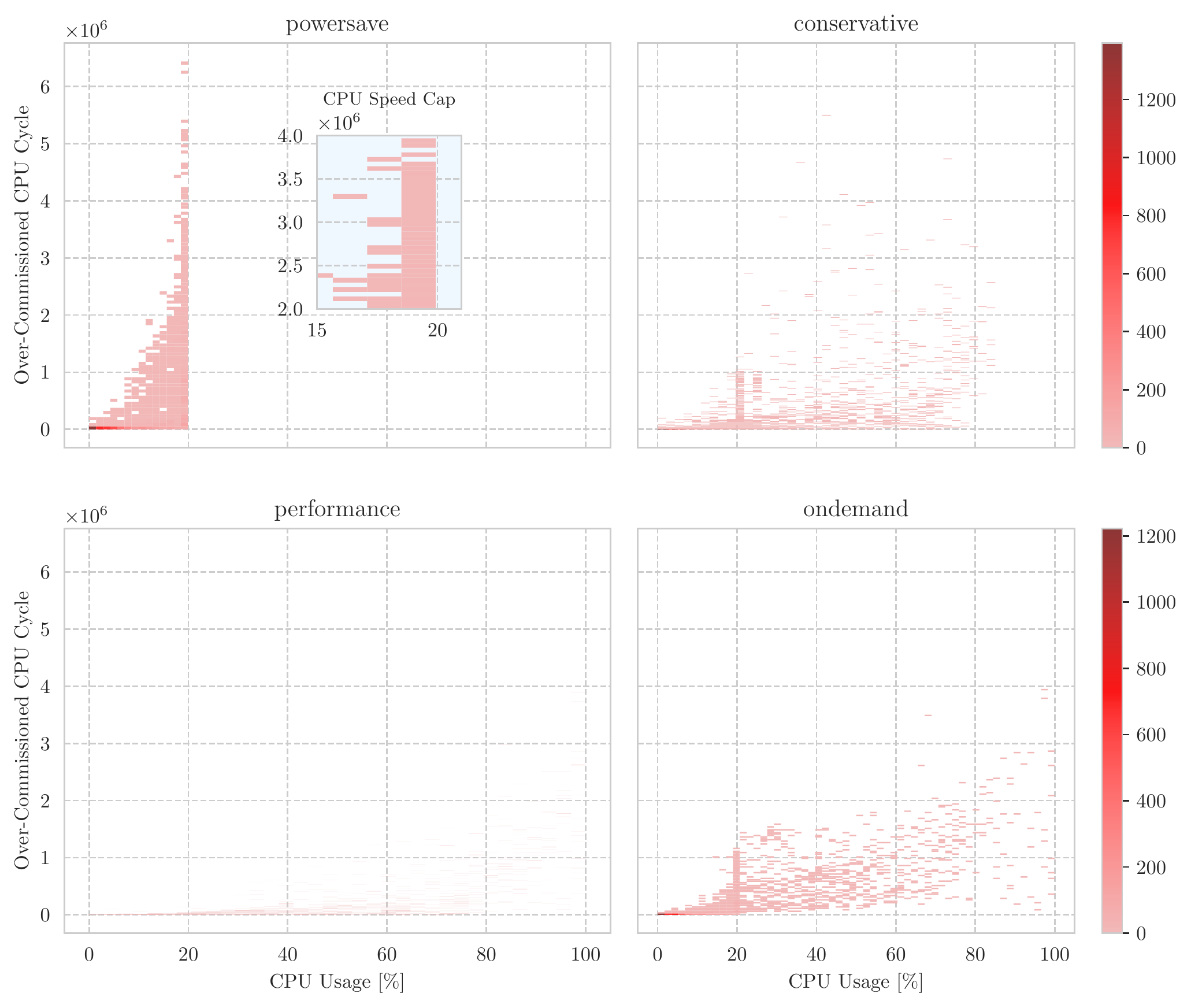}
    \end{adjustbox}
    \label{fig:dvfs_oc_usage}
\end{figure}

Firstly, Figure \ref{fig:dvfs_oc_usage} shows the relationship between CPU usage and over-commissioned CPU cycles. The \codeword{performance} governor barely exhibits any trace of over-commission across different levels of CPU usage, whilst other governors, especially, the \codeword{powersave} governor, demonstrate significant CPU over-commission. Due to the speed cap applied, the over-commission level of the \codeword{powersave} governor reaches a dramatic level at around 20\% of CPU usage. Furthermore, referring back to Figure \ref{fig:usage_box}, the \codeword{ondemand} governor has narrower interquartile range (IQR) then that of the \codeword{conservative}. Consequently, even though the overall range of CPU usage is larger in the case of \codeword{ondemand} governor than the \codeword{conservative}, \codeword{ondemand} exhibits higher levels of over-commission. Thus, the more concentrated the CPU usage (\ie the narrower the IQR), the higher level of CPU over-commission.

\begin{figure}[!ht]
    \centering
    \caption[Instant power draw at different over-commission levels]{Instant power draw at different over-commission levels of the four scaling governors.}
    \begin{adjustbox}{width=1.35\textwidth,center=\textwidth}
        \includegraphics[width=\textwidth]{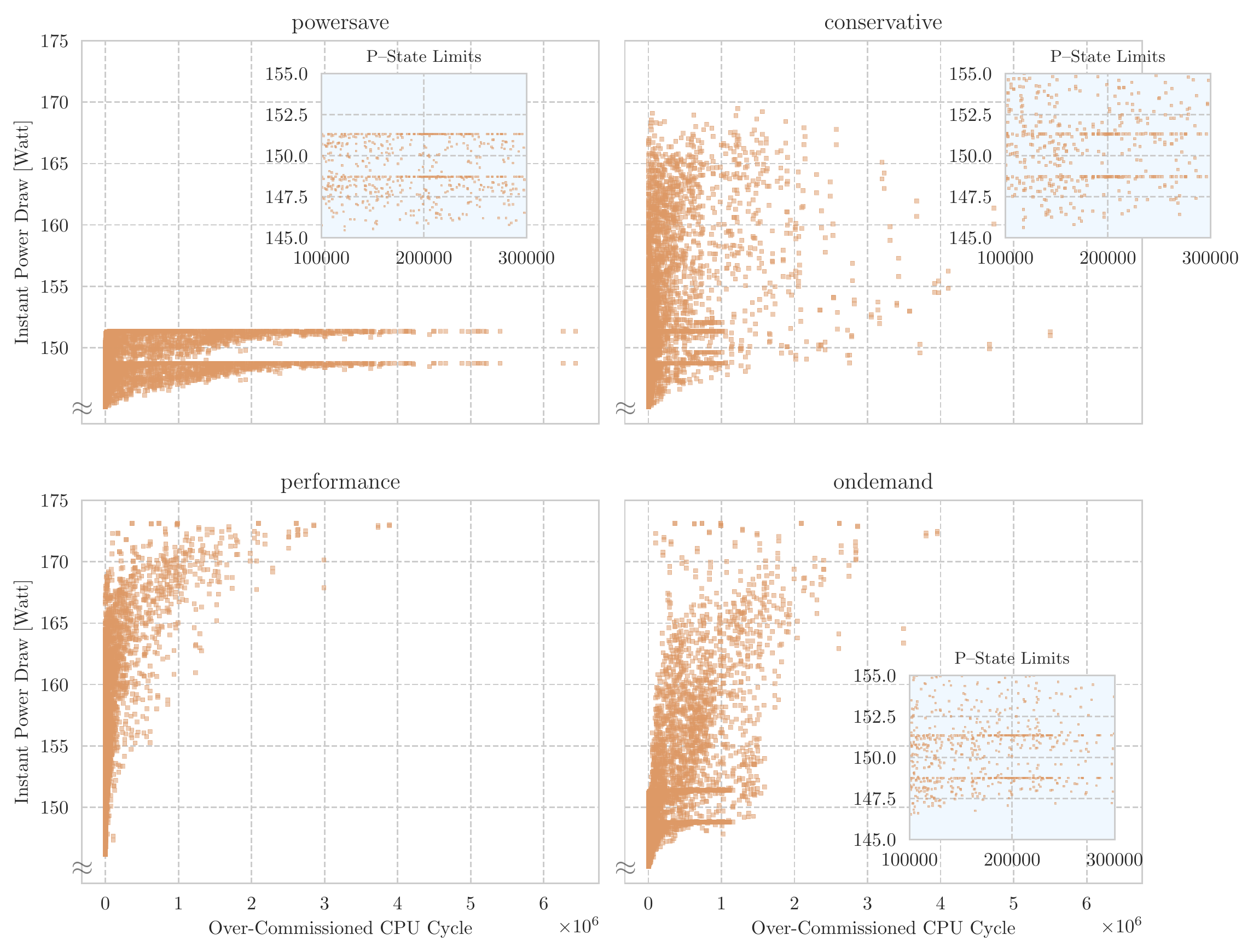}
    \end{adjustbox}
    \label{fig:dvfs_oc_power}
\end{figure}

Secondly, Figure \ref{fig:dvfs_oc_power} illustrates the relationship between the over-commission and the instant power draw. Although the \codeword{powersave} governor has a much higher over-commission level, its power is capped because of the frequency limit. Since both the \codeword{performance} and the \codeword{ondemand} governors exploit the full range of CPU usage (0--100\%), their instant power draw both reaches about 175 W. In contrast, as the CPU usage of the \codeword{conservative} is mostly below 80\% (Figure \ref{fig:usage_box}), its instant power is also restrained by about 170 W. Therefore, the level of power draw is \textit{not} directly associated with the CPU over-commission but depends upon the actual CPU usage. Additionally, due to the higher level of over-commission, governors other than the \codeword{performance} squeeze out the power steps caused by various discrete P-states.

\begin{figure}[!ht]
    \centering
    \caption[Comparison of granted work and instant power draw]{Comparison of granted work and instant power draw of the four scaling governors.}
    \begin{adjustbox}{width=1.35\textwidth,center=\textwidth}
        \includegraphics[width=\textwidth]{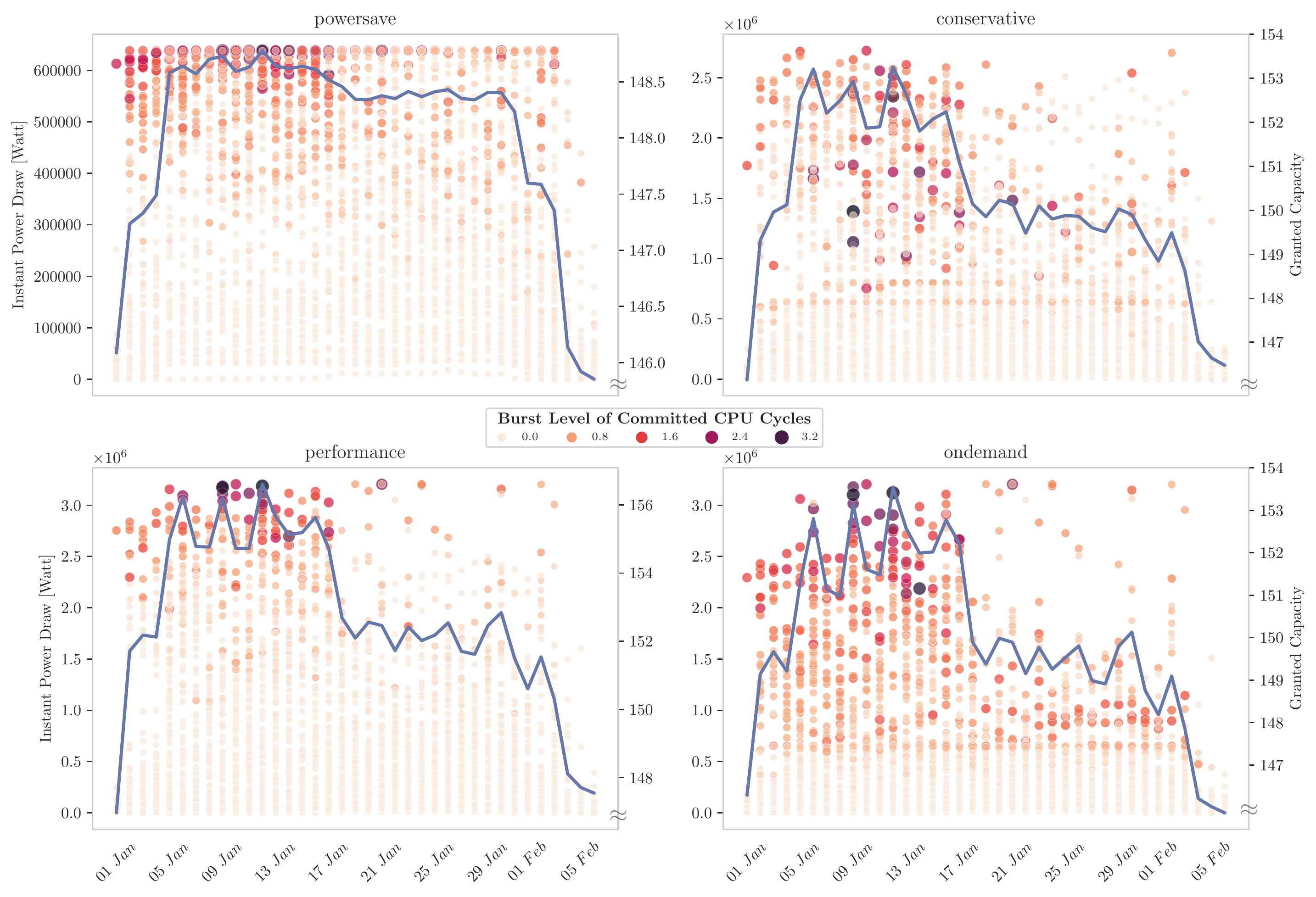}
    \end{adjustbox}
    \label{fig:dvfs_granted}
\end{figure}

Lastly, rather than looking at over-commission, we move on to the actual work committed by the CPU in Figure \ref{fig:dvfs_granted}. Different from over-commission, there is a clear connection between the amount of committed work and the level of instant power draw. In other words, the more work the CPU commits, the higher power the system requires.

\clearpage
\newpage
\newpage
\newpage

\subsection{Damping Factor} \label{sec:damping_factor}

\begin{figure}[!t]
    \centering
    \begin{adjustbox}{width=1\textwidth,center=\textwidth}
        \includegraphics[width=\textwidth]{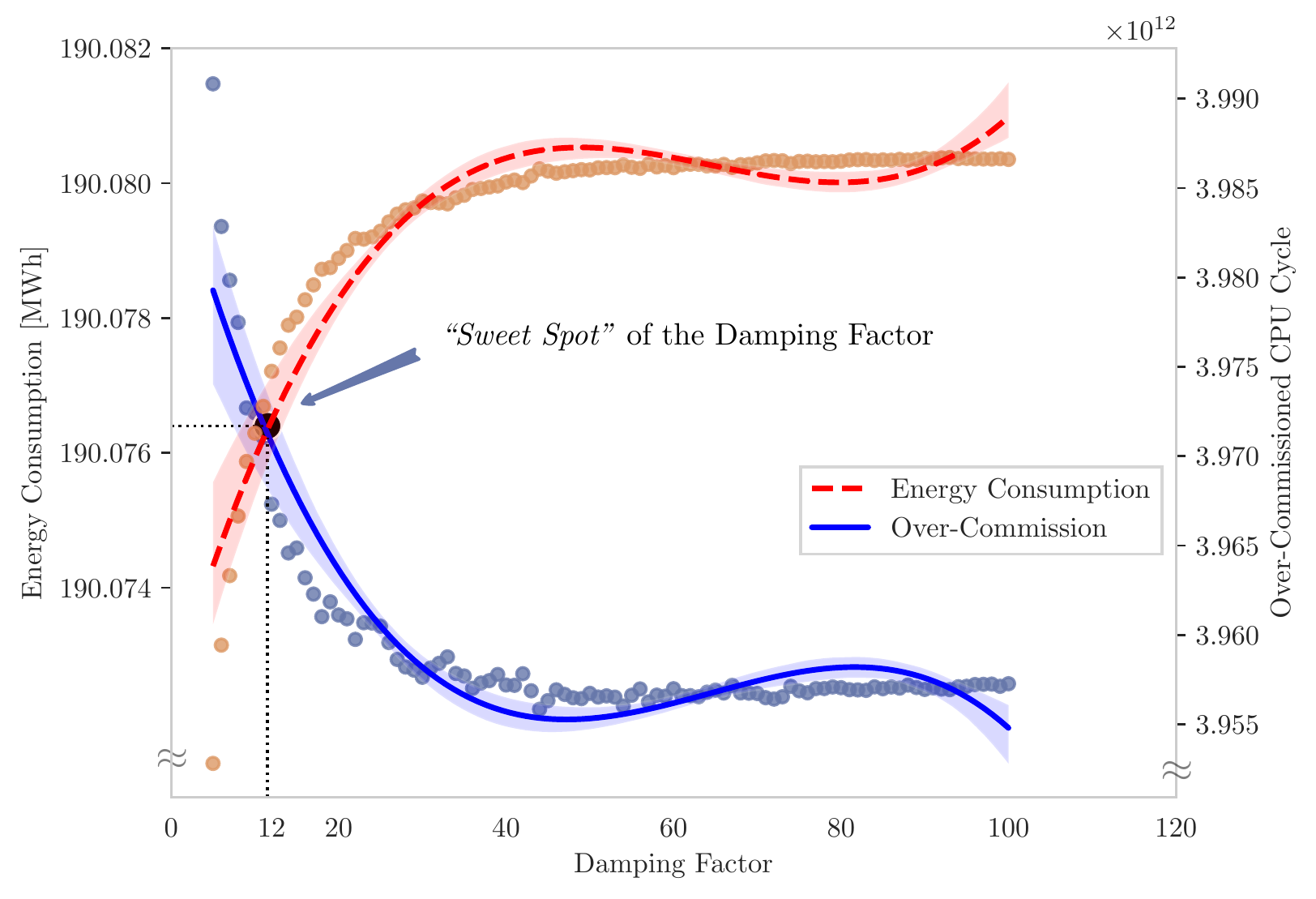}
    \end{adjustbox}
    \caption[Effect of the damping factor on energy use and over-commission]{Effect of the damping factor on energy use and CPU over-commission with third order regression.}
    \label{fig:damping_factor}
\end{figure}

Having studied the behaviours of the DVFS, we now focus on its proactive scheduler implemented in the EEMM extension. As described in Section \ref{sec:eemm_impl}, the damping factor is the hyperparameter responsible for ameliorating the stress imposed on the CPU. To this end, users can use the damping-factor plot shown in Figure \ref{fig:damping_factor} to strike a balance between the energy-saving and the level of CPU over-commission. Specifically, users determine the damping factor according to their desired level of energy saving and the acceptable level of CPU over-commission. In our case, the damping factors with which we have experimented run in the range of 0 to 110. As suggested in Figure \ref{fig:damping_factor}, both the energy consumption and the level of over-commission have converged when the damping factor is greater than $\sim$80. Then, we conduct a third-order regression on both the energy consumption (the red, dotted curve) and the CPU over-commission (the blue, solid curve). Since we do not assume a specific target for either of the two, we set the damping factor to the “Sweet Spot”, the intersection between the two regression lines marked by the black dot in the figure, which is about 12. To reiterate, finding the “Sweet Spot” is not the only way, certainly not necessarily the best way, for determining the damping factor. Instead, users ought to customize it according to their needs regarding energy saving and the CPU over-commission.

\subsection{Bounded Comparison} \label{sec:bounded}

\begin{figure}[!t]
    \centering
    \begin{adjustbox}{width=1.1\textwidth,center=\textwidth}
        \includegraphics[width=\textwidth]{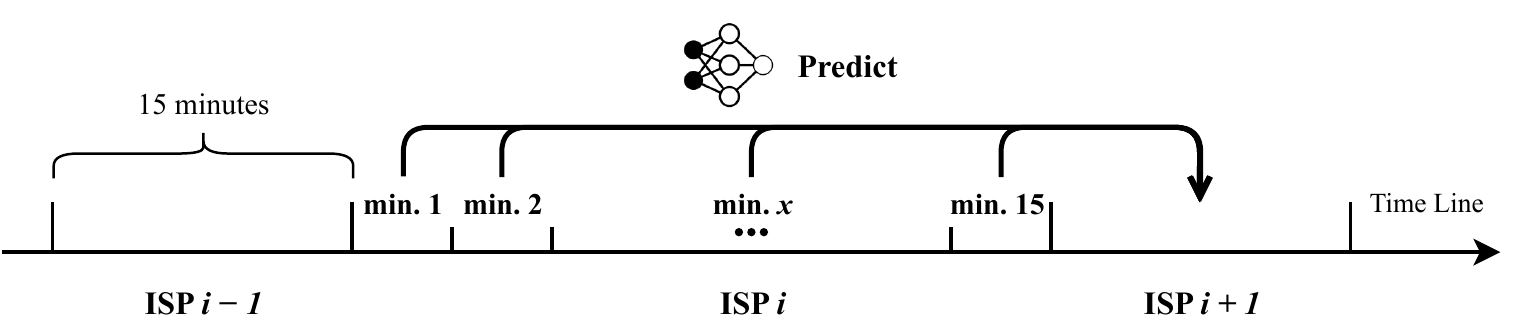}
    \end{adjustbox}
    \caption[Timeline of ML prediction]{Timeline of ML prediction (abbreviation: min. = minute).}
    \label{fig:ml_timeline}
\end{figure}

In this section, we describe the use of ML methods to further optimize the profit for datacenters when participating in the day-ahead and the balancing market. First and foremost, we elaborate on how the ML inferences are obtained. Referring back to Section \ref{sec:grid_markets}, unlike the day-ahead market, trading happens in the balancing market every 15 minutes. These trading intervals are often referred to as imbalance settlement periods (ISPs) and, in turn, there are 96 consecutive ISPs per trading day. As shown in Figure \ref{fig:ml_timeline}, we predict the energy price of the next ISP (\textbf{ISP $i\ +\ 1$}) during the current ISP (\textbf{ISP $i$}) on a minute-by-minute basis. In the following sections, the predictions produced in the first minute are referred to as the first ML inferences and the last ones are referred to as the last ML inferences. Also, the average predictions of every ISP are referred to as the average ML inferences. Furthermore, it is worth noting that the closer it gets to the next ISP, the more accurate the ML inferences are, but the shorter the time that allows datacenter operators to adjust their optional schedule accordingly. Thus, there is a trade-off between the performance of the ML methods and the operational leeway that datacenter operators have.

\subsubsection{Defining Metric}

\input{resources/tables/tab_sym_aa}

To measure the performance of the ML methods, we define a metric base upon the needs of the DVFS scheduler. Referring back to Algorithm \ref{algo:schedule}, there are two decision points at line \ref{line:1st_decision} and line \ref{line:2nd_decision} respectively. The first decision point is to check if the imbalance price is directly profitable. If it is not, in the second decision point, we compare the price level of the balancing market with that of the day-ahead market to determine whether or not to further suppress the CPU frequency. Based upon these two decision points, we define the agreement accuracy ($AA$) score by Equation \ref{eq:aa} as a measure of the performance of the ML methods. The $\mathbb{S}$ function  (Equation \ref{eq:sign}) checks the sign of the input value, and the $\mathds{1}$ is the indicator function. Table \ref{tab:sym_aa} present the meaning of the used symbols.

\begin{equation}
    \mathbb{S}(x) = 
        \begin{cases}
            +1 & x > 0\\
            0 & x = 0\\
            -1 & x <0
        \end{cases}
    \label{eq:sign}
\end{equation}

\begin{equation}
\mathlarger{
    AA = \cfrac{ \sum^{N_\text{ISP}}_{i} 
                \mathlarger{\mathds{1}\left\{
                    \mathds{1}\left[\mathbb{S}(p_i^B) = \mathbb{S}(p_i^F)\right]
                    = 
                    \mathds{1}\left[\mathbb{S}(p_i^B - p_i^S) = \mathbb{S}(p_i^F - p_i^S) \right]
                    \right\}}
            } 
             {N_\text{ISP}}
}
\label{eq:aa}
\end{equation}


\subsubsection{Synthetic Predictor} \label{sec:synthetic}

\begin{figure}[!t]
    \centering
    \begin{adjustbox}{width=0.9\textwidth,center=\textwidth}
        \includegraphics[width=\textwidth]{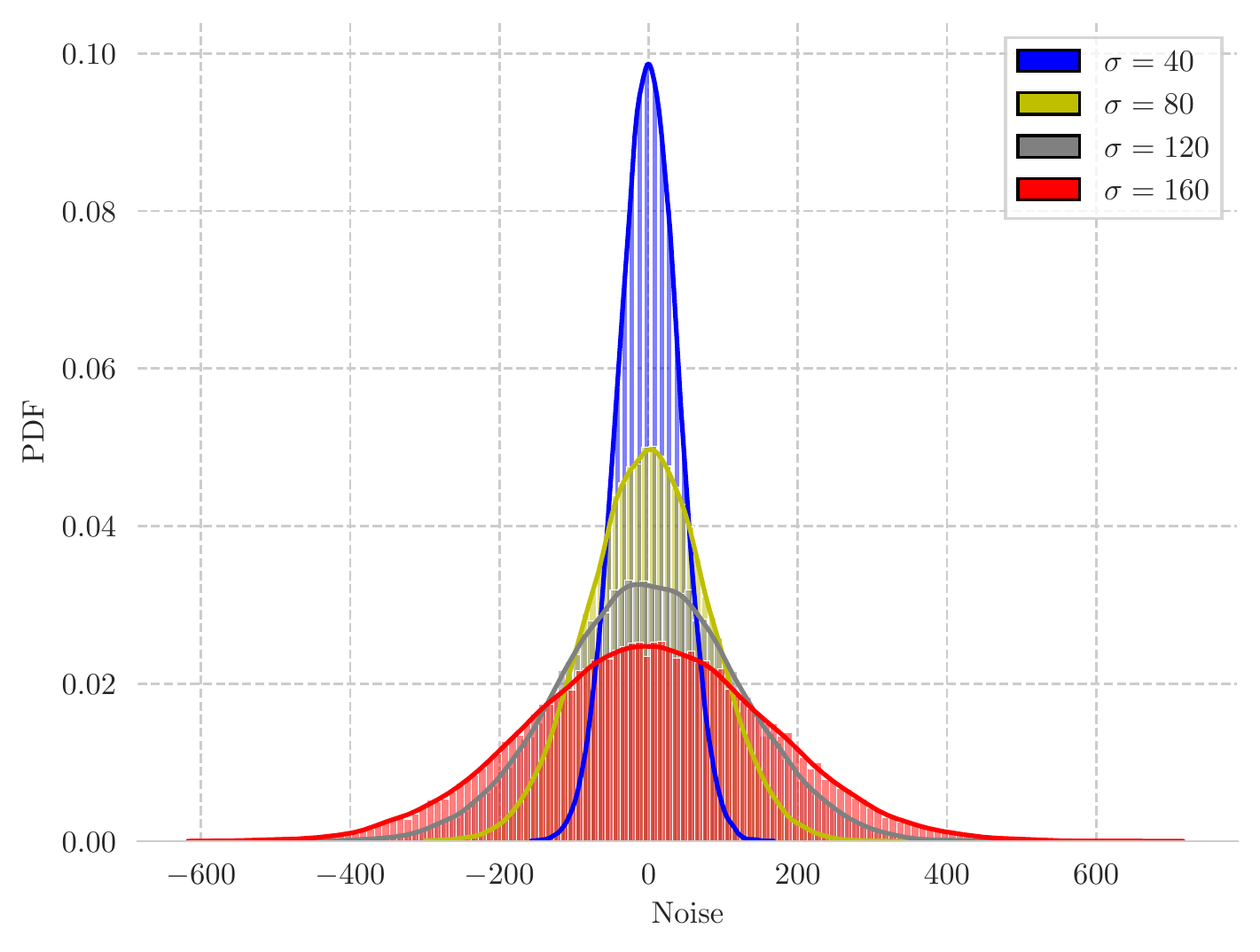}
    \end{adjustbox}
    \caption[Distributions of Gaussian noises with different $\sigma$ values]{Distributions of Gaussian noises with different $\sigma$ values.}
    \label{fig:noise_dist}
\end{figure}

To achieve bounded evaluation (\ref{nfr3}), we first construct a set of synthetic predictors by adding Gaussian noises to the \textit{actual} imbalance prices as follows:

\begin{equation}
\mathlarger{
        p^f = p^B + E
    }, \quad E \sim \mathlarger{\mathcal{N}}(0, \sigma) = \cfrac{1}{{\sigma \sqrt {2\pi } }}\cdot e^{-\frac{1}{2} \left(\frac{x}{\sigma}\right)^2}, 
\label{eq:gaussian}
\end{equation}

\noindent where $E$ is an addictive random variable representing the error term that follows the Gaussian distribution $\mathlarger{\mathcal{N}}(0, \sigma)$. Referring back to Figure \ref{fig:energy_markets}, the imbalance prices are capped by {800} \texteuro. Thus, we set $\sigma \in [0, 1200]$ in the following experiments to ensure that variation of errors is sufficiently large. 

\begin{figure}[!t]
    \centering
    \begin{adjustbox}{width=0.9\textwidth,center=\textwidth}
        \includegraphics[width=\textwidth]{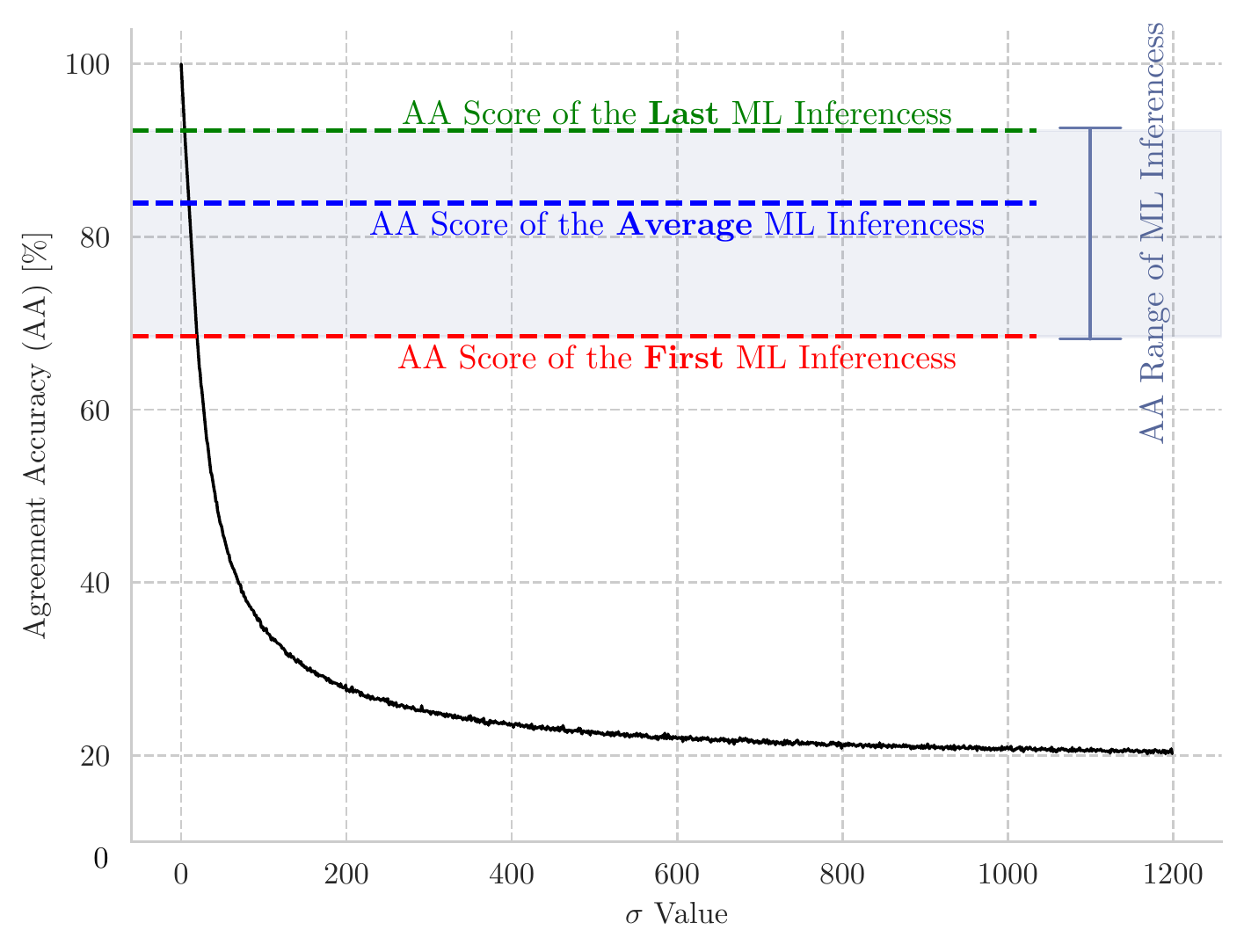}
    \end{adjustbox}
    \caption[$AA$ scores of ML and synthetic predictors]{Comparison of $AA$ scores of ML methods and synthetic predictors at different $\sigma$ levels.}
    \label{fig:sigma_aa}
\end{figure}

Figure \ref{fig:sigma_aa} shows the trending of the $AA$ scores of the ML methods in percentage and the synthetic predictors as the $\sigma$ value increases. As previously described, the later the ML inferences are produced, the better their performance. Indeed, the first ML inferences achieve the best $AA$ score, which is about 0.92. The $AA$ score of average ML inferences comes next (0.84), followed by the score of the first ML inferences, which is about 0.68. In respect of the synthetic predictors, when $\sigma = 0$, the $AA$ score is a perfect 1.0. It exhibits a plunge when $\sigma \in [0, 200]$. Then, as $\sigma$ continues to decrease, the $AA$ scores of the synthetic predictors level off and converge to 0.20 at last.

\begin{figure}[!t]
    \centering
    \begin{adjustbox}{width=1.1\textwidth,center=\textwidth}
        \includegraphics[width=\textwidth]{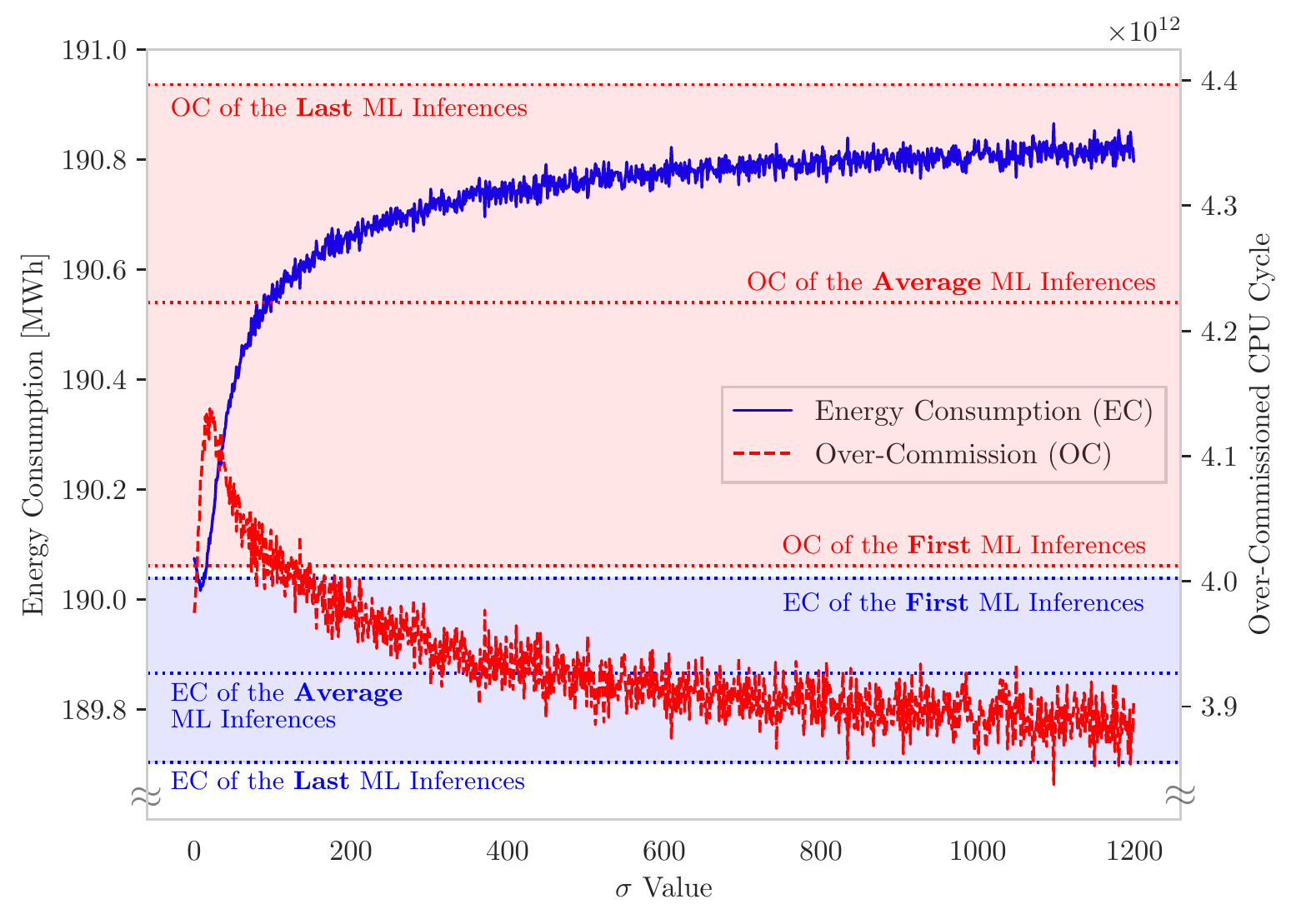}
    \end{adjustbox}
    \caption[Energy and over-commission of ML and synthetic predictors]{Comparison of energy consumption and over-commission between ML methods and synthetic predictors at different $\sigma$ levels.}
    \label{fig:sigma_perf}
\end{figure}

Next, we turn our attention to energy consumption and the level of CPU over-commission. Figure \ref{fig:sigma_perf} shows the comparison of energy consumption and over-commission between ML methods and synthetic predictors at different $\sigma$ levels. Regarding the ML methods, the last ML inferences optimize energy consumption over over-commission, which, in turn, leads to the lowest energy use but also the highest over-commission level amongst the ML methods. On the contrary, the first ML inferences leverage over-commission level over energy consumption, which results in the highest energy consumption and the lowest over-commission level. The ML method of the average inferences achieves average values in both metrics compared to the ML methods that use the first and the last inferences. Regarding the synthetic predictors, almost all of them, even the best synthetic predictor ($\sigma=0$), is not able to achieve the same level of energy-saving as that of the worst performed ML methods, the one using the first ML inferences. When it comes to the level of over-commission, however, these synthetic predictors perform particularly well, in the sense that when $\sigma >\ \sim 100$, none of the ML methods can reduce the over-commission to the same level as that of the synthetic predictors. Hence, we conclude that the scheduler powered by the ML methods is good at reducing the energy consumption but perform poorly in reducing the over-commission level, compared to using the synthetic predictors.

\clearpage

\begin{figure}[!t]
    \centering
    \begin{adjustbox}{width=0.9\textwidth,center=\textwidth}
        \includegraphics[width=\textwidth]{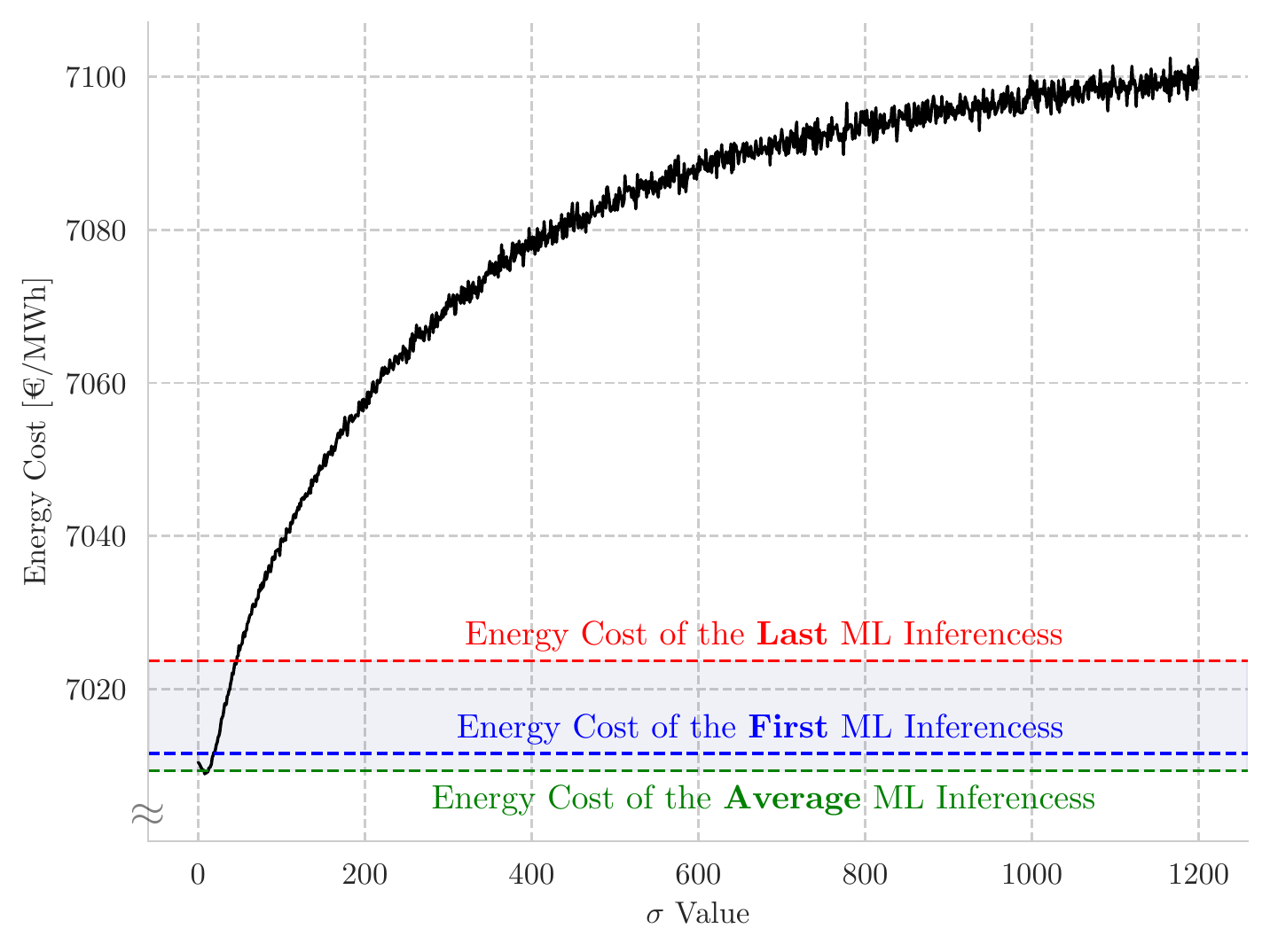}
    \end{adjustbox}
    \caption[Energy costs of ML methods and synthetic predictors]{Comparison of energy costs between ML methods and synthetic predictors at different $\sigma$ levels.}
    \label{fig:sigma_cost}
\end{figure}

Now, we focus on the total energy cost. Figure \ref{fig:sigma_cost} demonstrates the comparison of energy costs between ML methods and synthetic predictors at different $\sigma$ levels. Overall, compared to the synthetic predictors, the ML methods are excellent at leveraging the energy cost because even the best synthetic predictor ($\sigma=0$) is bounded by the best ML method. Furthermore, what counts most is that the best-performed ML method in saving cost is \textit{not} the one that saves the most energy, the one using the last ML inferences, but instead, the ML method that employs the average values achieves the lowest energy cost. This observation further hammers home the paramount importance of consuming the right amount of energy at the right time.

\clearpage

\subsubsection{Indirect Demand Response}

\begin{figure}[!t]
    \centering
    \begin{adjustbox}{width=1.1\textwidth,center=\textwidth}
        \includegraphics[width=\textwidth]{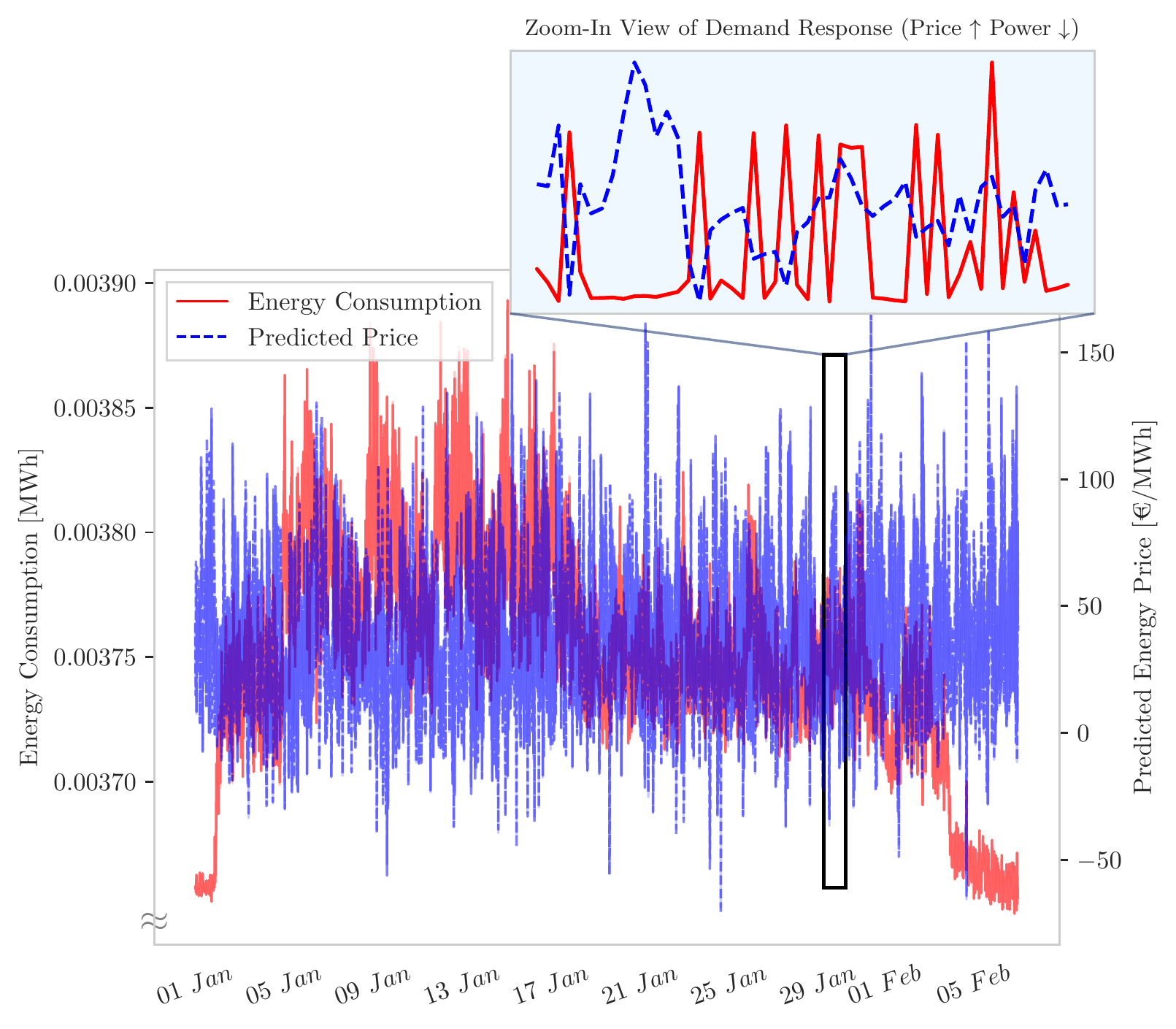}
    \end{adjustbox}
    \caption[Demonstration of indirect demand response]{Demonstration of indirect demand response, where the energy-consumption level is adjusted in response to the predicted energy cost from a synthetic predictor of $\sigma=50$.}
    \label{fig:dr}
\end{figure}

Last, but not certainly not least, we look into the indirect DR resulted from the proactive DVFS scheduling. In Figure \ref{fig:dr}, we visualize the energy consumption using the solid, red line together with the predicted energy price produced by a synthetic predictor of arbitrary $\sigma=50$ using a blue, dotted line. As we zoom in to a short timeframe, we can see the gaps between the two lines: when the energy is high, the scheduler tries to reduce the amount of energy consumed and vice versa. Note that, because of the actual demand and the effect of the damping factor, such gaps may not always be as obvious.

\subsection{Summary} \label{sec:scheduling_sum}

\begin{figure}[!t]
    \centering
    \begin{adjustbox}{width=0.9\textwidth,center=\textwidth}
        \includegraphics[width=\textwidth]{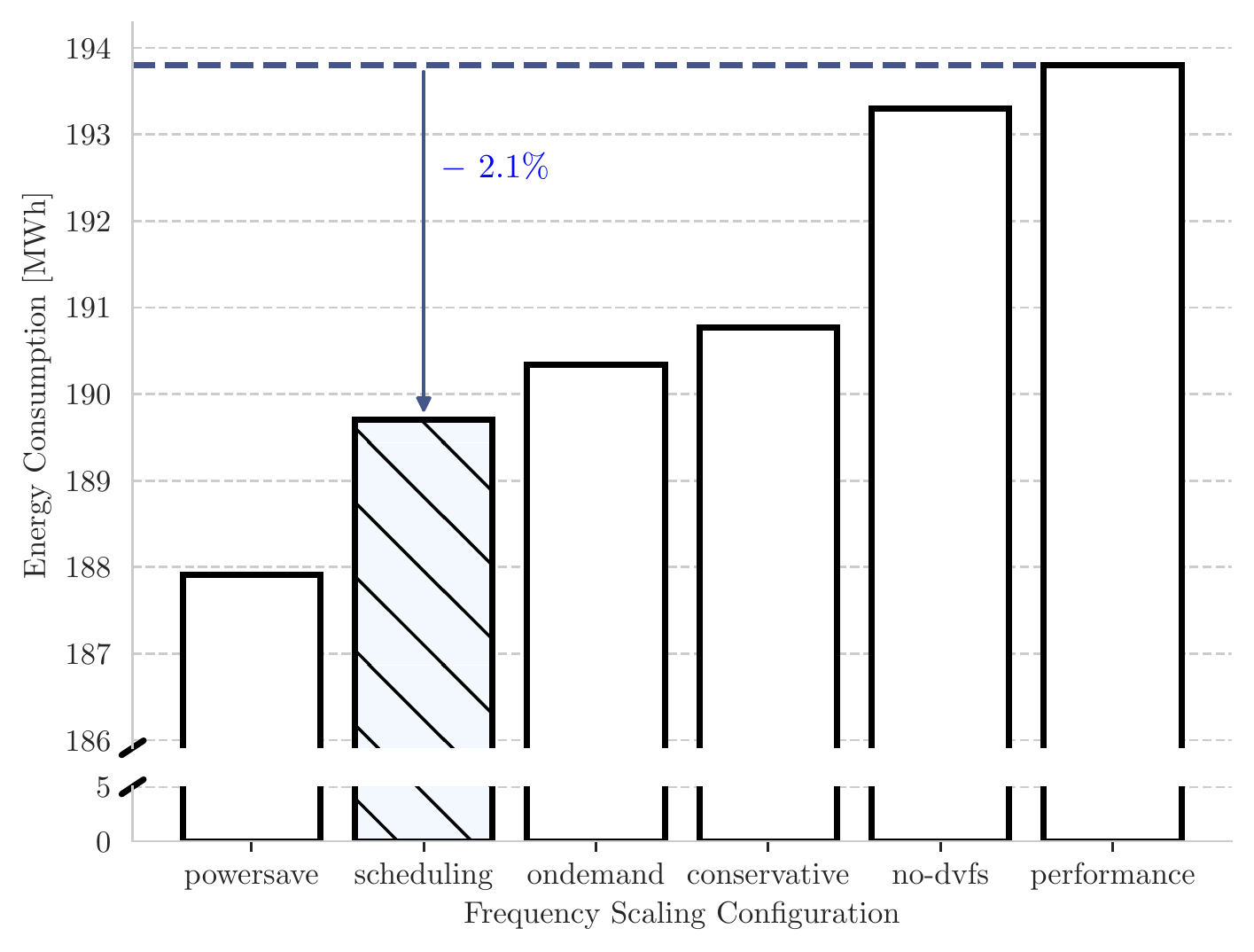}
    \end{adjustbox}
    \caption[Comparison of total energy consumption with DVFS scheduling]{Comparison of total energy consumption for DVFS scheduling.}
    \label{fig:dvfs_perf_energy}
\end{figure}

\begin{figure}[!t]
    \centering
    \begin{adjustbox}{width=0.9\textwidth,center=\textwidth}
        \includegraphics[width=\textwidth]{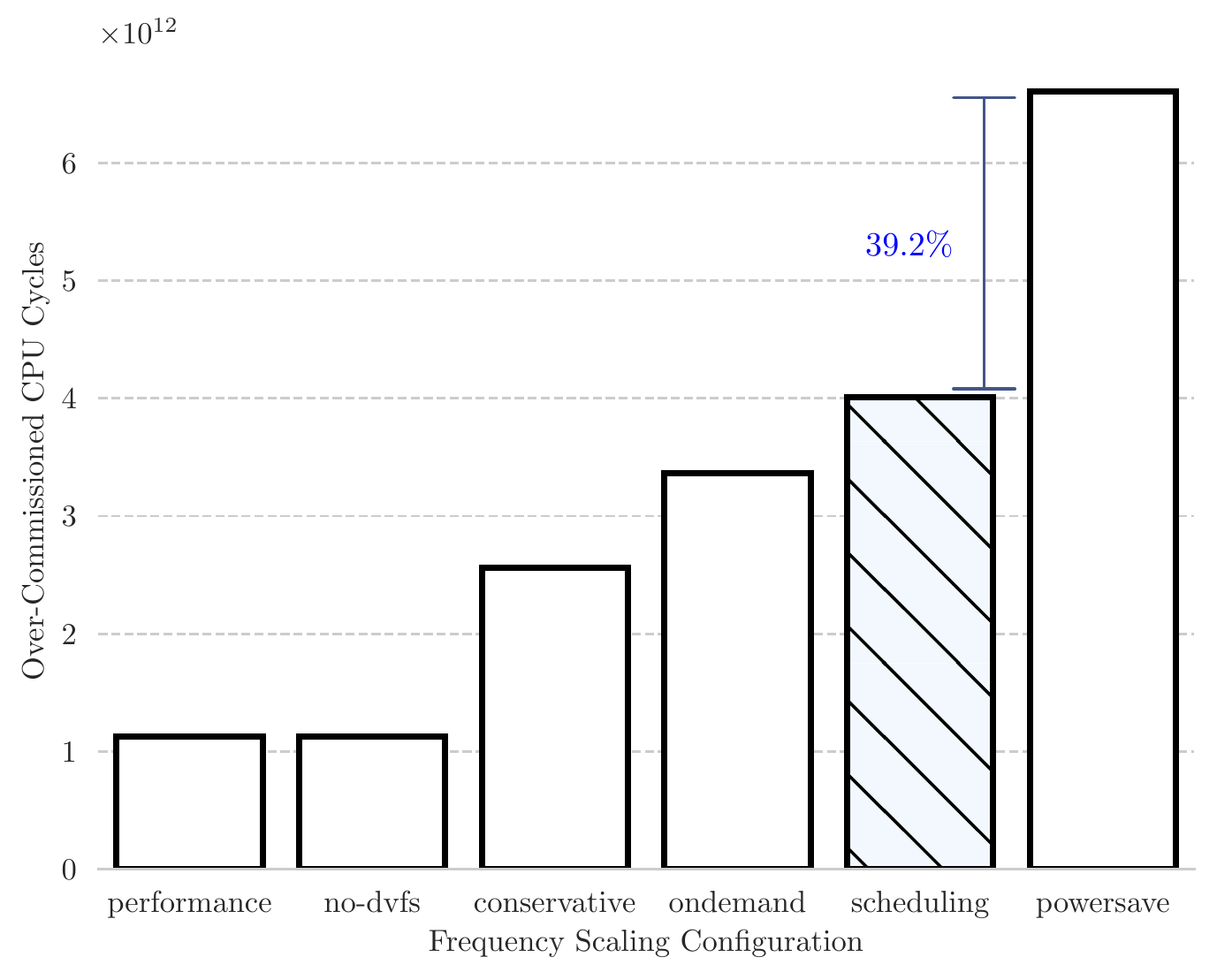}
    \end{adjustbox}
    \caption[Comparison of total CPU over-commission with DVFS scheduling]{Comparison of total CPU over-commission for DVFS scheduling.}
    \label{fig:dvfs_perf_oc}
\end{figure}

\begin{figure}[!t]
    \centering
    \begin{adjustbox}{width=1\textwidth,center=\textwidth}
        \includegraphics[width=\textwidth]{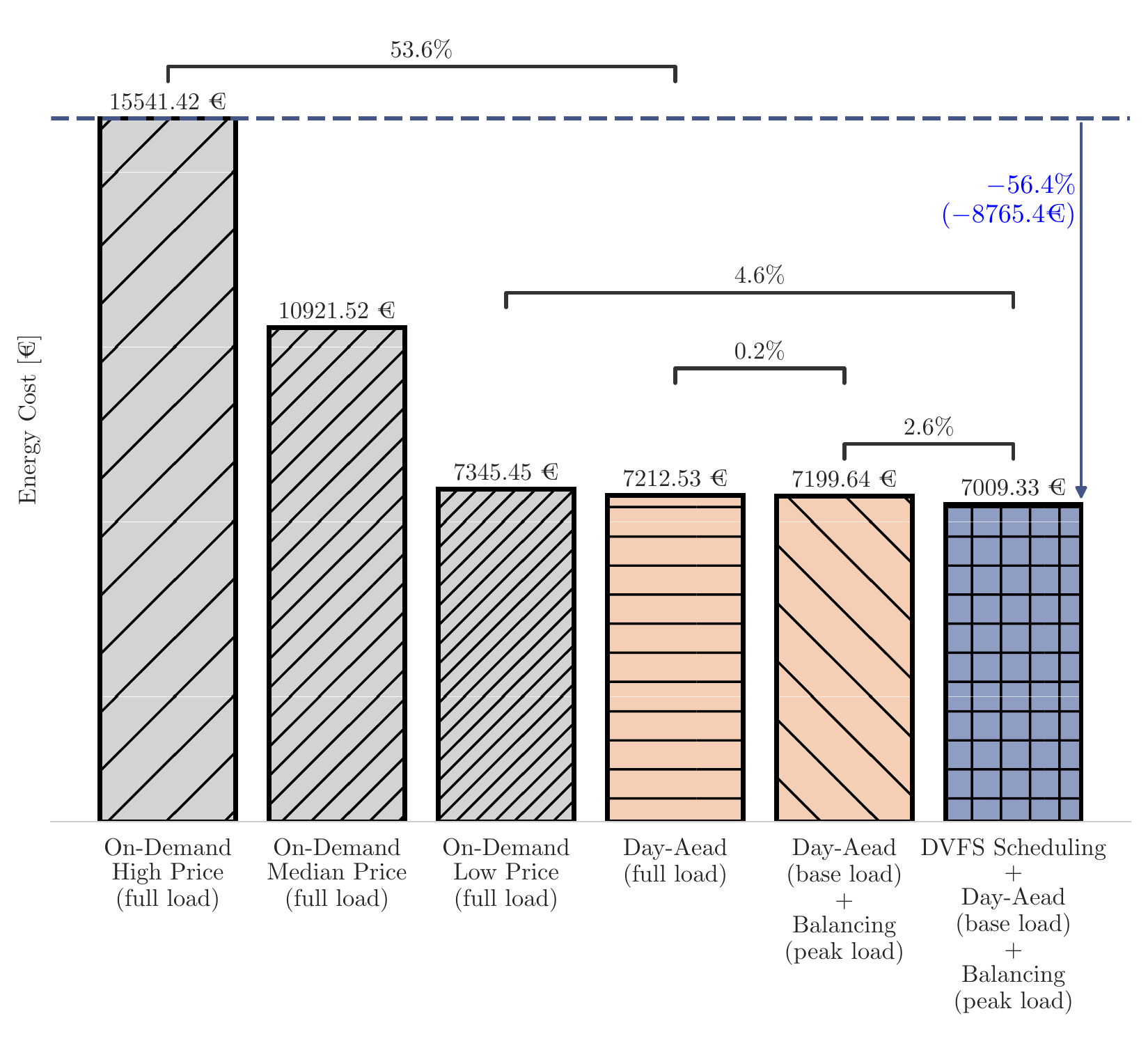}
    \end{adjustbox}
    \caption[Comparison of total energy costs for DVFS scheduling]{Comparison of total energy costs for DVFS scheduling.}
    \label{fig:dvfs_perf_costs}
\end{figure}

In this section, we demonstrate the behaviours of the implemented DVFS in Section \ref{sec:dvfs_behaviour} with a focus on CPU over-commission. Then, we determine the hyperparameter, the damping factor, of the scheduler using the damping factor plot (\S\ref{sec:damping_factor}). After that, by conducting bounded comparisons between the ML methods and the synthetic estimators, we show that the ML methods are good at optimizing the energy consumption and the cost, whereas they are not as performant in curbing the over-commission level. Lastly, we visualize the energy consumption level of the resulting DVFS schedule together with a randomly chosen synthetic predictor, illustrating the indirect DR.  

As shown in Figure \ref{fig:dvfs_perf_energy}, if we take the best results amongst the ML methods, we reduce about 2.1\% energy consumption compared to that of the \codeword{performance} governor. Such an energy saving on this single workload is roughly the equivalent of the annual energy consumption of two Dutch households \cite{cbs_save}. With regard to CPU over-commission, DVFS scheduling powered by the ML methods results in about 39\% improvement compared to that of the \codeword{powersave} governor (Figure \ref{fig:dvfs_perf_oc}). 

Finally, regarding the energy cost, together with the base-load procurement strategy (\ie schedule the bare-minimum energy in the day-ahead market and settle the peak load in the balancing market), the proactive DVFS scheduler powered by ML methods achieves a 2.6\% of reduction in energy cost compared to the case where it is disabled. Moreover, the cost reduction is about 56.4\% compared to the on-demand scheme of high price. 


In fact, we believe the results are rather \textit{conservative} because they are produced by the \textit{old} machine model whose peak load only takes up a tiny portion of its overall power load (Figure \ref{fig:new_costs}). Furthermore, the variations in power load between energy states of the old machine model are much smaller than that of the new machine model, as illustrated in Figure \ref{fig:new_loads}. In other words, the fraction of energy cost that can be leveraged is minuscule (\ref{fig:costs}). Therefore, the advantage of the proactive DVFS scheduling has not been fully exploited. Hence, if a newer machine model were available, the above results would be expected to be much more significant.

%% file: resources/tables/tab_sym_aa.tex
\begin{table}[H]
    \centering
    \begin{tabular}{ll}
    \toprule
        $N_\text{ISP}$ & Number of ISPs\\
        $p^S$ & Spot price in the day-ahead market \\
        $p^B$ & Shortage price in the balancing market \\
        $p^F$ & Forecasted imbalance price\\
    \bottomrule
    \end{tabular}
    \caption[Symbols used in defining the $AA$ score]{Symbols used in defining the $AA$ score.}
    \label{tab:sym_aa}
\end{table}

%% file: chapters/7-conclusion/7_conclusion.tex
\chapter{Conclusion} \label{cha:conclusion}

In this last chapter, we first summarize all conclusions drawn from previous chapters to provide answers to research questions \ref{rq1} to \ref{rq5} in Section \ref{sec:answers}. Then, we reflect on the limitations of this work (\S\ref{sec:limitations}) as well as lay out potential directions of future work (\S\ref{sec:future_work}).

\section{Answers to Research Questions} \label{sec:answers}

\paragraph*{Answering \ref{rq1}.} In this work, we extend the OpenDC simulator with advanced energy models so that it is able to model the whole power system of a typical datacenter in a flexible and highly customizable way. In Section \ref{sec:architecture}, we present the design of the energy management and modelling system, and in Section \ref{sec:sys_impl}, we describe the detailed implementations of the system and its subsystems.

\paragraph*{Answering \ref{rq2}.} To answer this and the following \textbf{RQ}s, we first develop a market extension of the energy management and modelling system. Then, by using this tool, we show that there is a strong financial incentive for datacenters to participate in both the day-ahead market and the balancing market, as substantial profit can be derived by doing so (\S\ref{sec:energy_costs}).

\paragraph*{Answering \ref{rq3}.} By simulating different load-forecast-based procurement strategies in Section \ref{sec:procurement}, we demonstrate that the most economical strategy is the base-load strategy, \ie scheduling only the bare-minimum amount of energy, the base load, in the day-ahead market and resolve the peak load in the balancing market.

\paragraph*{Answering \ref{rq4}.} Firstly, we develop a proactive DVFS scheduler powered by ML methods that can save energy whilst curbing the level of CPU over-commission (\S\ref{sec:eemm_impl}). In Section \ref{sec:scheduling}, we employ the DVFS scheduler to make fine-grained decisions in order to leverage the profit when participating in the energy market. Further, we carry out bounded comparisons between the ML methods and synthetic predictors (\S\ref{sec:synthetic}), showing that the ML methods are excellent at reducing both the energy consumption and the cost, whilst not as performant in curbing the CPU over-commission. Lastly, in Section \ref{sec:scheduling_sum}, we demonstrate that the DVFS scheduler is able to save about 2.1\% of energy compared to the \codeword{performance} governor and to improve about 39.2\% of CPU over-commission compared to the \codeword{powersave} governor. Such an energy saving on this single workload is roughly the equivalent of the annual energy consumption of two Dutch households (\S\ref{sec:scheduling_sum}). Moreover, together with the base-load procurement strategy, the scheduler is able to save about 56.4 \% energy cost compared to the on-demand scheme of high price. Furthermore, we conclude in Section \ref{sec:scheduling_sum} that these results are \textit{conservative} and are expected to be more significant on newer machine models.

\paragraph*{Answering \ref{rq5}.} To meet the requirements of this RQ, we follow a carefully designed development pipeline from the onset (\S\ref{sec:pipeline}) and conduct rigorous requirement engineering in Section \ref{sec:req_eng}. The software artefacts are created and maintained using strict software engineering methods. As a result, both the datacenter simulator and its extension EEMM can be run with a few clicks and/or simple shell commands, without requiring excessive prerequisite knowledge or any manual data preprocessing from users.

\section{Limitations} \label{sec:limitations}

Whilst acknowledging the promising results, we recognize the limitations thereof as well. In this section, we elaborate on the limitations and their corresponding mitigations.

\paragraph*{Internal Limitation.} In our experiments, we do not assume any specific machine types or computing platforms except for the new machine model. Consequently, the estimation of energy consumption may not be particularly representative. To mitigate this limitation, we use the knowledge gained from our previous studies, employing the linear model and the square-root model as the lower and upper bound respectively. In this way, the variation of the total energy consumption is bounded to a certain range.

\paragraph*{Construct Limitation.} As introduced in Section \ref{sec:power_grid}, the energy demand is inelastic. Thus, the market prices, albeit a good indicator and predictor, may not be able to satisfactorily capture the demand-supply balance in the power grid at all times. Nevertheless, as introduced in Chapter \ref{cha:intro}, this is the inherent disadvantage of the indirect DR approach. As the functions of the smart grid and the design of the energy market advance, we believe the energy prices will become increasingly responsive.

\paragraph*{External Limitations.} Although we strive to make our scientific tools more user-friendly and easy to use, users can still use them incorrectly and, in turn, produce invalid results. For example, the market data from ENTSO-E is in CET, whilst data from TenT is in GMT, so users may well feed the software with inputs that have mismatched timestamps. To mitigate this limitation, we write detailed documentation and make tutorials with examples \footnote{\url{http://opendc-eemm.rtfd.io}} \footnote{\url{https://github.com/atlarge-research/opendc\#documentation}}.

\section{Future Work \label{sec:future_work}}

According to our industry partners, energy planning in datacenters using simulation and modelling has been quickly gaining popularity as higher degrees of volatility caused by renewable energy sources are introduced to the market. Yet, the intersection between the energy market and datacenter simulation is still far from been fully explored. In this section, we identify potential directions of future research by means of questions; the following is a non-exhaustive list of such questions in no particular order.

\begin{enumerate}
    \item How feasible and beneficial is it for individual datacenters to serve as BSPs instead of BRPs?
    \item How can we develop a convenient and liable tool to measure P-state consumption levels, which will enable flexible experiments on more machine models?
    \item What is the impact of employing not only the energy price but also the frequency level of the power grid (with less frequent grid monitoring and communication) in datacenters’ decision-making on their market participation?
    \item How to improve the algorithms of resource allocation (\eg power distribution, VM scaling/placement, etc.) in response to market signals?
    \item What is the effect of core-level P-state frequency scaling on datacenters' participation in DR programmes?
    \item How can we improve the design of the energy market to incentivize datacenters' active participation?
    \item How can we optimize the direct participation in multiple markets across several, geographically distributed datacenters?
    \item How to orchestrate redundancies of datacenters (\eg PSU, UPS, etc.) to provide indirect DR?
    \item How can we tune the proactive scheduler base upon specifications in the SLA?
\end{enumerate}

\section{Summary}

In this work, we first model the entire power system of a quintessential datacenter. By conducting simulation on real-world traces, we then demonstrate the substantial financial incentive for individual datacenters to directly participate in the energy market, specifically, the day-ahead and the balancing markets. In turn, we suggest a new short-term, direct scheme of market participation for individual datacenters in place of the current long-term, inactive participation. Furthermore, we develop a novel proactive DVFS scheduling algorithm that is able to both reduce energy consumption and save the energy cost for datacenters when participating in the energy market. Also, in developing this scheduler, we propose an innovative combination of ML methods and the DVFS technology that is able to provide the power grid with indirect DR in an effort to combat the challenges brought by renewable energy sources.

Besides the aforementioned potential societal and economic impacts, we develop and open source scientific tools, bridging the gap between domain knowledge, and sharing our data and experimental results with the community without reservation. Last, but certainly not least, with all these efforts, we believe that we have opened a new research line; as such, we call for collaborations from both the industry and academia to help datacenters actively join the smart grid to tackle the climate crisis together.

%% file: main.bbl
\begin{thebibliography}{186}
\providecommand{\natexlab}[1]{#1}
\providecommand{\url}[1]{\texttt{#1}}
\expandafter\ifx\csname urlstyle\endcsname\relax
  \providecommand{\doi}[1]{doi: #1}\else
  \providecommand{\doi}{doi: \begingroup \urlstyle{rm}\Url}\fi

\bibitem[acp()]{acpi}
\textbf{ACPICA: ACPI Component Architecture}.
\newblock Available at \url{https://www.acpica.org/}.

\bibitem[Aalami et~al.(2008)Aalami, Yousefi, and Moghadam]{aalami2008demand}
H~Aalami, GR~Yousefi, and M~Parsa Moghadam.
\newblock \textbf{Demand response model considering EDRP and TOU programs}.
\newblock In \emph{2008 IEEE/PES Transmission and Distribution Conference and
  Exposition}, pages 1--6. IEEE, 2008.

\bibitem[Abbasi et~al.(2010)Abbasi, Varsamopoulos, and
  Gupta]{Abbasi2010ThermalAS}
Zahra Abbasi, Georgios Varsamopoulos, and Sandeep K.~S. Gupta.
\newblock \textbf{Thermal aware server provisioning and workload distribution
  for internet data centers}.
\newblock In \emph{HPDC '10}, 2010.

\bibitem[Andreadis et~al.(2021)Andreadis, Mastenbroek, van Beek, and
  Iosup]{Andreadis2021CapelinDC}
George Andreadis, F.~Mastenbroek, Vincent van Beek, and A.~Iosup.
\newblock \textbf{Capelin: Data-Driven Compute Capacity Procurement for Cloud
  Datacenters Using Portfolios of Scenarios}.
\newblock \emph{IEEE Transactions on Parallel and Distributed Systems},
  33:\penalty0 26--39, 2021.

\bibitem[Bag and Bassiouni(2006)]{Bag2006EnergyET}
Anirban Bag and Mostafa~A. Bassiouni.
\newblock \textbf{Energy Efficient Thermal Aware Routing Algorithms for
  Embedded Biomedical Sensor Networks}.
\newblock \emph{2006 IEEE International Conference on Mobile Ad Hoc and Sensor
  Systems}, pages 604--609, 2006.

\bibitem[Bajracharya et~al.(2018)Bajracharya, Khan, Michael, and
  Tonkoski]{Bajracharya2018ForecastingDC}
Abhilasha Bajracharya, Md~Riaz~Ahmed Khan, Semhar Michael, and R.~Tonkoski.
\newblock \textbf{Forecasting Data Center Load Using Hidden Markov Model}.
\newblock \emph{2018 North American Power Symposium (NAPS)}, pages 1--5, 2018.

\bibitem[Baldick(2018)]{Baldick2018IncentivePO}
R.~Baldick.
\newblock \textbf{Incentive properties of coincident peak pricing}.
\newblock \emph{Journal of Regulatory Economics}, 54:\penalty0 165--194, 2018.

\bibitem[Bambrik(2020)]{Bambrik2020ASO}
Ilyas Bambrik.
\newblock \textbf{A Survey on Cloud Computing Simulation and Modeling}.
\newblock \emph{SN Comput. Sci.}, 1:\penalty0 249, 2020.

\bibitem[Barroso et~al.(2018)Barroso, H{\"o}lzle, and
  Ranganathan]{Barroso2018TheDA}
L.~Barroso, Urs H{\"o}lzle, and P.~Ranganathan.
\newblock \textbf{The Datacenter as a Computer: Designing Warehouse-Scale
  Machines, Third Edition}.
\newblock In \emph{The Datacenter as a Computer}, 2018.

\bibitem[Barroso and H{\"o}lzle(2007)]{Barroso2007TheCF}
Luiz~Andr{\'e} Barroso and Urs H{\"o}lzle.
\newblock \textbf{The Case for Energy-Proportional Computing}.
\newblock \emph{Computer}, 40, 2007.

\bibitem[Barroso and H{\"o}lzle(2008)]{Barroso2008TheDA}
Luiz~Andr{\'e} Barroso and Urs H{\"o}lzle.
\newblock \textbf{The Datacenter as a Computer: An Introduction to the Design
  of Warehouse-Scale Machines}.
\newblock In \emph{The Datacenter as a Computer: An Introduction to the Design
  of Warehouse-Scale Machines}, 2008.

\bibitem[Bates et~al.(2014)Bates, Ghatikar, Abdulla, Koenig, Bhalachandra,
  Sheikhalishahi, Patki, Rountree, and Poole]{Bates2014ElectricalGA}
N.~Bates, G.~Ghatikar, G.~Abdulla, et~al.
\newblock \textbf{Electrical Grid and Supercomputing Centers: An Investigative
  Analysis of Emerging Opportunities and Challenges}.
\newblock \emph{Informatik-Spektrum}, 38:\penalty0 111--127, 2014.

\bibitem[Bellosa(2000)]{Bellosa2000TheBO}
Frank Bellosa.
\newblock \textbf{The benefits of event: driven energy accounting in
  power-sensitive systems}.
\newblock In \emph{EW 9}, 2000.

\bibitem[Beloglazov et~al.(2012)Beloglazov, Abawajy, and
  Buyya]{Beloglazov2012EnergyawareRA}
A.~Beloglazov, J.~Abawajy, and R.~Buyya.
\newblock \textbf{Energy-aware resource allocation heuristics for efficient
  management of data centers for Cloud computing}.
\newblock \emph{Future Gener. Comput. Syst.}, 28:\penalty0 755--768, 2012.

\bibitem[Beloglazov et~al.(2010)Beloglazov, Buyya, Lee, and
  Zomaya]{Beloglazov2010ATA}
Anton Beloglazov, Rajkumar Buyya, Young~Choon Lee, and Albert~Y. Zomaya.
\newblock \textbf{A Taxonomy and Survey of Energy-Efficient Data Centers and
  Cloud Computing Systems}.
\newblock \emph{Advances in Computers}, 82:\penalty0 47--111, 2010.

\bibitem[Bezjak et~al.(2018)Bezjak, Clyburne-Sherin, Conzett, Fernandes,
  G{\"o}r{\"o}gh, Helbig, Kramer, Labastida, Niemeyer, Psomopoulos,
  et~al.]{bezjak2018open}
Sonja Bezjak, April Clyburne-Sherin, Philipp Conzett, et~al.
\newblock \textbf{The open science training handbook}.
\newblock Technical report, [sn], 2018.

\bibitem[Boru et~al.(2013)Boru, Kliazovich, Granelli, Bouvry, and
  Zomaya]{Boru2013EnergyefficientDR}
Dejene Boru, Dzmitry Kliazovich, Fabrizio Granelli, et~al.
\newblock \textbf{Energy-efficient data replication in cloud computing
  datacenters}.
\newblock In \emph{GLOBECOM Workshops}, 2013.

\bibitem[Brooks et~al.(2000)Brooks, Tiwari, and Martonosi]{Brooks2000WattchAF}
D.~Brooks, V.~Tiwari, and M.~Martonosi.
\newblock \textbf{Wattch: a framework for architectural-level power analysis
  and optimizations}.
\newblock \emph{Proceedings of 27th International Symposium on Computer
  Architecture (IEEE Cat. No.RS00201)}, pages 83--94, 2000.

\bibitem[Brown et~al.(2008{\natexlab{a}})Brown, Incorporated, and
  Incorporated]{Brown2008ReportTC}
R.~Brown, Icf Incorporated, and Erg Incorporated.
\newblock \textbf{Report to Congress on Server and Data Center Energy
  Efficiency: Public Law 109-431}.
\newblock \emph{Lawrence Berkeley National Laboratory}, 2008{\natexlab{a}}.

\bibitem[Brown et~al.(2008{\natexlab{b}})Brown, Masanet, Nordman, Tschudi,
  Shehabi, Stanley, Koomey, Sartor, and Chan]{58670}
Richard~E. Brown, Eric~R. Masanet, Bruce Nordman, et~al.
\newblock \textbf{Report to Congress on Server and Data Center Energy
  Efficiency: Public Law 109-431}.
\newblock Technical report, Berkeley, CA, 06/2008 2008{\natexlab{b}}.

\bibitem[Bucek et~al.(2018)Bucek, Lange, and Kistowski]{Bucek2018SPECCN}
James Bucek, K.~Lange, and J.~Kistowski.
\newblock \textbf{SPEC CPU2017: Next-Generation Compute Benchmark}.
\newblock \emph{Companion of the 2018 ACM/SPEC International Conference on
  Performance Engineering}, 2018.

\bibitem[Butler(2018)]{Butler2018ClimateCH}
C.~Butler.
\newblock \textbf{Climate Change, Health and Existential Risks to Civilization:
  A Comprehensive Review (1989–2013)}.
\newblock \emph{International Journal of Environmental Research and Public
  Health}, 15, 2018.

\bibitem[B.V.()]{low_od_price}
NieuweStroom B.V.
\newblock \textbf{Tarieven zakelijk grootverbruik}.
\newblock Available at
  \url{https://www.nieuwestroom.nl/producten/fullflex-stroom-en-gas-grootverbruik/tarieven-zakelijk-grootverbruik/}.

\bibitem[Calheiros et~al.(2011)Calheiros, Ranjan, Beloglazov, Rose, and
  Buyya]{Calheiros2011CloudSimAT}
Rodrigo~N. Calheiros, Rajiv Ranjan, Anton Beloglazov, et~al.
\newblock \textbf{CloudSim: a toolkit for modeling and simulation of cloud
  computing environments and evaluation of resource provisioning algorithms}.
\newblock \emph{Softw., Pract. Exper.}, 41:\penalty0 23--50, 2011.

\bibitem[Casanova et~al.(2014)Casanova, Giersch, Legrand, Quinson, and
  Suter]{Casanova2014VersatileSA}
H.~Casanova, A.~Giersch, Arnaud Legrand, et~al.
\newblock \textbf{Versatile, scalable, and accurate simulation of distributed
  applications and platforms}.
\newblock \emph{J. Parallel Distributed Comput.}, 74:\penalty0 2899--2917,
  2014.

\bibitem[Casta{\~n}{\'e} et~al.(2013)Casta{\~n}{\'e}, N{\'u}{\~n}ez, Llopis,
  and Carretero]{Casta2013Emc2AF}
G.~G. Casta{\~n}{\'e}, Alberto N{\'u}{\~n}ez, P.~Llopis, and J.~Carretero.
\newblock \textbf{E-mc2: A formal framework for energy modelling in cloud
  computing}.
\newblock \emph{Simul. Model. Pract. Theory}, 39:\penalty0 56--75, 2013.

\bibitem[Chang et~al.(2002)Chang, Farkas, and
  Ranganathan]{Chang2002EnergyDrivenSS}
Fay Chang, K.~Farkas, and P.~Ranganathan.
\newblock \textbf{Energy-Driven Statistical Sampling: Detecting Software
  Hotspots}.
\newblock In \emph{PACS}, 2002.

\bibitem[Chen et~al.(2013{\natexlab{a}})Chen, Coskun, and
  Caramanis]{Chen2013RealtimePC}
H.~Chen, A.~Coskun, and M.~Caramanis.
\newblock \textbf{Real-time power control of data centers for providing
  Regulation Service}.
\newblock \emph{52nd IEEE Conference on Decision and Control}, pages
  4314--4321, 2013{\natexlab{a}}.

\bibitem[Chen et~al.(2013{\natexlab{b}})Chen, Hankendi, Caramanis, and
  Coskun]{Chen2013DynamicSP}
H.~Chen, Can Hankendi, M.~Caramanis, and A.~Coskun.
\newblock \textbf{Dynamic server power capping for enabling data center
  participation in power markets}.
\newblock \emph{2013 IEEE/ACM International Conference on Computer-Aided Design
  (ICCAD)}, pages 122--129, 2013{\natexlab{b}}.

\bibitem[Chen et~al.(2014)Chen, Caramanis, and Coskun]{Chen2014TheDC}
H.~Chen, M.~Caramanis, and A.~Coskun.
\newblock \textbf{The data center as a grid load stabilizer}.
\newblock \emph{2014 19th Asia and South Pacific Design Automation Conference
  (ASP-DAC)}, pages 105--112, 2014.

\bibitem[Chen et~al.(2019)Chen, Zhang, Caramanis, and
  Coskun]{Chen2019EnergyQAREQD}
H.~Chen, Yijia Zhang, M.~Caramanis, and A.~Coskun.
\newblock \textbf{EnergyQARE: QoS-Aware Data Center Participation in Smart Grid
  Regulation Service Reserve Provision}.
\newblock \emph{ACM Trans. Model. Perform. Evaluation Comput. Syst.},
  4:\penalty0 2:1--2:31, 2019.

\bibitem[Chen et~al.(2013{\natexlab{c}})Chen, Hankendi, Caramanis, and
  Coskun]{chen2013dynamic}
Hao Chen, Can Hankendi, Michael~C Caramanis, and Ayse~K Coskun.
\newblock \textbf{Dynamic server power capping for enabling data center
  participation in power markets}.
\newblock In \emph{2013 IEEE/ACM International Conference on Computer-Aided
  Design (ICCAD)}, pages 122--129. IEEE, 2013{\natexlab{c}}.

\bibitem[Chen and Li(2015)]{Chen2015OnTI}
Lijun Chen and N.~Li.
\newblock \textbf{On the Interaction Between Load Balancing and Speed Scaling}.
\newblock \emph{IEEE Journal on Selected Areas in Communications}, 33:\penalty0
  2567--2578, 2015.

\bibitem[Chen et~al.(2010)Chen, Gmach, Hyser, Wang, Bash, Hoover, and
  Singhal]{Chen2010IntegratedMO}
Y.~Chen, D.~Gmach, C.~Hyser, et~al.
\newblock \textbf{Integrated management of application performance, power and
  cooling in data centers}.
\newblock \emph{2010 IEEE Network Operations and Management Symposium - NOMS
  2010}, pages 615--622, 2010.

\bibitem[Chien et~al.(2015)Chien, Wolski, and Yang]{Chien2015ZeroCarbonCA}
Andrew~A. Chien, Richard Wolski, and Fan Yang.
\newblock \textbf{Zero-Carbon Cloud: A Volatile Resource for High-Performance
  Computing}.
\newblock \emph{2015 IEEE International Conference on Computer and Information
  Technology; Ubiquitous Computing and Communications; Dependable, Autonomic
  and Secure Computing; Pervasive Intelligence and Computing}, pages
  1997--2001, 2015.

\bibitem[Cignetti et~al.(2000)Cignetti, Komarov, and
  Ellis]{Cignetti2000EnergyET}
Todd~L. Cignetti, K.~Komarov, and C.~Ellis.
\newblock \textbf{Energy estimation tools for the Palm}.
\newblock In \emph{MSWIM '00}, 2000.

\bibitem[Con~Kolivas()]{linux_load}
Pavel~Machek Con~Kolivas.
\newblock \textbf{Linux CPU Load}.
\newblock Available at
  \url{https://www.kernel.org/doc/html/latest//admin-guide/cpu-load.html}.

\bibitem[Conejo et~al.(2010)Conejo, Morales, and Baringo]{Conejo2010RealTimeDR}
A.~Conejo, J.~Morales, and L.~Baringo.
\newblock \textbf{Real-Time Demand Response Model}.
\newblock \emph{IEEE Transactions on Smart Grid}, 1:\penalty0 236--242, 2010.

\bibitem[Contreras and Martonosi(2005)]{Contreras2005PowerPF}
Gilberto Contreras and M.~Martonosi.
\newblock \textbf{Power prediction for Intel XScale/spl reg/ processors using
  performance monitoring unit events}.
\newblock \emph{ISLPED '05. Proceedings of the 2005 International Symposium on
  Low Power Electronics and Design, 2005.}, pages 221--226, 2005.

\bibitem[Corbet()]{linux_plt}
Jonathan Corbet.
\newblock \textbf{Linux Per-entity load tracking}.
\newblock Available at \url{https://lwn.net/Articles/531853/}.

\bibitem[Corporation()]{ibm}
International Business~Machines Corporation.
\newblock \textbf{IBM CPU utilization}.
\newblock Available at
  \url{https://www.ibm.com/docs/en/informix-servers/12.10?topic=performance-cpu-utilization}.

\bibitem[da~Silva et~al.(2019)da~Silva, Orgerie, Casanova, Tanaka, Deelman, and
  Suter]{Silva2019AccuratelySE}
Rafael~Ferreira da~Silva, Anne-C{\'e}cile Orgerie, H.~Casanova, et~al.
\newblock \textbf{Accurately Simulating Energy Consumption of I/O-Intensive
  Scientific Workflows}.
\newblock In \emph{ICCS}, 2019.

\bibitem[(DDA)()]{dda}
Dutch Data Center~Association (DDA).
\newblock \textbf{Data Center Factsheet}.
\newblock Available at \url{https://www.dutchdatacenters.nl/en/factsheet/}.

\bibitem[De~Bono(2017)]{de2017six}
Edward De~Bono.
\newblock \emph{Six thinking hats}.
\newblock Penguin uk, 2017.

\bibitem[Demirbas(2005)]{Demirbas2005PotentialAO}
A.~Demirbas.
\newblock \textbf{Potential applications of renewable energy sources, biomass
  combustion problems in boiler power systems and combustion related
  environmental issues}.
\newblock \emph{Progress in Energy and Combustion Science}, 31:\penalty0
  171--192, 2005.

\bibitem[Eaton()]{ups_as_reserve}
Eaton.
\newblock \textbf{Eaton Launches Industry First UPS-as-a-Reserve Service to
  Support the Power Grid in Frequency Containment Reserve}.
\newblock Available at
  \url{http://powerquality.eaton.com/emea/about-us/news-events/2017/pr031017.asp}
  (2017/10/03).

\bibitem[Economou et~al.(2006)Economou, Rivoire, Kozyrakis, and
  Ranganathan]{Economou2006FullSystemPA}
D.~Economou, S.~Rivoire, C.~Kozyrakis, and P.~Ranganathan.
\newblock \textbf{Full-System Power Analysis and Modeling for Server
  Environments}.
\newblock 2006.

\bibitem[Fan et~al.(2007)Fan, Weber, and Barroso]{Fan2007PowerPF}
Xiaobo Fan, W.~Weber, and L.~Barroso.
\newblock \textbf{Power provisioning for a warehouse-sized computer}.
\newblock In \emph{ISCA '07}, 2007.

\bibitem[Fang et~al.(2012)Fang, Misra, Xue, and Yang]{Fang2012SmartG}
X.~Fang, S.~Misra, G.~Xue, and D.~Yang.
\newblock \textbf{Smart Grid — The New and Improved Power Grid: A Survey}.
\newblock \emph{IEEE Communications Surveys \& Tutorials}, 14:\penalty0
  944--980, 2012.

\bibitem[(Firm)(2015)]{ashrae2015thermal}
ASHRAE (Firm).
\newblock \emph{Thermal guidelines for Data processing environments}.
\newblock ASHRAE, 2015.

\bibitem[Flinn and Satyanarayanan(1999)]{Flinn1999PowerScopeAT}
J.~Flinn and M.~Satyanarayanan.
\newblock \textbf{PowerScope: a tool for profiling the energy usage of mobile
  applications}.
\newblock \emph{Proceedings WMCSA'99. Second IEEE Workshop on Mobile Computing
  Systems and Applications}, pages 2--10, 1999.

\bibitem[Fu et~al.(2020{\natexlab{a}})Fu, Han, Baker, and
  Zuo]{Fu2020AssessmentsOD}
Y.~Fu, X.~Han, K.~Baker, and W.~Zuo.
\newblock \textbf{Assessments of data centers for provision of frequency
  regulation}.
\newblock \emph{Applied Energy}, 277:\penalty0 115621, 2020{\natexlab{a}}.

\bibitem[Fu et~al.(2020{\natexlab{b}})Fu, Zuo, and Baker]{Fu2020MultimarketOO}
Yang-Yang Fu, W.~Zuo, and K.~Baker.
\newblock \textbf{Multi-market optimization of a data center without storage
  systems}.
\newblock 2020{\natexlab{b}}.

\bibitem[Fulpagare et~al.(2017)Fulpagare, Joshi, and
  Bhargav]{Fulpagare2017RackLF}
Yogesh Fulpagare, Y.~Joshi, and A.~Bhargav.
\newblock \textbf{Rack level forecasting model of data center}.
\newblock \emph{2017 16th IEEE Intersociety Conference on Thermal and
  Thermomechanical Phenomena in Electronic Systems (ITherm)}, pages 824--829,
  2017.

\bibitem[Gandhi et~al.(2009)Gandhi, Harchol-Balter, Das, and
  Lefurgy]{Gandhi2009OptimalPA}
Anshul Gandhi, Mor Harchol-Balter, Rajarshi Das, and C.~Lefurgy.
\newblock \textbf{Optimal power allocation in server farms}.
\newblock In \emph{SIGMETRICS '09}, 2009.

\bibitem[Gandhi et~al.(2011)Gandhi, Chen, Gmach, Arlitt, and
  Marwah]{Gandhi2011MinimizingDC}
Anshul Gandhi, Y.~Chen, D.~Gmach, et~al.
\newblock \textbf{Minimizing data center SLA violations and power consumption
  via hybrid resource provisioning}.
\newblock \emph{2011 International Green Computing Conference and Workshops},
  pages 1--8, 2011.

\bibitem[Ghamkhari and Rad(2013)]{Ghamkhari2013EnergyAP}
Mahdi Ghamkhari and H.~Rad.
\newblock \textbf{Energy and Performance Management of Green Data Centers: A
  Profit Maximization Approach}.
\newblock \emph{IEEE Transactions on Smart Grid}, 4:\penalty0 1017--1025, 2013.

\bibitem[Ghatikar(2012)]{Ghatikar2012DemandRO}
G.~Ghatikar.
\newblock \textbf{Demand Response Opportunities and Enabling Technologies for
  Data Centers: Findings From Field Studies}.
\newblock 2012.

\bibitem[Ghemawat et~al.(2003)Ghemawat, Gobioff, and Leung]{Ghemawat2003TheGF}
Sanjay Ghemawat, H.~Gobioff, and Shun-Tak Leung.
\newblock \textbf{The Google file system}.
\newblock In \emph{SOSP '03}, 2003.

\bibitem[Glanz(2012)]{glanz2012power}
James Glanz.
\newblock \textbf{Power, pollution and the internet}.
\newblock \emph{The New York Times}, 22, 2012.

\bibitem[Gu{\'e}rout et~al.(2013)Gu{\'e}rout, Monteil, Costa, Calheiros, Buyya,
  and Alexandru]{Gurout2013EnergyawareSW}
T.~Gu{\'e}rout, T.~Monteil, Georges~Da Costa, et~al.
\newblock \textbf{Energy-aware simulation with DVFS}.
\newblock \emph{Simul. Model. Pract. Theory}, 39:\penalty0 76--91, 2013.

\bibitem[Gupta et~al.(2011)Gupta, Gilbert, Banerjee, Abbasi, Mukherjee, and
  Varsamopoulos]{Gupta2011GDCSimAT}
Sandeep K.~S. Gupta, Rose~Robin Gilbert, Ayan Banerjee, et~al.
\newblock \textbf{GDCSim: A tool for analyzing Green Data Center design and
  resource management techniques}.
\newblock \emph{2011 International Green Computing Conference and Workshops},
  pages 1--8, 2011.

\bibitem[Gurumurthi et~al.(2002)Gurumurthi, Sivasubramaniam, Irwin,
  Vijaykrishnan, Kandemir, Li, and John]{Gurumurthi2002UsingCM}
S.~Gurumurthi, A.~Sivasubramaniam, M.~J. Irwin, et~al.
\newblock \textbf{Using complete machine simulation for software power
  estimation: the SoftWatt approach}.
\newblock \emph{Proceedings Eighth International Symposium on High Performance
  Computer Architecture}, pages 141--150, 2002.

\bibitem[Guyon et~al.(2017)Guyon, Orgerie, Morin, and Agarwal]{guyon2017much}
David Guyon, Anne-C{\'e}cile Orgerie, Chrtistine Morin, and Deb Agarwal.
\newblock \textbf{How Much Energy can Green HPC Cloud Users Save?}
\newblock In \emph{2017 25th Euromicro International Conference on Parallel,
  Distributed and Network-based Processing (PDP)}, pages 416--420. IEEE, 2017.

\bibitem[Hackenberg et~al.(2015)Hackenberg, Sch{\"o}ne, Ilsche, Molka,
  Schuchart, and Geyer]{Hackenberg2015AnEE}
Daniel Hackenberg, R.~Sch{\"o}ne, T.~Ilsche, et~al.
\newblock \textbf{An Energy Efficiency Feature Survey of the Intel Haswell
  Processor}.
\newblock \emph{2015 IEEE International Parallel and Distributed Processing
  Symposium Workshop}, pages 896--904, 2015.

\bibitem[Hamming(1997)]{Hamming1997TheAO}
Richard~R. Hamming.
\newblock \textbf{The Art of Doing Science and Engineering: Learning to Learn}.
\newblock 1997.

\bibitem[He(2020)]{hongyu_hp}
Hongyu He.
\newblock \textbf{Modelling Energy Consumption in the OpenDC Datacenter
  Simulator for Analysing Energy-Aware Cloud Infrastructure}.
\newblock Technical report, 2020.

\bibitem[Heinrich et~al.(2017)Heinrich, Cornebize, Degomme, Legrand,
  Carpen-Amarie, Hunold, Orgerie, and Quinson]{Heinrich2017PredictingTE}
F.~Heinrich, Tom Cornebize, A.~Degomme, et~al.
\newblock \textbf{Predicting the Energy-Consumption of MPI Applications at
  Scale Using Only a Single Node}.
\newblock \emph{2017 IEEE International Conference on Cluster Computing
  (CLUSTER)}, pages 92--102, 2017.

\bibitem[Heiser(2020)]{benchmarkcrimes}
Gernot Heiser.
\newblock \textbf{Systems Benchmarking Crimes}.
\newblock Available at
  \url{http://www.cse.unsw.edu.au/~Gernot/benchmarking-crimes.html}, 2020.

\bibitem[Hemmati et~al.(2017)Hemmati, Saboori, and
  Jirdehi]{Hemmati2017StochasticPA}
R.~Hemmati, H.~Saboori, and M.~A. Jirdehi.
\newblock \textbf{Stochastic planning and scheduling of energy storage systems
  for congestion management in electric power systems including renewable
  energy resources}.
\newblock \emph{Energy}, 133:\penalty0 380--387, 2017.

\bibitem[Heo et~al.(2011)Heo, Jayachandran, Shin, Wang, Abdelzaher, and
  Liu]{Heo2011OptiTunerOP}
Jin-Man Heo, P.~Jayachandran, I.~Shin, et~al.
\newblock \textbf{OptiTuner: On Performance Composition and Server Farm Energy
  Minimization Application}.
\newblock \emph{IEEE Transactions on Parallel and Distributed Systems},
  22:\penalty0 1871--1878, 2011.

\bibitem[Hui et~al.(2020)Hui, Ding, Shi, Li, Song, and Yan]{Hui20205GNI}
Hongxun Hui, Yi~Ding, Qingxin Shi, et~al.
\newblock \textbf{5G network-based Internet of Things for demand response in
  smart grid: A survey on application potential}.
\newblock \emph{Applied Energy}, 257:\penalty0 113972, 2020.

\bibitem[Inc.()]{aws_spot}
Amazon Inc.
\newblock \textbf{Amazon EC2 Spot Instances}.
\newblock Available at
  \url{https://aws.amazon.com/ec2/spot/?cards.sort-by=item.additionalFields.startDateTime&cards.sort-order=asc}.

\bibitem[.Inc({\natexlab{a}})]{intel_manual}
Intel .Inc.
\newblock \textbf{Intel® 64 and IA-32 Architectures Optimization Reference
  Manual}.
\newblock Available at
  \url{https://www.intel.com/content/dam/www/public/us/en/documents/manuals/64-ia-32-architectures-optimization-manual.pdf}
  (2016/06), {\natexlab{a}}.

\bibitem[.Inc({\natexlab{b}})]{intel_powerarch}
Intel .Inc.
\newblock \textbf{Power Management in Intel® Architecture Servers}.
\newblock Available at
  \url{https://www.intel.com/content/dam/support/us/en/documents/motherboards/server/sb/power_management_of_intel_architecture_servers.pdf}
  (2009/04), {\natexlab{b}}.

\bibitem[Inc.()]{vmware_metrics}
VMware Inc.
\newblock \textbf{VMware vSphere Documentation}.
\newblock Available at
  \url{https://docs.vmware.com/en/VMware-vSphere/index.html?topic=\%2Fcom.vmware.wssdk.apiref.doc\%2Fcpu_counters.html}.

\bibitem[Institute()]{uptime_pue}
Uptime Institute.
\newblock \textbf{Data center PUEs flat since 2013}.
\newblock Available at
  \url{https://journal.uptimeinstitute.com/data-center-pues-flat-since-2013/}.

\bibitem[Intel()]{dvfs_twofold}
Intel.
\newblock \textbf{What exactly is a P-state?}
\newblock Available at
  \url{https://software.intel.com/content/www/us/en/develop/blogs/what-exactly-is-a-p-state-pt-1.html?language=en}.

\bibitem[Iosup et~al.(2019)Iosup, Versluis, Trivedi, Eyk, Toader, Beek,
  Frascaria, Musaafir, and Talluri]{Iosup2019TheAV}
A.~Iosup, L.~Versluis, A.~Trivedi, et~al.
\newblock \textbf{The AtLarge Vision on the Design of Distributed Systems and
  Ecosystems}.
\newblock \emph{2019 IEEE 39th International Conference on Distributed
  Computing Systems (ICDCS)}, pages 1765--1776, 2019.

\bibitem[Iosup et~al.(2017)Iosup, Andreadis, Van~Beek, Bijman, Van~Eyk, Neacsu,
  Overweel, Talluri, Versluis, and Visser]{iosup2017opendc}
Alexandru Iosup, Georgios Andreadis, Vincent Van~Beek, et~al.
\newblock \textbf{The OpenDC vision: Towards collaborative datacenter
  simulation and exploration for everybody}.
\newblock In \emph{2017 16th International Symposium on Parallel and
  Distributed Computing (ISPDC)}, pages 85--94. IEEE, 2017.

\bibitem[ISO()]{cal_iso}
California ISO.
\newblock \textbf{California ISO website}.
\newblock Available at \url{http://www.caiso.com/Pages/default.aspx}.

\bibitem[Jain(1991)]{Jain1991TheAO}
R.~Jain.
\newblock \textbf{The art of computer systems performance analysis - techniques
  for experimental design, measurement, simulation, and modeling}.
\newblock In \emph{Wiley professional computing}, 1991.

\bibitem[Jain(1990)]{Jain1990TheAO}
Ray Jain.
\newblock \textbf{The Art of Computer Systems Performance Analysis}.
\newblock In \emph{Int. CMG Conference}, 1990.

\bibitem[Jindal et~al.(2020)Jindal, Aujla, Kumar, and
  Villari]{Jindal2020GUARDIANBS}
Anish Jindal, G.~Aujla, N.~Kumar, and M.~Villari.
\newblock \textbf{GUARDIAN: Blockchain-Based Secure Demand Response Management
  in Smart Grid System}.
\newblock \emph{IEEE Transactions on Services Computing}, 13:\penalty0
  613--624, 2020.

\bibitem[Johari and Tsitsiklis(2011)]{Johari2011ParameterizedSF}
R.~Johari and J.~Tsitsiklis.
\newblock \textbf{Parameterized Supply Function Bidding: Equilibrium and
  Efficiency}.
\newblock \emph{Oper. Res.}, 59:\penalty0 1079--1089, 2011.

\bibitem[Karabinaoglu and G{\"o}zel(2017)]{Karabinaoglu2017LoadFM}
Murat~Salim Karabinaoglu and T.~G{\"o}zel.
\newblock \textbf{Load forecasting modelling of data centers and IT systems by
  using artificial neural networks}.
\newblock \emph{2017 10th International Conference on Electrical and
  Electronics Engineering (ELECO)}, pages 62--66, 2017.

\bibitem[Kecskemeti(2015)]{Kecskemeti2015DISSECTCFAS}
Gabor Kecskemeti.
\newblock \textbf{DISSECT-CF: A simulator to foster energy-aware scheduling in
  infrastructure clouds}.
\newblock \emph{Simul. Model. Pract. Theory}, 58:\penalty0 188--218, 2015.

\bibitem[Kecskemeti et~al.(2017)Kecskemeti, Hajji, and
  Tso]{Kecskemeti2017ModellingLP}
Gabor Kecskemeti, Wajdi Hajji, and Fung~Po Tso.
\newblock \textbf{Modelling Low Power Compute Clusters for Cloud Simulation}.
\newblock \emph{2017 25th Euromicro International Conference on Parallel,
  Distributed and Network-based Processing (PDP)}, pages 39--45, 2017.

\bibitem[Kempton and Tomic(2005)]{Kempton2005VehicletogridPF}
W.~Kempton and J.~Tomic.
\newblock \textbf{Vehicle-to-grid power fundamentals: Calculating capacity and
  net revenue}.
\newblock \emph{Journal of Power Sources}, 144:\penalty0 268--279, 2005.

\bibitem[kernel~development community({\natexlab{a}})]{intelp}
The kernel~development community.
\newblock \textbf{\textit{intel\_pstate} CPU Performance Scaling Driver}.
\newblock Available at
  \url{https://www.kernel.org/doc/html/latest/admin-guide/pm/intel_pstate.html},
  {\natexlab{a}}.

\bibitem[kernel~development community({\natexlab{b}})]{linux_kernel}
The kernel~development community.
\newblock \textbf{The Linux Kernel Documentations}.
\newblock Available at \url{https://www.kernel.org/doc/html/latest/index.html},
  {\natexlab{b}}.

\bibitem[Kheir(1995)]{Kheir1995SystemsMA}
N.~Kheir.
\newblock \textbf{Systems Modeling and Computer Simulation}.
\newblock 1995.

\bibitem[Khemakhem et~al.(2019)Khemakhem, Rekik, and
  Krichen]{Khemakhem2019DoubleLH}
Siwar Khemakhem, M.~Rekik, and L.~Krichen.
\newblock \textbf{Double layer home energy supervision strategies based on
  demand response and plug-in electric vehicle control for flattening power
  load curves in a smart grid}.
\newblock \emph{Energy}, 167:\penalty0 312--324, 2019.

\bibitem[Kliazovich et~al.(2010)Kliazovich, Bouvry, and
  Khan]{Kliazovich2010GreenCloudAP}
Dzmitry Kliazovich, Pascal Bouvry, and Samee~Ullah Khan.
\newblock \textbf{GreenCloud: a packet-level simulator of energy-aware cloud
  computing data centers}.
\newblock \emph{The Journal of Supercomputing}, 62:\penalty0 1263--1283, 2010.

\bibitem[Koomey et~al.(2011)]{koomey2011growth}
Jonathan Koomey et~al.
\newblock \textbf{Growth in data center electricity use 2005 to 2010}.
\newblock \emph{A report by Analytical Press, completed at the request of The
  New York Times}, 9\penalty0 (2011):\penalty0 161, 2011.

\bibitem[Koomey et~al.(2007)]{koomey2007estimating}
Jonathan~G Koomey et~al.
\newblock \textbf{Estimating total power consumption by servers in the US and
  the world}, 2007.

\bibitem[Kurowski et~al.(2013)Kurowski, Oleksiak, Piatek, Piontek,
  Przybyszewski, and Weglarz]{Kurowski2013DCwormsA}
K.~Kurowski, A.~Oleksiak, W.~Piatek, et~al.
\newblock \textbf{DCworms - A tool for simulation of energy efficiency in
  distributed computing infrastructures}.
\newblock \emph{Simul. Model. Pract. Theory}, 39:\penalty0 135--151, 2013.

\bibitem[Kusic et~al.(2008)Kusic, Kephart, Hanson, Kandasamy, and
  Jiang]{Kusic2008PowerAP}
D.~Kusic, J.~Kephart, James~E. Hanson, et~al.
\newblock \textbf{Power and performance management of virtualized computing
  environments via lookahead control}.
\newblock \emph{Cluster Computing}, 12:\penalty0 1--15, 2008.

\bibitem[Le~Sueur and Heiser(2010)]{le2010dynamic}
Etienne Le~Sueur and Gernot Heiser.
\newblock \textbf{Dynamic voltage and frequency scaling: The laws of
  diminishing returns}.
\newblock In \emph{Proceedings of the 2010 international conference on Power
  aware computing and systems}, pages 1--8, 2010.

\bibitem[Li et~al.(2019)Li, Zhao, and Fang]{Li2019CSLdrivenAE}
Hongjian Li, Yuyan Zhao, and Shuyong Fang.
\newblock \textbf{CSL-driven and energy-efficient resource scheduling in cloud
  data center}.
\newblock \emph{The Journal of Supercomputing}, 76:\penalty0 481 -- 498, 2019.

\bibitem[Li et~al.(2011)Li, Chen, and Low]{Li2011OptimalDR}
N.~Li, Lijun Chen, and S.~Low.
\newblock \textbf{Optimal demand response based on utility maximization in
  power networks}.
\newblock \emph{2011 IEEE Power and Energy Society General Meeting}, pages
  1--8, 2011.

\bibitem[Li et~al.(2013)Li, Chiu, Liu, Phan, Gill, Aggarwal, Zhang, Loo, Maier,
  and McManus]{Li2013TowardsDP}
Yang Li, David Chiu, Changbin Liu, et~al.
\newblock \textbf{Towards dynamic pricing-based collaborative optimizations for
  green data centers}.
\newblock \emph{2013 IEEE 29th International Conference on Data Engineering
  Workshops (ICDEW)}, pages 272--278, 2013.

\bibitem[Liang(2017)]{Liang2017EmergingPQ}
X.~Liang.
\newblock \textbf{Emerging Power Quality Challenges Due to Integration of
  Renewable Energy Sources}.
\newblock \emph{IEEE Transactions on Industry Applications}, 53:\penalty0
  855--866, 2017.

\bibitem[Lim et~al.(2009)Lim, Sharma, Nam, Kim, and Das]{Lim2009MDCSimAM}
Seung-Hwan Lim, Bikash Sharma, Gunwoo Nam, et~al.
\newblock \textbf{MDCSim: A multi-tier data center simulation, platform}.
\newblock \emph{2009 IEEE International Conference on Cluster Computing and
  Workshops}, pages 1--9, 2009.

\bibitem[Lin et~al.(2011)Lin, Wierman, Andrew, and Thereska]{Lin2011DynamicRF}
Minghong Lin, A.~Wierman, L.~Andrew, and Eno Thereska.
\newblock \textbf{Dynamic right-sizing for power-proportional data centers}.
\newblock \emph{2011 Proceedings IEEE INFOCOM}, pages 1098--1106, 2011.

\bibitem[Lin et~al.(2012)Lin, Liu, Wierman, and Andrew]{Lin2012OnlineAF}
Minghong Lin, Zhenhua Liu, A.~Wierman, and L.~Andrew.
\newblock \textbf{Online algorithms for geographical load balancing}.
\newblock \emph{2012 International Green Computing Conference (IGCC)}, pages
  1--10, 2012.

\bibitem[Liu et~al.(2012)Liu, Ren, Loo, Mao, and Basu]{Liu2012CologneAD}
Changbin Liu, Lu~Ren, B.~T. Loo, et~al.
\newblock \textbf{Cologne: A Declarative Distributed Constraint Optimization
  Platform}.
\newblock \emph{Proc. VLDB Endow.}, 5:\penalty0 752--763, 2012.

\bibitem[Liu et~al.(2013)Liu, Wierman, Chen, Razon, and Chen]{Liu2013DataCD}
Zhenhua Liu, A.~Wierman, Y.~Chen, et~al.
\newblock \textbf{Data center demand response: avoiding the coincident peak via
  workload shifting and local generation}.
\newblock In \emph{SIGMETRICS '13}, 2013.

\bibitem[Liu et~al.(2014)Liu, Liu, Low, and Wierman]{Liu2014PricingDC}
Zhenhua Liu, Iris Liu, S.~Low, and A.~Wierman.
\newblock \textbf{Pricing data center demand response}.
\newblock In \emph{SIGMETRICS '14}, 2014.

\bibitem[Liu et~al.(2015)Liu, Lin, Wierman, Low, and Andrew]{Liu2015GreeningGL}
Zhenhua Liu, Minghong Lin, A.~Wierman, et~al.
\newblock \textbf{Greening Geographical Load Balancing}.
\newblock \emph{IEEE/ACM Transactions on Networking}, 23:\penalty0 657--671,
  2015.

\bibitem[Louis et~al.(2015)Louis, Mitra, Saguna, and
  {\AA}hlund]{Louis2015CloudSimDiskES}
B.~Louis, K.~Mitra, S.~Saguna, and C.~{\AA}hlund.
\newblock \textbf{CloudSimDisk: Energy-Aware Storage Simulation in CloudSim}.
\newblock \emph{2015 IEEE/ACM 8th International Conference on Utility and Cloud
  Computing (UCC)}, pages 11--15, 2015.

\bibitem[Malik et~al.(2017)Malik, Bilal, Malik, Anwar, Aziz, Kliazovich, Ghani,
  Khan, and Buyya]{Malik2017CloudNetSimAG}
A.~Malik, Kashif Bilal, S.~Malik, et~al.
\newblock \textbf{CloudNetSim++: A GUI Based Framework for Modeling and
  Simulation of Data Centers in OMNeT++}.
\newblock \emph{IEEE Transactions on Services Computing}, 10:\penalty0
  506--519, 2017.

\bibitem[Malone and Belady(2006)]{malone2006metrics}
Christopher Malone and Christian Belady.
\newblock \textbf{Metrics to characterize data center \& IT equipment energy
  use}.
\newblock In \emph{Proceedings of the Digital Power Forum, Richardson, TX},
  volume~68. sn, 2006.

\bibitem[Mare(2010)]{Mare2010DemandRA}
K.~Mare.
\newblock \textbf{Demand Response and Open Automated Demand Response
  Opportunities for Data Centers}.
\newblock \emph{Lawrence Berkeley National Laboratory}, 2010.

\bibitem[M{\'a}rkus et~al.(2017)M{\'a}rkus, Kert{\'e}sz, and
  Kecskemeti]{Mrkus2017CostAwareIE}
Andr{\'a}s M{\'a}rkus, A.~Kert{\'e}sz, and Gabor Kecskemeti.
\newblock \textbf{Cost-Aware IoT Extension of DISSECT-CF}.
\newblock \emph{Future Internet}, 9:\penalty0 47, 2017.

\bibitem[Marr et~al.(2002)Marr, Binns, Hill, Hinton, Koufaty, Miller, and
  Upton]{marr2002hyper}
Deborah~T Marr, Frank Binns, David~L Hill, et~al.
\newblock \textbf{Hyper-Threading Technology Architecture and
  Microarchitecture.}
\newblock \emph{Intel Technology Journal}, 6\penalty0 (1), 2002.

\bibitem[Marsh(2009)]{marsh2009intermittent}
George Marsh.
\newblock \textbf{From intermittent to variable: can we manage the wind?}
\newblock \emph{Renewable energy focus}, 10\penalty0 (5):\penalty0 42--47,
  2009.

\bibitem[Masanet et~al.(2020)Masanet, Shehabi, Lei, Smith, and
  Koomey]{masanet2020recalibrating}
Eric Masanet, Arman Shehabi, Nuoa Lei, et~al.
\newblock \textbf{Recalibrating global data center energy-use estimates}.
\newblock \emph{Science}, 367\penalty0 (6481):\penalty0 984--986, 2020.

\bibitem[Mastenbroek et~al.(2021)Mastenbroek, Andreadis, Jounaid, Lai, Burley,
  Bosch, van Eyk, Versluis, van Beek, and Iosup]{mastenbroek2021opendc}
Fabian Mastenbroek, Georgios Andreadis, Soufiane Jounaid, et~al.
\newblock \textbf{OpenDC 2.0: Convenient modeling and simulation of emerging
  technologies in cloud datacenters}.
\newblock CCGRID, 2021.

\bibitem[Mastroleon et~al.(2005)Mastroleon, Bambos, Kozyrakis, and
  Economou]{Mastroleon2005AutomaticPM}
Lykomidis Mastroleon, Nicholas Bambos, Christoforos~E. Kozyrakis, and Dimitris
  Economou.
\newblock \textbf{Automatic power management schemes for Internet servers and
  data centers}.
\newblock \emph{GLOBECOM '05. IEEE Global Telecommunications Conference,
  2005.}, 2:\penalty0 5 pp.--, 2005.

\bibitem[Meisner et~al.(2011)Meisner, Sadler, Barroso, Weber, and
  Wenisch]{Meisner2011PowerMO}
David Meisner, Christopher~M. Sadler, L.~Barroso, et~al.
\newblock \textbf{Power management of online data-intensive services}.
\newblock \emph{2011 38th Annual International Symposium on Computer
  Architecture (ISCA)}, pages 319--330, 2011.

\bibitem[Mukherjee et~al.(2009)Mukherjee, Banerjee, Varsamopoulos, Gupta, and
  Rungta]{Mukherjee2009SpatiotemporalTJ}
Tridib Mukherjee, Ayan Banerjee, Georgios Varsamopoulos, et~al.
\newblock \textbf{Spatio-temporal thermal-aware job scheduling to minimize
  energy consumption in virtualized heterogeneous data centers}.
\newblock \emph{Comput. Networks}, 53:\penalty0 2888--2904, 2009.

\bibitem[NATIONS()]{paris_agreement}
UNITED NATIONS.
\newblock \textbf{Paris Agreement}.
\newblock Available at
  \url{https://unfccc.int/sites/default/files/english_paris_agreement.pdf}.

\bibitem[N{\'u}{\~n}ez et~al.(2012)N{\'u}{\~n}ez, V{\'a}zquez-Poletti,
  Caminero, Casta{\~n}{\'e}, Carretero, and Llorente]{Nez2012iCanCloudAF}
Alberto N{\'u}{\~n}ez, J.~L. V{\'a}zquez-Poletti, A.~Caminero, et~al.
\newblock \textbf{iCanCloud: A Flexible and Scalable Cloud Infrastructure
  Simulator}.
\newblock \emph{Journal of Grid Computing}, 10:\penalty0 185--209, 2012.

\bibitem[N.V.()]{high_od_price}
Essent N.V.
\newblock \textbf{ENERGIE BESPAREN VIA ESSENT}.
\newblock Available at
  \url{https://www.essent.nl/content/particulier/energie-besparen/index.html}.

\bibitem[of~State()]{climate_summit}
U.S.~Department of~State.
\newblock \textbf{Leaders Summit on Climate}.
\newblock Available at \url{https://www.state.gov/leaders-summit-on-climate/}.

\bibitem[Ousterhout(2018)]{Ousterhout2018AlwaysMO}
J.~Ousterhout.
\newblock \textbf{Always measure one level deeper}.
\newblock \emph{Communications of the ACM}, 61:\penalty0 74 -- 83, 2018.

\bibitem[Oyj()]{1_2_price}
Fingrid Oyj.
\newblock \textbf{Two-price and one-price system}.
\newblock Available at
  \url{https://www.fingrid.fi/en/electricity-market/balance-service/description-of-balance-model/two-price-and-one-price-system/}.

\bibitem[Paterakis et~al.(2015)Paterakis, Erdinç, Bakirtzis, and
  Catal{\~a}o]{Paterakis2015OptimalHA}
N.~Paterakis, O.~Erdinç, A.~Bakirtzis, and J.~Catal{\~a}o.
\newblock \textbf{Optimal Household Appliances Scheduling Under Day-Ahead
  Pricing and Load-Shaping Demand Response Strategies}.
\newblock \emph{IEEE Transactions on Industrial Informatics}, 11:\penalty0
  1509--1519, 2015.

\bibitem[Patterson et~al.(2013)Patterson, Poole, Hsu, Maxwell, Tschudi, Coles,
  Martinez, and Bates]{patterson2013tue}
Michael~K Patterson, Stephen~W Poole, Chung-Hsing Hsu, et~al.
\newblock \textbf{TUE, a new energy-efficiency metric applied at ORNL’s
  Jaguar}.
\newblock In \emph{International Supercomputing Conference}, pages 372--382.
  Springer, 2013.

\bibitem[Pedram(2012)]{pedram2012energy}
Massoud Pedram.
\newblock \textbf{Energy-efficient datacenters}.
\newblock \emph{IEEE Transactions on Computer-Aided Design of Integrated
  Circuits and Systems}, 31\penalty0 (10):\penalty0 1465--1484, 2012.

\bibitem[Peffers et~al.(2008)Peffers, Tuunanen, Rothenberger, and
  Chatterjee]{Peffers2008ADS}
K.~Peffers, T.~Tuunanen, M.~Rothenberger, and S.~Chatterjee.
\newblock \textbf{A Design Science Research Methodology for Information Systems
  Research}.
\newblock \emph{Journal of Management Information Systems}, 24:\penalty0 45 --
  77, 2008.

\bibitem[Pipattanasomporn et~al.(2014)Pipattanasomporn, Kuzlu, Rahman, and
  Teklu]{Pipattanasomporn2014LoadPO}
M.~Pipattanasomporn, M.~Kuzlu, S.~Rahman, and Y.~Teklu.
\newblock \textbf{Load Profiles of Selected Major Household Appliances and
  Their Demand Response Opportunities}.
\newblock \emph{IEEE Transactions on Smart Grid}, 5:\penalty0 742--750, 2014.

\bibitem[PricewaterhouseCoopers()]{med_od_price}
PricewaterhouseCoopers.
\newblock \textbf{Vergelijking van gas-en elektriciteitsprijzen 2017}.
\newblock Available at
  \url{https://zoek.officielebekendmakingen.nl/blg-850506.pdf/}.

\bibitem[Qureshi et~al.(2009)Qureshi, Weber, Balakrishnan, Guttag, and
  Maggs]{Qureshi2009CuttingTE}
Asfandyar Qureshi, Rick Weber, H.~Balakrishnan, et~al.
\newblock \textbf{Cutting the electric bill for internet-scale systems}.
\newblock In \emph{SIGCOMM '09}, 2009.

\bibitem[Rad and Leon-Garcia(2010)]{Rad2010OptimalRL}
H.~Rad and A.~Leon-Garcia.
\newblock \textbf{Optimal Residential Load Control With Price Prediction in
  Real-Time Electricity Pricing Environments}.
\newblock \emph{IEEE Transactions on Smart Grid}, 1:\penalty0 120--133, 2010.

\bibitem[Raghavendra et~al.(2008)Raghavendra, Ranganathan, Talwar, Wang, and
  Zhu]{Raghavendra2008NoS}
R.~Raghavendra, P.~Ranganathan, V.~Talwar, et~al.
\newblock \textbf{No "power" struggles: coordinated multi-level power
  management for the data center}.
\newblock In \emph{ASPLOS}, 2008.

\bibitem[Rao et~al.(2010)Rao, Liu, Xie, and Liu]{Rao2010MinimizingEC}
Lei Rao, X.~Liu, Le~Xie, and Wenyu Liu.
\newblock \textbf{Minimizing Electricity Cost: Optimization of Distributed
  Internet Data Centers in a Multi-Electricity-Market Environment}.
\newblock \emph{2010 Proceedings IEEE INFOCOM}, pages 1--9, 2010.

\bibitem[Rasmussen(2007)]{Rasmussen2007ElectricalEM}
N.~Rasmussen.
\newblock \textbf{Electrical Efficiency Modeling for Data Centers}.
\newblock 2007.

\bibitem[Rawson(2004)]{Rawson2004MEMPOWERAS}
F.~Rawson.
\newblock \textbf{MEMPOWER: A Simple Memory Power Analysis Tool Set}.
\newblock 2004.

\bibitem[Rogers(2002)]{rogers2002google}
Ian Rogers.
\newblock \textbf{The Google Pagerank algorithm and how it works}.
\newblock 2002.

\bibitem[Root et~al.(2017)Root, Presume, Proudfoot, Willis, and
  Masiello]{Root2017UsingBE}
C.~Root, H.~Presume, D.~Proudfoot, et~al.
\newblock \textbf{Using battery energy storage to reduce renewable resource
  curtailment}.
\newblock \emph{2017 IEEE Power \& Energy Society Innovative Smart Grid
  Technologies Conference (ISGT)}, pages 1--5, 2017.

\bibitem[Ruepp et~al.(2017)Ruepp, Pilimon, Thrane, Galili, Berger, and
  Dittmann]{Ruepp2017CombiningHA}
S.~Ruepp, Artur Pilimon, Jakob Thrane, et~al.
\newblock \textbf{Combining hardware and simulation for datacenter scaling
  studies}.
\newblock \emph{2017 International Conference on Optical Network Design and
  Modeling (ONDM)}, pages 1--6, 2017.

\bibitem[Ryckbosch et~al.(2011)Ryckbosch, Polfliet, and
  Eeckhout]{ryckbosch2011trends}
Frederick Ryckbosch, Stijn Polfliet, and Lieven Eeckhout.
\newblock \textbf{Trends in server energy proportionality}.
\newblock \emph{Computer}, 44\penalty0 (9):\penalty0 69--72, 2011.

\bibitem[S{\ae}le and Grande(2011)]{Sle2011DemandRF}
H.~S{\ae}le and O.~S. Grande.
\newblock \textbf{Demand Response From Household Customers: Experiences From a
  Pilot Study in Norway}.
\newblock \emph{IEEE Transactions on Smart Grid}, 2:\penalty0 102--109, 2011.

\bibitem[Samson(2009)]{samson2009power}
Ted Samson.
\newblock \textbf{Power capping yields savings and floor space}, 2009.

\bibitem[Seta et~al.(1995)Seta, Hara, Kuroda, Kakumu, and
  Sakurai]{Seta199550AS}
K.~Seta, Hideki Hara, Tomonori Kuroda, et~al.
\newblock \textbf{50\% active-power saving without speed degradation using
  standby power reduction (SPR) circuit}.
\newblock \emph{Proceedings ISSCC '95 - International Solid-State Circuits
  Conference}, pages 318--319, 1995.

\bibitem[Shaw et~al.(2017)Shaw, Kumar, and Singh]{Shaw2017UseOT}
Subhadra~Bose Shaw, C.~Kumar, and Anil~Kumar Singh.
\newblock \textbf{Use of time-series based forecasting technique for balancing
  load and reducing consumption of energy in a cloud data center}.
\newblock \emph{2017 International Conference on Intelligent Computing and
  Control (I2C2)}, pages 1--6, 2017.

\bibitem[Shen et~al.(2015)Shen, Beek, and Iosup]{Shen2015StatisticalCO}
S.~Shen, V.~V. Beek, and A.~Iosup.
\newblock \textbf{Statistical Characterization of Business-Critical Workloads
  Hosted in Cloud Datacenters}.
\newblock \emph{2015 15th IEEE/ACM International Symposium on Cluster, Cloud
  and Grid Computing}, pages 465--474, 2015.

\bibitem[Siddha(2007)]{Siddha2007ProcessSC}
Suresh Siddha.
\newblock \textbf{Process Scheduling Challenges in the Era of Multi-core
  Processors}.
\newblock 2007.

\bibitem[Siqi~Shen()]{traces}
Alexandru~Iosup Siqi~Shen, Vincent van~Beek.
\newblock \textbf{The Grid Workloads Archive}.
\newblock Available at
  \url{http://gwa.ewi.tudelft.nl/datasets/gwa-t-12-bitbrains}.

\bibitem[Soeleman and Roy(1999)]{Soeleman1999UltralowPD}
Hendrawan Soeleman and Kairshik Roy.
\newblock \textbf{Ultra-low power digital subthreshold logic circuits}.
\newblock \emph{Proceedings. 1999 International Symposium on Low Power
  Electronics and Design (Cat. No.99TH8477)}, pages 94--96, 1999.

\bibitem[Stadler(2008)]{stadler2008power}
Ingo Stadler.
\newblock \textbf{Power grid balancing of energy systems with high renewable
  energy penetration by demand response}.
\newblock \emph{Utilities Policy}, 16\penalty0 (2):\penalty0 90--98, 2008.

\bibitem[Tian et~al.(2015)Tian, Zhao, Xu, Zhong, and Sun]{Tian2015ATF}
Wenhong Tian, Yong Zhao, Minxian Xu, et~al.
\newblock \textbf{A Toolkit for Modeling and Simulation of Real-Time Virtual
  Machine Allocation in a Cloud Data Center}.
\newblock \emph{IEEE Transactions on Automation Science and Engineering},
  12:\penalty0 153--161, 2015.

\bibitem[Tighe et~al.(2012)Tighe, Keller, Bauer, and Lutfiyya]{tighe2012dcsim}
Michael Tighe, Gaston Keller, Michael Bauer, and Hanan Lutfiyya.
\newblock \textbf{DCSim: A data centre simulation tool for evaluating dynamic
  virtualized resource management}.
\newblock In \emph{2012 8th international conference on network and service
  management (cnsm) and 2012 workshop on systems virtualiztion management
  (svm)}, pages 385--392. IEEE, 2012.

\bibitem[Toosi et~al.(2011)Toosi, Calheiros, Thulasiram, and
  Buyya]{Toosi2011ResourcePP}
Adel~Nadjaran Toosi, Rodrigo~N. Calheiros, Ruppa~K. Thulasiram, and Rajkumar
  Buyya.
\newblock \textbf{Resource Provisioning Policies to Increase IaaS Provider's
  Profit in a Federated Cloud Environment}.
\newblock \emph{2011 IEEE International Conference on High Performance
  Computing and Communications}, pages 279--287, 2011.

\bibitem[TSO()]{tennet}
TenneT TSO.
\newblock \textbf{Imbalance Pricing System}.
\newblock Available at
  \url{https://www.tennet.eu/fileadmin/user_upload/SO_NL/Imbalance_pricing_system.pdf}.

\bibitem[Utilities()]{peak_bill}
Fort~Collins Utilities.
\newblock \textbf{Coincident Peak}.
\newblock Available at
  \url{https://www.fcgov.com/utilities/business/manage-your-account/rates/electric/coincident-peak}.

\bibitem[Uv and Pillai(2018)]{Uv2018EnergyMO}
N.~Uv and Kishore Kumar~G Pillai.
\newblock \textbf{Energy Management of Cloud Data Center Using Neural
  Networks}.
\newblock \emph{2018 IEEE International Conference on Cloud Computing in
  Emerging Markets (CCEM)}, pages 85--89, 2018.

\bibitem[Vasques et~al.(2019)Vasques, Moura, and Almeida]{Vasques2019ARO}
Thiago~Lara Vasques, Pedro~S. Moura, and A.~D. Almeida.
\newblock \textbf{A review on energy efficiency and demand response with focus
  on small and medium data centers}.
\newblock \emph{Energy Efficiency}, 12:\penalty0 1399--1428, 2019.

\bibitem[Verma et~al.(2008)Verma, Ahuja, and Neogi]{Verma2008pMapperPA}
Akshat Verma, Puneet Ahuja, and A.~Neogi.
\newblock \textbf{pMapper: Power and Migration Cost Aware Application Placement
  in Virtualized Systems}.
\newblock In \emph{Middleware}, 2008.

\bibitem[Vesa et~al.(2020)Vesa, Cioara, Anghel, Antal, Pop, Iancu, Salomie,
  et~al.]{vesa2020energy}
Andreea~Valeria Vesa, Tudor Cioara, Ionut Anghel, et~al.
\newblock \textbf{Energy Flexibility Prediction for Data Center Engagement in
  Demand Response Programs}.
\newblock \emph{Sustainability}, 12\penalty0 (4):\penalty0 1417, 2020.

\bibitem[Vijaykrishnan et~al.(2000)Vijaykrishnan, Kandemir, Irwin, Kim, and
  Ye]{Vijaykrishnan2000EnergydrivenIH}
N.~Vijaykrishnan, M.~Kandemir, M.~J. Irwin, et~al.
\newblock \textbf{Energy-driven integrated hardware-software optimizations
  using SimplePower}.
\newblock \emph{Proceedings of 27th International Symposium on Computer
  Architecture (IEEE Cat. No.RS00201)}, pages 95--106, 2000.

\bibitem[VMware()]{vmware_vm}
Inc. VMware.
\newblock \textbf{Difference between cpu.usage and cpu.utilization counters for
  HostSystem object}.
\newblock Available at \url{https://kb.vmware.com/s/article/2055995}.

\bibitem[voor~de Statistiek()]{cbs_save}
Centraal~Bureau voor~de Statistiek.
\newblock \textbf{Energy consumption private dwellings; type of dwelling and
  regions}.
\newblock Available at
  \url{https://www.cbs.nl/en-gb/figures/detail/81528ENG?q=parts\%20of\%20the\%20country}.

\bibitem[Vrettos et~al.(2014)Vrettos, Oldewurtel, Zhu, and
  Andersson]{Vrettos2014RobustPO}
E.~Vrettos, F.~Oldewurtel, Fengtian Zhu, and G.~Andersson.
\newblock \textbf{Robust Provision of Frequency Reserves by Office Building
  Aggregations}.
\newblock \emph{IFAC Proceedings Volumes}, 47:\penalty0 12068--12073, 2014.

\bibitem[Wang et~al.(2002)Wang, Zhu, Peh, and Malik]{Wang2002OrionAP}
Hangsheng Wang, X.~Zhu, L.~Peh, and S.~Malik.
\newblock \textbf{Orion: a power-performance simulator for interconnection
  networks}.
\newblock \emph{35th Annual IEEE/ACM International Symposium on
  Microarchitecture, 2002. (MICRO-35). Proceedings.}, pages 294--305, 2002.

\bibitem[Wang et~al.(2016)Wang, Huang, Lin, and Rad]{Wang2016ProactiveDR}
Hao Wang, Jianwei Huang, X.~Lin, and H.~Rad.
\newblock \textbf{Proactive Demand Response for Data Centers: A Win-Win
  Solution}.
\newblock \emph{IEEE Transactions on Smart Grid}, 7:\penalty0 1584--1596, 2016.

\bibitem[Wang et~al.(2017)Wang, Gu, He, Guo, Sun, Vinel, and
  Shen]{Wang2017DistributedEM}
K.~Wang, Liqiu Gu, Xiaoming He, et~al.
\newblock \textbf{Distributed Energy Management for Vehicle-to-Grid Networks}.
\newblock \emph{IEEE Network}, 31:\penalty0 22--28, 2017.

\bibitem[Wang et~al.(2012)Wang, Rao, Liu, and Qi]{Wang2012DProDD}
Peijian Wang, Lei Rao, X.~Liu, and Yong Qi.
\newblock \textbf{D-Pro: Dynamic Data Center Operations With Demand-Responsive
  Electricity Prices in Smart Grid}.
\newblock \emph{IEEE Transactions on Smart Grid}, 3:\penalty0 1743--1754, 2012.

\bibitem[Wang et~al.(2019)Wang, Abdolrashidi, Yu, and
  Wong]{Wang2019FrequencyRS}
Weiming Wang, AmirAli Abdolrashidi, N.~Yu, and Daniel Wong.
\newblock \textbf{Frequency regulation service provision in data center with
  computational flexibility}.
\newblock \emph{Applied Energy}, 251:\penalty0 113304--113304, 2019.

\bibitem[Wierman et~al.(2014)Wierman, Liu, Liu, and
  Rad]{Wierman2014OpportunitiesAC}
A.~Wierman, Zhenhua Liu, Iris Liu, and H.~Rad.
\newblock \textbf{Opportunities and challenges for data center demand
  response}.
\newblock \emph{International Green Computing Conference}, pages 1--10, 2014.

\bibitem[Wikipedia()]{wiki}
Wikipedia.
\newblock \textbf{CPU utilization}.
\newblock Available at
  \url{https://en.wikipedia.org/wiki/Load_\%28computing\%29}.

\bibitem[Wilkinson et~al.(2016)Wilkinson, Dumontier, Aalbersberg, Appleton,
  Axton, Baak, Blomberg, Boiten, da~Silva~Santos, Bourne, Bouwman, Brookes,
  Clark, Crosas, Dillo, Dumon, Edmunds, Evelo, Finkers,
  Gonz{\'a}lez-Beltr{\'a}n, Gray, Groth, Goble, Grethe, Heringa, ‘t Hoen,
  Hooft, Kuhn, Kok, Kok, Lusher, Martone, Mons, Packer, Persson, Rocca-Serra,
  Roos, van Schaik, Sansone, Schultes, Sengstag, Slater, Strawn, Swertz,
  Thompson, van~der Lei, van Mulligen, Velterop, Waagmeester, Wittenburg,
  Wolstencroft, Zhao, and Mons]{Wilkinson2016TheFG}
M.~Wilkinson, M.~Dumontier, I.~J. Aalbersberg, et~al.
\newblock \textbf{The FAIR Guiding Principles for scientific data management
  and stewardship}.
\newblock \emph{Scientific Data}, 3, 2016.

\bibitem[Xu and Li(2012)]{Xu2012CostED}
H.~Xu and B.~Li.
\newblock \textbf{Cost efficient datacenter selection for cloud services}.
\newblock \emph{2012 1st IEEE International Conference on Communications in
  China (ICCC)}, pages 51--56, 2012.

\bibitem[Xu et~al.(2016)Xu, Li, and Low]{Xu2016DemandRW}
Yunjian Xu, N.~Li, and S.~Low.
\newblock \textbf{Demand Response With Capacity Constrained Supply Function
  Bidding}.
\newblock \emph{IEEE Transactions on Power Systems}, 31:\penalty0 1377--1394,
  2016.

\bibitem[Xu and Liang(2013)]{Xu2013MinimizingTO}
Zichuan Xu and Weifa Liang.
\newblock \textbf{Minimizing the Operational Cost of Data Centers via
  Geographical Electricity Price Diversity}.
\newblock \emph{2013 IEEE Sixth International Conference on Cloud Computing},
  pages 99--106, 2013.

\bibitem[Yadav et~al.(2020)Yadav, Zhang, Li, Liu, Shafiq, and
  Karn]{Yadav2020AnAH}
Rahul Yadav, Weizhe Zhang, K.~Li, et~al.
\newblock \textbf{An adaptive heuristic for managing energy consumption and
  overloaded hosts in a cloud data center}.
\newblock \emph{Wireless Networks}, 26:\penalty0 1905--1919, 2020.

\bibitem[Yao et~al.(2012)Yao, Huang, Sharma, Golubchik, and
  Neely]{Yao2012DataCP}
Y.~Yao, Longbo Huang, A.~Sharma, et~al.
\newblock \textbf{Data centers power reduction: A two time scale approach for
  delay tolerant workloads}.
\newblock \emph{2012 Proceedings IEEE INFOCOM}, pages 1431--1439, 2012.

\bibitem[Yu et~al.(2014)Yu, Wang, Yin, and Ling]{Yu2014ReviewerRO}
Y.~Yu, H.~Wang, Gang Yin, and C.~Ling.
\newblock \textbf{Reviewer Recommender of Pull-Requests in GitHub}.
\newblock \emph{2014 IEEE International Conference on Software Maintenance and
  Evolution}, pages 609--612, 2014.

\bibitem[Zedlewski et~al.(2003)Zedlewski, Sobti, Garg, Zheng, Krishnamurthy,
  and Wang]{Zedlewski2003ModelingHP}
John Zedlewski, Sumeet Sobti, Nitin Garg, et~al.
\newblock \textbf{Modeling Hard-Disk Power Consumption}.
\newblock In \emph{FAST}, 2003.

\bibitem[Zhang et~al.(2015)Zhang, Ren, Wu, and Li]{Zhang2015ATI}
Linquan Zhang, Shaolei Ren, C.~Wu, and Z.~Li.
\newblock \textbf{A truthful incentive mechanism for emergency demand response
  in colocation data centers}.
\newblock \emph{2015 IEEE Conference on Computer Communications (INFOCOM)},
  pages 2632--2640, 2015.

\bibitem[Zhang et~al.(2012)Zhang, Zhani, Zhang, Zhu, Boutaba, and
  Hellerstein]{Zhang2012DynamicEC}
Q.~Zhang, M.~Zhani, S.~Zhang, et~al.
\newblock \textbf{Dynamic energy-aware capacity provisioning for cloud
  computing environments}.
\newblock In \emph{ICAC '12}, 2012.

\bibitem[Zhang et~al.(2009)Zhang, Wang, Fu, and Wang]{zhang2009smart}
Qin Zhang, Xifan Wang, Min Fu, and Jianxue Wang.
\newblock \textbf{Smart grid from the perspective of demand response}.
\newblock \emph{Automation of Electric Power Systems}, 17:\penalty0 49--55,
  2009.

\bibitem[Zheng et~al.(2020)Zheng, Geng, Ciais, Davis, Martin, Meng, Wu,
  Chevallier, Broquet, Boersma, et~al.]{zheng2020satellite}
Bo~Zheng, Guannan Geng, Philippe Ciais, et~al.
\newblock \textbf{Satellite-based estimates of decline and rebound in China’s
  CO2 emissions during COVID-19 pandemic}.
\newblock \emph{Science Advances}, 6\penalty0 (49):\penalty0 eabd4998, 2020.

\bibitem[Zhou et~al.(2016)Zhou, Yi, Cui, Jin, Guo, and Yang]{Zhou2016DemandRC}
Y.~Zhou, Y.~Yi, Gaoying Cui, et~al.
\newblock \textbf{Demand response control strategy of groups of central
  air-conditionings for power grid energy saving}.
\newblock \emph{2016 IEEE International Conference on Power and Renewable
  Energy (ICPRE)}, pages 323--327, 2016.

\end{thebibliography}
